\documentclass[useAMS,usenatbib]{mn2e}

\usepackage[dvips]{graphicx}
\usepackage{subfigure}
\usepackage{amsmath}
\usepackage{amssymb}

\newcommand{\aap}{A\&A}

\newcommand{\apj}{ApJ}
\newcommand{\apjl}{ApJ}
\newcommand{\mnras}{MNRAS}

\newcommand{\pasp}{PASP}

\newcommand{\apjs}{ApJS}

\newcommand{\aj}{AJ}

\newcommand{\hb}{{H$\beta$}}
\newcommand{\ha}{{H$\alpha$}}
\newcommand{\ppak}{{\sc Ppak}}
\newcommand{\spak}{{\sc SparsePak}}

\def\Msun{$\mbox{M}_\odot$}

\def\deg{^\circ}

\def\farcs{\hbox{$.\!\!^{\prime\prime}$}} 

\title[\ha\ kinematics of bulgeless disks]{Two-dimensional \ha\ kinematics of bulgeless disk 
galaxies}

\author[Neumayer et al.]{Nadine Neumayer$^{1,2}$\thanks{E-mail:
nadine.neumayer@universe-cluster.de}, Carl~Jakob~Walcher$^{3,4}$, David~Andersen$^{5}$, Sebastian~F. S\'anchez$^{6}$, 
\newauthor Torsten B\"oker$^{3}$, Hans-Walter Rix$^{7}$\\
$^{1}$European Southern Observatory, Karl-Schwarzschild-Str. 2, 85748 Garching bei M\"unchen, Germany\\
$^{2}$Excellence Cluster Universe, Technische Universit\"at M\"unchen, Boltzmannstr. 2, 85748, Garching bei M\"unchen, Germany\\
$^{3}$Research and Scientific Support Department, European Space Agency, Keplerlaan1, 2200AG Noordwijk, The Netherlands\\
$^{4}$Astrophysikalisches Institut Potsdam, An der Sternwarte 16, 14482 Potsdam, Germany\\
$^{5}$NRC Herzberg Institute of Astrophysics, 5071 W Saanich Road, Victoria, BC V9E 2E7\\
$^{6}$Centro Astron\'omico Hispano Alem\'an, Calar Alto, (CSIC-MPG),
  C/Jes\'us Durb\'an Rem\'on 2-2, E-04004 Almeria, Spain\\
$^{7}$Max-Planck Institute for Astronomy, K\"onigstuhl 17, 69117 Heidelberg, Germany}

\begin{document}
\bibliographystyle{mn2e}

\date{Accepted 2010 December 24. Received 2010 December 23 ; in original form 2010 October 29}

\pagerange{\pageref{firstpage}--\pageref{lastpage}} \pubyear{2010}

\maketitle

\label{firstpage}

\begin{abstract}
We present two-dimensional \ha\ velocity fields for 20 late-type, disk-dominated 
spiral galaxies, the largest sample to date with high-resolution \ha\ 
velocity fields for bulgeless disks. From these data we derive 
rotation curves and the location of the kinematic centers. The galaxy sample was 
selected to contain nucleated and non-nucleated galaxies 
(as determined from prior HST imaging), which allows us to investigate what impact the gas 
kinematics in the host disk have on the presence (or absence) of a nuclear star cluster. 
In general, we find that the velocity fields span a broad range of morphologies. While some 
galaxies show regular rotation, most have some degree of irregular gas motions, which in 
nearly all cases can be either attributed to the presence of a bar or is connected to a rather 
patchy distribution of the \ha\ emission and the stellar light. There appears to be no systematic 
difference in the kinematics of nucleated and non-nucleated disks. Due to the large fields of view 
of the integral field units we use, we are able to observe the flattening of the rotation curve 
in almost all of our sample galaxies. This makes modeling of the velocity fields relatively 
straight-forward.

Due to the complexities of the velocity fields, we obtain reliable determinations of the kinematic center 
for only 6 of our 20 sample galaxies. For all of these the locations of the nuclear star cluster/photometric center 
and the kinematic center agree within the uncertainties. These locations also agree for 7 more objects, despite 
considerably larger uncertainties as to the accuracy of the kinematic center. If we disregard all kinematically 
irregular galaxies, our study concludes that nuclear star clusters truly occupy the nuclei, or dynamical 
centers, of their hosts. 
Our results are thus consistent with in-situ formation of nuclear star clusters.
Yet, many well-motivated formation scenarios for nuclear clusters invoke
off-center cluster formation and subsequent ÒsinkingÓ of clusters due to
dynamical friction. In that case, our results imply that dynamical friction in
the centers of bulgeless galaxies must be very effective in driving massive
clusters to the kinematic center.\end{abstract}

\begin{keywords}
galaxies: bulges; galaxies: nuclei; galaxies: star clusters; galaxies: spiral; galaxies: kinematics and dynamics
\end{keywords}

\section{Introduction}

What defines the center of a galaxy? This question is not merely
academic, because throughout the last decade, a number of studies 
have found tight correlations between the global properties of galaxies and 
the properties of their nuclei  \citep[e.g.~][]{ferrarese00,gebhardt00,marconi03,haring04}. 
These global-to-nuclear scaling relations can be interpreted such that the
mass assembly of a galaxy is pre-determined by its nuclear properties,
or alternatively, that galaxy nuclei evolve in a way that is governed by the 
assembly of the entire galaxy.

Either way, characterizing the nuclear properties has become an important 
diagnostic tool in constraining the formation mechanism(s) of galaxies.
The question of where the galaxy nucleus - i.e. its center of mass -  is 
located seems obvious in ellipticals and bulge-dominated spirals, where very 
often a luminous active galactic nucleus (AGN) marks the location of 
a super-massive black hole (SMBH) which almost certainly marks the
bottom of the potential well.

However, in the latest Hubble-types, i.e. in bulgeless, ``pure'' disk galaxies,
(luminous) AGN are rare \citep{satyapal09}, and it is less obvious whether the galaxy disk rotates around 
a nucleus that follows in any way the scaling relations mentioned above. 
A number of recent studies have suggested that in late-type spirals, the 
nucleus is marked by a massive stellar cluster \citep{phillips96,carollo98,matthews99,boker02,boker04}. 
Such Nuclear Clusters (NCs) are also present in earlier type galaxies 
\citep[e.g. recently][]{balcells03,lotz04,cote06}, but the exact relation between NCs in 
galaxies of different Hubble types remains unclear to date. NCs have masses 
of $\sim 10^6-10^7 M_{\odot}$ \citep{walcher05} and show stellar populations of 
multiple ages \citep{walcher06, rossa06,seth06}, pointing towards them having a complex 
formation history. 

On average, the location of NCs appears to coincide 
with the photometric center (PC) as derived from isophotal fits \citep{boker02}.
However, the often irregular and asymmetric shape of late-type disk galaxies causes
rather large uncertainties in defining the PC, and doubts have 
been raised on whether NCs actually define the bottom of the potential 
well \citep{matthews02,andersen08}. 

Settling this question is important in order to rule out a number of suggested
formation mechanisms for NCs. For example, if migration and/or merging 
of massive clusters is the dominant formation mechanism of NCs, as suggested by
\cite{bekki07}, \cite{capuzzo08} and \cite{agarwal10}, one would expect to find a number of NCs
displaced from the nucleus, as e.g. \cite{georgiev09} find for dwarf irregular galaxies.

Since not all late-type spirals harbor an obvious NC \citep{boker02}, one may also ask 
whether there is a galaxy property that {\it prevents} the formation of a NC. Nuclear star
formation may be suppressed in galaxies with irregular gas kinematics, and hence there 
may be systematic differences in the gas rotation patterns of galaxies with and without NCs.

In order to address both these questions, we have obtained two-dimensional 
velocity fields of the ionized gas (as traced by the \ha\ line) for a sample of 
20 late-type spiral galaxies. 
The use of \ha\ as a tracer for the general gas kinematics is required for an accurate
comparison of the position of the kinematic center (KC) and the PC, because it can be 
observed at high spatial resolution. While \ha\ may be affected by stellar winds,
supernovae, or other deviations from the pure disk rotation, in general, it has been 
found to represent well the overall rotation field of the neutral (i.e. HI) gas 
\citep{swaters09}. 

Similar studies have been performed in the past \citep[e.g.][]{ganda06,bershady10}
but this study is the first to focus exclusively on bulgeless spirals. 
In addition, our analysis includes parameterised descriptions of the \ha\ rotation 
curve for all galaxies, thus enabling a direct comparison with the kinematics of
other gas components.

This paper is structured as follows: following this introduction, we describe in \S~\ref{sec:obs}
the galaxy sample, the observations, and data reduction methods. In \S~\ref{sec:analysis},
we detail the methods used to fit the \ha\ velocity fields, to derive the \ha\ rotation curves,
and to extract the location of the KC. We discuss our results in \S~\ref{sec:discussion} and conclude
in \S~\ref{sec:conclusions}.

\section{Galaxy sample and Observations}
\label{sec:obs}

\subsection{Sample Selection}
The galaxy sample discussed here was selected from the sample of \citet{boker02},
and thus consists of spirals with late Hubble-type (Scd or later) and low inclination ($< 40^{\deg}$). 
In order to gauge the importance of a NC for the \ha\ kinematics, we selected galaxies
with and without a NC in roughly equal parts. 
Given the visibility constraints from Calar Alto\footnote{The German-Spanish Astronomical 
Center, Calar Alto, is jointly operated by the Max-Planck-Institut fŸr Astronomie Heidelberg 
and the Instituto de Astrof'sica de Andaluc'a (CSIC).} and WIYN\footnote{The WIYN Observatory is a joint facility 
of the University of Wisconsin-Madison, Indiana University, Yale University, 
and the National Optical Astronomy Observatories.}, we identified a sample of 20 
objects, summarized in Table~\ref{table:sample}. The object distances are between 6 and 28 Mpc, 
which implies that the field of view of the \ppak\ and \spak\ instruments cover between 
2 and 9.5~kpc in radius. The main instrumental parameters of \ppak\ and \spak\ are summarised in 
Table~\ref{table:inst.params}.

\begin{table*}
\caption{Galaxy Sample}
\begin{tabular}{r c c c r@{.}l r@{.}l c r@{.}l r@{.}l}
\hline
\hline
Galaxy & Hubble type & RA & DEC & \multicolumn{2}{c} {D} & \multicolumn{2}{c} {$L_I$} & $v_{max}$ & \multicolumn{2}{c} {$M_I^{\rm NC}$} & \multicolumn{2}{c} {$\mu_0$} \\
            & 			   & (hh:mm:ss.ss) & (dd:mm:ss.s) & \multicolumn{2}{c} {(Mpc)} & \multicolumn{2}{c} {($10^9$ L$_{\odot}$)}  & (km s$^{-1}$)  & \multicolumn{2}{c} {(mag)}  & \multicolumn{2}{c} {(mag/arcsec$^2$)} \\

\hline
UGC\,3574$^2$      & SAcd     & 06:53:10.4 & +57:10:40   &  20&7$\pm$3.0   & 10&0  & 167  & $-11$&90$\pm0.18$	& 18&9$\pm$0.1 \\
NGC\,2552$^2$      & SAm      & 08:19:20.5 & +50:00:35   &  11&9$\pm$2.0   & 2&3    & 62    & $ -12$&04$\pm0.01 $ & 20&15$\pm$0.1 \\
UGC\,4499$^2$      & SABdm    & 08:37:41.5 & +51:39:09   &  12&8$\pm$2.0   & 0&23  & 51    & $  -8$&59$\pm0.63 $   & 19&85$\pm0.05$  \\
UGC\,5288$^1$      & Sdm      & 09:51:17.0 & +07:49:39   &    6&0$\pm$1.2  & 0&072 & 49    &  \multicolumn{2}{c} {--}  &  19&8 $\pm$0.1 \\
NGC\,3206$^2$      & SBcd     & 10:21:47.6 & +56:55:50   &   20&7$\pm$0.4  & 3&0     & 78    & \multicolumn{2}{c} {--} & 18&78$\pm$0.05  \\
NGC\,3346$^1$      & SBcd     & 10:43:38.9 & +14:52:19   &   22&4$\pm$3.0  & 8&8     & 124  & $-11$&78 $\pm 0.01$ & 18&3$\pm$0.2 \\
NGC\,3423$^2$      & SAcd     & 10:51:14.3 & +05:50:24   &  11&3$\pm$0.6   & 1&4     & 127  & $-11$&84$\pm0.05$     & 17&9$\pm$0.15 \\
NGC\,3445$^1$      & SABm     & 10:54:35.5 & +56:59:26   &  17&5$\pm$2.0   &  3&4    & 148  & $-13$&42$\pm0.10$     & 18&1$\pm$0.1  \\
NGC\,4204$^1$      & SBdm     & 12:15:14.3 & +20:39:32   &    7&9$\pm$2.0  &  0&25  & --       & $-10$&26$\pm0.02 $ & 19&6$\pm$0.1  \\
NGC\,4299$^1$      & SABdm    & 12:21:40.9 & +11:30:12   &  16&8$\pm$2.0   &  3&9    & 109  & $-11$&73$\pm0.04 $ & 19&0$\pm$0.5 \\
NGC\,4496a$^1$     & SBm      & 12:31:39.2 & +03:56:22   &  15&6$\pm$1.2   & 5&8     & 94    & $-11$&99$\pm0.02 $ & 18&9$\pm$0.1  \\
NGC\,4517a$^1$     & SBdm     & 12:32:28.1 & +00:23:23   &  24&5$\pm$3.0   &  1&4    & 69    & \multicolumn{2}{c} {--} & 19&8$\pm$0.1 \\
NGC\,4540$^2$      & SABcd    & 12:34:50.8 & +15:33:05   &  16&8$\pm$2.0   &  4&8    & 83    & $-12$&29$\pm0.02 $ & 18&05$\pm$0.05 \\
NGC\,4625$^1$      & SABm     & 12:41:52.7 & +41:16:26   &    8&2$\pm$2.0  & 0&51   & 39    & $-10$&61$\pm0.08 $ & 17&5$\pm$0.2  \\
NGC\,4904$^{1,2}$  & SBcd     & 13:00:58.6 & $-00$:01:40 &  20&0$\pm$2.0   & 2&1    & 105  &  \multicolumn{2}{c} {--}& 17&3$\pm$0.2 \\
UGC\,8516$^1$      & Scd      & 13:31:52.6 & +20:00:04   &  20&4$\pm$2.0   &  1&6    & 60    & $-10$&97$\pm0.09 $ & 18&7$\pm$0.1 \\
NGC\,5669$^2$      & SABcd    & 14:32:43.5 & +09:53:26   &  18&0$\pm$2.0   & 8&6      & 98   & $-10$&03$\pm0.01 $ & 18&5$\pm$0.1 \\
NGC\,5789$^{1,2}$  & Sdm      & 14:56:35.5 & +30:14:03   &  33&0$\pm$3.0   &  4&3     & 123 & \multicolumn{2}{c} {--}	    & 19&9$\pm$0.05 \\
NGC\,5964$^1$      & SBd      & 15:37:36.3 & +05:58:24   &  26&5$\pm$2.0   & 21&6    & 121 & $-12$&62$\pm0.06$      & 18&9$\pm$0.1  \\
NGC\,6509$^1$      & Sd       & 17:59:25.3 & +06:17:13   &  28&2$\pm$3.0   & 6&3      & 218 & $-13$&08$\pm0.07$     & 17&8$\pm$0.1 \\
\end{tabular}
\label{table:sample}
\medskip

Note. -- Galaxies with $^1$ were observed with CAHA/\ppak, those with $^2$ with WIYN/\spak.
RA, DEC and Hubble types are from the RC3 \citep{devauc91} through NED.
Distances have been assembled through NEDs summary statistics and are from redshift independent 
measurements, where possible. If only one measurement was available we assumed 
a default uncertainty of max(2Mpc, D*0.1). For UGC\,8516 and NGC\,5789 the distance comes 
from the redshift, corrected for Virgo, Great Attractor and Shapley cluster infall, following the formulae 
of \citet{mould00}.  $L_I$ and $v_c$ have been assembled through Leda and use their {\tt itc} and 
{\tt $v_{max}$} parameters.  Central surface brightnesses $\mu_0$ have been derived from the HST 
I-band surface brightness profiles (SBP) of \cite{boker02}. For galaxies without a NC the SBPs are flat in all 
cases and $\mu_0$ is well defined. For galaxies with a NC it is unclear to what extend the disk extends further 
into the NC, we therefore chose to use the surface brightness just outside the NC as a measure of $\mu_0$. 
Error bars give our estimate of the systematic uncertainty of this number.

\end{table*}

\subsection{\ppak\ Data}
During the nights of May 7th and 8th, 2007, thirteen galaxies of our sample were 
observed at the 3.5 m telescope of the Calar Alto observatory, using the Potsdam 
Multi Aperture Spectrograph \citep[PMAS, ][]{roth05} in its \ppak\ mode \citep{kelz06}.  
The \ppak\ science fibre bundle consists of
382 fibres of 2.7 arcsec diameter each, of which 331 (the science fibres) are
concentrated in a hexagonal bundle covering a field-of-view of
72''$\times$64'' with a filling factor of $\sim$\,65\%. The sky background is
sampled by 36 additional fibres, distributed in 6 bundles of 6 fibres each,
distributed along a circle $\sim$\,90 arcsec from the center of the instrument
FOV. The sky-fibres are distributed among the science fibres within the
pseudo-slit in order to have a good characterization of the sky; the remaining
15 fibres are used for calibration purposes. Cross-talk between adjacent
fibres is estimated to be less than 5\% when using a simple aperture extraction
\citep{sanchez06}. Adjacent fibres in the pseudo-slit may cover very 
different locations on the sky, thus further reducing the effect of cross-talk.

The J1200 grating, mounted backwards in 2nd order, was used for all 
observations. It covers the wavelength range $\sim$\,6350-6690 \AA\ with a
spectral resolution of $FWHM\sim$\,0.5 \AA ($R\sim$11000). For each galaxy, 
two exposures were taken, with exposure times between 900s and 1500s,
depending on the target brightness. 
The nights were clear, with a slightly elevated extinction ($A_V\sim$0.18
mag), but stable in both cases. The seeing was variable, ranging between
0.8$\arcsec$ and 1.3$\arcsec$. A spectrophotometric standard star was observed 
each night, in order to correct for the transmission curve of the instrument. 
Note though that spectrophotometric accuracy is not required for our analysis.

The data were reduced using {\sc R3D} \citep{sanchez06}, in combination
with {\sc IRAF}\footnote{IRAF is distributed by the National Optical Astronomy
Observatories, which are operated by the Association of Universities for
Research in Astronomy, Inc., under cooperative agreement with the National
Science Foundation.} packages and {\sc E3D} \citep{sanchez04}. The
reduction consists of the standard steps for fibre-based integral-field
spectroscopy. A master bias frame (created by averaging all bias frames
observed during the night) was subtracted from the science frames. Exposures
of a given sky position were median-combined using {\sc IRAF} routines,
thus clipping any cosmic rays. The locations of the spectra on the CCD were 
traced using an exposure of a continuum lamp taken before the science exposures. 
Each spectrum was then extracted from the science frames, and stored in a 
row-stacked-spectrum file \citep{sanchez04}. 
Wavelength calibration was performed using the position of Ne lines in 
lamp exposures obtained before and after each pointing, yielding an accuracy
of rms$\sim$\,0.15 \AA. Differences in the relative fibre-to-fibre
transmission throughput were corrected by comparing the wavelength-calibrated
science frames with the corresponding frames derived from sky exposures
taken during twilight. Then, the data was corrected for the average instrument
throughput curve, by comparing the observed spectrum of the spectrophotometric 
calibration star with a flux-calibrated one. Finally, a contemporaneous (average) 
night sky spectrum was obtained by combining the spectra of the 36 sky fibers, 
and subtracted from the science spectra. The relative location of the final science
spectra on the sky were obtained via the standard \ppak\ position table.

\begin{table}
\caption{Instrumental parameters}
\begin{tabular}{l c c}
\hline
\hline
 & \ppak & \spak\\
\hline
\# science fibres & 331& 75 \\
\# sky fibres  & 36 & 7 \\
size of fibres  & 2$\farcs$7 & 4$\farcs$675\\
field-of-view  & $72\arcsec \times 64\arcsec $ & $70\arcsec \times 70\arcsec$  \\
filling factor & $\sim 57\%$ & $\sim 27 \%$ \\
wavelength range  & $6350 -6690 \rm{\AA}$ & $6500 - 6900 \rm{\AA}$ \\
instrumental FWHM  & $0.5\rm{\AA}$ & $0.65\rm{\AA}$ \\
\hline
\end{tabular}
\label{table:inst.params}
\end{table}%

\subsection{\spak\ Data}
Nine galaxies in our sample were observed using \spak\ \citep{bershady04}
on the 3.5m WIYN telescope during the course of
four nights, namely March 27--30, 2007.  

\spak\ is a fiber optic--array that feeds light from the WIYN Nasmyth f/6.3 focus 
imaging port to the Bench Spectrograph. The \spak\ IFU consists of 82 fibers 
with a diameter of $4.675\arcsec$ each, arranged in a sparse grid with a field of 
regard of $\sim 70\arcsec \times 70\arcsec$, thus offering good coverage
and sampling of our target galaxies. The spectrograph was configured with the
Bench Spectrograph Camera (BSC) and an Echelle grating with 316 lines/mm, 
used in 8th order. It covers the wavelength range 6500\,\AA $< \lambda < 6900$\,\AA 
with a dispersion of 0.195\,\AA/pix (8.6 km s$^{-1}$/pix) and an instrumental 
FWHM of 0.67\,\AA~(30.5 km s$^{-1}$).  

This rather high resolution is mandatory for our purpose: fitting velocity field models 
to galaxies with observed rotation velocities of roughly 100 km s$^{-1}$ requires 
a centroiding accuracy of $\approx$ 5 km s$^{-1}$. 
The BSC images the spectrograph onto a T2KA CCD with 2048x2048 pixels. 
The spectra are aligned along the columns of the 
CCD. The chip has a read noise of 4 e$^-$ and was used with the standard gain of 2.1
e$^-$/ADU.  The system throughput for this setup is roughly 4\% \citep{bershady05}.

The sparse grid can be filled in with 3 pointings, and we observed 8 of the 9
galaxies with 3 pointings (NGC\,5789 was only observed with two pointings). 
When WIYN first points to a target, the slit-viewing camera is used to put the
PC roughly coincident with fiber 52 at the center of the
\spak\ array (if the target has a surface brightness that is too low, as is
the case for many of these targets, we trust that the WIYN pointing is accurate
enough to deliver the science target onto the fiber array after offsetting from a nearby star). After $2\times20$
minute exposures (we use 2 exposures to be able to better reject cosmic rays), 
guiding is stopped, and the telescope is offset by $5.6\arcsec$ towards the
South and guiding is resumed.  After the next $2\times20$ minute 
exposures, guiding is paused, the telescope is offset by $4.9\arcsec$ 
West and $2.8\arcsec$ North from the second position. Guiding is resumed,
and the final $2\times20$ minute exposures for a galaxy are taken.

\begin{figure*}
\begin{center}
\includegraphics[width=17cm]{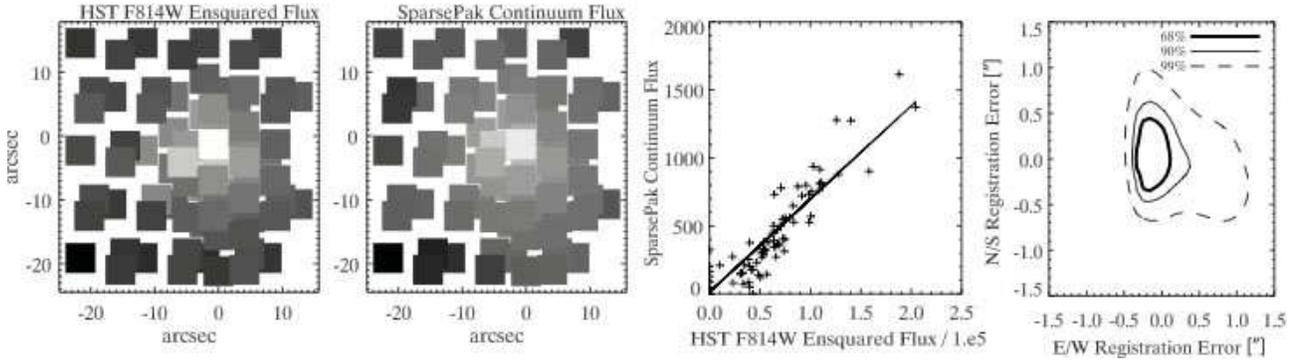}
\end{center}
\caption{Registration of the IFU data to the HST images: {\it Left:} HST F814W image convolved with the \spak\ PSF and sampled to the 
\spak\ pixelsize in comparison with the continuum levels from the \spak\ spaxels 
{\it (second panel)} for NGC3423. {\it Third panel:} Spaxel-ensquared HST WFPC2 
F814W flux vs. \spak\ continuum flux with best-fit linear regression through (0,0). {\it Right:} Error 
contours of the image IFU registration based on the $\chi^2$ map, indicating an astrometric accuracy of
the \spak\ data of $\pm 0\farcs3$. }
\label{fig:registration}
\end{figure*}

Data were overscan- and bias-corrected and trimmed using the NOAO {\sc
IRAF} package {\it ccdproc}. Cosmic ray rejection was performed before
spectral extraction, using the method described in \cite{andersen06}.
Following cosmic-ray cleaning, basic spectral extraction, flattening, and
wavelength calibration were done using the {\sc IRAF} package {\it dohydra}.
During this process, we made use of bias frames, dome flats, and Thorium Argon 
emission spectra that were taken each night. Finally, the emission from the night
sky was subtracted by averaging the spectra of the seven sky fibers, and 
subtracting the result from each of the 75 source spectra.

\subsection{Image Registration}
\label{subsec:registration}
In order to correlate the results of the kinematical analysis with the galaxy morphology,
and in particular the positions of KC, PC, and NC, we use 
archival HST/WFPC2 F814W images (roughly corresponding to the Johnson I-band) 
from the snapshot survey of \cite{boker02}.  
For each galaxy, we first created a continuum image from the IFU data cube 
by fitting the continuum in a spectral window free of emission lines.
This should  
allow a fair comparison to the I-band images. We then smoothed the HST image with a 
Gaussian beam whose width matches the seeing-limited resolution of the IFU
observations (typically 1.0\arcsec ).  

We cross-correlated both images as follows: for a given position of the 
(smoothed) HST image, we extracted its flux within the footprint of each IFU spaxel.  
We then fit a linear relation between the IFU and the HST spaxel fluxes and 
tabulated $\chi^2$. We repeated this process over a grid of offsets between
both images, thus ``mapping'' $\chi^2$. The offset grid has a granularity corresponding
to five times the pixel size of the WFPC2 data, i.e. 0.232\arcsec . The location of the $\chi^2$ 
minimum within the offset grid was used as the best registration, and the minimum
$\chi^2$ value to estimate the uncertainty in the registration (see Fig.~\ref{fig:registration} for 
an illustration).

\subsection{Line Fitting and Extraction of Velocity Fields}

\begin{figure}
\begin{center}
\includegraphics[height=7cm]{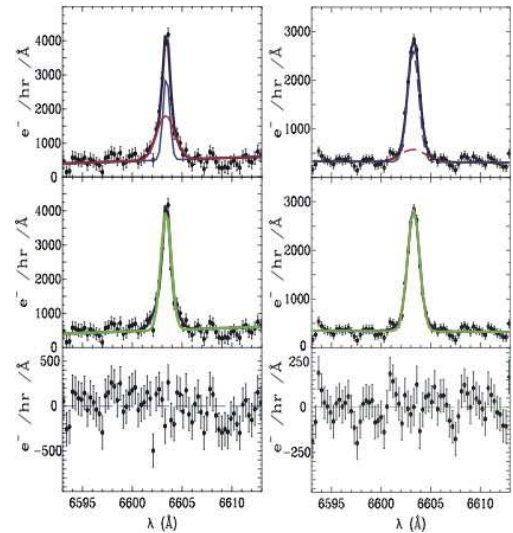}
\end{center}
\caption{\ha\ emission lines from two \ppak\ spaxels around the center of NGC6509 
comparing the double-Gaussian fits ({\it top}) to the single-Gaussian fits ({\it middle}). 
The bottom panels show the fit residuals to the fit in the middle panel. \ha\ was detected 
in 3078 of 4303 \ppak\ spaxels and 1220 of 1950 \spak\ spaxels. Most lines can be 
well fit by a single Gaussian, as shown in the plots. }
\label{fig:linefits}
\end{figure}

Once spectra were processed as described in \S~\ref{sec:obs}, 
we identified \ha\ emission-lines and
measured fluxes, widths, centers and the corresponding errors
for lines in a given spectral window.
We detected \ha\ emission lines and measured their fluxes, widths and
centroids in 3078 of 4303 \ppak\ and 1220 of 1950 \spak\ object spectra 
(72\% and 63\% detection rate, respectively).  While the
galaxy described in Andersen et al. (2008), NGC\,2139, contained multiple 
kinematic components in each spectra, most emission lines
in this sample were best fit by a single Gaussian line. Furthermore, typically 10
spectra per galaxy were best fit by two Gaussians with centers that were close
to being coincident, i.e., the lines appeared to have a strong core with 
broader, low-level wings. Still, the center positions and widths of these
lines can be very well fit by a single Gaussian (see Fig.~\ref{fig:linefits}). 
Only about 2-3 lines per galaxy exhibited two de-coupled 
Gaussian profiles. Almost all of these double-line features occur at the 
very center of the galaxies. In case of double line features, we used the more highly
peaked, dominant component of the line fit to derive the velocity map.

\section{Analysis}
\label{sec:analysis}

\subsection{Kinematic Modelling}
\label{subsec:kinmod}

In order to interpret the \ha\ velocity fields that result from the analysis described in the last
section, we fit them with a kinematic model. While there are arguments for fitting a more flexible 
tilted ring model, which allows for warps in the gas disks, the sometimes sparse sampling
of the overall velocity field
of our data did not allow this approach for all galaxies. In the interest of a uniform analysis
for the entire sample, we therefore chose to follow the approach of \citet{courteau97} who
proposed the following parametric form for the (de-projected) rotation curve:

\begin{equation}
\label{eq:rotcurve}
v(r) = v_0 + \frac{2}{\pi} v_c \mbox{arctan}(\frac{r-r_0}{r_t})
\end{equation}

\citet{courteau97} shows that this function describes the rotation curves of spiral galaxies well. Its specific 
advantage in our context is that it has only two intrinsic free parameters, i.e. scale radius and velocity scale, 
which is the minimum number of parameters possible for a rotation curve. The modeled two-dimensional 
velocity fields thus have seven free parameters: the systemic velocity $v_0$,the asymptotic velocity 
$v_c$, the center positions $x_0$ and $y_0$, the scale radius $r_t$, the inclination $i$ and the position angle $PA$. 

The fits are carried out via a custom-written Markov Chain Monte Carlo code. In addition
to the best fit parameters, we also obtain their uncertainties. As in \citet{andersen08} we find that the formal, 
statistical errors on the \ha\ velocities are too small to yield a useful measure of the quality of the fit. 
We thus add a second, additive error term $\sigma_{\rm mod}$ to make sure that the reduced $\chi^2$ of the best 
fit is not unreasonably high. We found that although our best-fit center does not depend on $\sigma_{\rm mod}$, 
the uncertainties in the position of the KC do. Also, this ad-hoc additional error term, 
hampers our capability to independently assess the quality of the final fit. We chose to set $\sigma_{\rm mod} = 8$ km/s 
for all galaxies \citep[compare e.g.~][and references therein]{kamphuis93,andersen06,sellwood10}. 

\begin{table*}
\caption{Final Kinematic Parameters}
\begin{tabular}{r r@{$\pm$}l c r@{$\pm$}l  r@{$\pm$}l  c c c r@{$\pm$}l  r@{$\pm$}l }
\hline
\hline
Galaxy &  \multicolumn{2}{|c|}{V$_{\mathrm sys}$} & V$_{\mathrm{sys, NED}}$ & \multicolumn{2}{|c|}{PA} &  \multicolumn{2}{|c|}{i} & Center RA & Center DEC & 
Center error & \multicolumn{2}{|c|}{v$_{\mathrm circ}$}  & \multicolumn{2}{|c|}{r$_{t}$} \\ 
            & \multicolumn{2}{|c|}{(km s$^{-1}$ )} & (km s$^{-1}$ ) & \multicolumn{2}{|c|}{($\deg$)} & \multicolumn{2}{|c|}{($\deg$)} & (hh:mm:ss.ss) & (dd:mm:ss.s) &  (arcsec) & \multicolumn{2}{|c|}{(km s$^{-1}$)} & \multicolumn{2}{|c|}{arcsec}    \\
 (1)  & \multicolumn{2}{|c|}{(2)} & (3) & \multicolumn{2}{|c|}{(4)} & \multicolumn{2}{|c|}{(5)} & (6) & (7) & (8) & \multicolumn{2}{|c|}{(9)} & \multicolumn{2}{|c|}{(10)} \\
\hline
UGC\,3574$^2$  &  1438&2   &  1441    &   98.4&0.5   &  25.3&0.2    & 06:53:10.43 & +57:10:37.9 &   0.9     &  167.2&0.3   &   198.4&0.4   \\
NGC\,2552$^2$  &   511&3   &  524     &   62.4&0.9   &  25.4&0.4    & 08:19:19.44 & +50:00:48.7 &   2.3     &	62.2&1.6   &   221.0&1.9   \\
UGC\,4499$^2$  &   685&2   &  691     &  141.9&1.7   &  29.0&0.2    & 08:37:41.49 & +51:39:13.6 &   6.8     &	50.9&1.6   &   200.2&2.7   \\
UGC\,5288$^1$  &   576&6   &  556     &   56.7&0.5   &  29.9&0.2    & 09:51:16.88 & +07:49:47.3 &   4.3     &	48.6&0.4   &   179.0&0.4   \\
NGC\,3206$^2$  &  1151&1   &  1150    &  183.0&0.5   &  26.5&0.1    & 10:21:47.81 & +56:55:49.8 &   0.6     &	78.4&0.4   &   229.3&1.2   \\
NGC\,3346$^1$  &  1274&1   &  1260    &  -67.1&0.6   &  29.4&0.3    & 10:43:38.85 & +14:52:16.6 &   0.6     &  123.6&0.8   &   207.1&1.3   \\
NGC\,3423$^2$  &  1004&2   &  1011    &   45.1&0.8   &  19.0&0.2    & 10:51:14.36 & +05:50:24.3 &   1.1     &  127.0&0.5   &   295.6&1.2   \\
NGC\,3445$^1$  &  2048&1   &  2069    &   -9.8&1.1   &  27.6&0.4    & 10:54:35.21 & +56:59:23.7 &   1.9     &  148.1&2.3   &   132.1&4.9   \\
NGC\,4204$^1$  &   870&3   &  856     &  240.2&2.5   &  13.6&1.7    & 12:15:14.36 & +20:39:29.6 &   2.6     &	50.0&2.8   &   238.0&19.3   \\
NGC\,4299$^1$  &   237&2   &  232     & -110.3&1.0   &  28.7&1.2    & 12:21:40.59 & +11:30:11.4 &   0.9     &  109.1&1.1   &   189.9&4.8   \\
NGC\,4496a$^1$ &  1747&3   &  1730    &   47.5&0.3   &  30.4&0.1    & 12:31:39.75 & +03:56:18.9 &   1.7     &   94.3&0.2   &   221.3&0.5  \\
NGC\,4517a$^1$ &  1525&2   &  1509    & -156.6&1.5   &  33.7&1.2    & 12:32:28.32 & +00:23:28.7 &   1.9     &   68.6&0.8   &   175.5&5.8  \\
NGC\,4540$^2$  &  1291&2   &  1286    &   13.9&1.7   &  27.9&1.1    & 12:34:50.91 & +15:33:06.3 &   1.4     &	83.4&1.5   &   201.0&2.6   \\
NGC\,4625$^1$  &   621&1   &  609     &  -55.8&1.1   &  13.3&0.3    & 12:41:53.06 & +41:16:24.0 &   1.3     &	38.7&0.2   &   188.3&1.3   \\
NGC\,4904$^1$  &  1180&1   &  1189    & -134.0&0.5   &  38.5&0.6    & 13:00:58.62 & -00:01:37.8 &   0.3     &  105.2&0.2   &   235.1&1.8   \\
NGC\,4904$^2$  &  1162&2   &  1189    &  227.2&0.9   &  39.0&0.5    & 13:00:58.54 & -00:01:37.9 &   0.4     &  105.2&0.9   &   247.5&2.8   \\
UGC\,8516$^1$  &  1026&3   &  1023    &   14.5&0.7   &  43.3&0.3    & 13:31:52.60 & +20:00:03.9 &   1.4     &	60.2&0.9   &   154.5&1.2   \\
NGC\,5669$^2$  &  1368&2   &  1371    &   69.1&0.7   &  35.6&0.0    & 14:32:44.06 & +09:53:29.5 &   0.5     &	98.4&0.5   &   186.6&0.6   \\
NGC\,5789$^1$  &  1811&1   &  1805    &  150.3&0.8   &  27.4&0.1    & 14:56:35.58 & +30:14:02.4 &   1.2     &  122.9&1.0   &   140.3& .6   \\
NGC\,5789$^2$  &  1809&2   &  1805    &  151.7&2.9   &  27.0&0.2    & 14:56:35.56 & +30:14:01.1 &   2.2     &  122.9&1.3   &   148.8&4.5   \\
NGC\,5964$^1$  &  1457&4   &  1447    &  131.8&0.9   &  38.4&0.2    & 15:37:36.94 & +05:58:17.3 &   5.1     &  120.8&0.8   &   220.1&1.2   \\
NGC\,6509$^1$  &  1780&1   &  1813    &  -81.6&0.4   &  49.3&0.7    & 17:59:25.46 & +06:17:10.5 &   0.3     &  218.1&0.4   &   213.7&0.7   \\
\end{tabular}		   
\label{table:kinpar}	   
\medskip		   
			   
Note. -- Galaxies with $^1$ were observed with CAHA/\ppak, those with $^2$ with WIYN/\spak. The kinematic parameters as derived in this paper are (2) systemic velocity , (4) Position to the major axis, (5) inclination, (6) and (7) RA and DEC of the KC, (8) error on the kinematic center, (9) Circular velocity, defined as the maximum velocity in the rotation curve fits, and (10) the scale radius of the rotation curve. 
For comparison to the modelled systemic velocity (2), we give the systemic velocity assembled through NED (3). 
\end{table*}

\begin{table*}
\caption{Compilation of Kinematic Centers, Nuclear Cluster Positions, and Photometric Centers}
\begin{tabular}{lccccccccc}
\hline
\hline
	    & \multicolumn{2}{|c|}{Kinematic Center} & offset &  \multicolumn{2}{|c|}{Nuclear Cluster Position} & offset & \multicolumn{2}{|c|}{Photometric Center}\\
Galaxy & RA & DEC & (KC-NC) & RA & DEC &  (PC-KC) & RA & DEC & $q$ \\ 
            & hh:mm:ss.ss & dd:mm:ss.s & $\arcsec$/pc & hh:mm:ss.ss & dd:mm:ss.s & $\arcsec$/pc & hh:mm:ss.ss & dd:mm:ss.s \\
\hline
UGC\,3574$^2$   & 06:53:10.43 & +57:10:37.9 &   1.7/170.6   &06:53:10.39  & +57:10:39.5 &   5.7/ 572.0 & 06:53:10.39 & +57:10:39.5 & 1 \\
NGC\,2552$^2$   & 08:19:19.44 & +50:00:48.7 &  21.3/1228.9  &08:19:20.38  & +50:00:32.8 &  20.6/1188.5 & 08:19:20.28 & +50:00:32.4 &  1\\
UGC\,4499$^2$   & 08:37:41.49 & +51:39:13.6 &   4.5/279.3   &08:37:41.43  & +51:39:09.2 &   5.1/316.5 & 08:37:41.44 & +51:39:08.6 &  1 \\
UGC\,5288$^1$   & 09:51:16.88 & +07:49:47.3 & --            &--	  	  & --	  	&   8.8/256.0 & 09:51:17.09 & +07:49:39.1 &  0 \\
NGC\,3206$^2$   & 10:21:47.81 & +56:55:49.8 & --            &--	   	  & --	  	&   0.2/20.1 & 10:21:47.79 & +56:55:49.7 &  2 \\
NGC\,3346$^1$   & 10:43:38.85 & +14:52:16.6 &   0.6/65.2    &10:43:38.84  & +14:52:17.2 &   0.9/97.7 & 10:43:38.82 & +14:52:17.4 &  2 \\
NGC\,3423$^2$   & 10:51:14.36 & +05:50:24.3 &   3.1/169.8   &10:51:14.31  & +05:50:24.3 &   3.1/169.8 & 10:51:14.31 & +05:50:24.3 &  2 \\
NGC\,3445$^1$   & 10:54:35.21 & +56:59:23.7 &   5.8/492.1   &10:54:35.55  & +56:59:26.4 &   6.1/517.5 & 10:54:35.59 & +56:59:25.9 &  1 \\
NGC\,4204$^1$   & 12:15:14.36 & +20:39:29.6 &   1.4/53.6    &12:15:14.43  & +20:39:28.8 &   1.4/53.6 & 12:15:14.40 & +20:39:29.6 &  0 \\
NGC\,4299$^1$   & 12:21:40.59 & +11:30:11.4 &   2.9/236.2   &12:21:40.39  & +11:30:11.8 &   2.3/187.3 & 12:21:40.44 & +11:30:10.6 &  2\\
NGC\,4496a$^1$  & 12:31:39.75 & +03:56:18.9 &   7.8/589.9   &12:31:39.24  & +03:56:20.5 &   7.6/574.8 & 12:31:39.26 & +03:56:20.9 &  1 \\
NGC\,4517a$^1$  & 12:32:28.32 & +00:23:28.7 &  --           &--	    	  & --	  	&   7.0/831.5 & 12:32:28.15 & +00:23:22.2 &  1 \\
NGC\,4540$^2$   & 12:34:50.91 & +15:33:06.3 &   2.7/219.9   &12:34:50.89  & +15:33:06.9 &   0.7/57.0 & 12:34:50.89 & +15:33;06.9 &  1 \\
NGC\,4625$^1$   & 12:41:53.06 & +41:16:24.0 &   1.6/63.6    &12:41:53.06  & +41:16:25.6 &   1.6/63.6 & 12:41:53.06 & +41:16:25.6 &  1 \\
NGC\,4904$^1$   & 13:00:58.62 & -00:01:37.8 &   --          &--	    	  & --	  	&   2.8/271.5 & 13:00:58.59 & -00:01:37.9 &  2 \\
NGC\,4904$^2$   & 13:00:58.54 & -00:01:37.9 &  --           &--	    	  & --	  	&   0.8/77.6 & 13:00:58.59 & -00:01:37.9 &  2 \\
UGC\,8516$^1$   & 13:31:52.60 & +20:00:03.9 &   0.6/59.3    &13:31:52.56  & +20:00:03.8 &   0.9/89.0 & 13:31:52.54 & +20:00:03.6 &  1 \\
NGC\,5669$^2$   & 14:32:44.06 & +09:53:29.5 &   1.2/104.7   &14:32:44.09  & +09:53:30.6 &   1.0/87.3 & 14:32:44.09 & +09:53:30.4 &  2\\
NGC\,5789$^1$   & 14:56:35.58 & +30:14:02.4 &  --           &--	    	  & --	  	&   4.0/640.0 & 14:56:35.84 & +30:14:03.1 &  1 \\
NGC\,5789$^2$   & 14:56:35.56 & +30:14:01.1 &  --           &--	    	  & --	  	&   4.8/767.9 & 14:56:35.84 & +30:14:03.1 &  1 \\
NGC\,5964$^1$   & 15:37:36.94 & +05:58:17.3 &  15.0/1927.1  &15:37:36.11  & +05:58:25.9 &  14.7/1888.6 & 15:37:36.13 & +05:58:25.7  &  0 \\
NGC\,6509$^1$   & 17:59:25.46 & +06:17:10.5 &   2.5/341.8   &17:59:25.36  & +06:17:12.5 &   2.6/ 355.5 & 17:59:25.34 & +06:17:12.4 &  1 \\
\end{tabular}
\label{table:pos}
\medskip

Note. -- Galaxies with $^1$ were observed with CAHA/\ppak, those with $^2$ with WIYN/\spak. The position of 
the kinematic center as derived in this paper is given in columns RA$_{KC}$ and DEC$_{KC}$. The Nuclear Cluster
coordinates (RA$_{NC}$/DEC$_{NC}$) and coordinates of the photometric center (RA$_{PC}$/DEC$_{PC}$) are taken
from \cite{boker02}. The offsets between nuclear cluster and kinematic centre (KC-NC) and photometric centre and kinematic centre (PC-KC) are given in arcseconds and parsecs (using the distances given in Table~\ref{table:sample}. The parameter $q$ is a measure for how well the velocity fields can be modelled, where 0 denotes a 
bad fit to the data, 1 denotes models with low trust and 2 denote models of high fidelity.
\end{table*}

The best-fit kinematic models are presented in column~4 of Fig.~\ref{fig:mc1}, and the derived rotation curves,
including the best fit to Eq.~\ref{eq:rotcurve}, are shown in Figure \ref{fig:rotcurv}. The Figures show that our model velocity fields
seem to be a fair representation of the observations, confirming that also the latest Hubble types can be 
fit reasonably well with this simple functional form. The resulting best-fit kinematic parameters are summarised in Table~\ref{table:kinpar}. 

Since we are mostly interested in the position of the KC, we have checked the dependence of the 
fit results on the chosen starting values for the center position, for $i$, $PA$, $v_0$, etc. We find that the fit always 
converges to the same solution, within $\approx$30\% of the nominal error bar, demonstrating that our results 
are insensitive to the choice of starting values. The derived coordinates of the KCs are given in Table~\ref{table:pos}.

\begin{figure}
\begin{center}
\includegraphics[width=8cm]{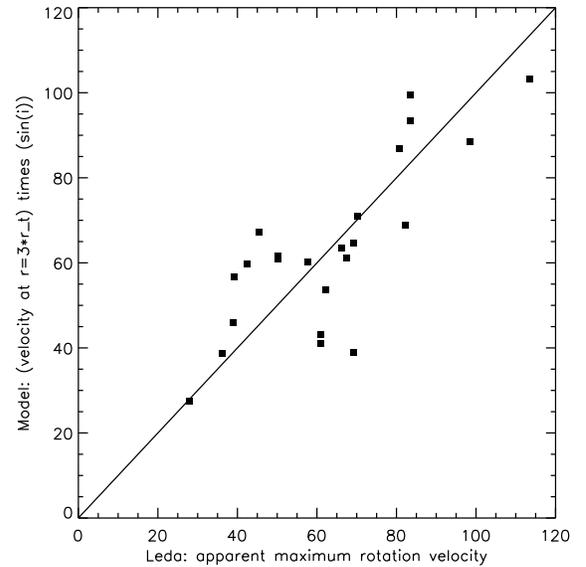}
\end{center}
\caption{ The projected maximal rotation velocity of HI gas from the Hyperleda database against 
the projected velocity of the model at 3 times the scale radius $r_t$. We find a very acceptable 
correlation. The line indicates the expected one-to-one correlation.   }
\label{fig:vmax}
\end{figure}

\subsection{Uncertainties in the kinematic modelling}

For an independent check, we compared our results to the apparent maximal rotation velocity of the gas (parameter 
$vmaxg$) from the HyperLEDA database\footnote{http://leda.univ-lyon1.fr} \citep{paturel03}. As pointed out by \citet{courteau97}, the parameter $v_c$ in Eq.~\ref{eq:rotcurve} 
is in itself not a good measure of the maximal velocity. Instead, we use $v_{3t}$, i.e. the model velocity at three times $r_t$.
Figure \ref{fig:vmax} shows that $v_{3t}$ correlates well with the $v_{max}$ parameter from LEDA, after correction for 
the galaxy inclination as derived from our fits. There are a number of uncertainties underlying the plotted values giving 
rise to the scatter in the plotted relation, like irregularities in the velocity fields for both HI and \ha\ or the fact that we 
do not sample the full velocity field with our \ha data. Given these uncertainties, the scatter is surprisingly low. 
This demonstrates the reliability of our kinematic models\footnote{We note 
that it is doubtful whether further exploration of our data for Tully-Fisher relation purposes 
is reasonable, due to the many kinematic uncertainties we identify.}.

  \begin{figure*}
  \begin{center}
  \begin{tabular}{c}
  \hspace*{-1.3cm} \includegraphics[height=4.5cm]{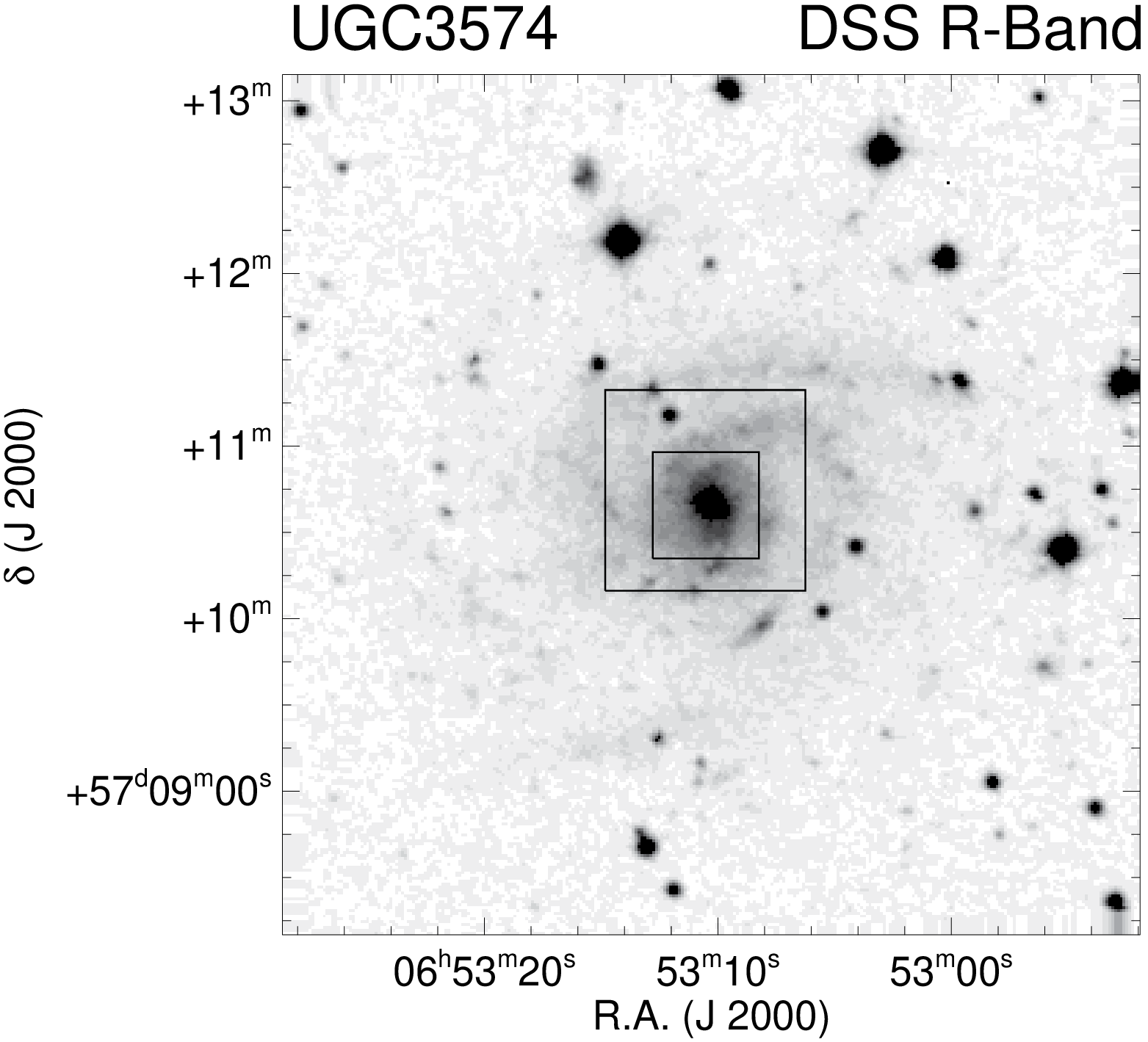}
   \hspace*{-0.25cm}  \includegraphics[width=4.1cm]{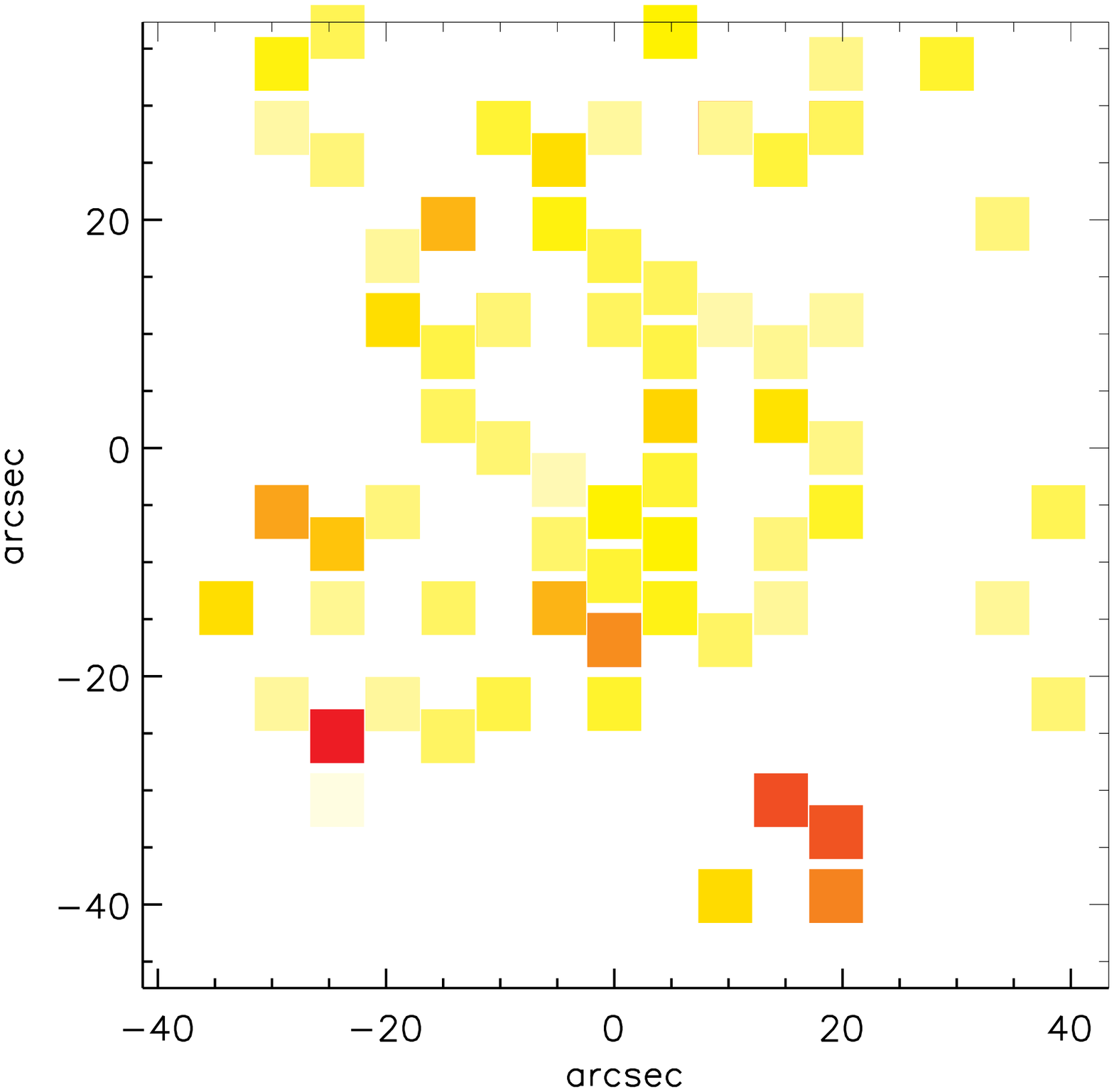}
    \hspace*{-0.25cm}  \includegraphics[height=4.1cm]{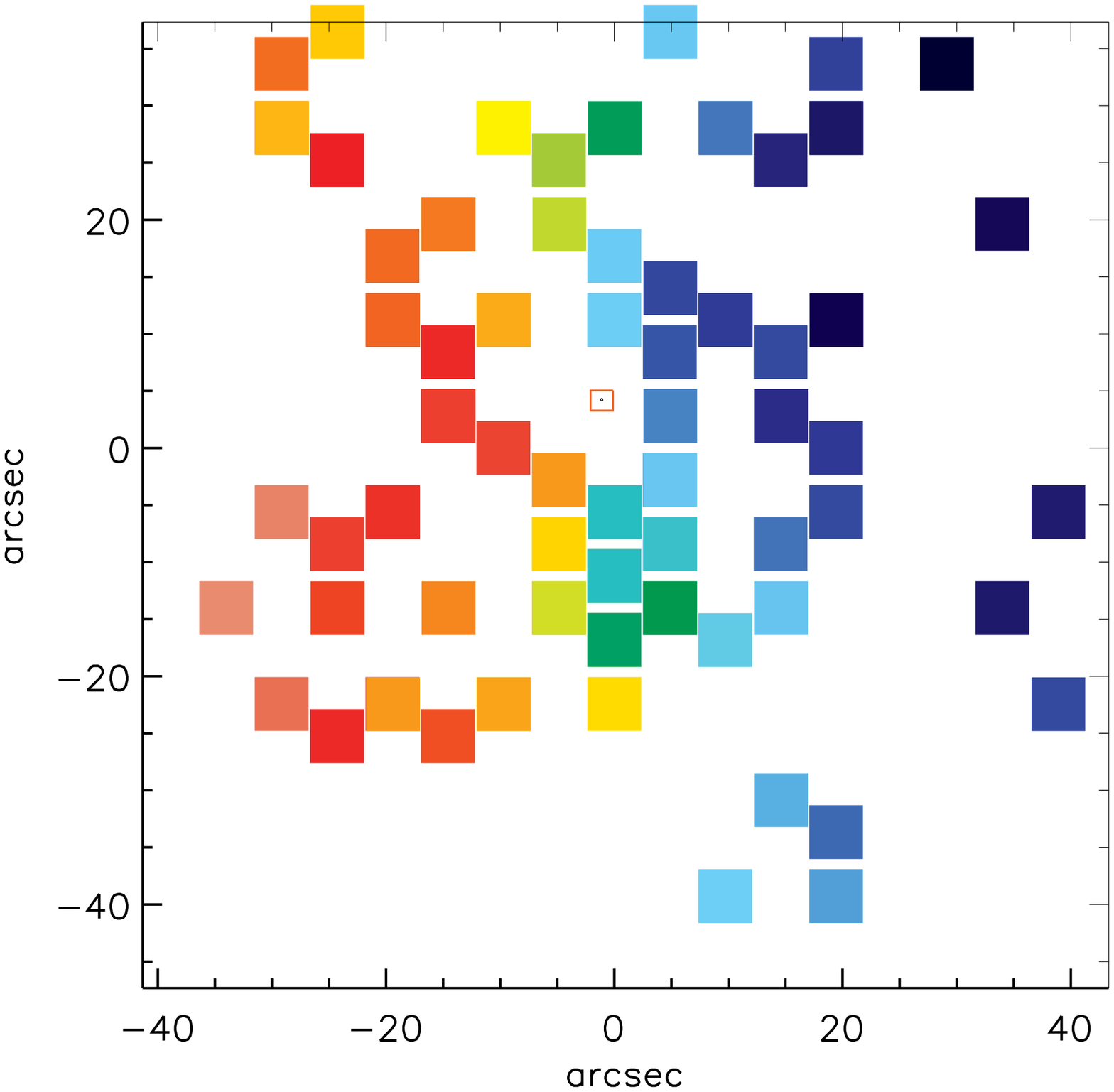}
    \hspace*{-0.25cm}  \includegraphics[height=4.1cm]{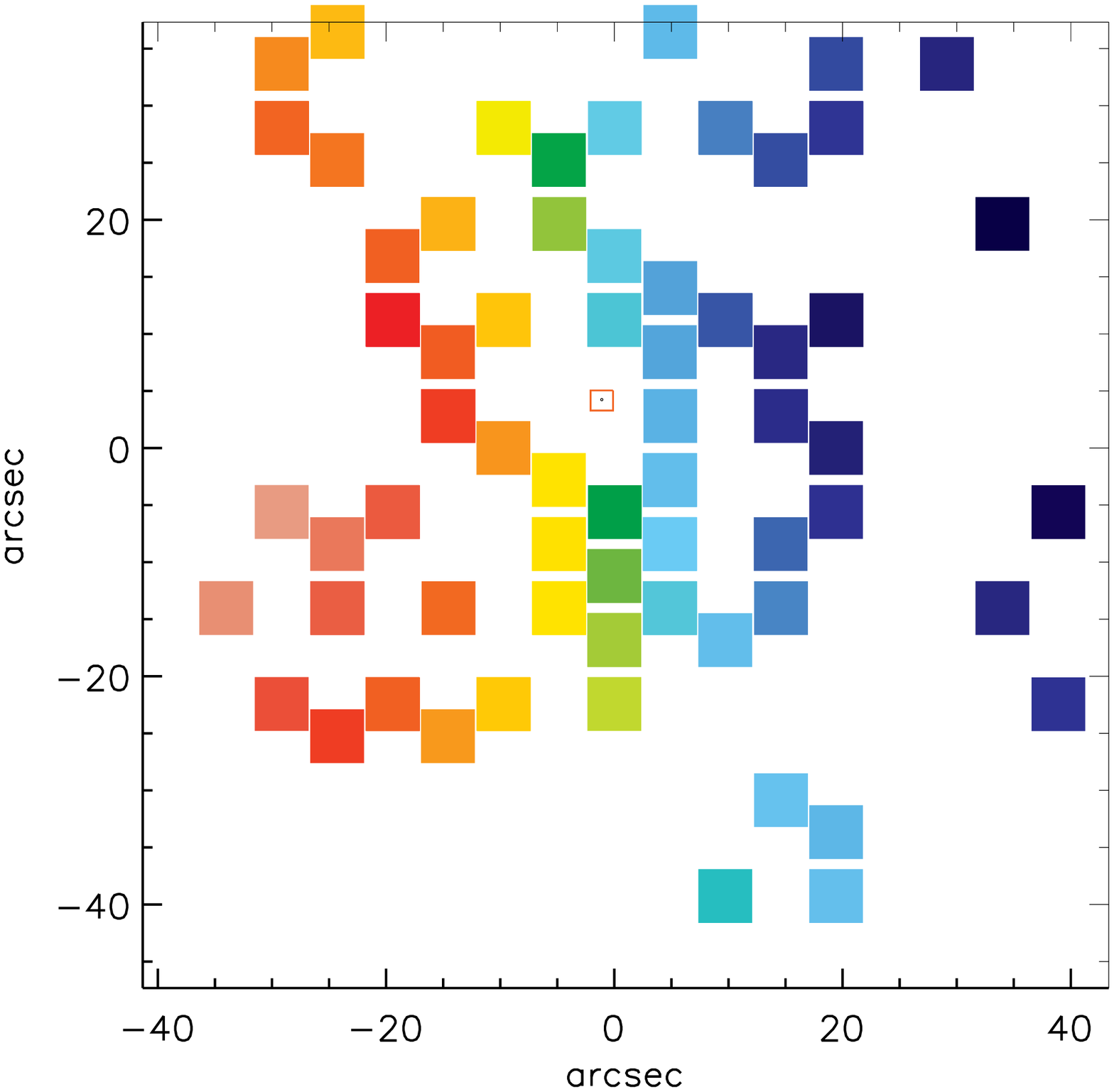}
    \hspace*{-0.25cm} \includegraphics[height=3.82cm]{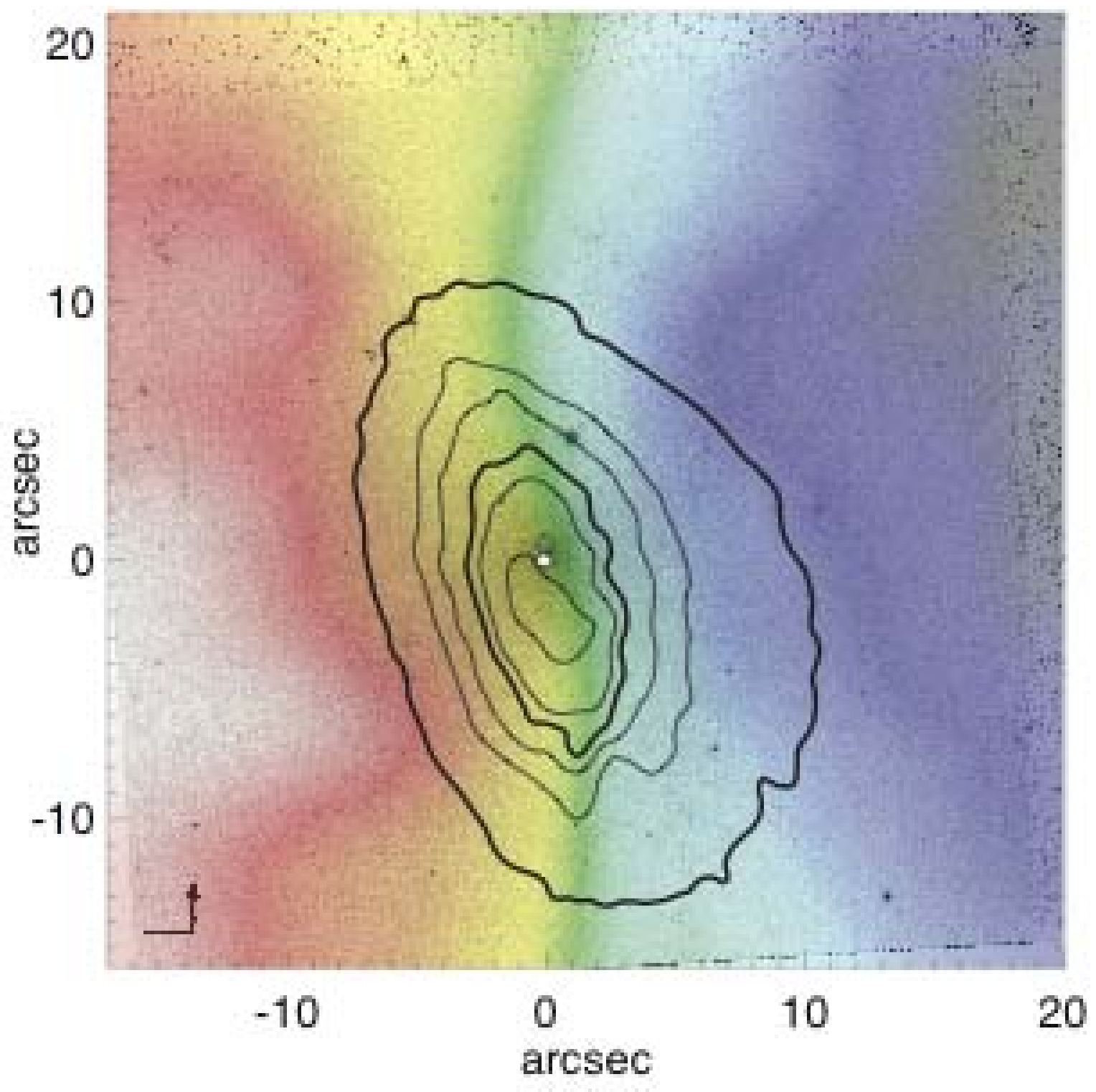}
    \vspace{-0.6cm}
  \end{tabular}
  \end{center}

  \begin{center}
  \begin{tabular}{cc}
  \hspace*{-1.3cm} \includegraphics[height=4.5cm]{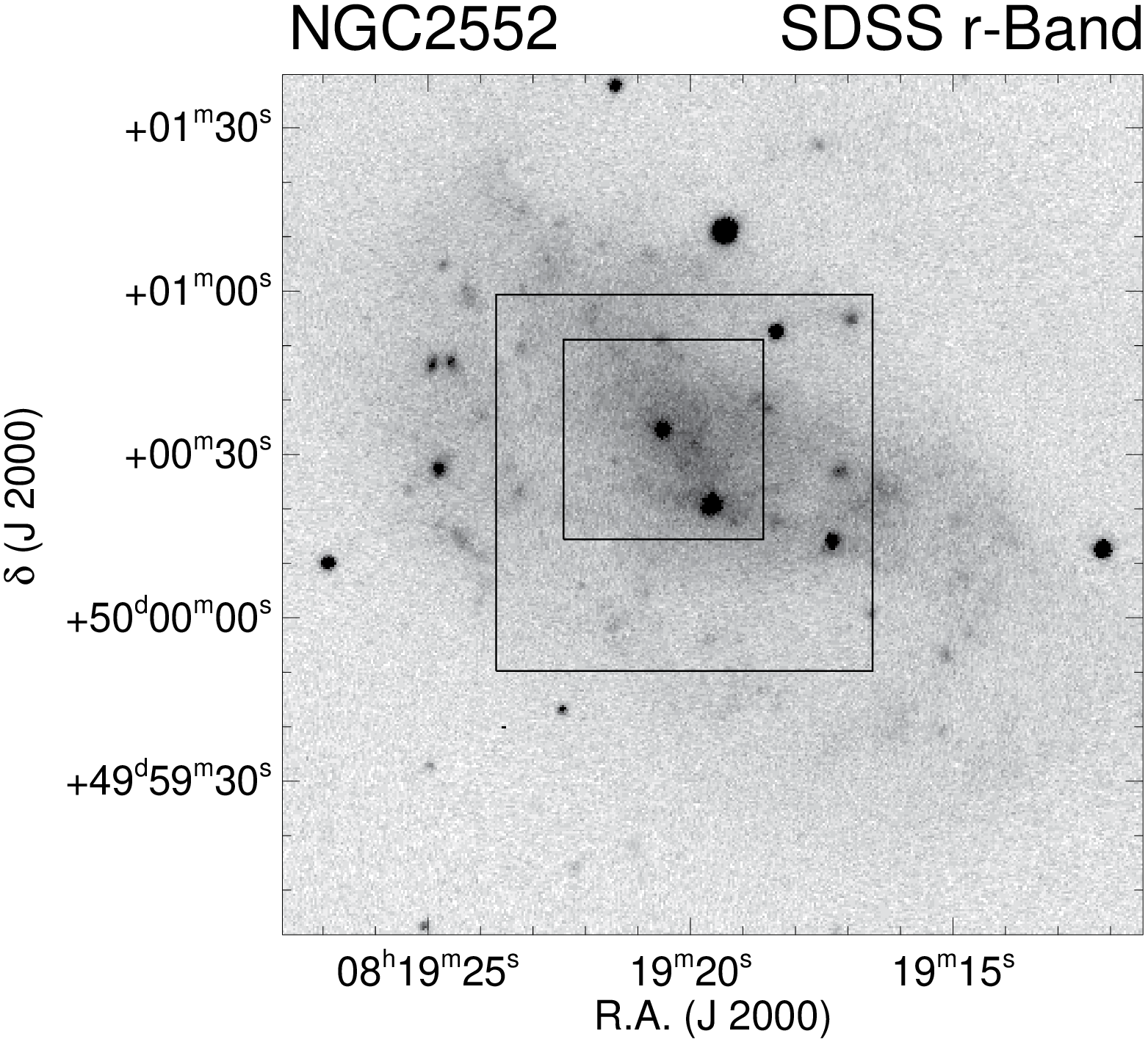}
   \hspace*{-0.25cm} \includegraphics[height=4.1cm]{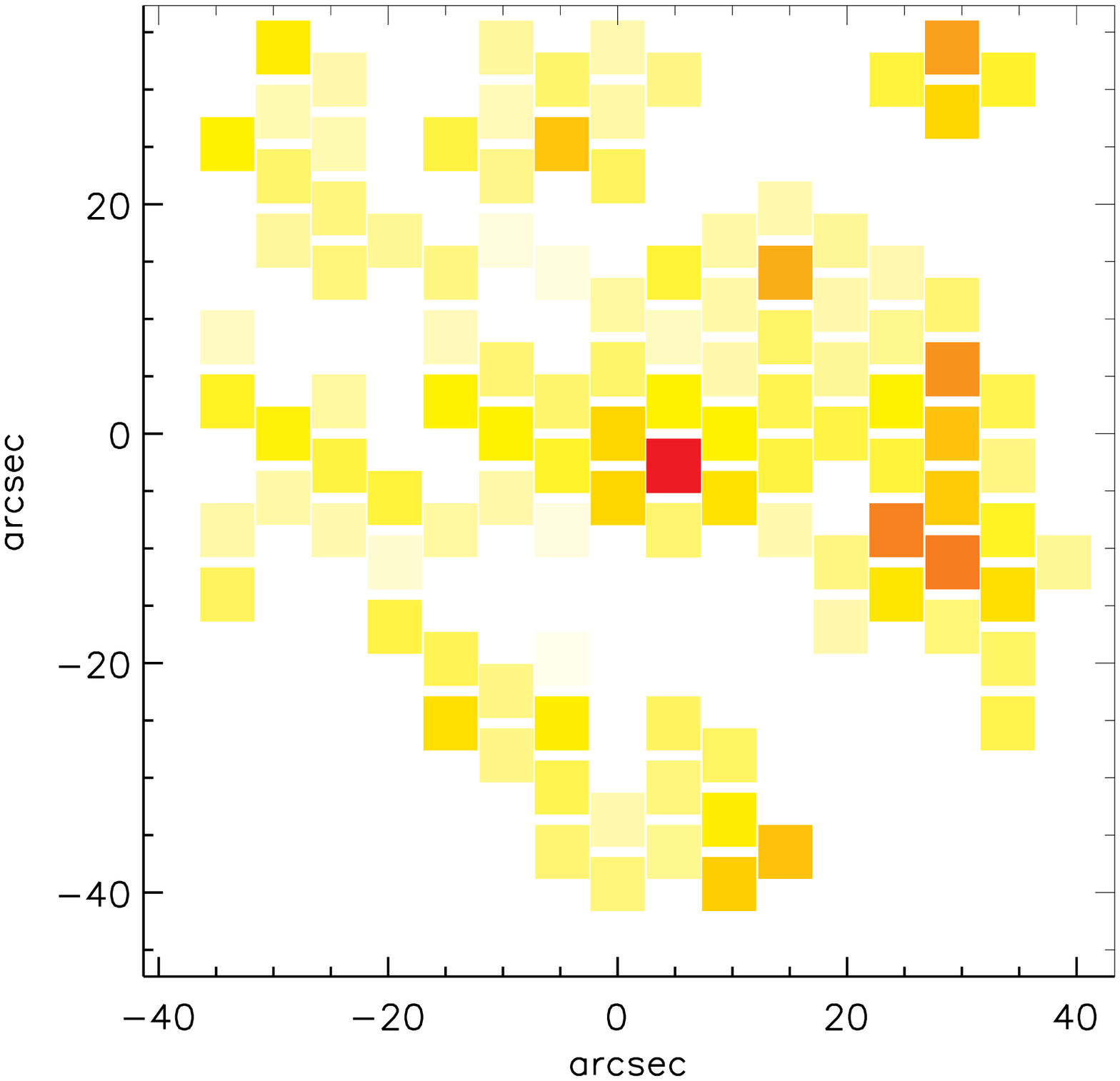}
   \hspace*{-0.25cm}   \includegraphics[height=4.1cm]{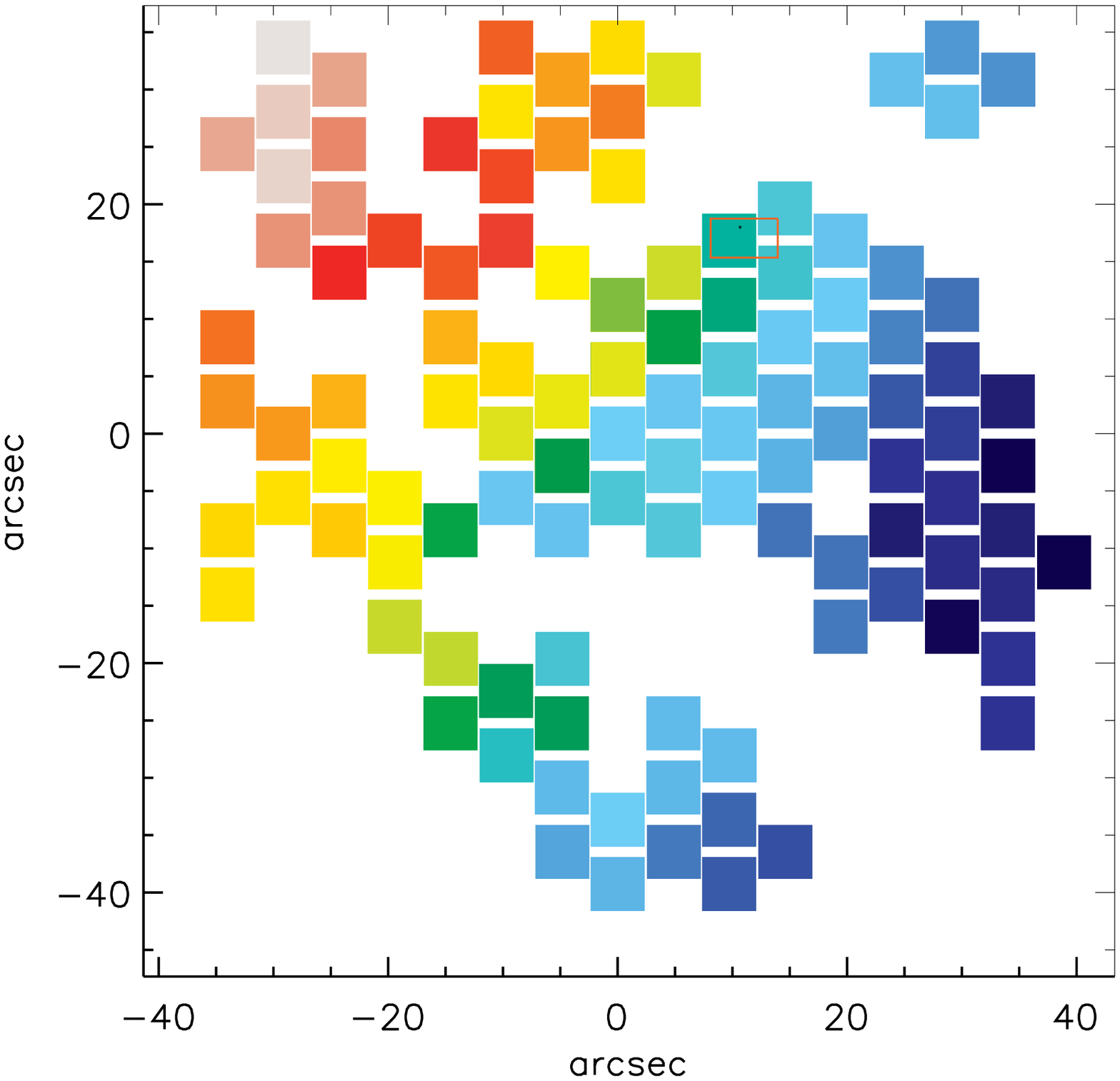}
   \hspace*{-0.25cm}   \includegraphics[height=4.1cm]{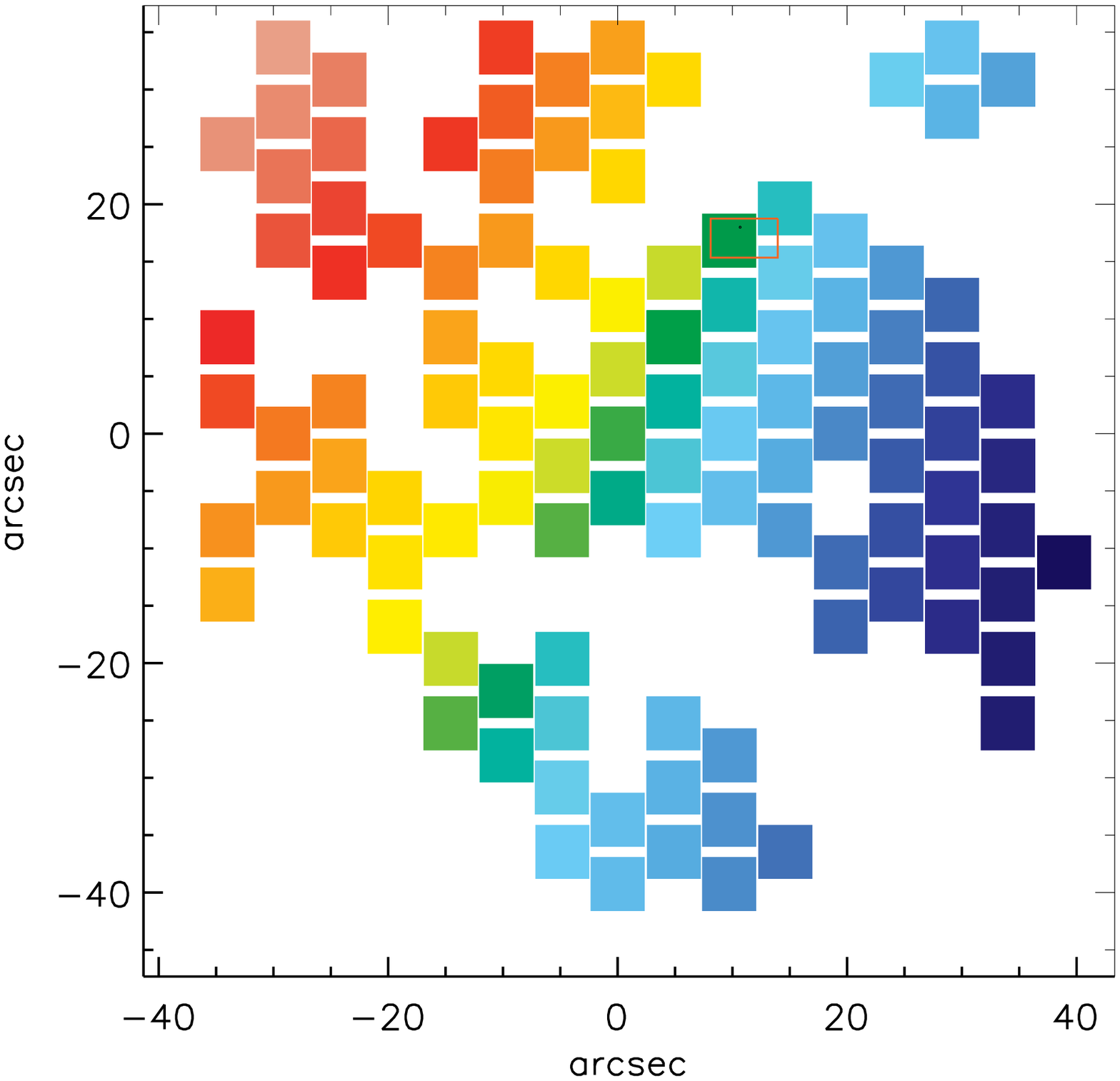} 
   \hspace*{-0.25cm}  \includegraphics[height=3.82cm]{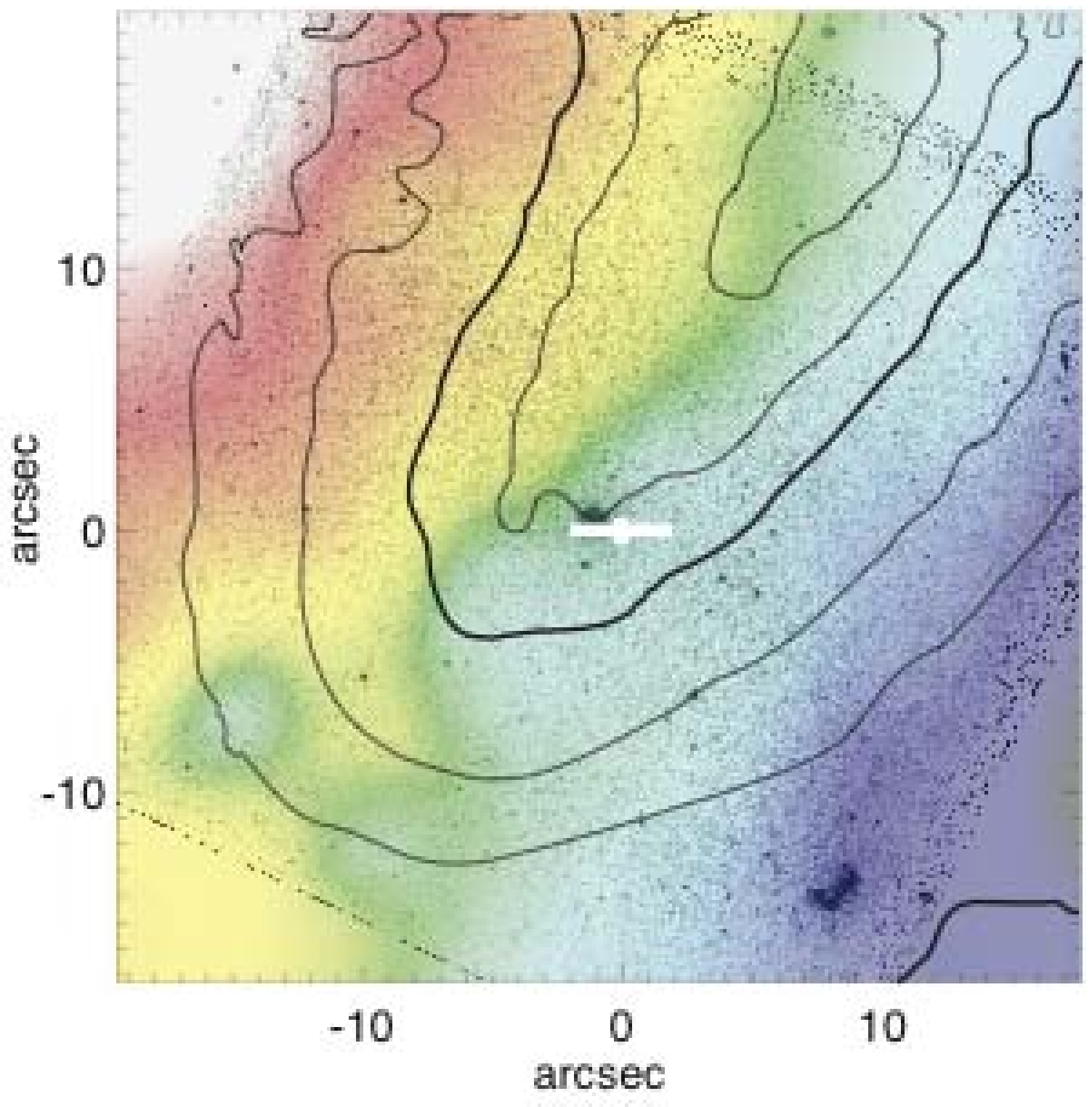}
   \vspace{-0.6cm}
  \end{tabular}
  \end{center}

  \begin{center}
  \begin{tabular}{cc}
  \hspace*{-1.3cm} \includegraphics[height=4.5cm]{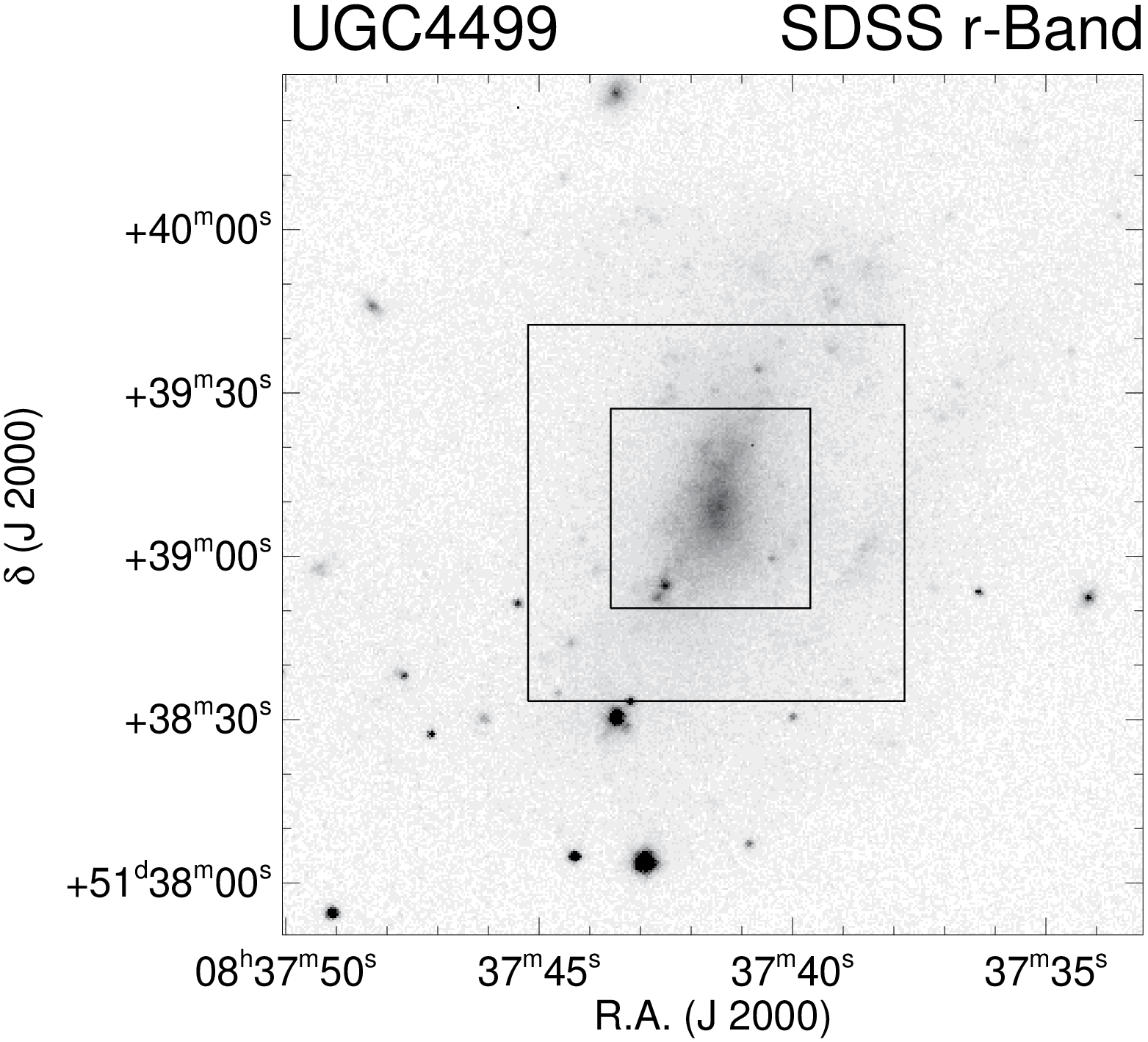}
   \hspace*{-0.25cm} \includegraphics[height=4.1cm]{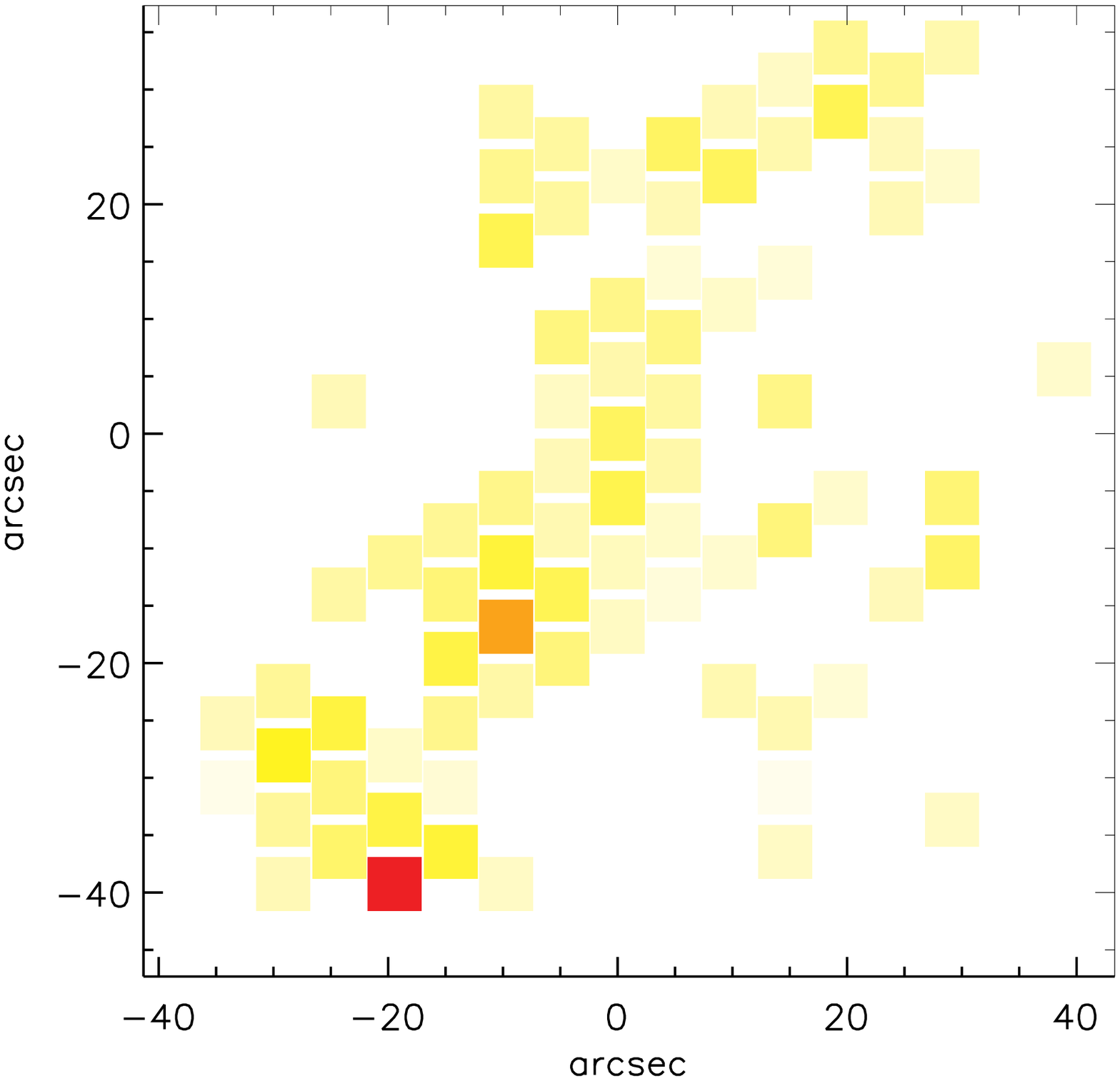}
   \hspace*{-0.25cm}   \includegraphics[height=4.1cm]{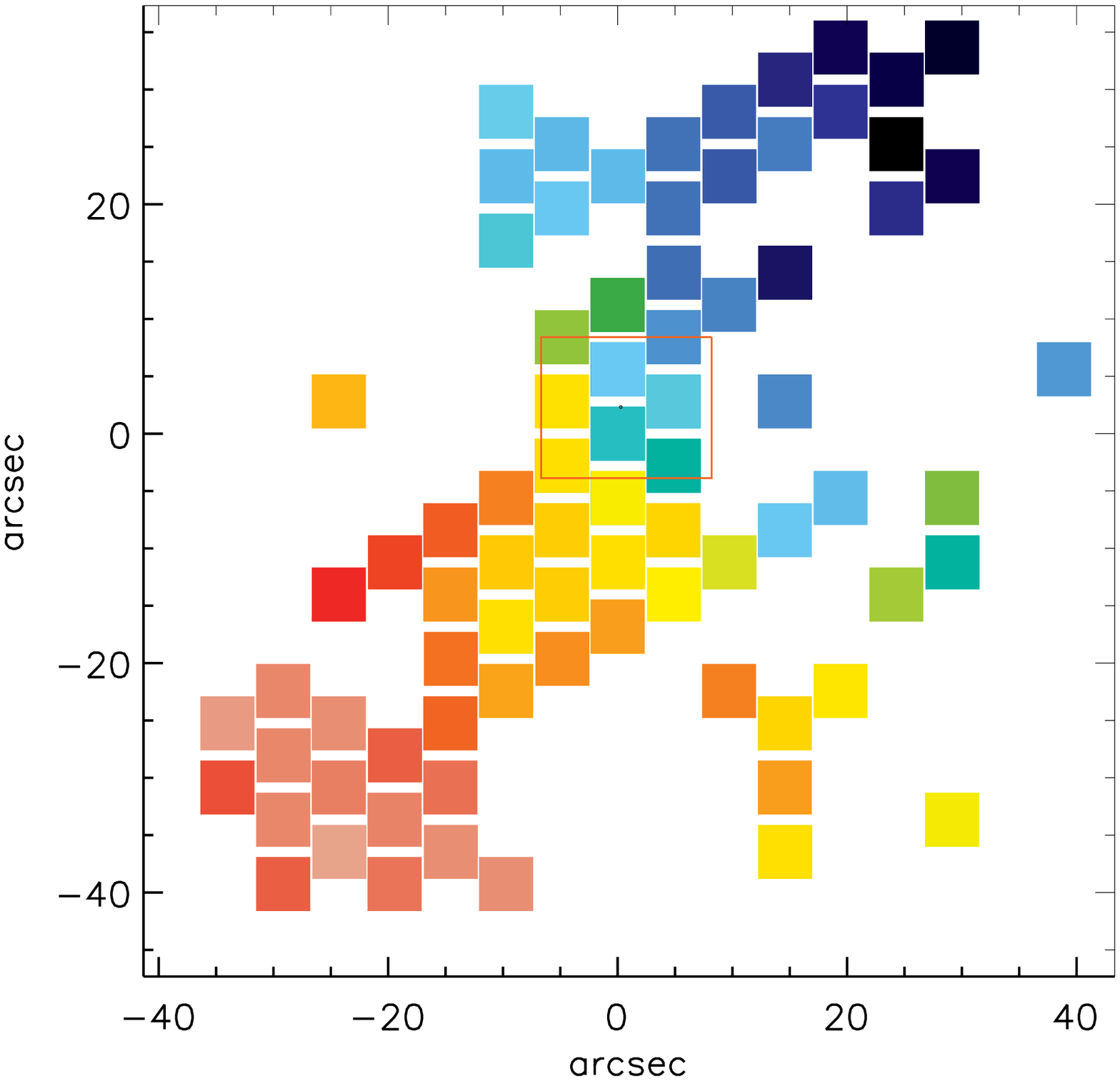}
   \hspace*{-0.25cm}   \includegraphics[height=4.1cm]{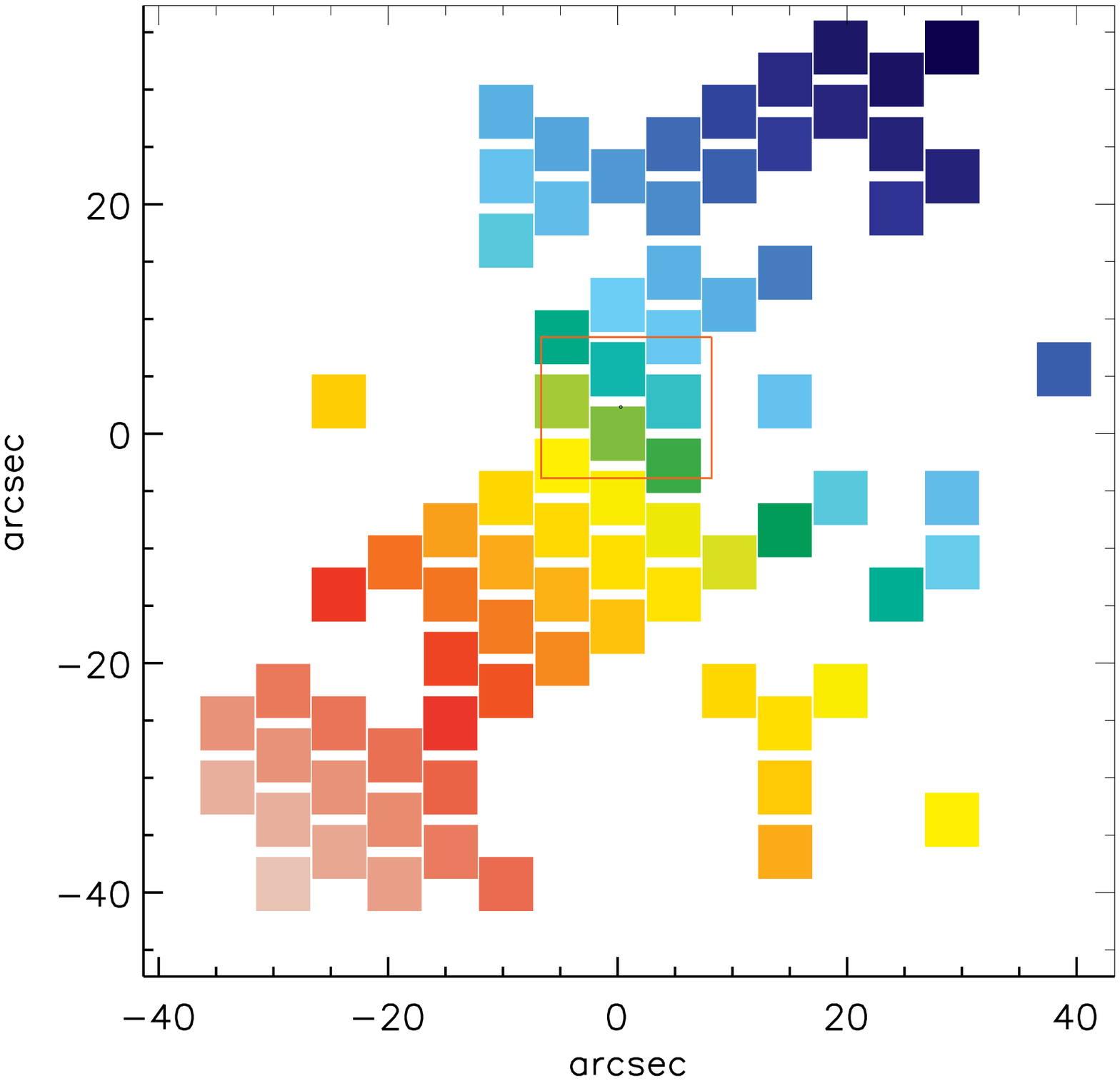} 
  \hspace*{-0.25cm} \includegraphics[height=3.82cm]{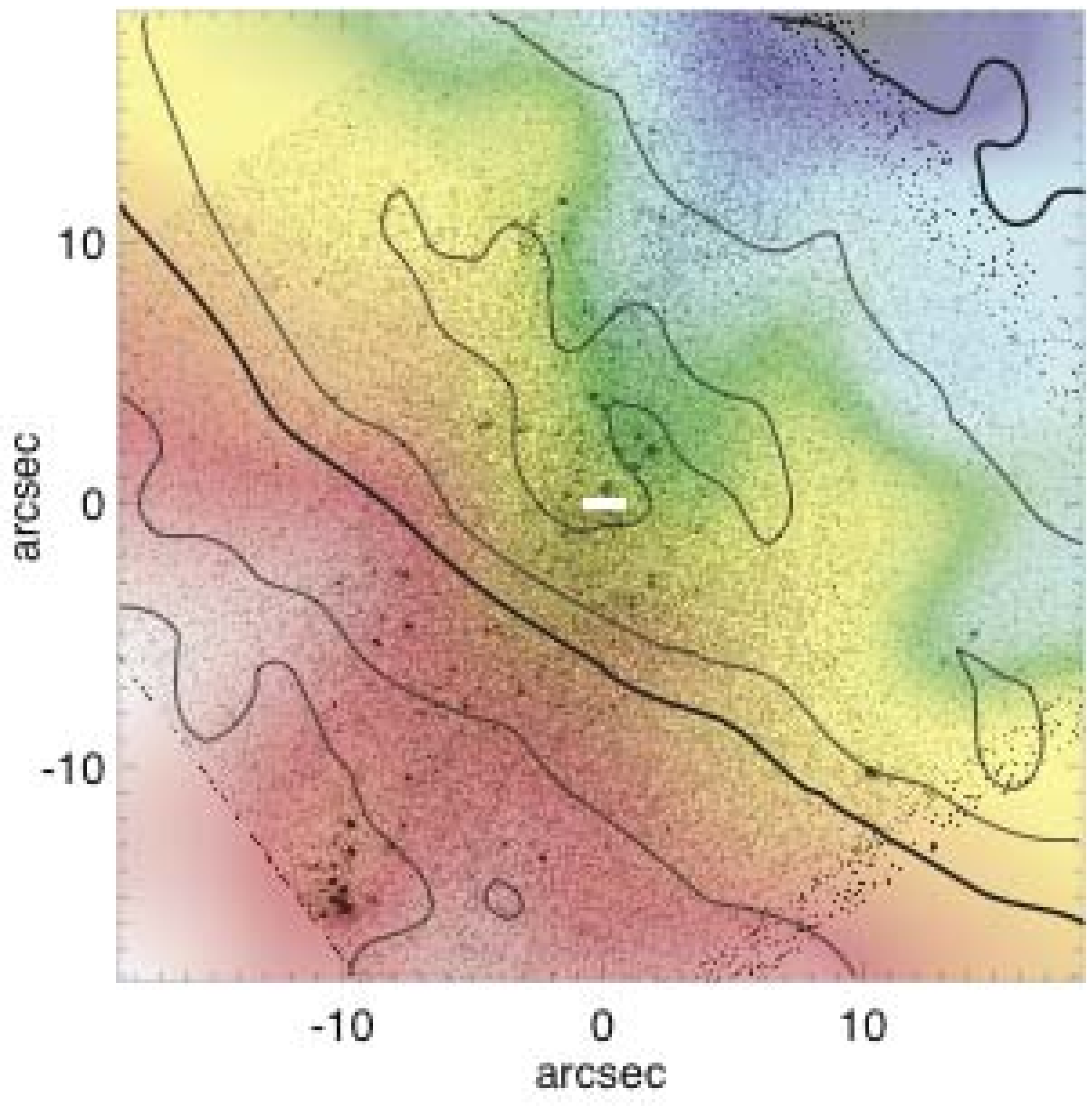}
   \vspace{-0.6cm}
  \end{tabular}
  \end{center}

 \begin{center}
  \begin{tabular}{cc}
  \hspace*{-1.3cm} \includegraphics[height=4.5cm]{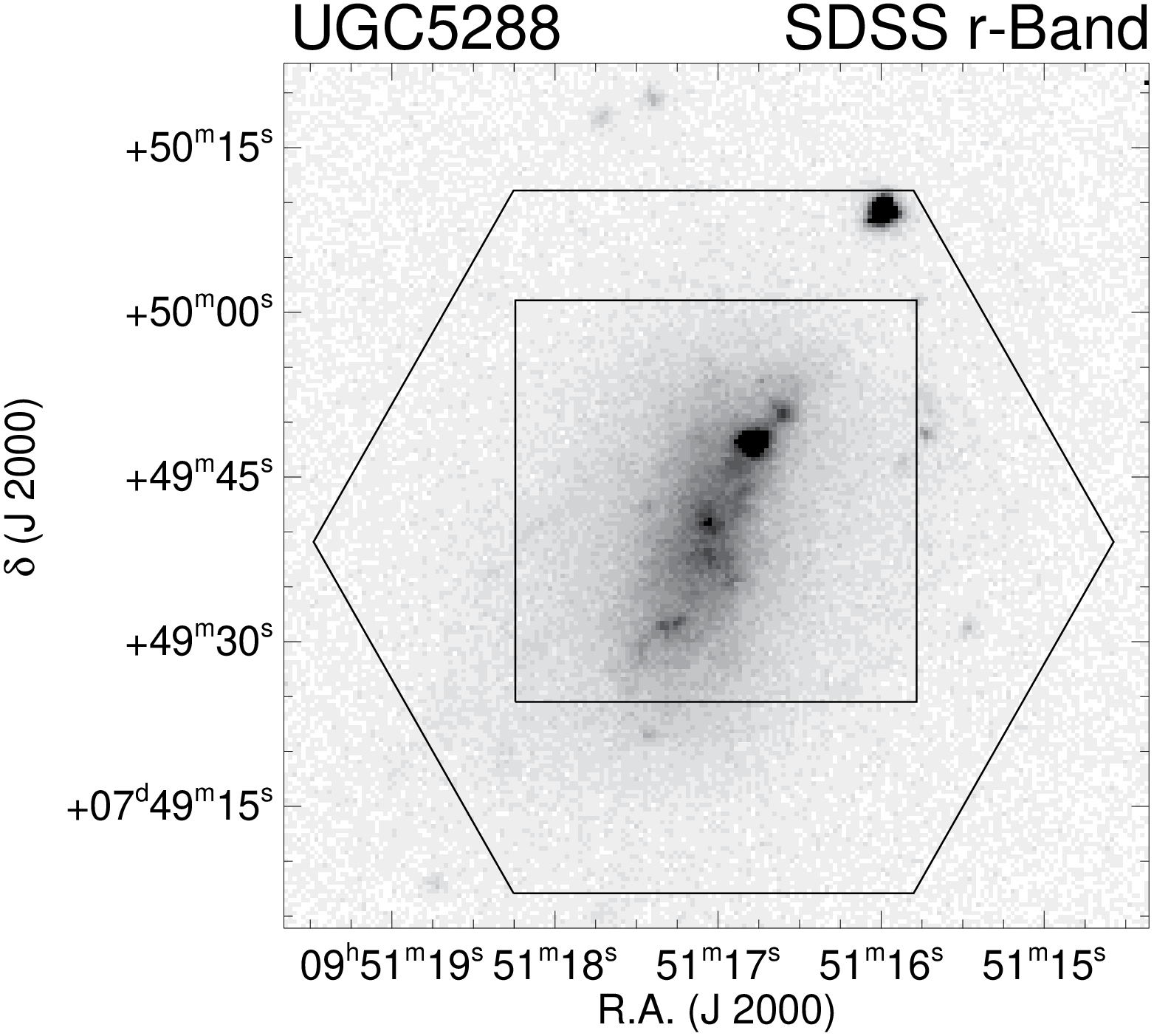}
   \hspace*{-0.25cm} \includegraphics[height=4.1cm]{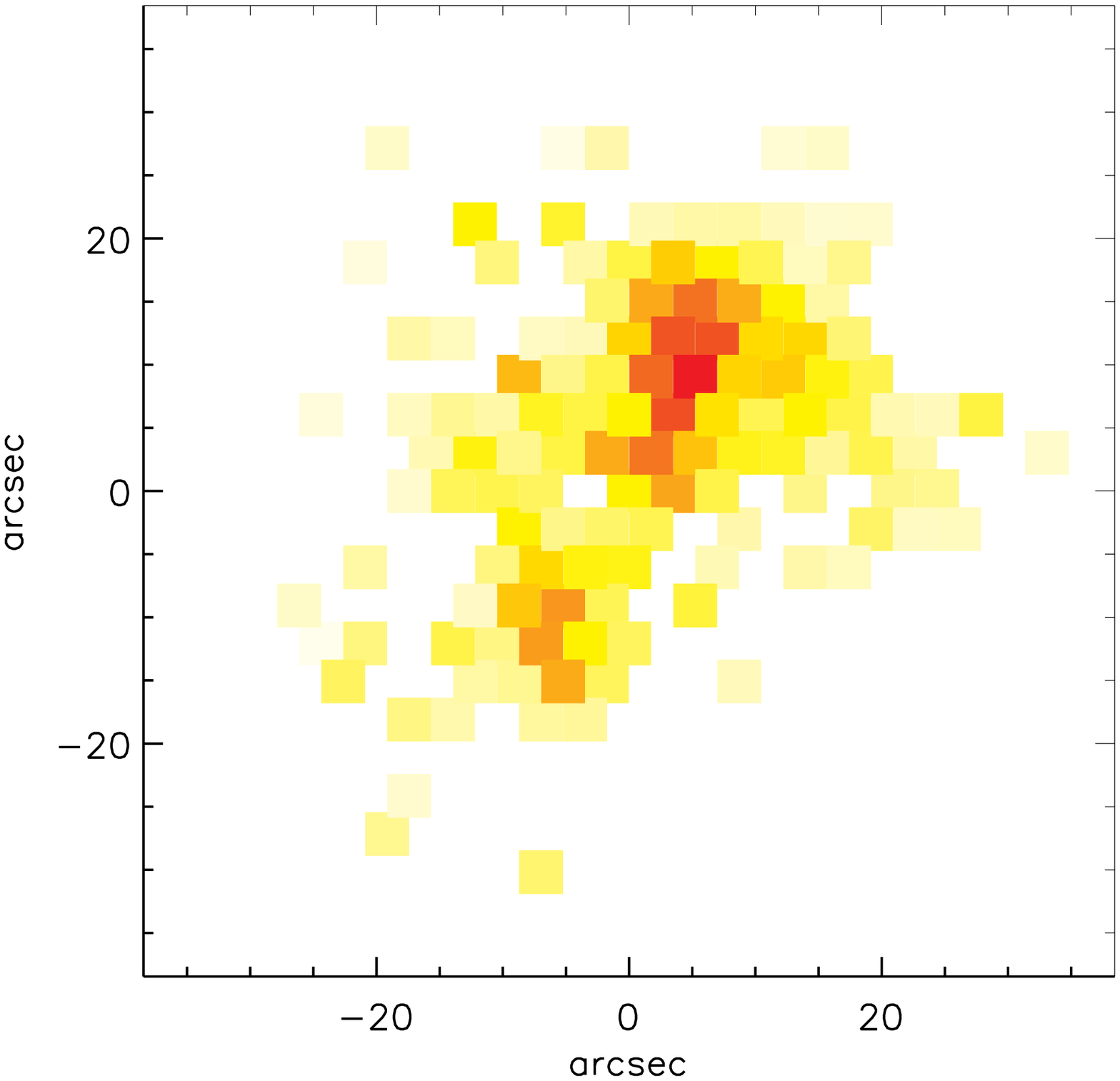}
   \hspace*{-0.25cm}   \includegraphics[height=4.1cm]{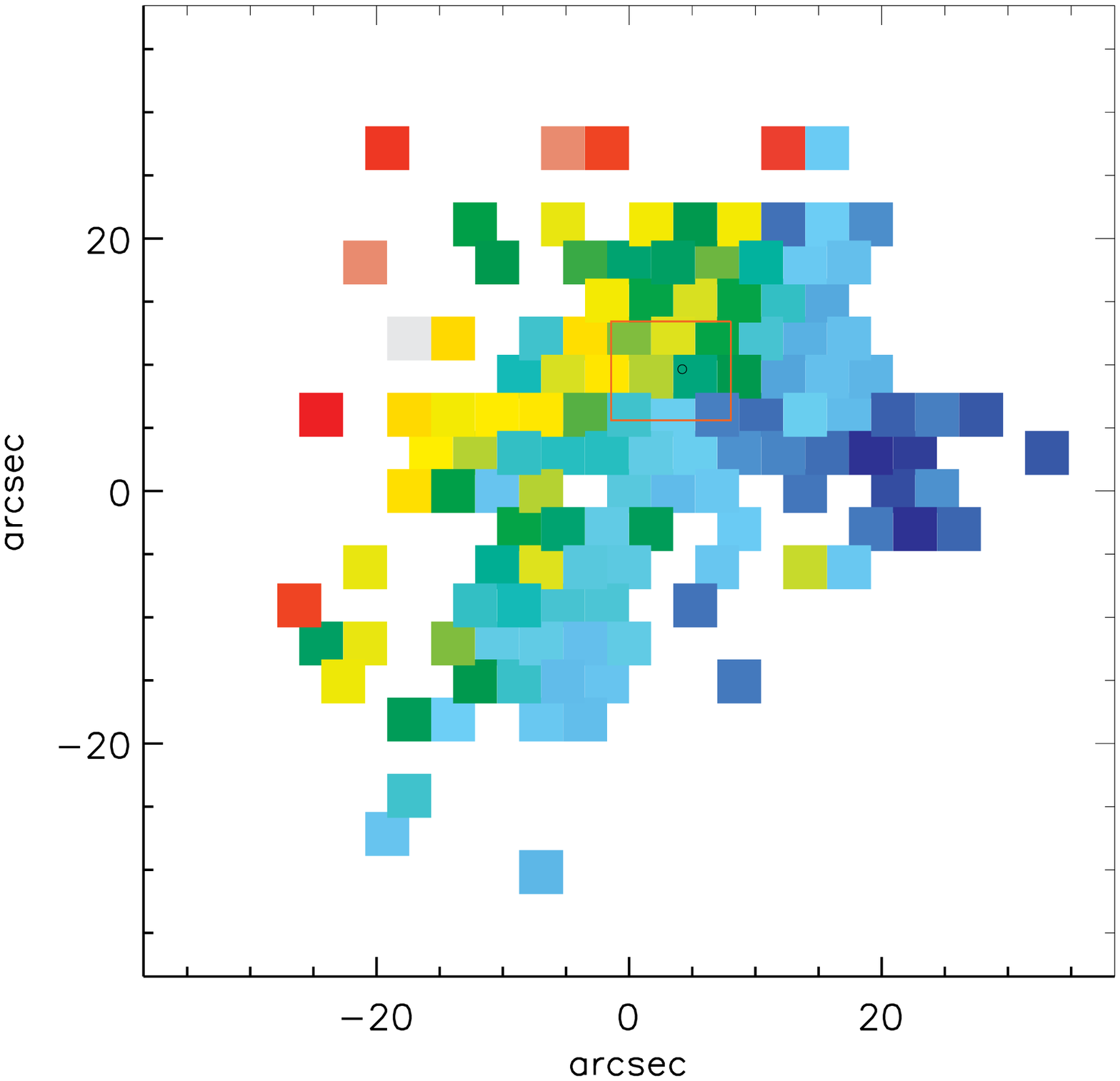}
   \hspace*{-0.25cm}   \includegraphics[height=4.1cm]{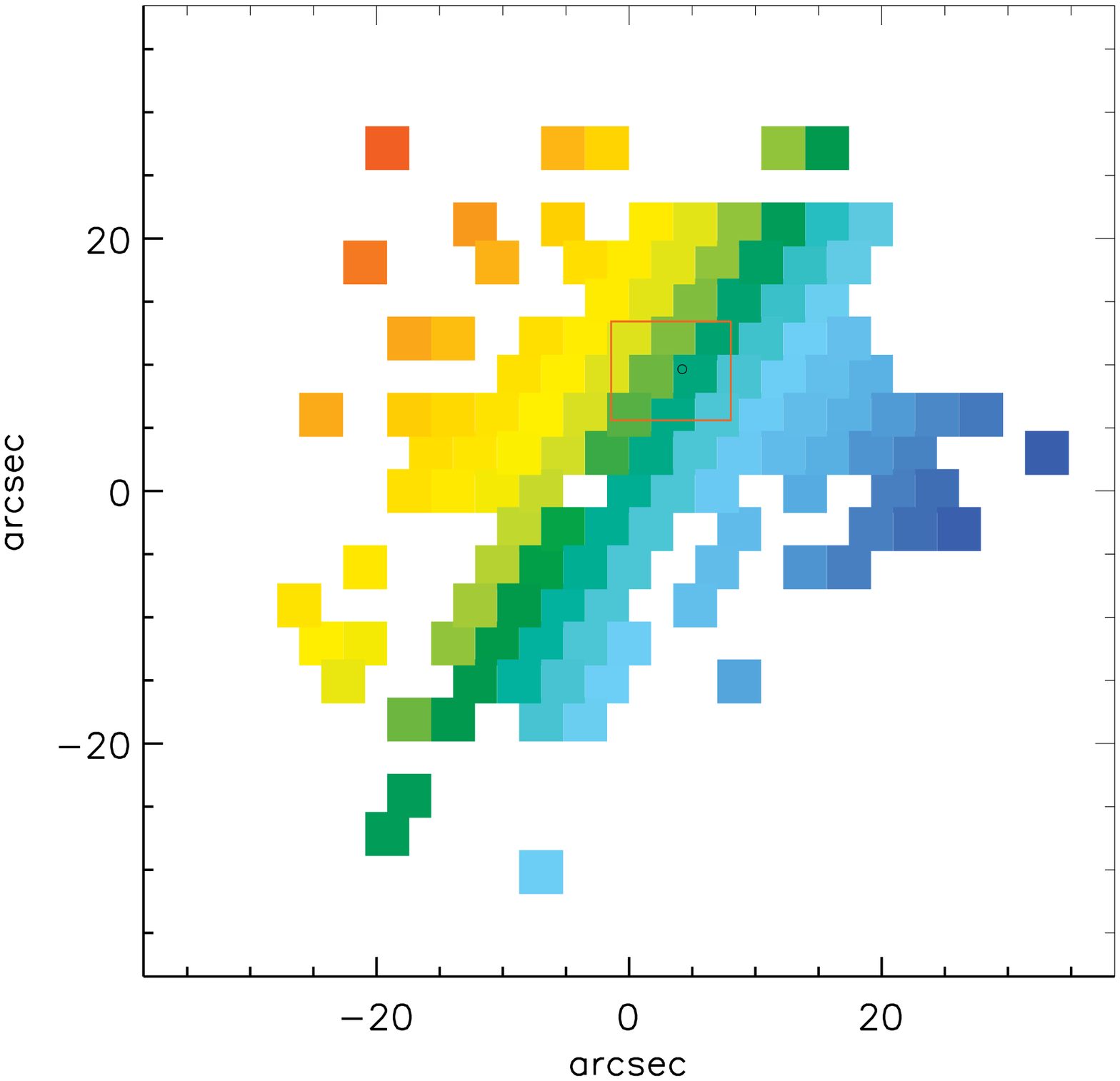} 
  \hspace*{-0.25cm}   \includegraphics[height=3.82cm]{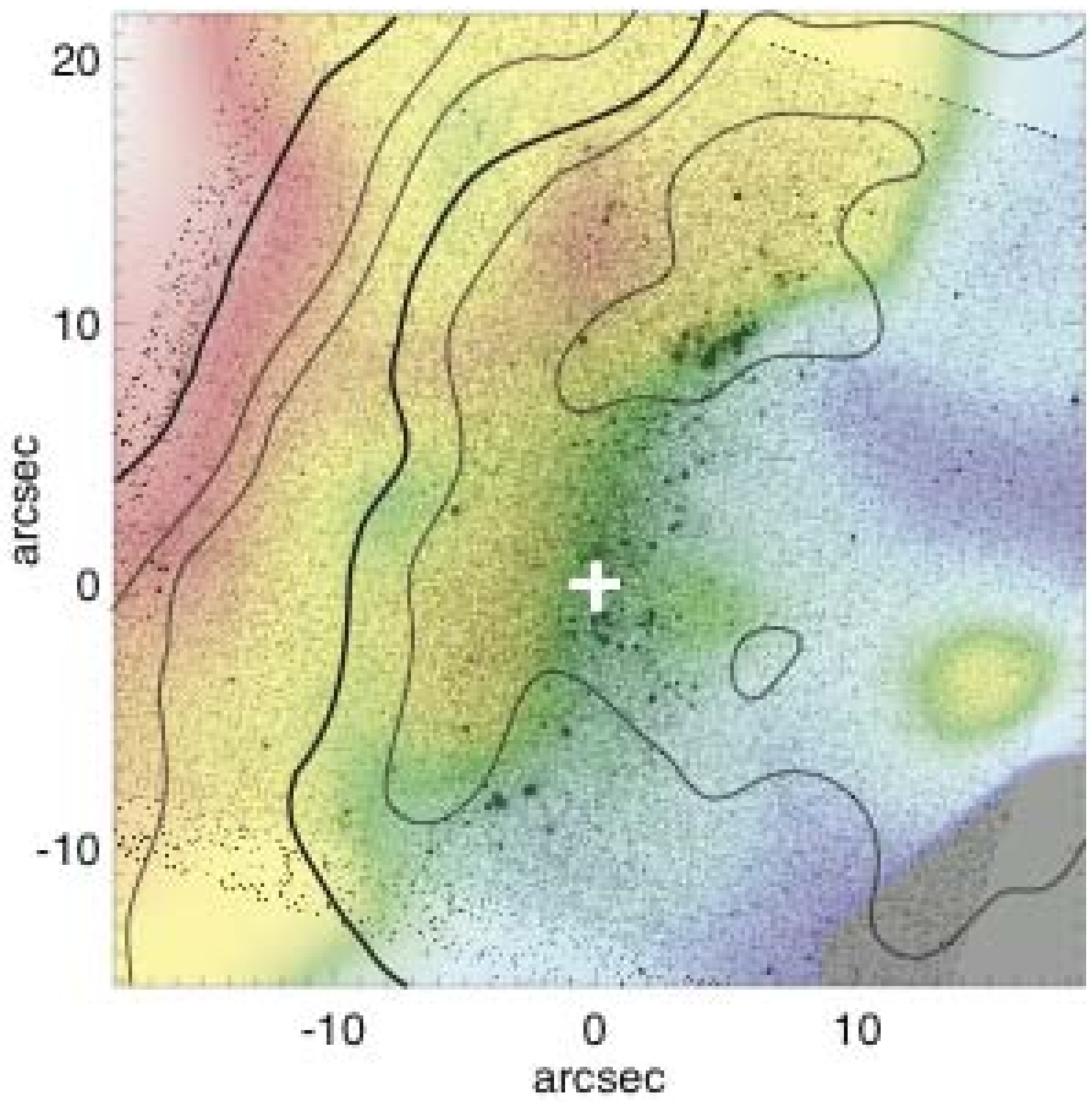}
   \vspace{-0.6cm}
  \end{tabular}
  \end{center}

 \begin{center}
  \begin{tabular}{cc}
  \hspace*{-1.3cm} \includegraphics[height=4.5cm]{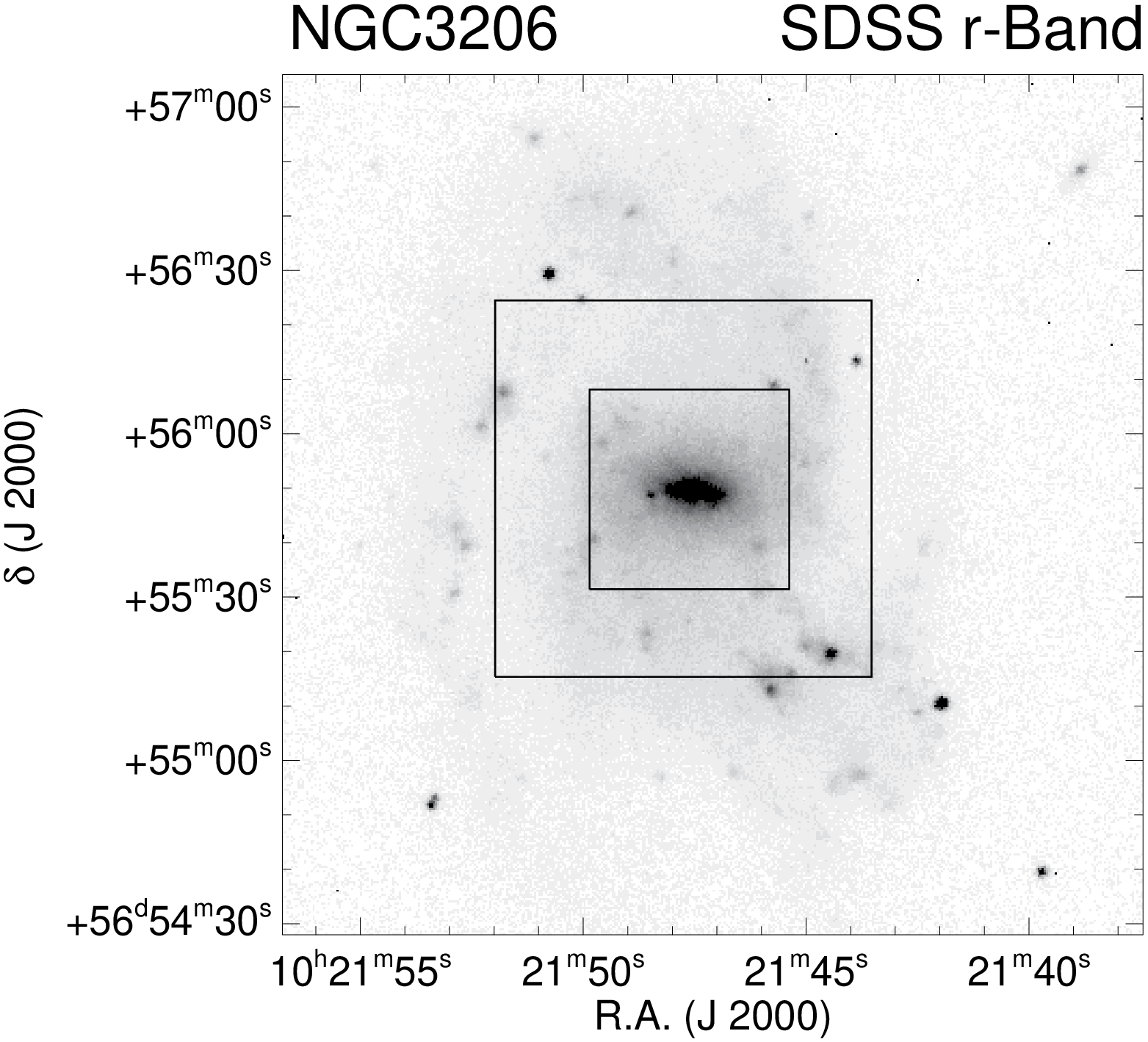}
   \hspace*{-0.25cm} \includegraphics[height=4.1cm]{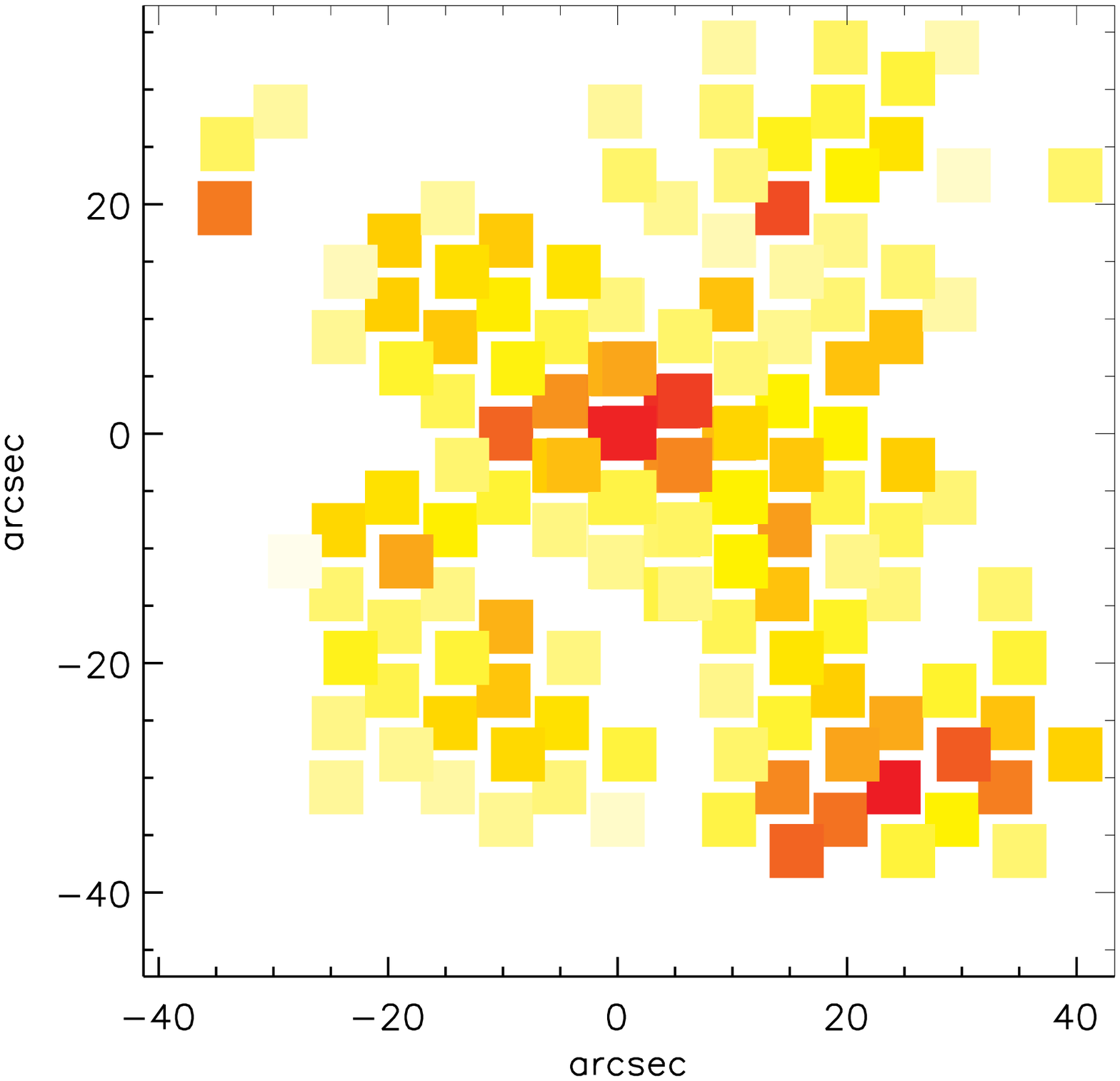}
   \hspace*{-0.25cm}   \includegraphics[height=4.1cm]{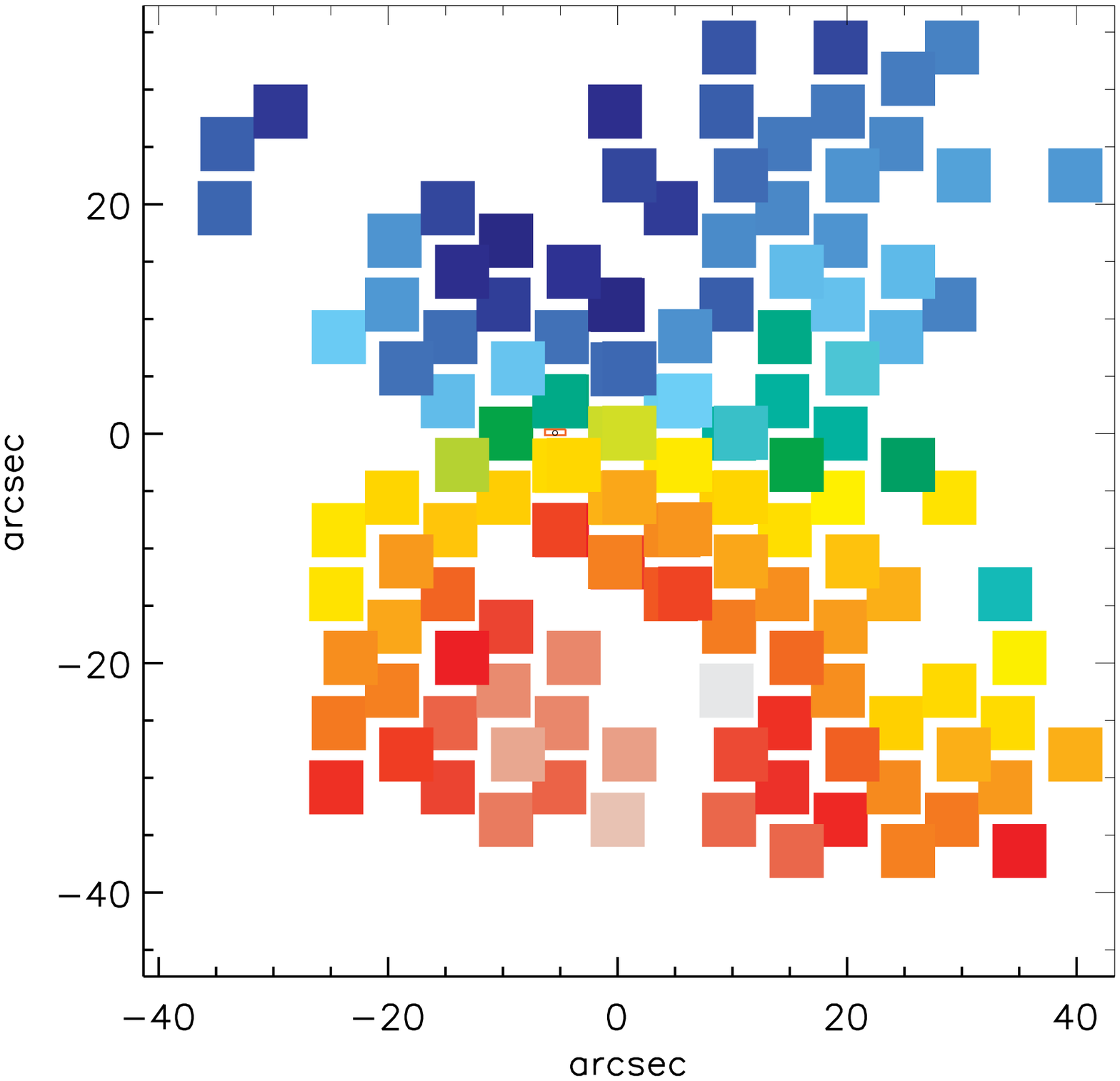}
   \hspace*{-0.25cm}   \includegraphics[height=4.1cm]{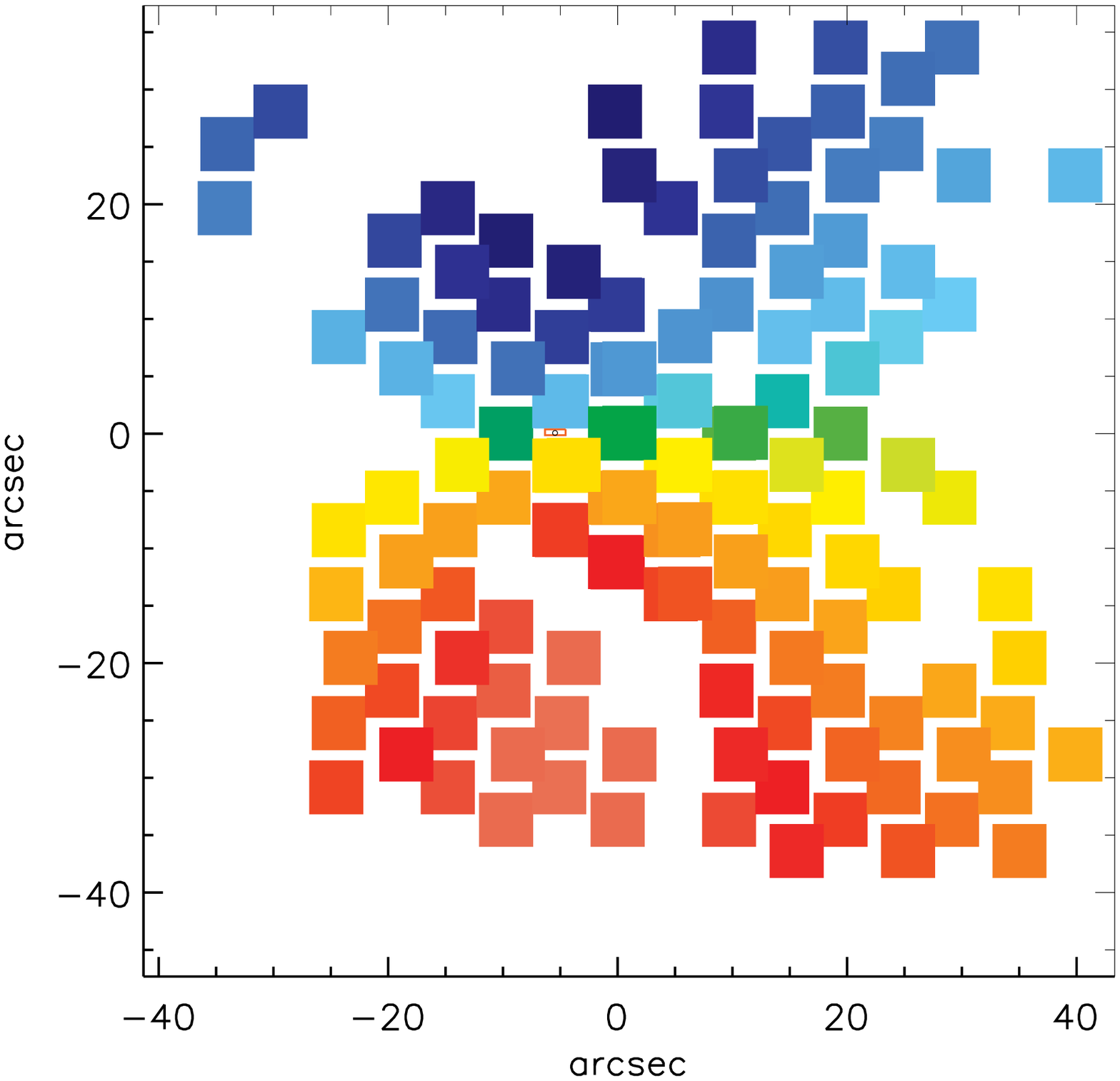} 
  \hspace*{-0.25cm}   \includegraphics[height=3.82cm]{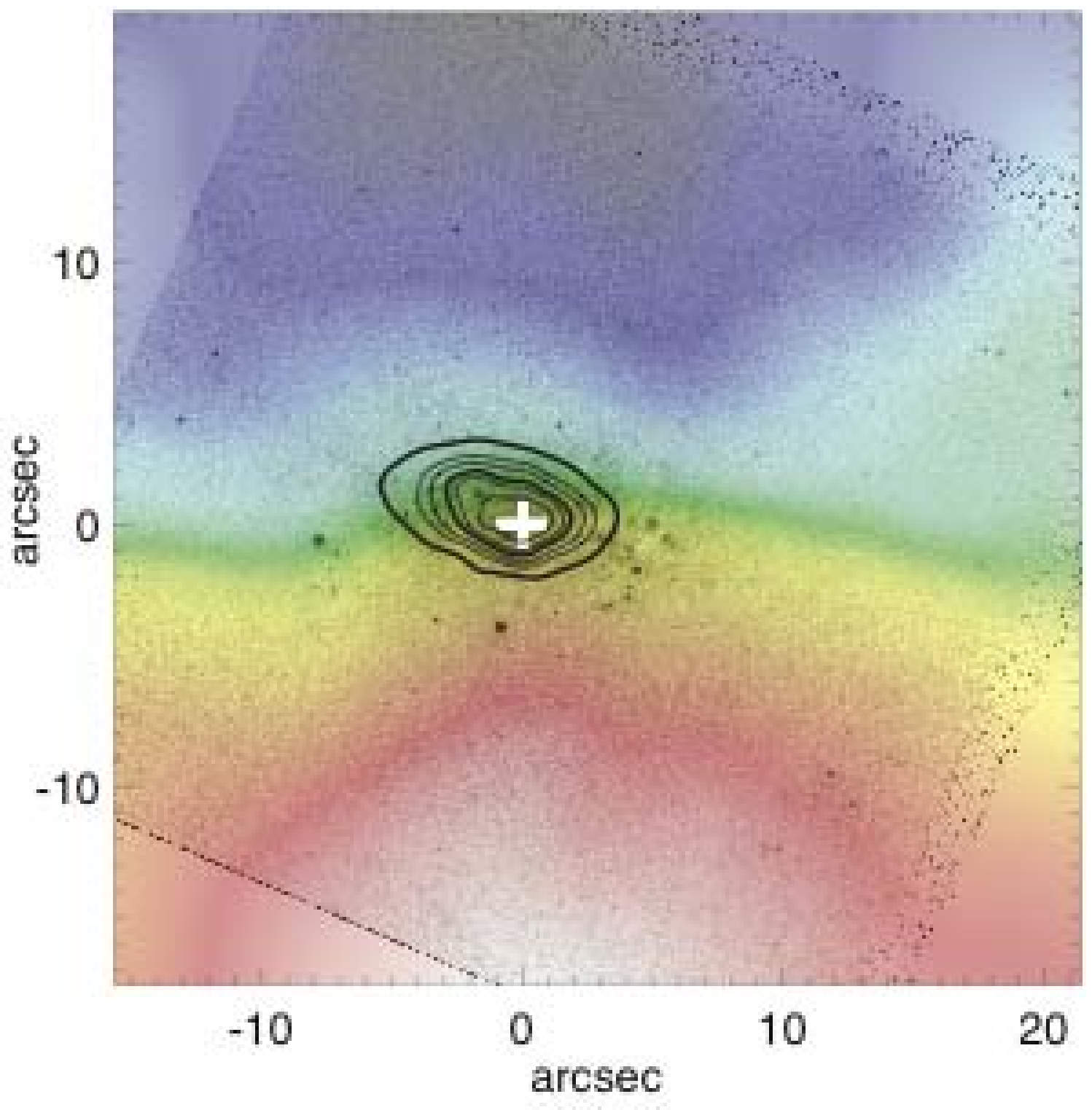}
  \end{tabular}
  \end{center}

\caption{ \label{fig:mc1}
Montage of the sample galaxies sorted by right ascension, one line per galaxy. Panels from 
left to right: (i) SDSS r-Band (when available) or DSS R-Band image, with two footprints per 
galaxy overplotted. The small square footprint indicates the HST WFPC2 field-of-view, the 
larger square gives the \spak\ and the hexagon the \ppak\ footprint. (ii) \ha\ flux map and 
(iii) observed \ha\ velocity map in comparison to (iv) the modelled velocity field (masked 
with the observed flux map). The right panel shows the HST WFPC2 F814W image with 
a transparent overlay of the velocity map. Gaps in the velocity map are filled in with model 
values. The white cross indicates the position and uncertainty of the PC 
\citep[as derived by][]{boker02}, black contours indicate the position and uncertainty of the 
KC (from 1 $\sigma$ to 6 $\sigma$, marginalized over all other parameters), 
derived in this paper.   }
\end{figure*}
  
  \begin{figure*}
  \begin{center}
  \begin{tabular}{c}
  \hspace*{-1.3cm} \includegraphics[height=4.5cm]{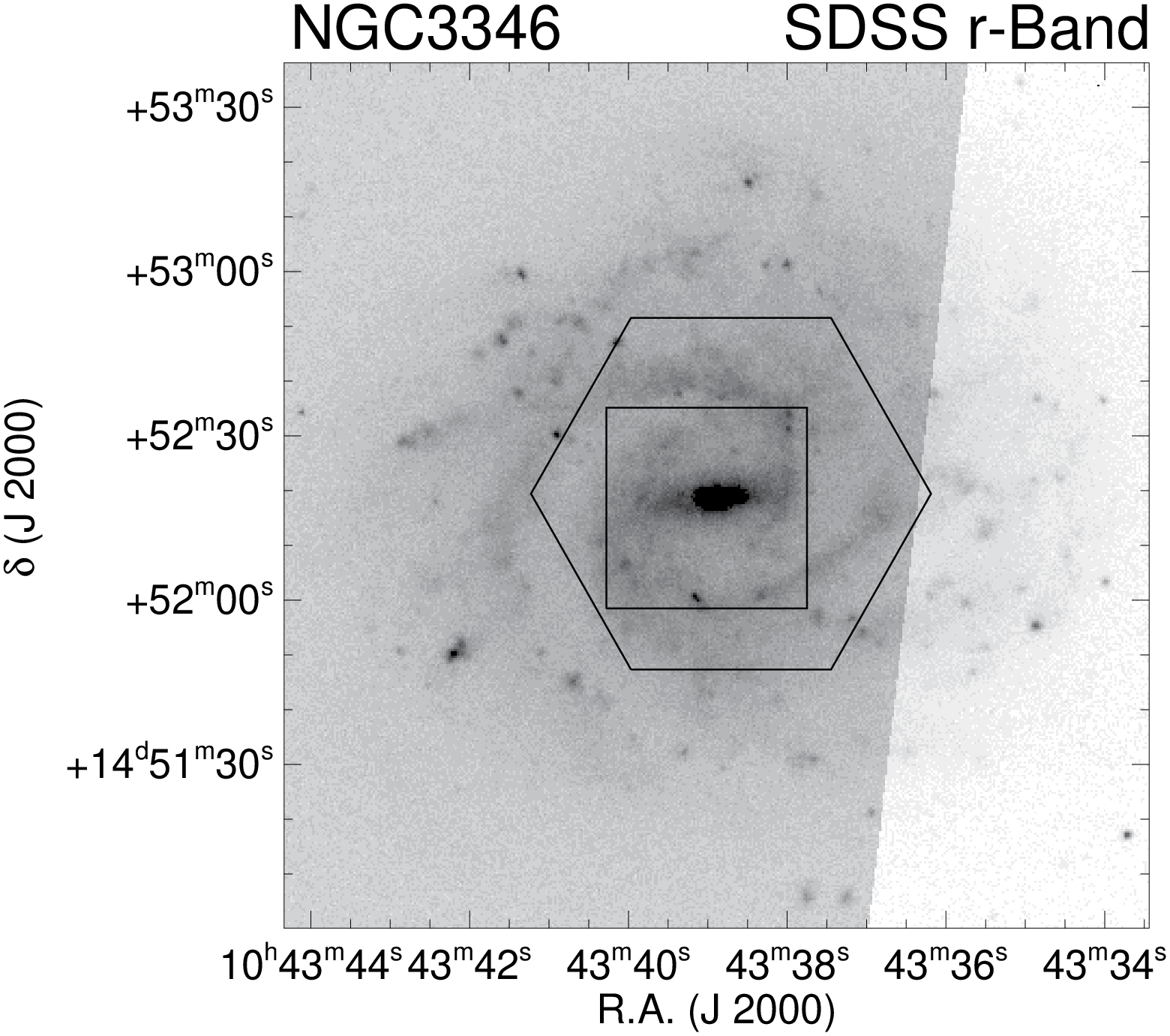}
   \hspace*{-0.25cm}  \includegraphics[width=4.1cm]{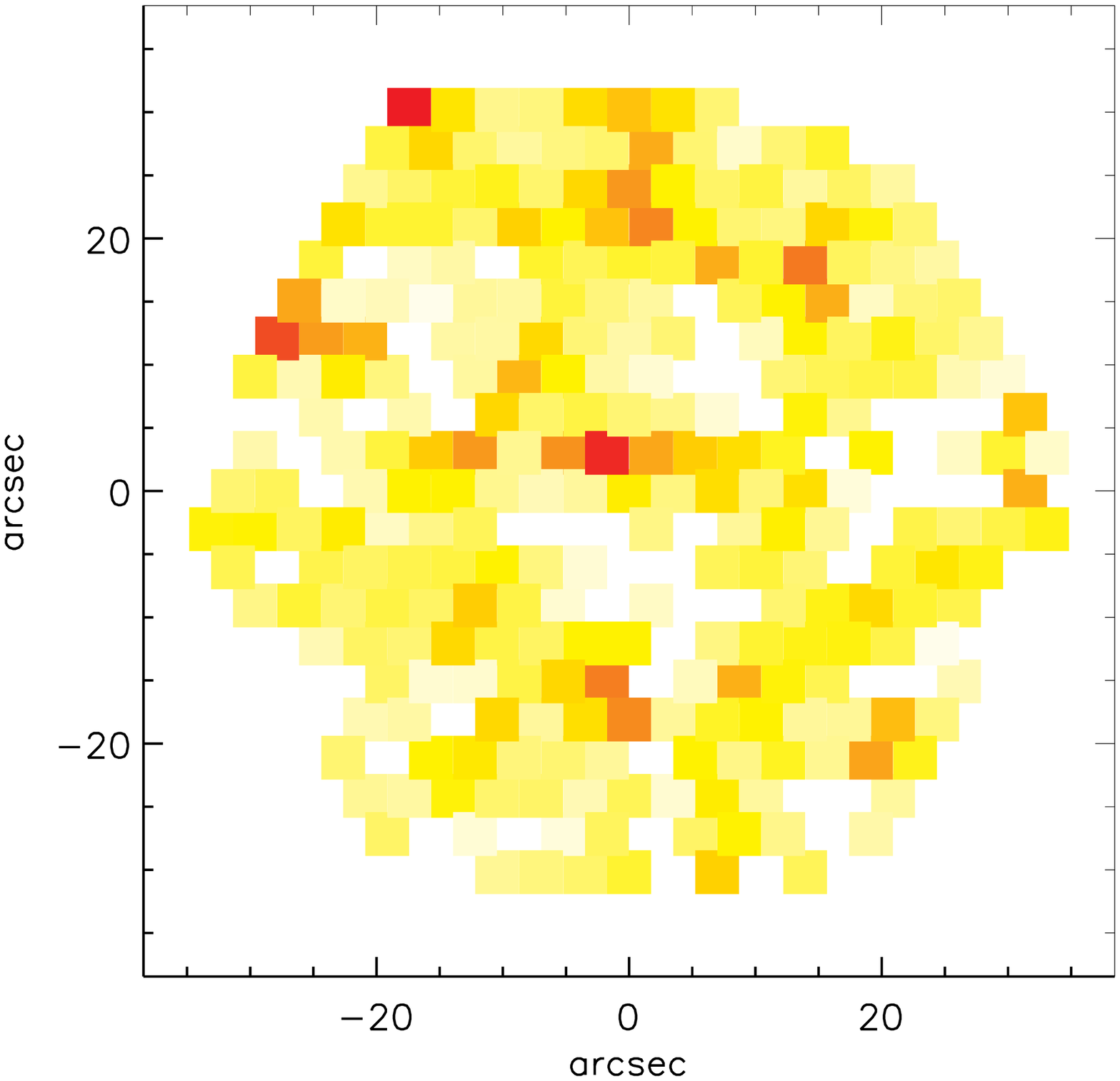}
    \hspace*{-0.25cm}  \includegraphics[height=4.1cm]{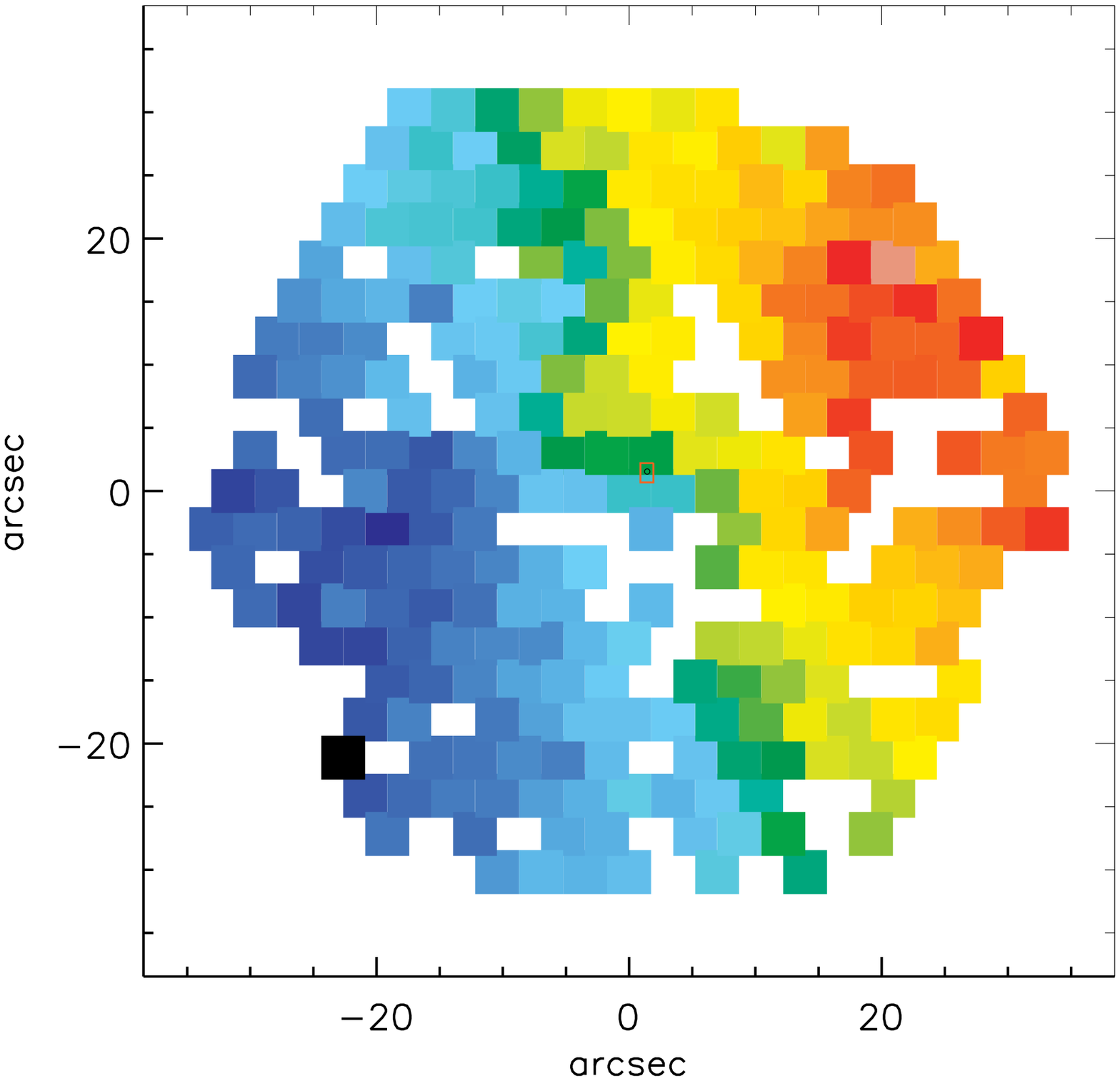}
    \hspace*{-0.25cm}  \includegraphics[height=4.1cm]{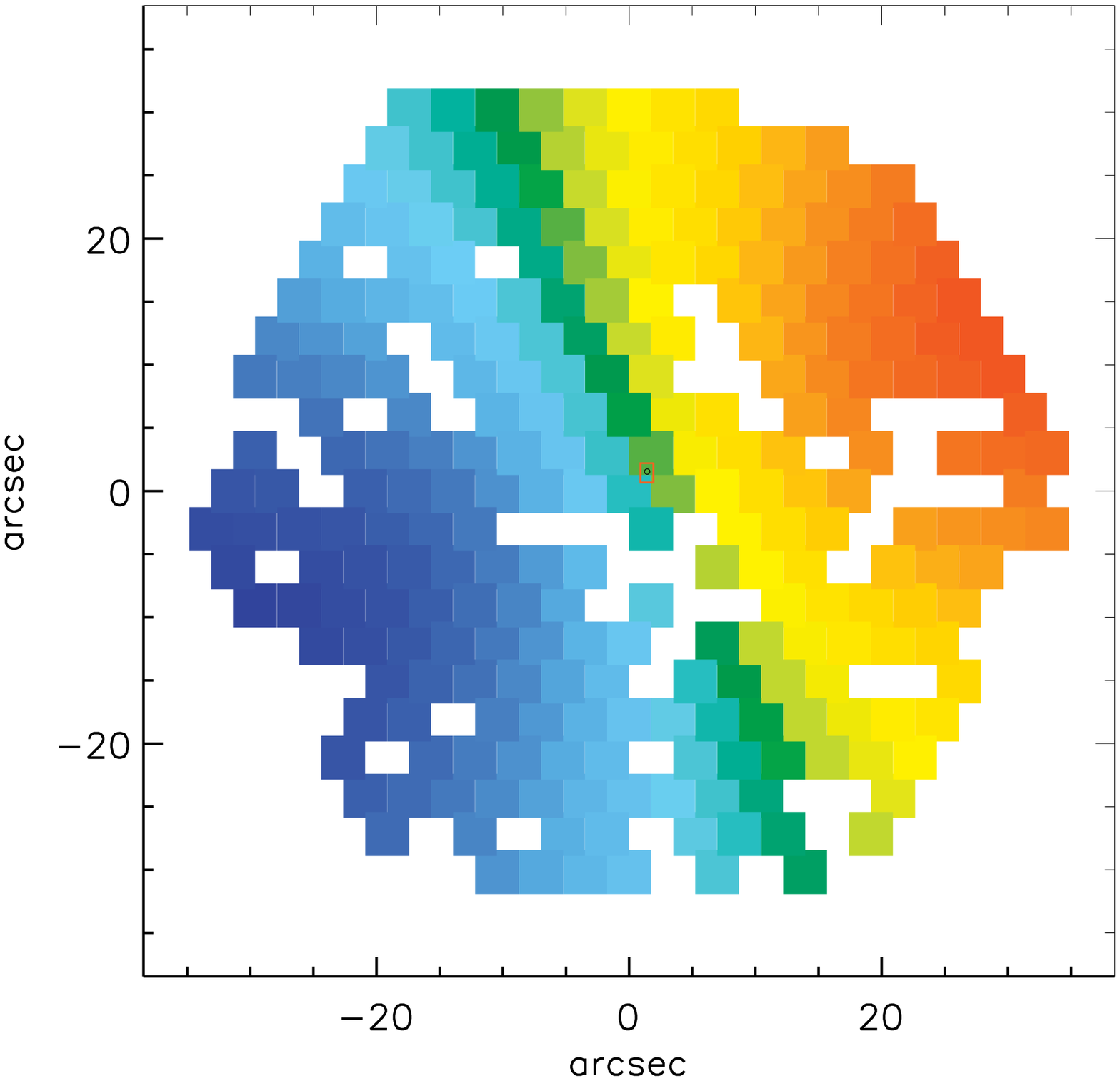}
 \hspace*{-0.25cm} \includegraphics[height=3.82cm]{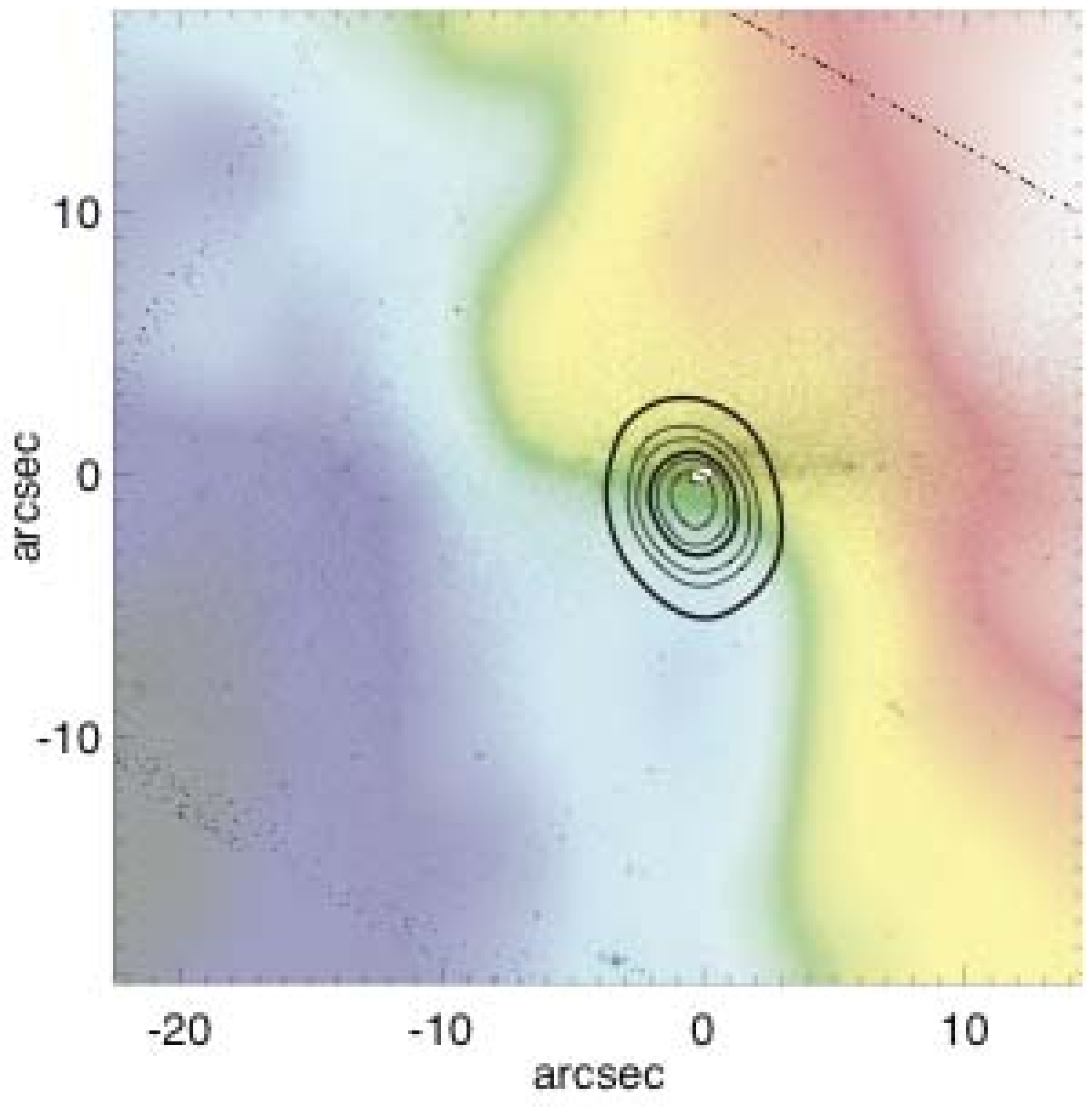}
    \vspace{-0.6cm}
  \end{tabular}
  \end{center}

  \begin{center}
  \begin{tabular}{cc}
  \hspace*{-1.3cm} \includegraphics[height=4.5cm]{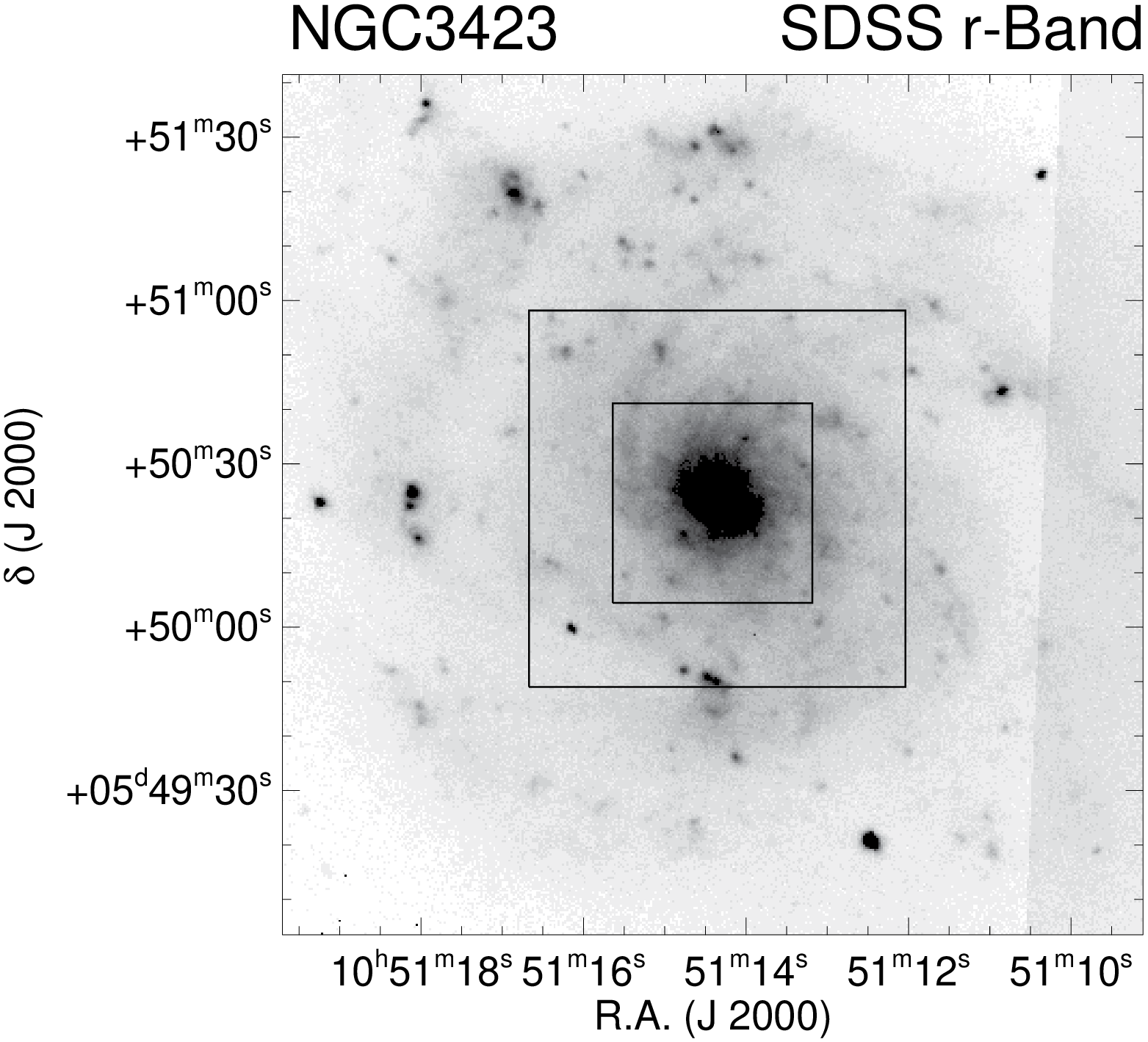}
   \hspace*{-0.25cm} \includegraphics[height=4.1cm]{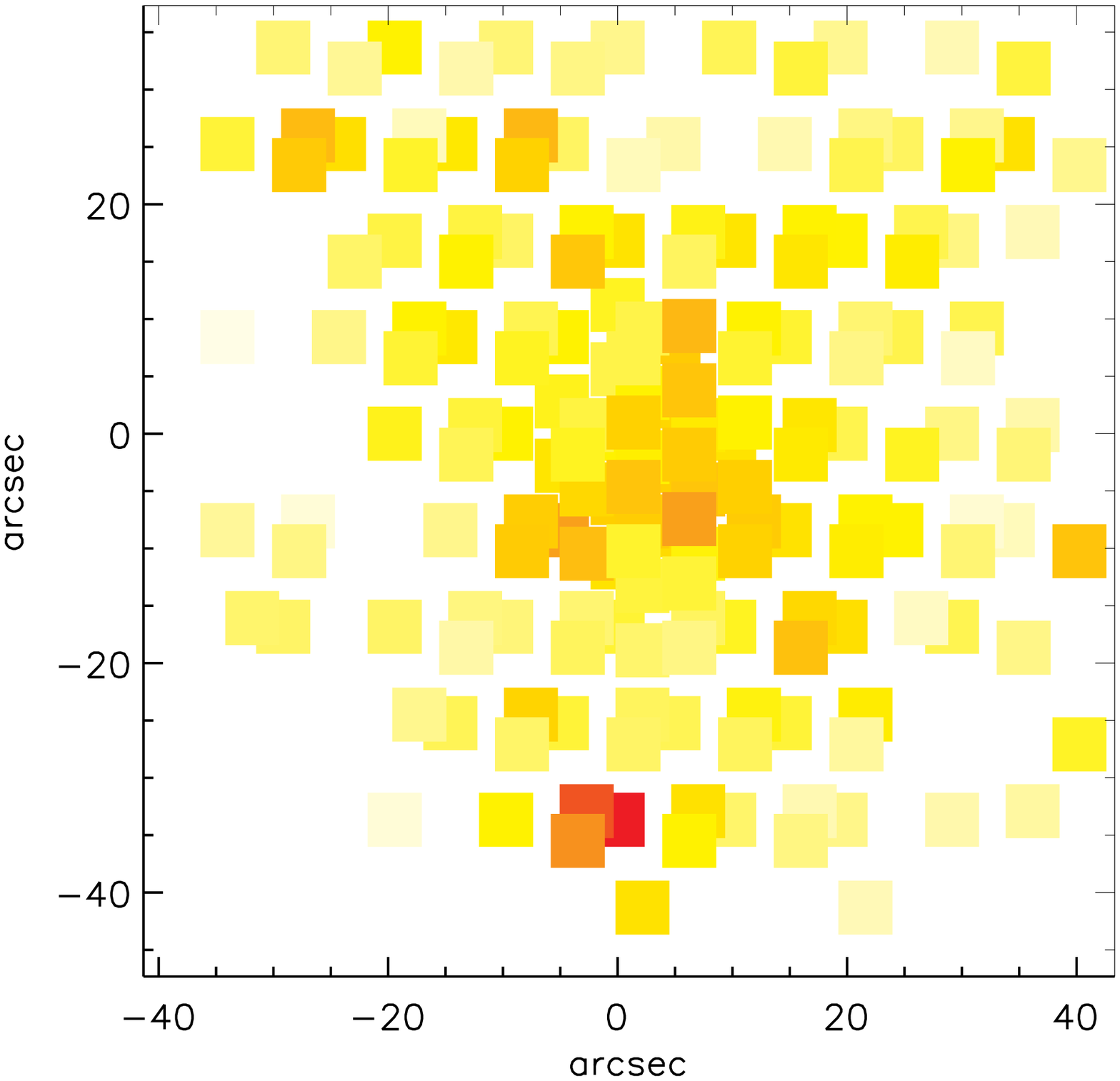}
   \hspace*{-0.25cm}   \includegraphics[height=4.1cm]{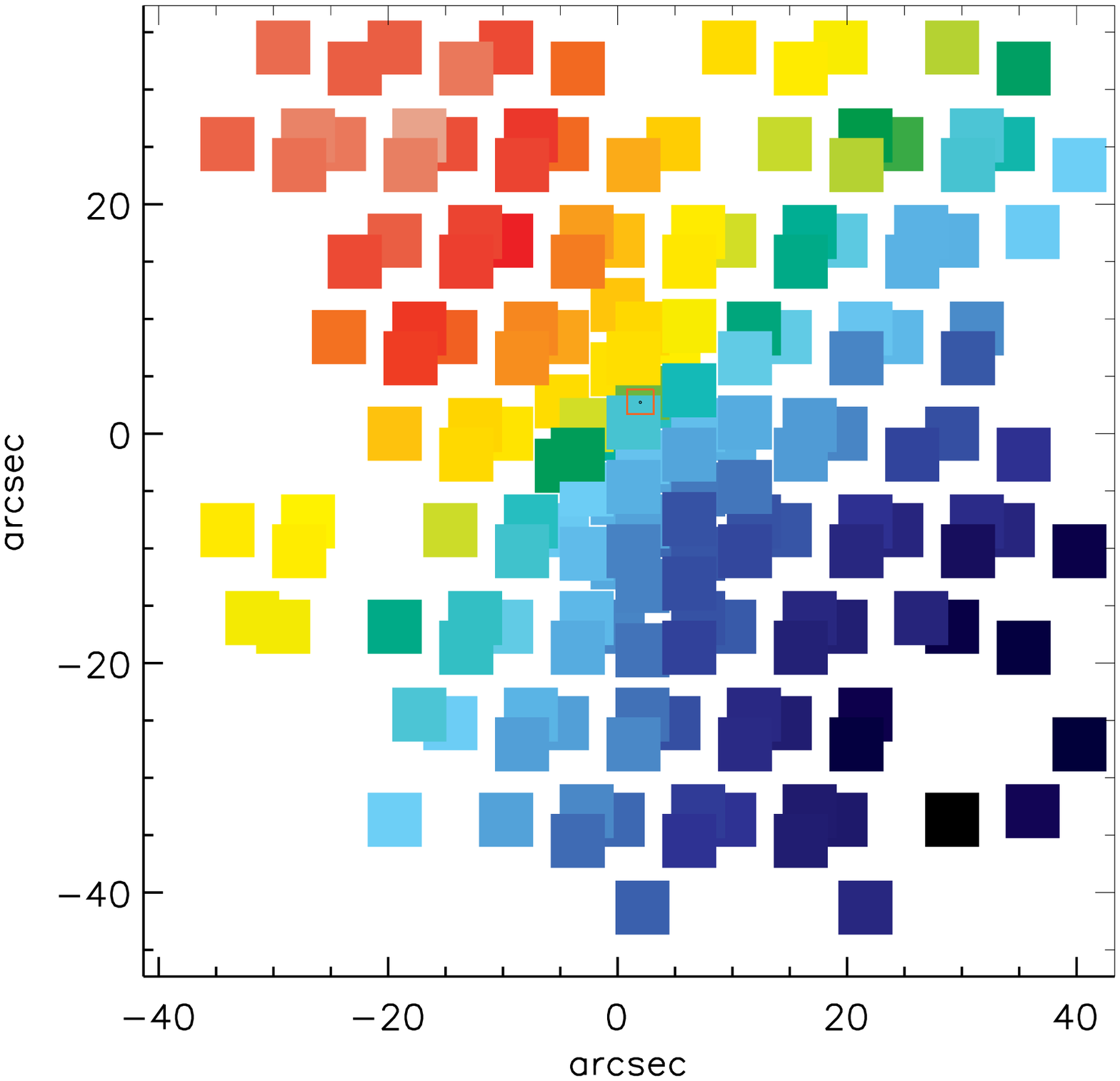}
   \hspace*{-0.25cm}   \includegraphics[height=4.1cm]{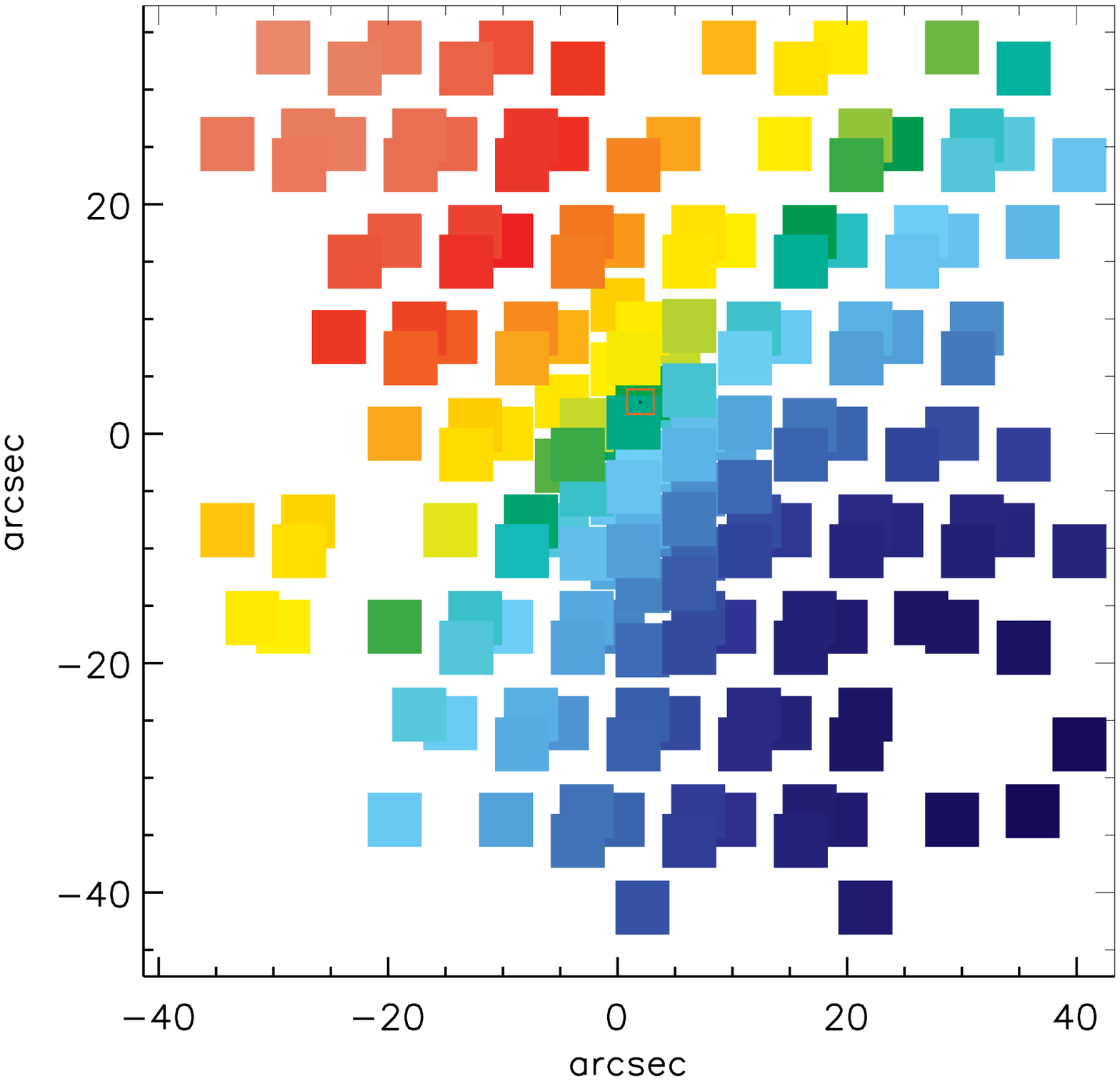} 
 \hspace*{-0.25cm}  \includegraphics[height=3.82cm]{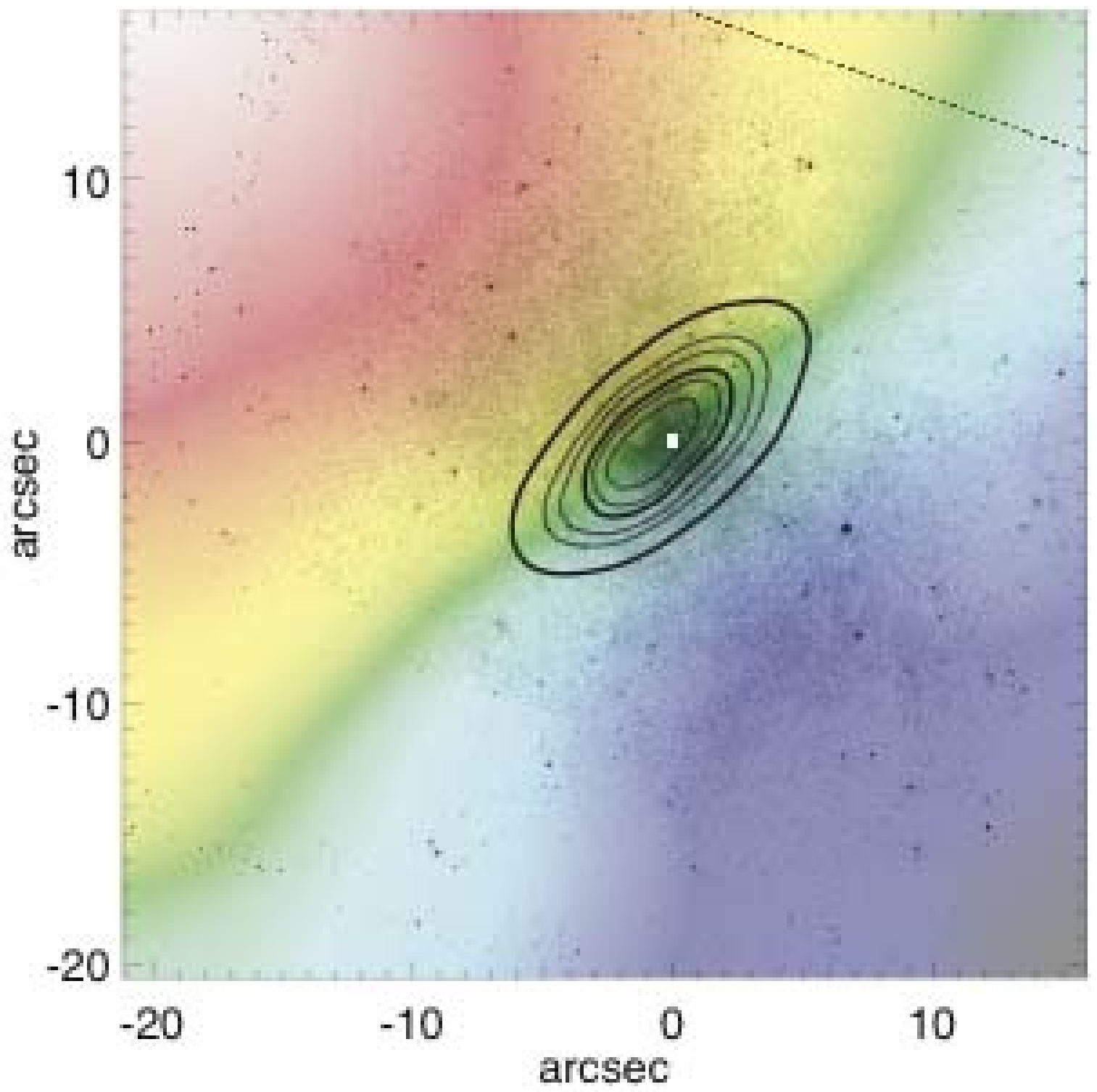}
   \vspace{-0.6cm}
  \end{tabular}
  \end{center}

  \begin{center}
  \begin{tabular}{cc}
  \hspace*{-1.3cm} \includegraphics[height=4.5cm]{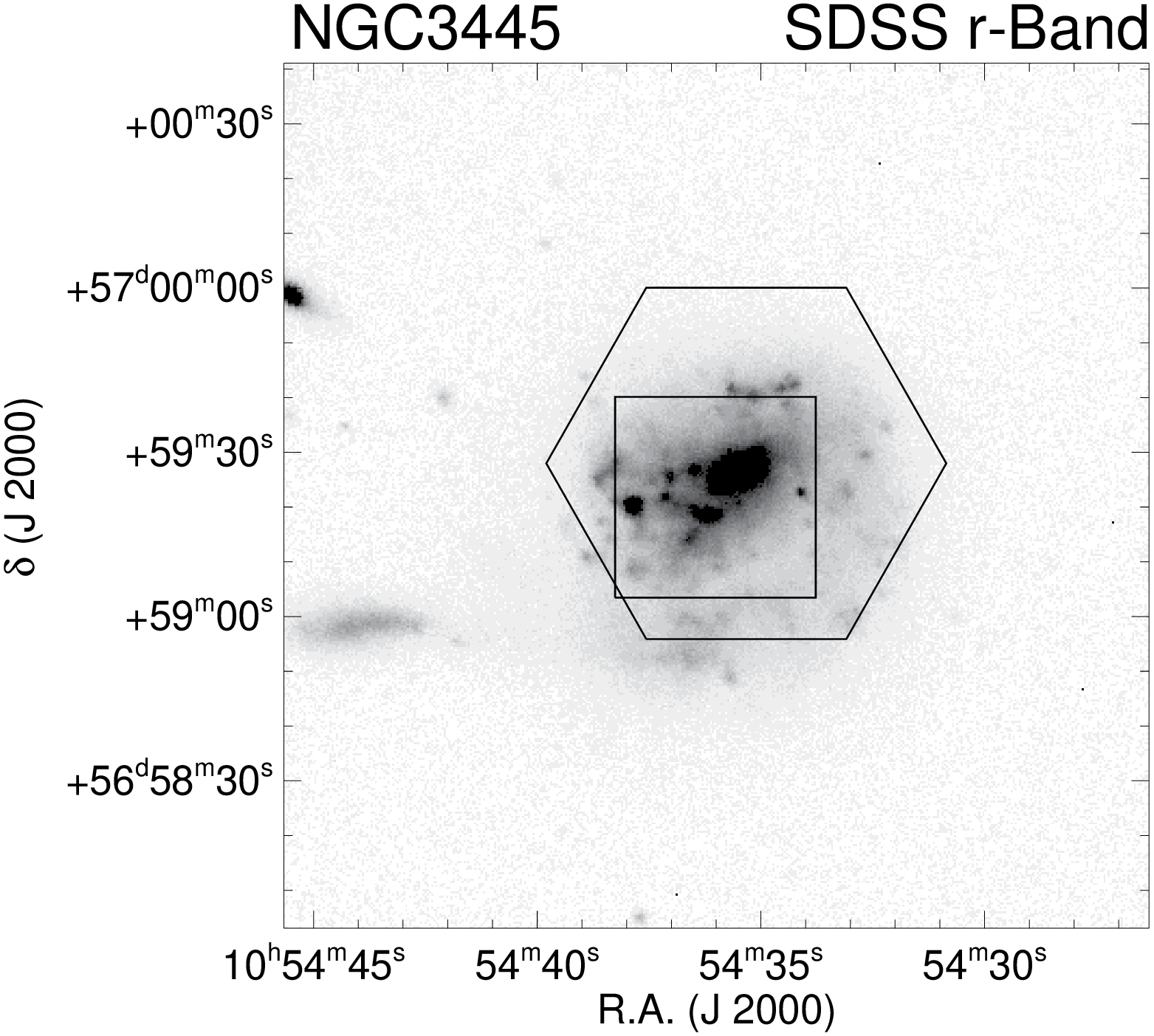}
   \hspace*{-0.25cm} \includegraphics[height=4.1cm]{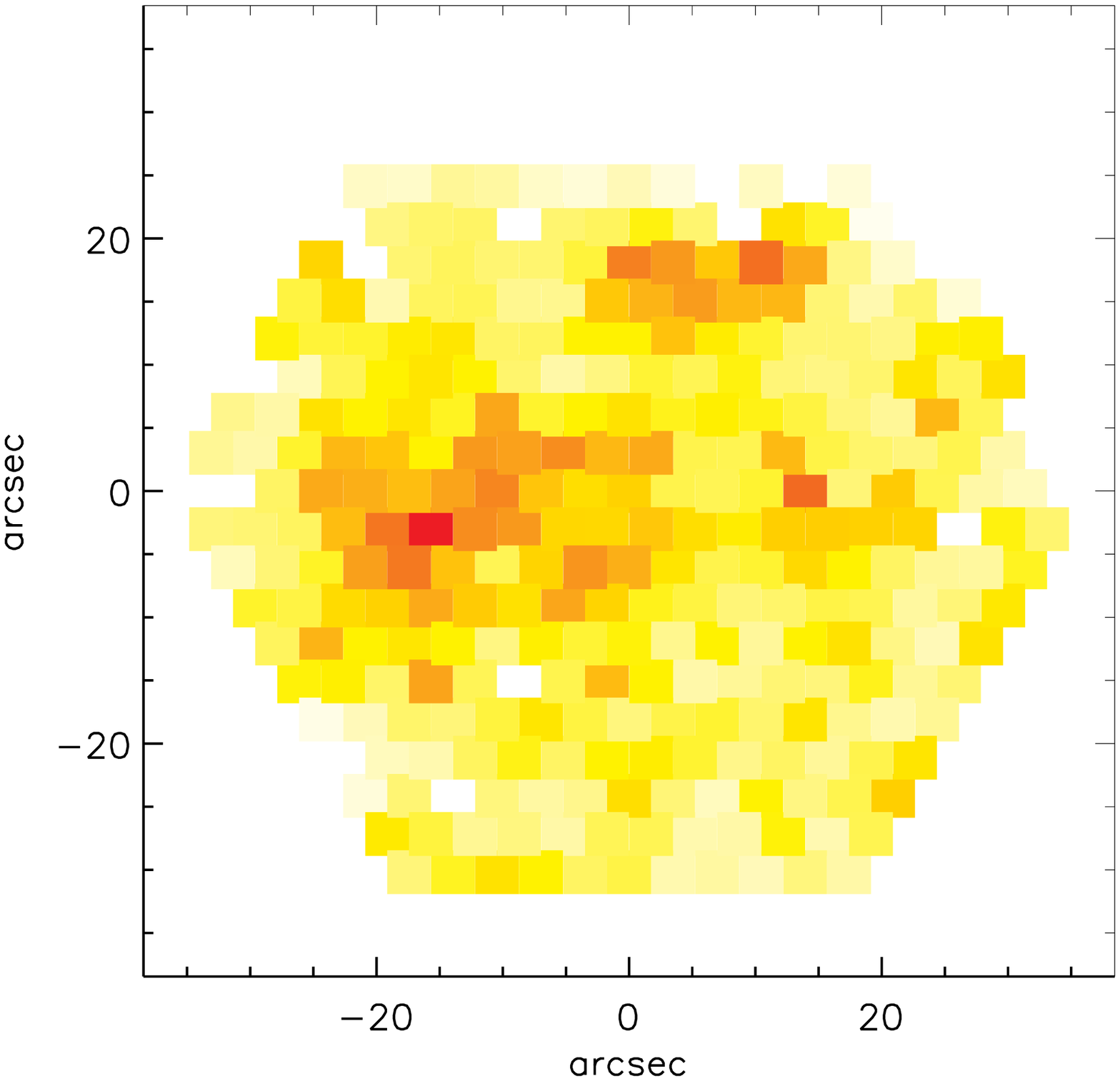}
   \hspace*{-0.25cm}   \includegraphics[height=4.1cm]{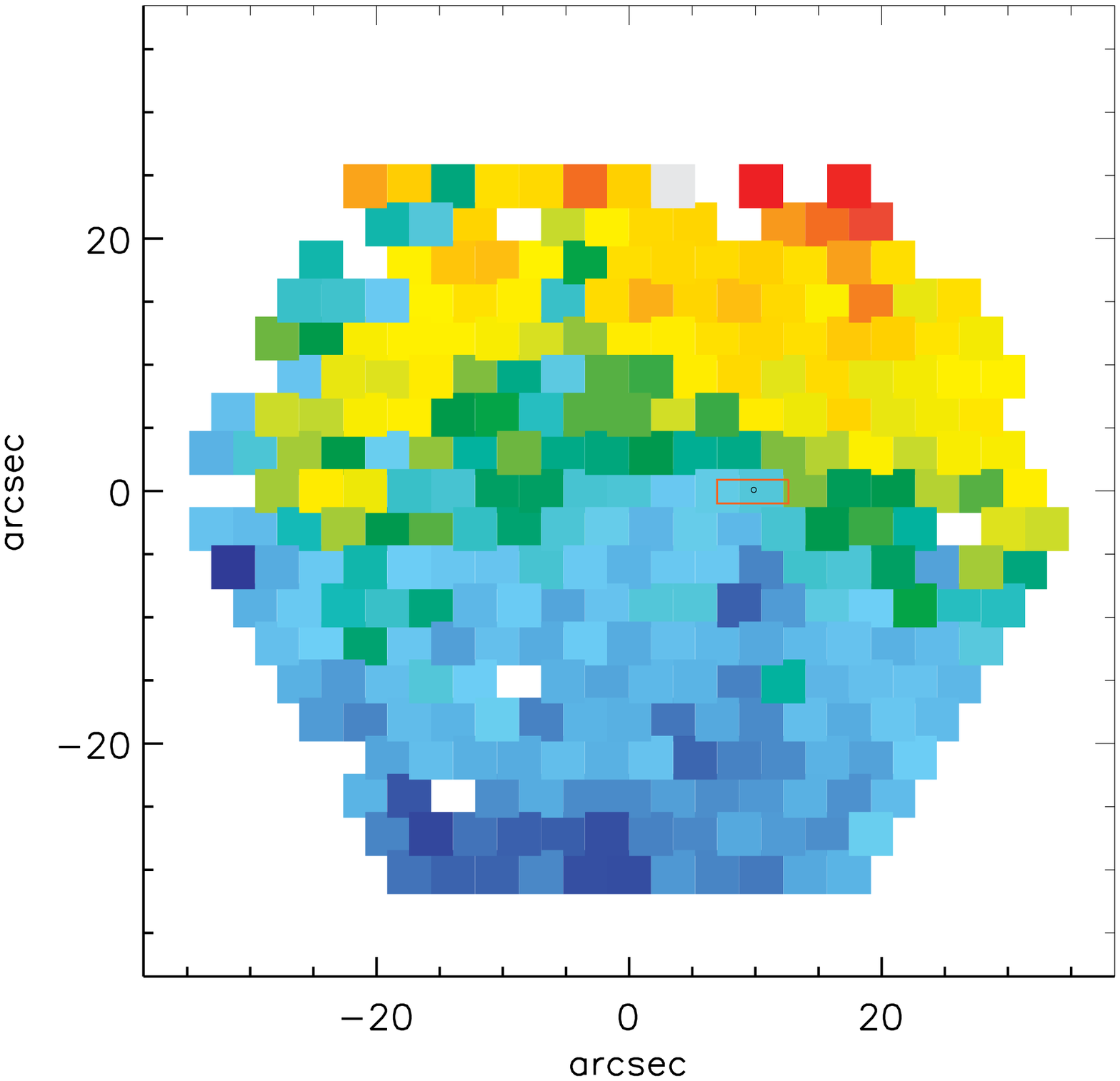}
   \hspace*{-0.25cm}   \includegraphics[height=4.1cm]{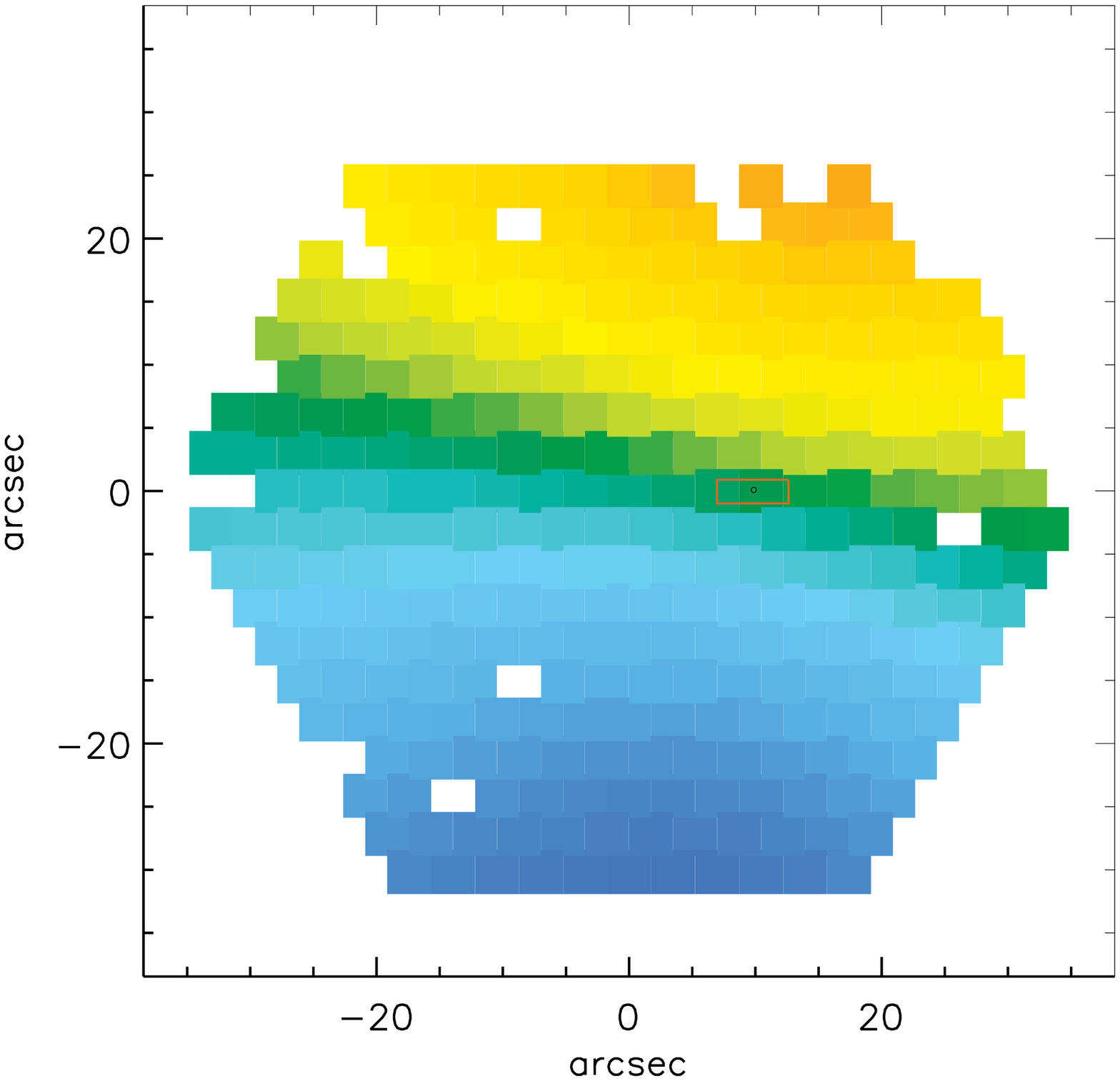} 
  \hspace*{-0.25cm} \includegraphics[height=3.82cm]{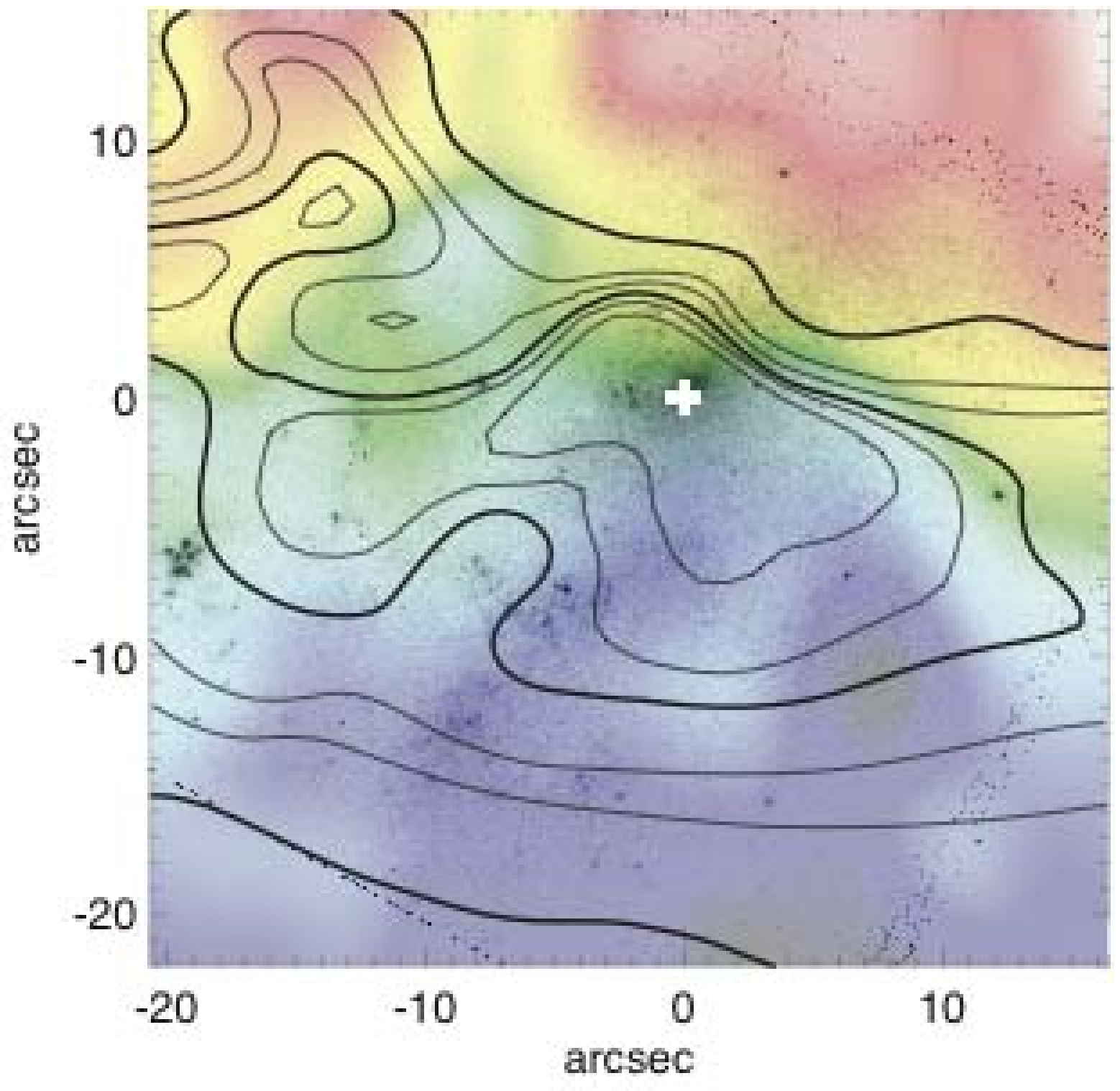}
   \vspace{-0.6cm}
  \end{tabular}
  \end{center}

 \begin{center}
  \begin{tabular}{cc}
  \hspace*{-1.3cm} \includegraphics[height=4.5cm]{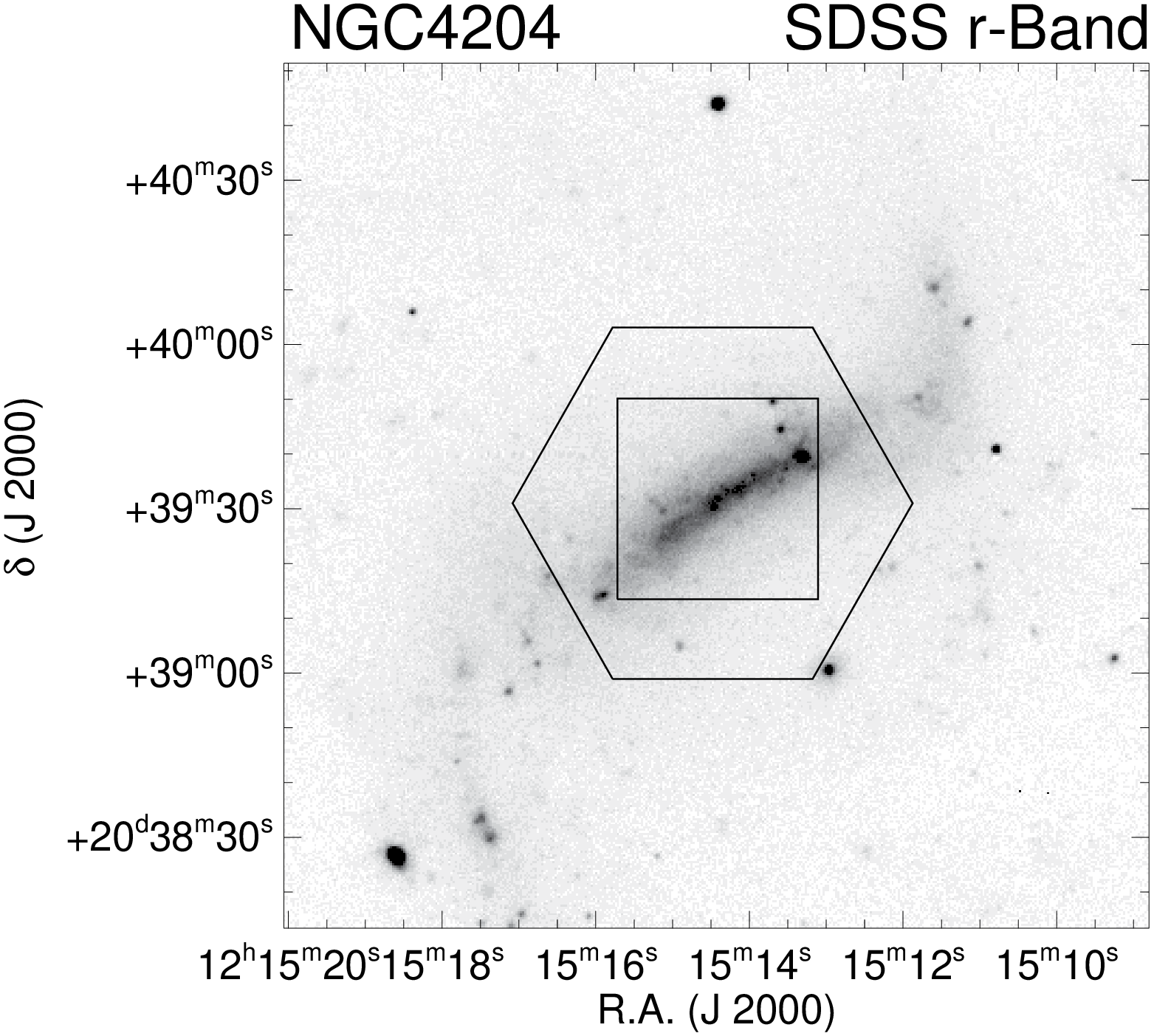}
   \hspace*{-0.25cm} \includegraphics[height=4.1cm]{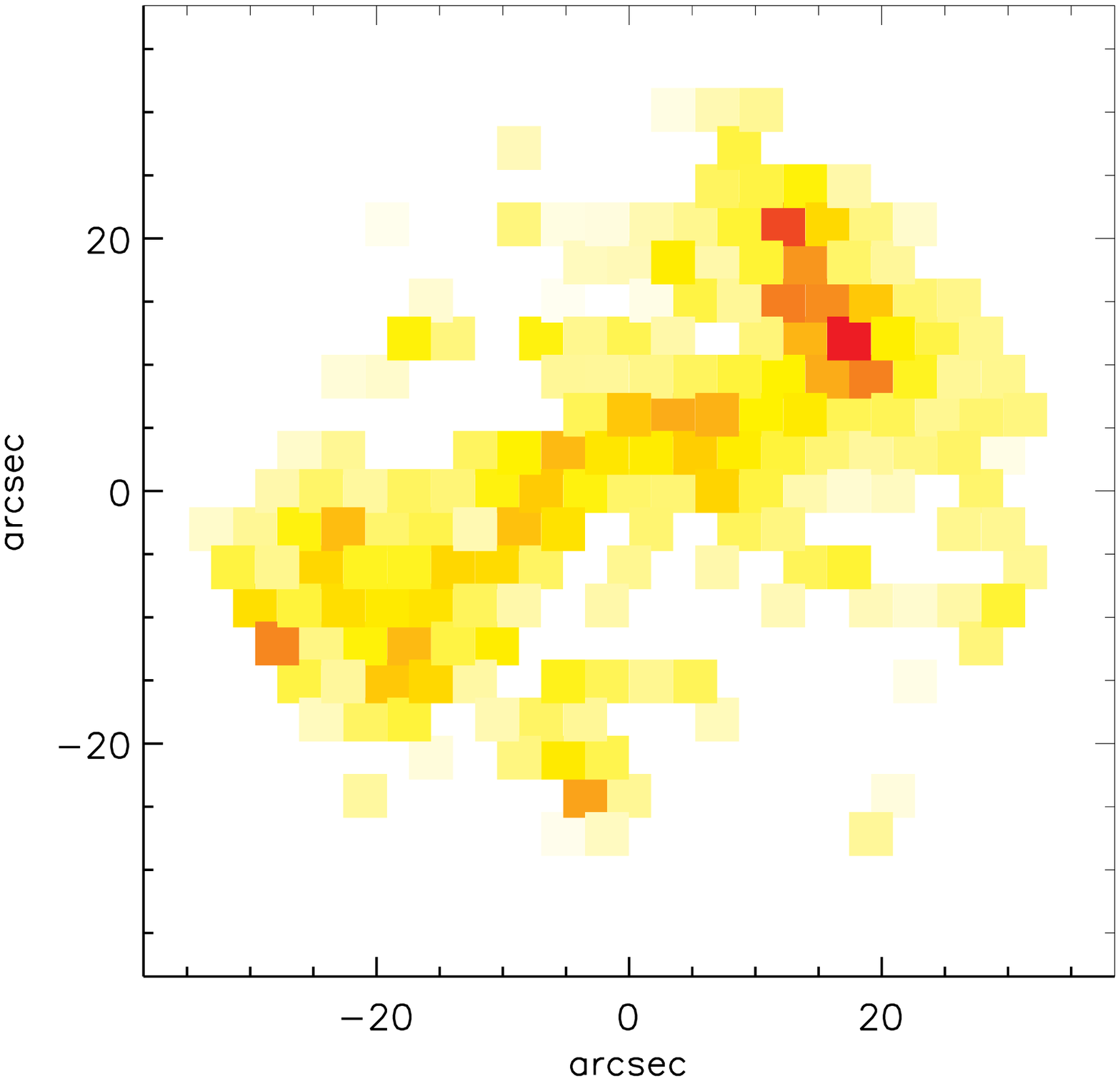}
   \hspace*{-0.25cm}   \includegraphics[height=4.1cm]{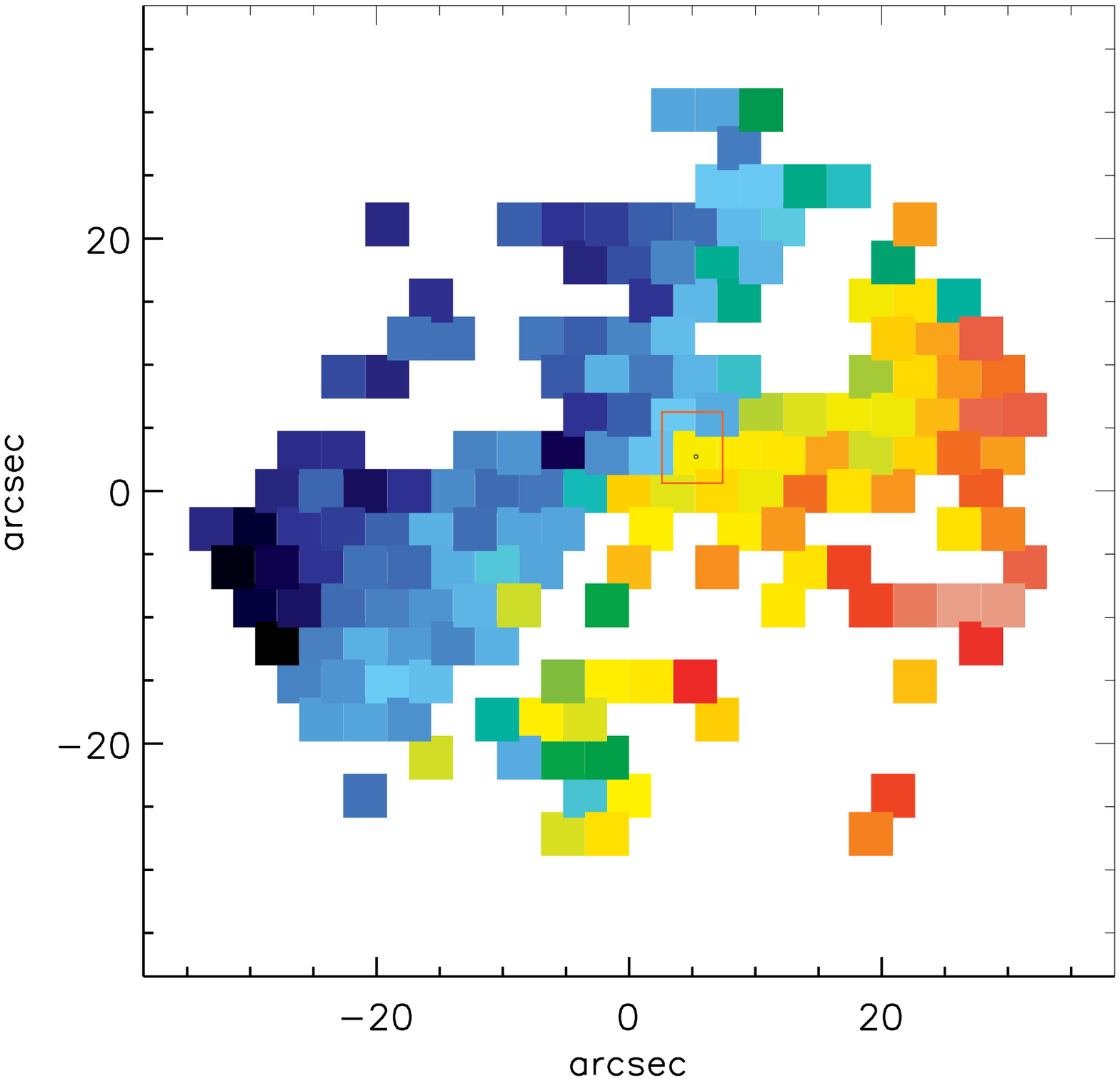}
   \hspace*{-0.25cm}   \includegraphics[height=4.1cm]{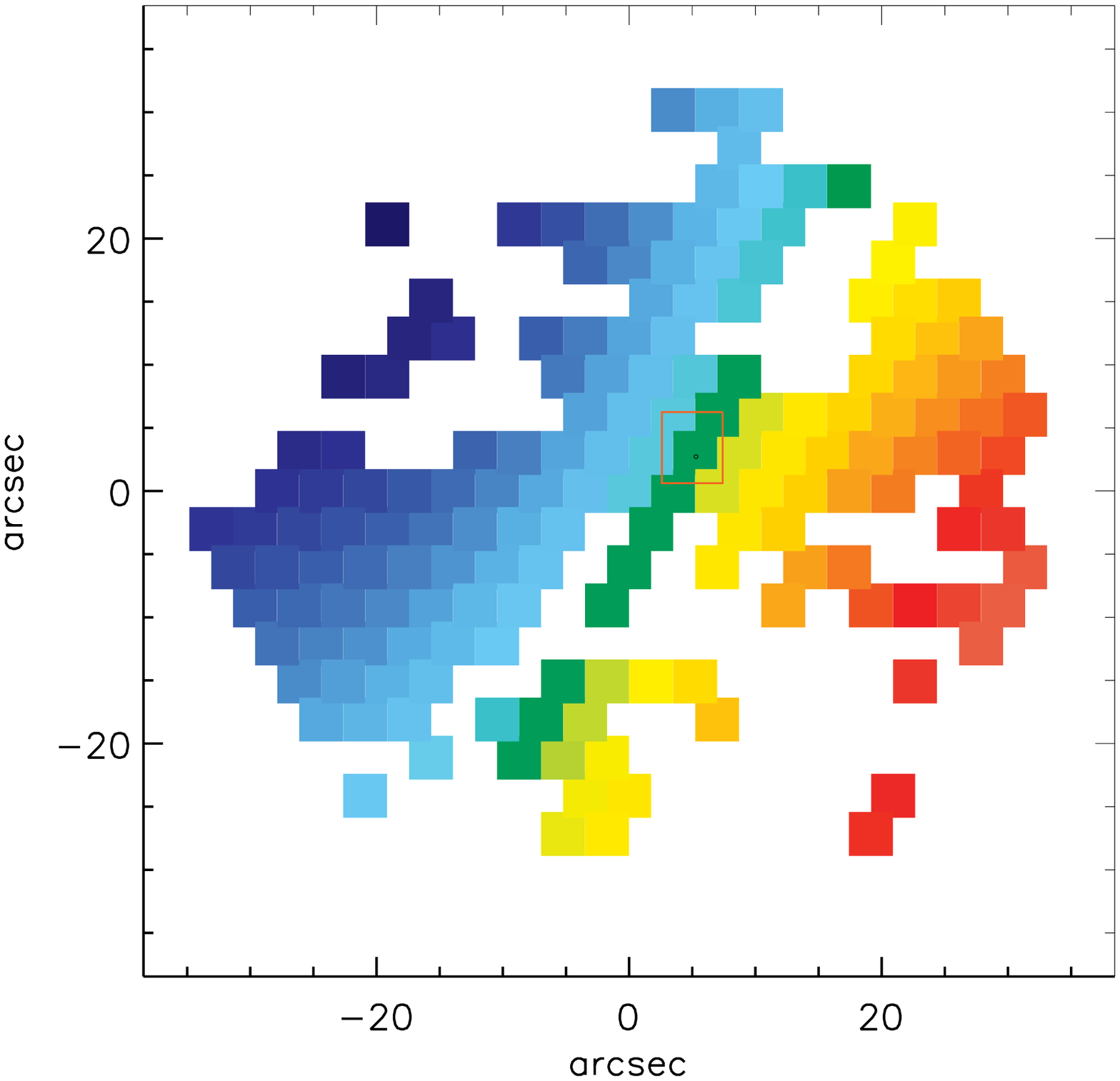} 
  \hspace*{-0.25cm}   \includegraphics[height=3.82cm]{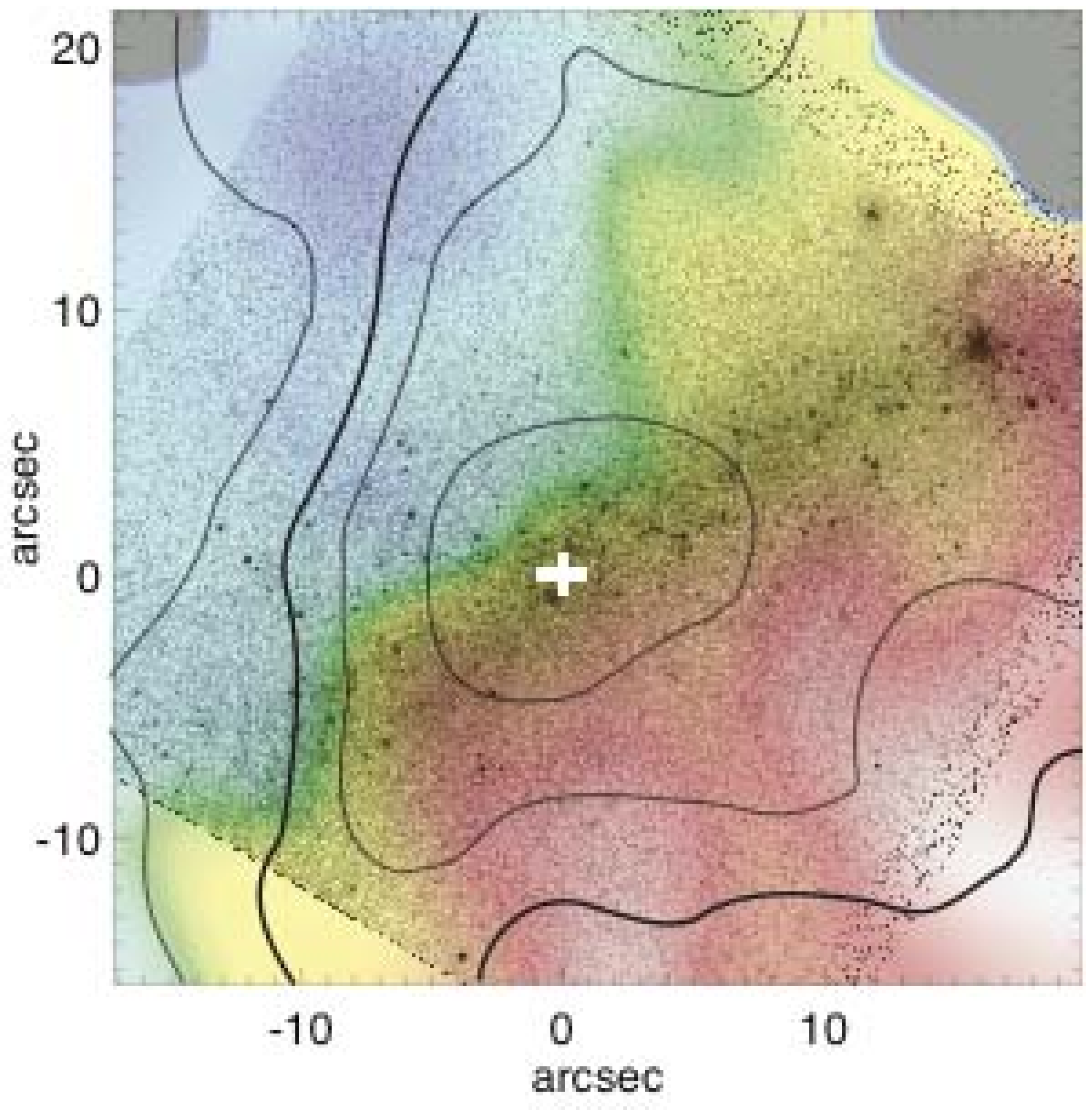}
   \vspace{-0.6cm}
  \end{tabular}
  \end{center}

 \begin{center}
  \begin{tabular}{cc}
  \hspace*{-1.3cm} \includegraphics[height=4.5cm]{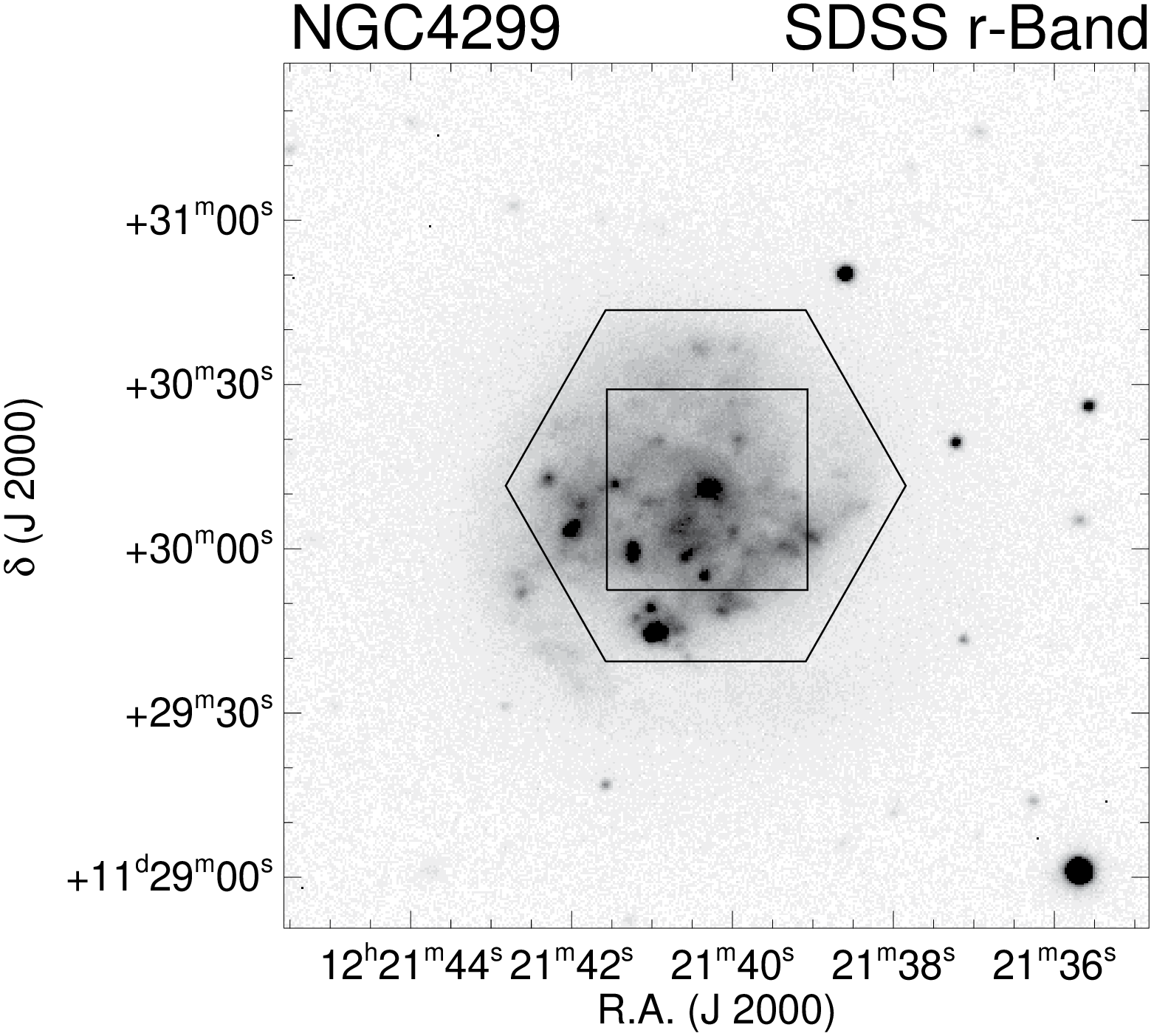}
   \hspace*{-0.25cm} \includegraphics[height=4.1cm]{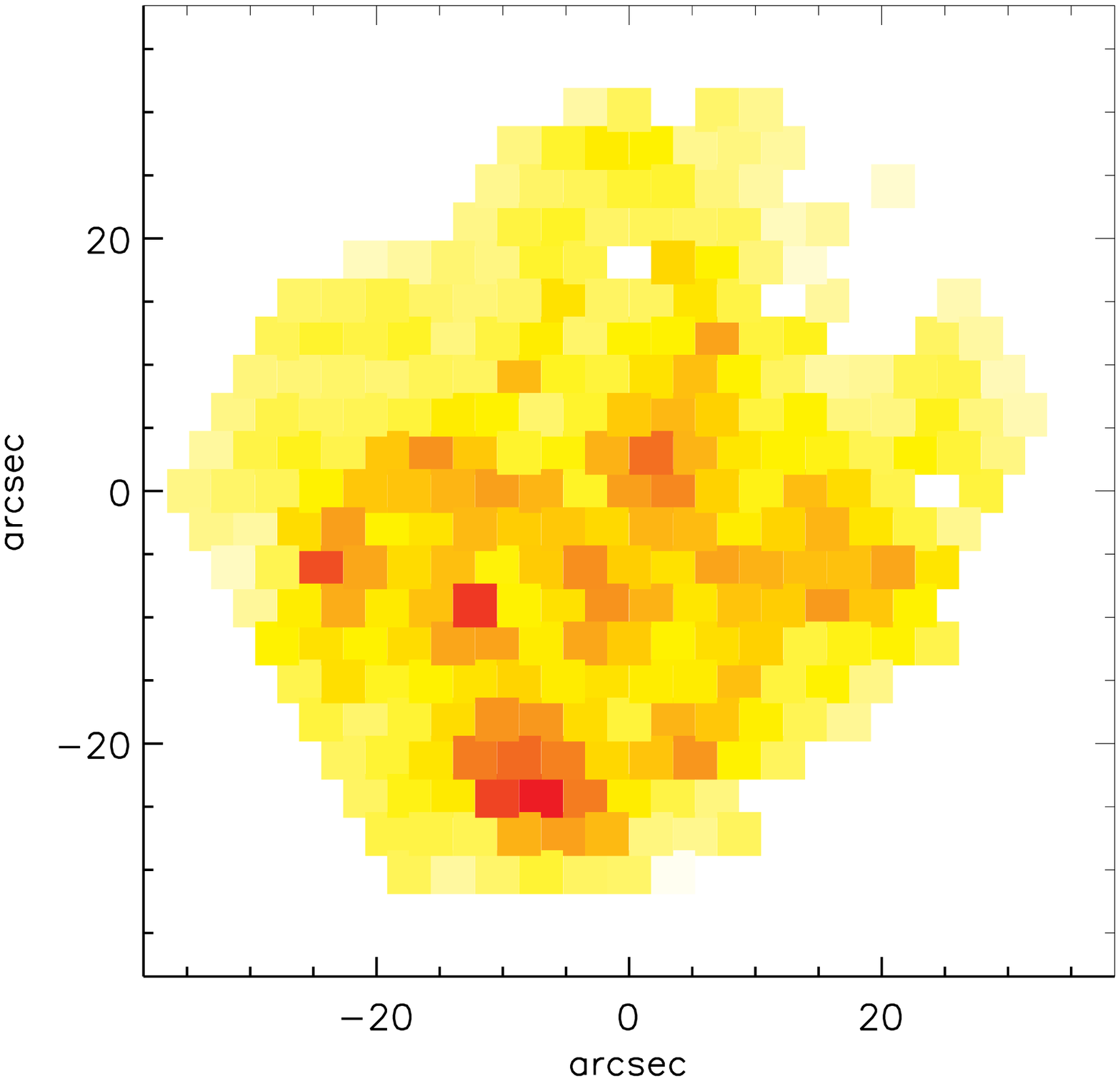}
   \hspace*{-0.25cm}   \includegraphics[height=4.1cm]{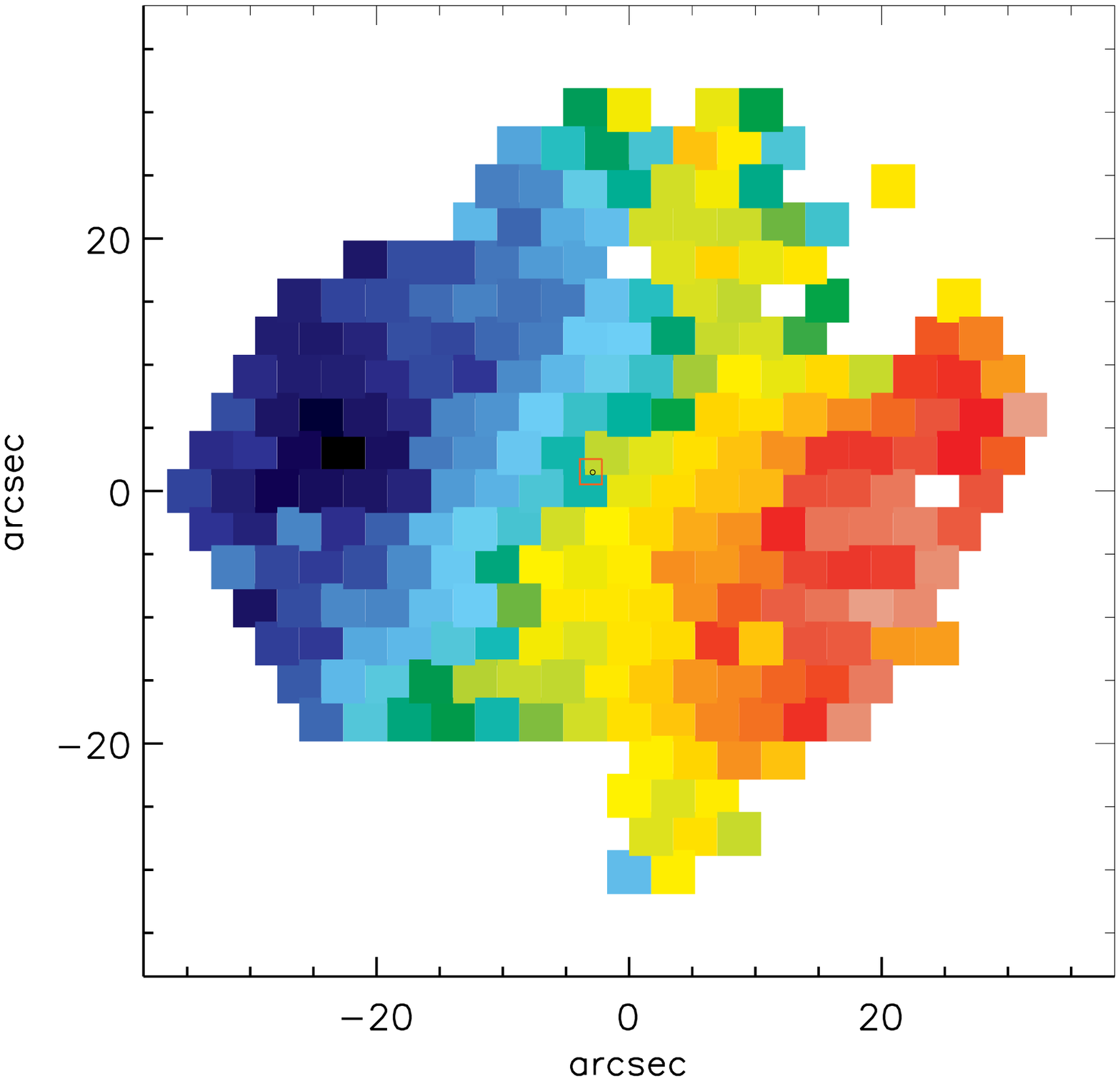}
   \hspace*{-0.25cm}   \includegraphics[height=4.1cm]{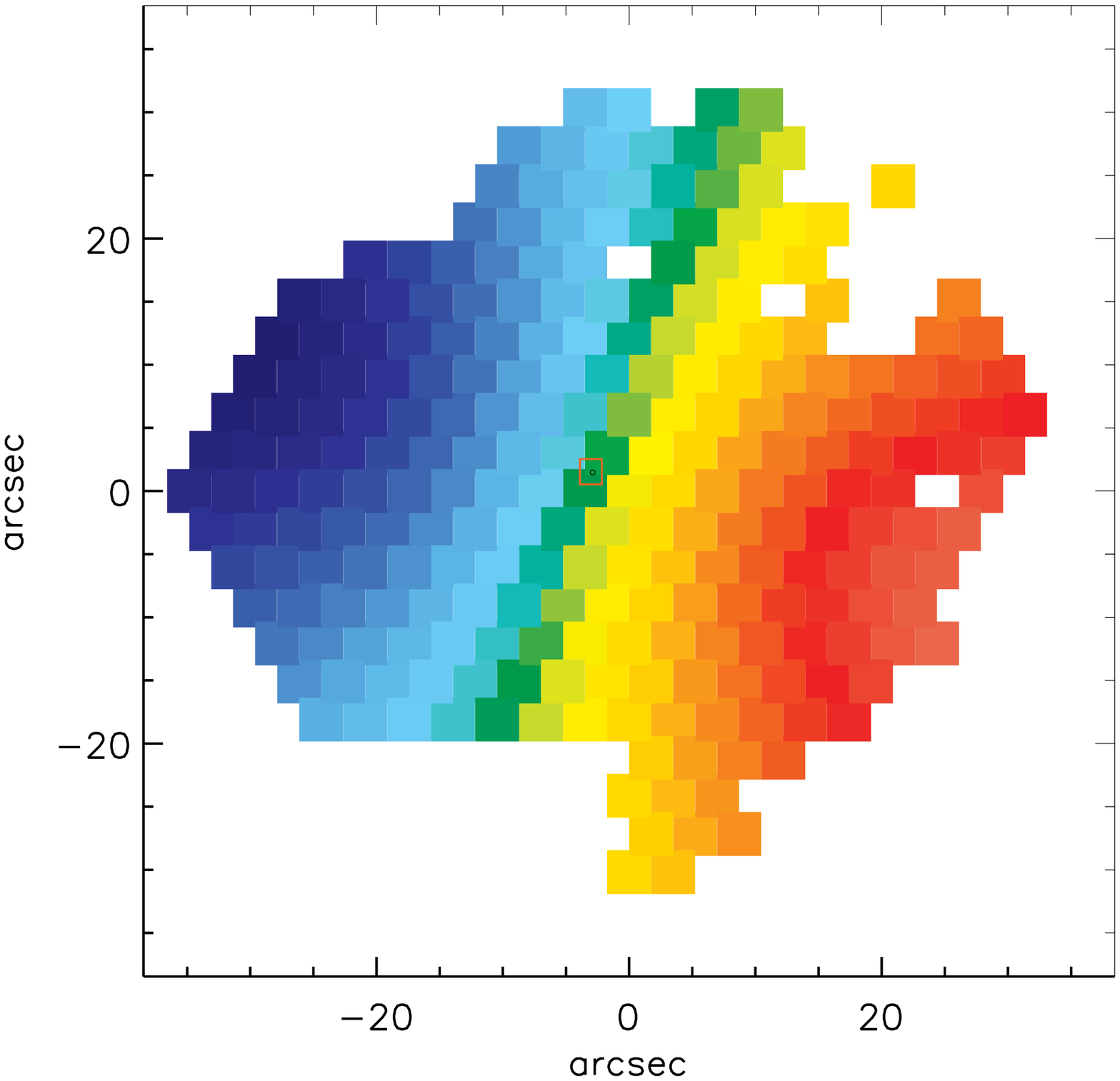} 
  \hspace*{-0.25cm}   \includegraphics[height=3.82cm]{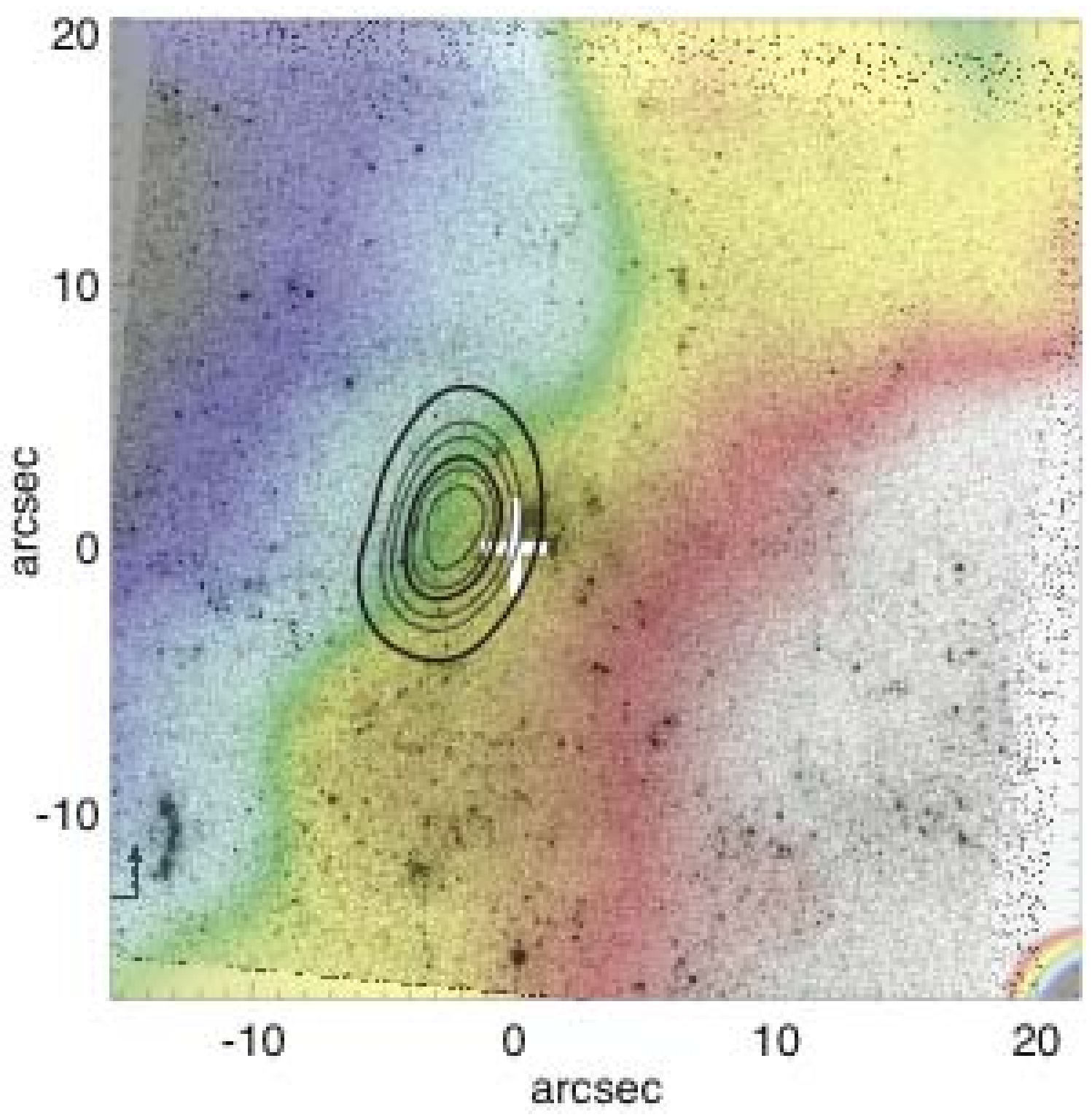}
  \end{tabular}
  \end{center}

  \contcaption{
}
  \end{figure*}
  
  \begin{figure*}
  \begin{center}
  \begin{tabular}{c}
  \hspace*{-1.3cm} \includegraphics[height=4.5cm]{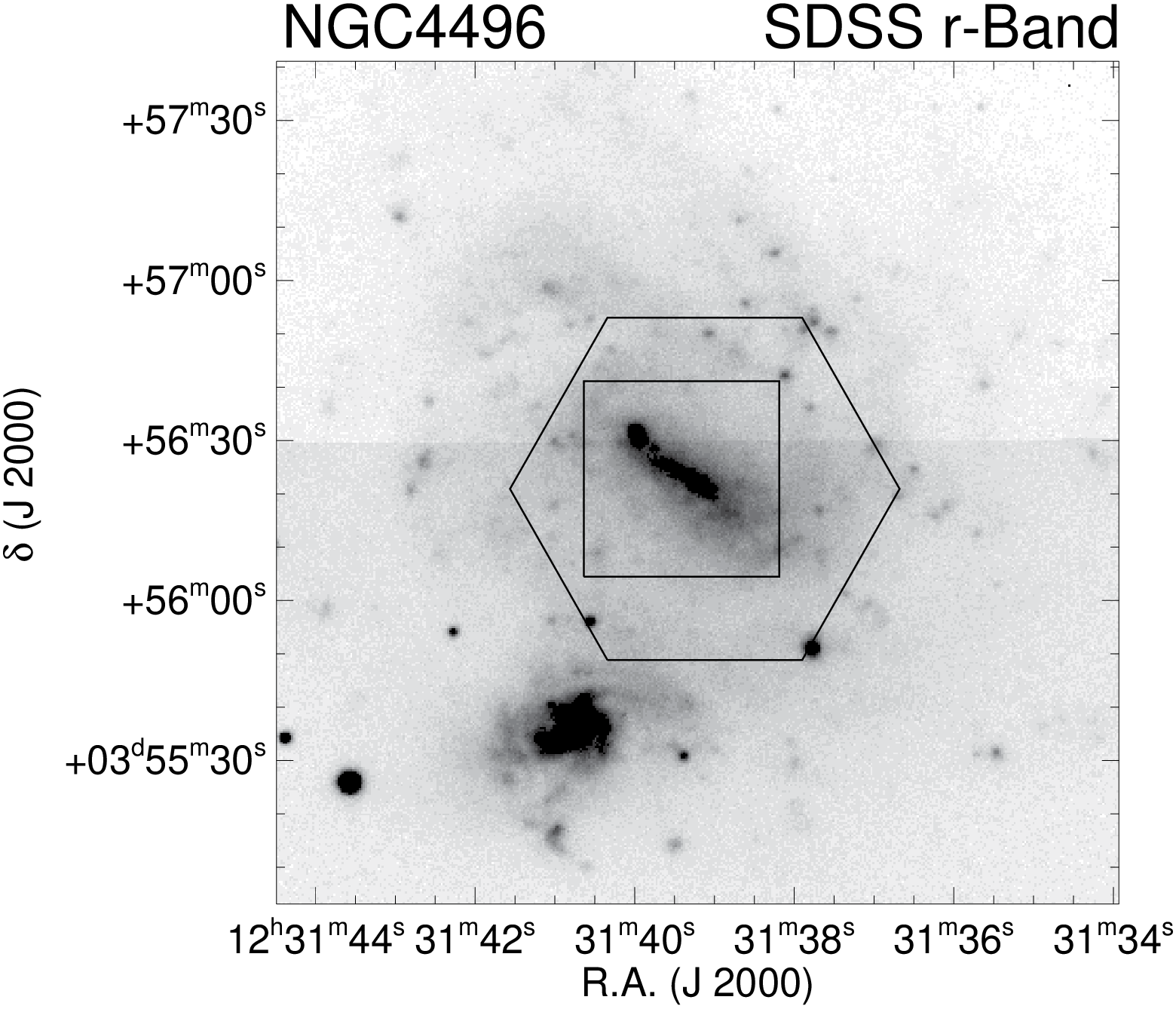}
   \hspace*{-0.25cm}  \includegraphics[width=4.1cm]{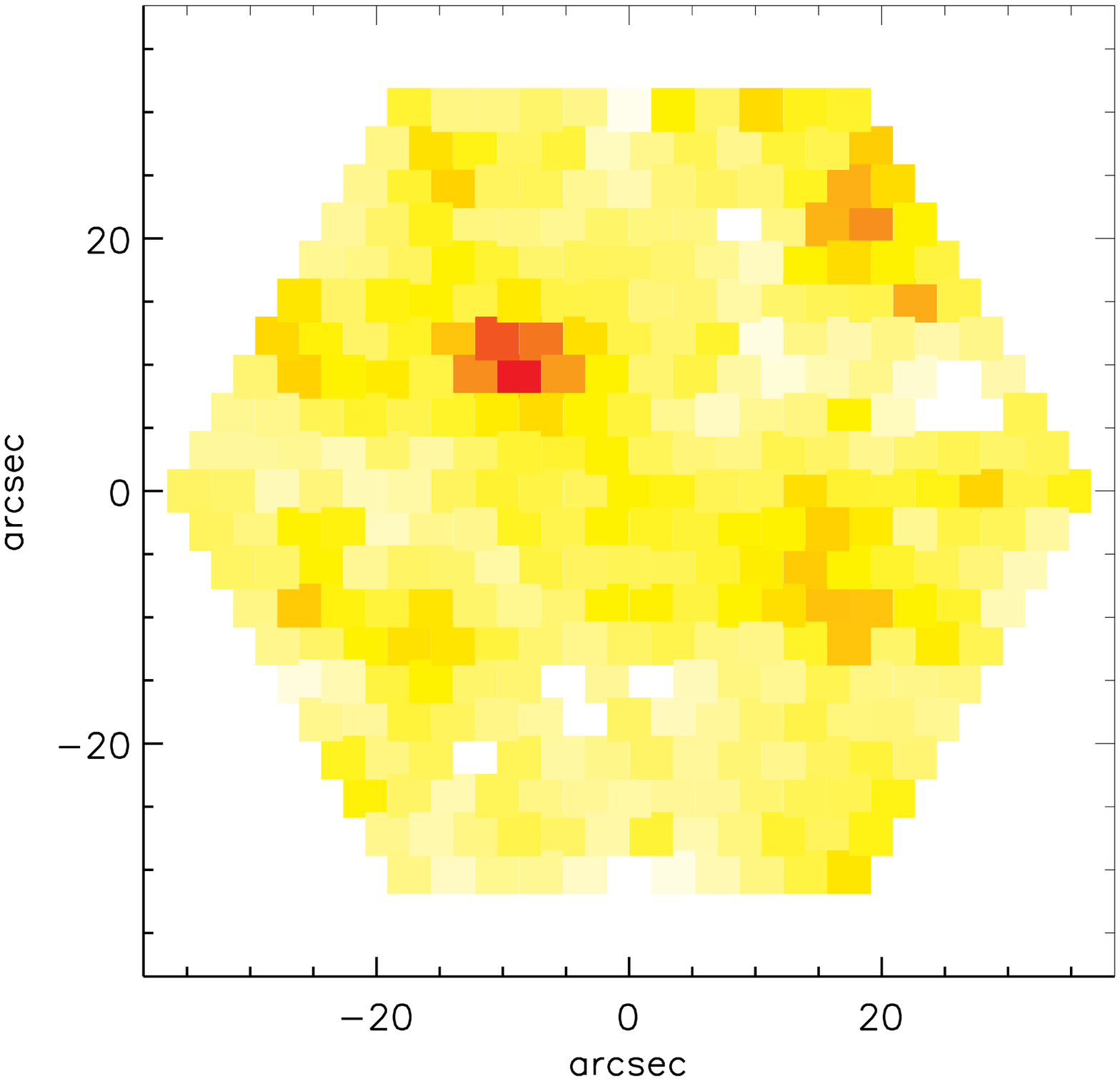}
    \hspace*{-0.25cm}  \includegraphics[height=4.1cm]{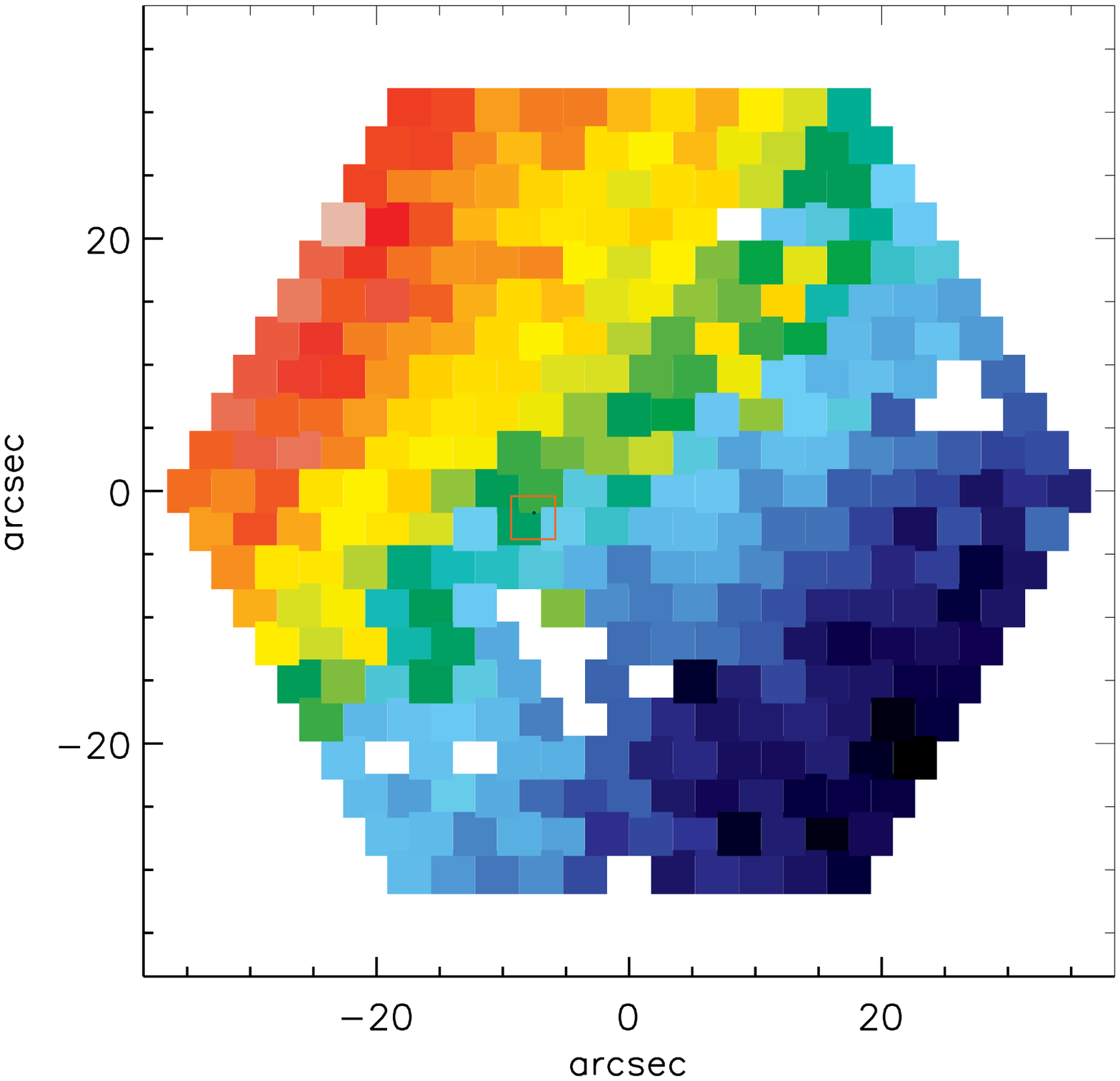}
    \hspace*{-0.25cm}  \includegraphics[height=4.1cm]{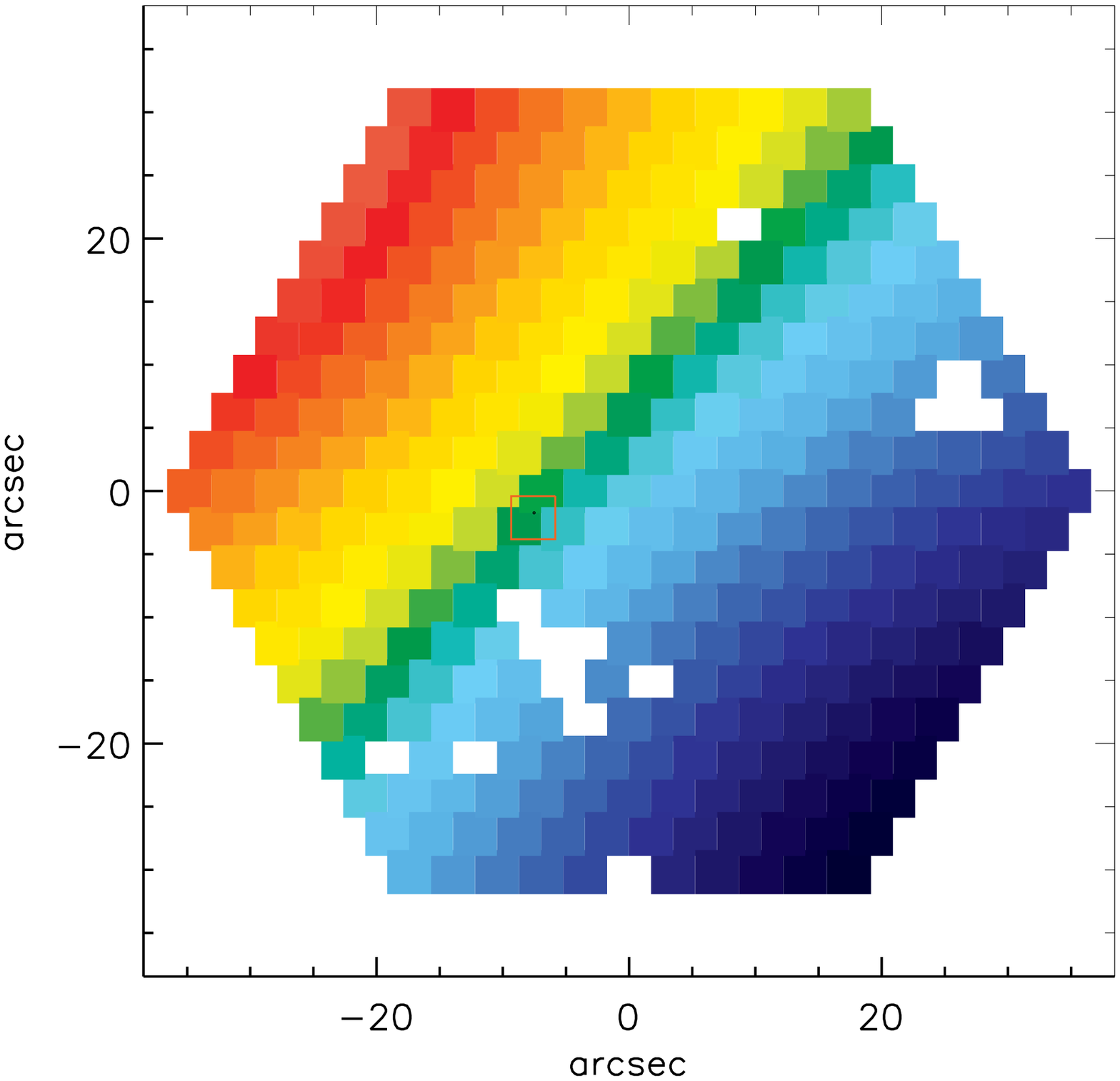}
  \hspace*{-0.25cm} \includegraphics[height=3.82cm]{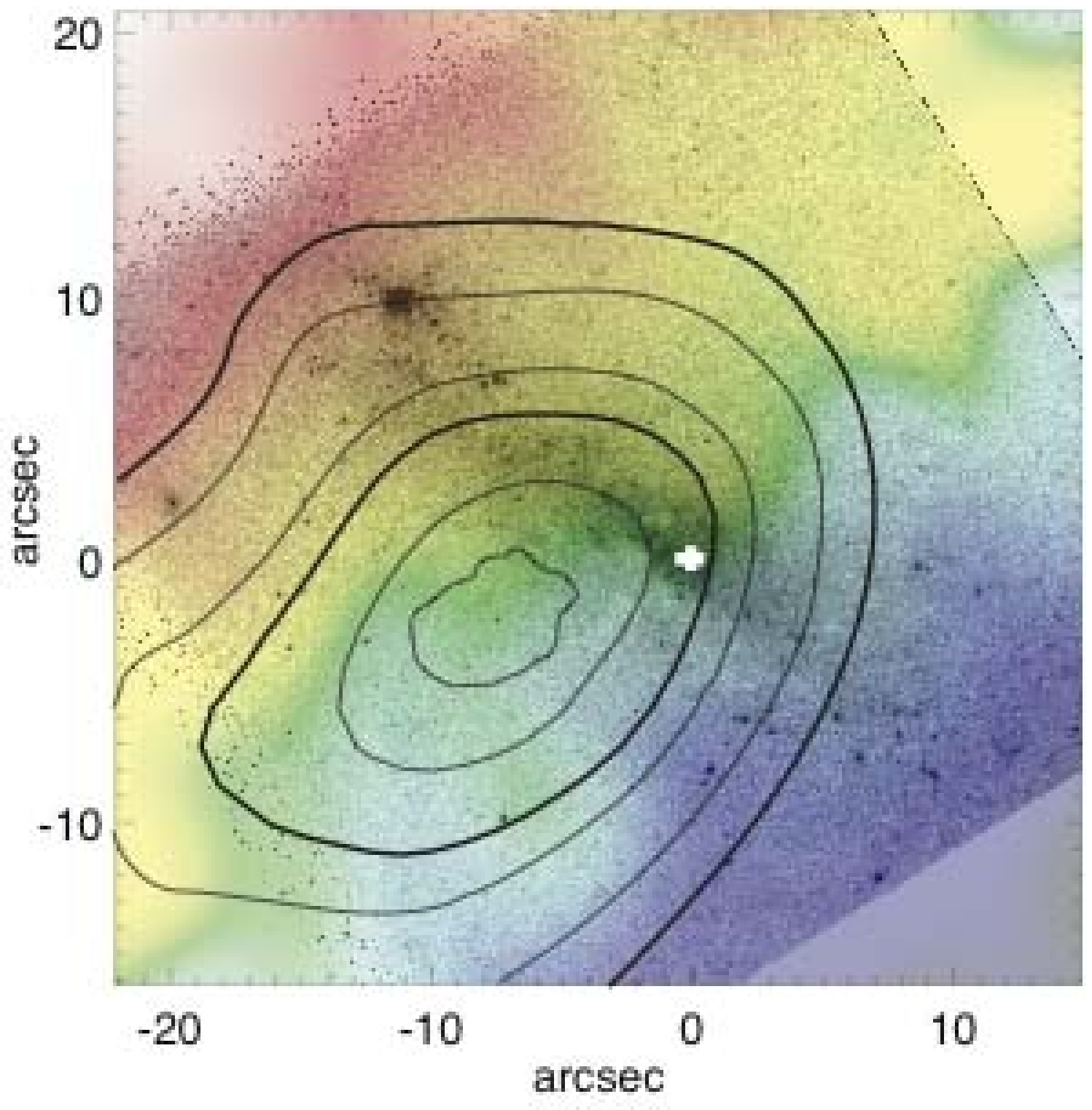}
    \vspace{-0.6cm}
  \end{tabular}
  \end{center}

  \begin{center}
  \begin{tabular}{cc}
  \hspace*{-1.3cm} \includegraphics[height=4.5cm]{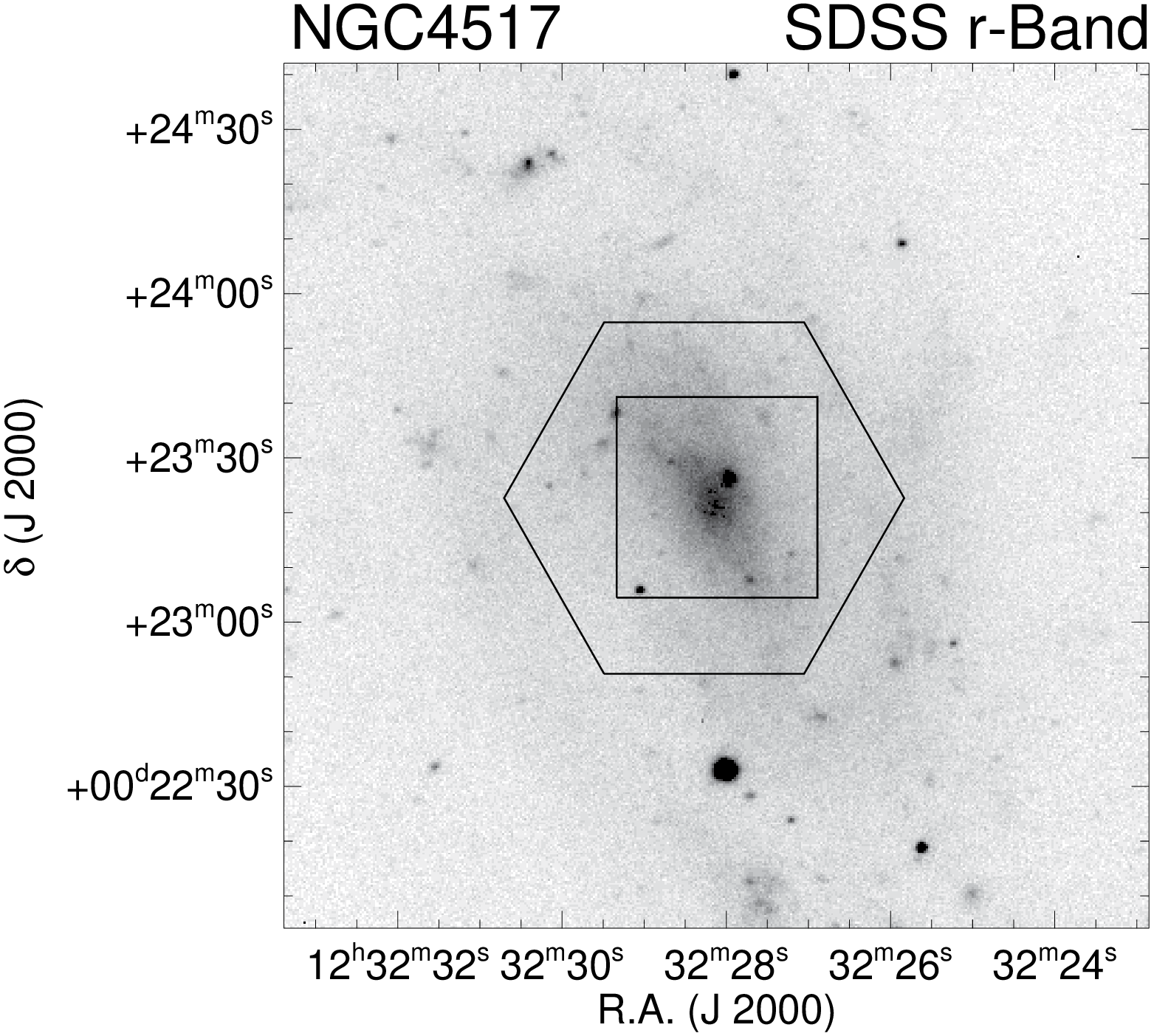}
   \hspace*{-0.25cm} \includegraphics[height=4.1cm]{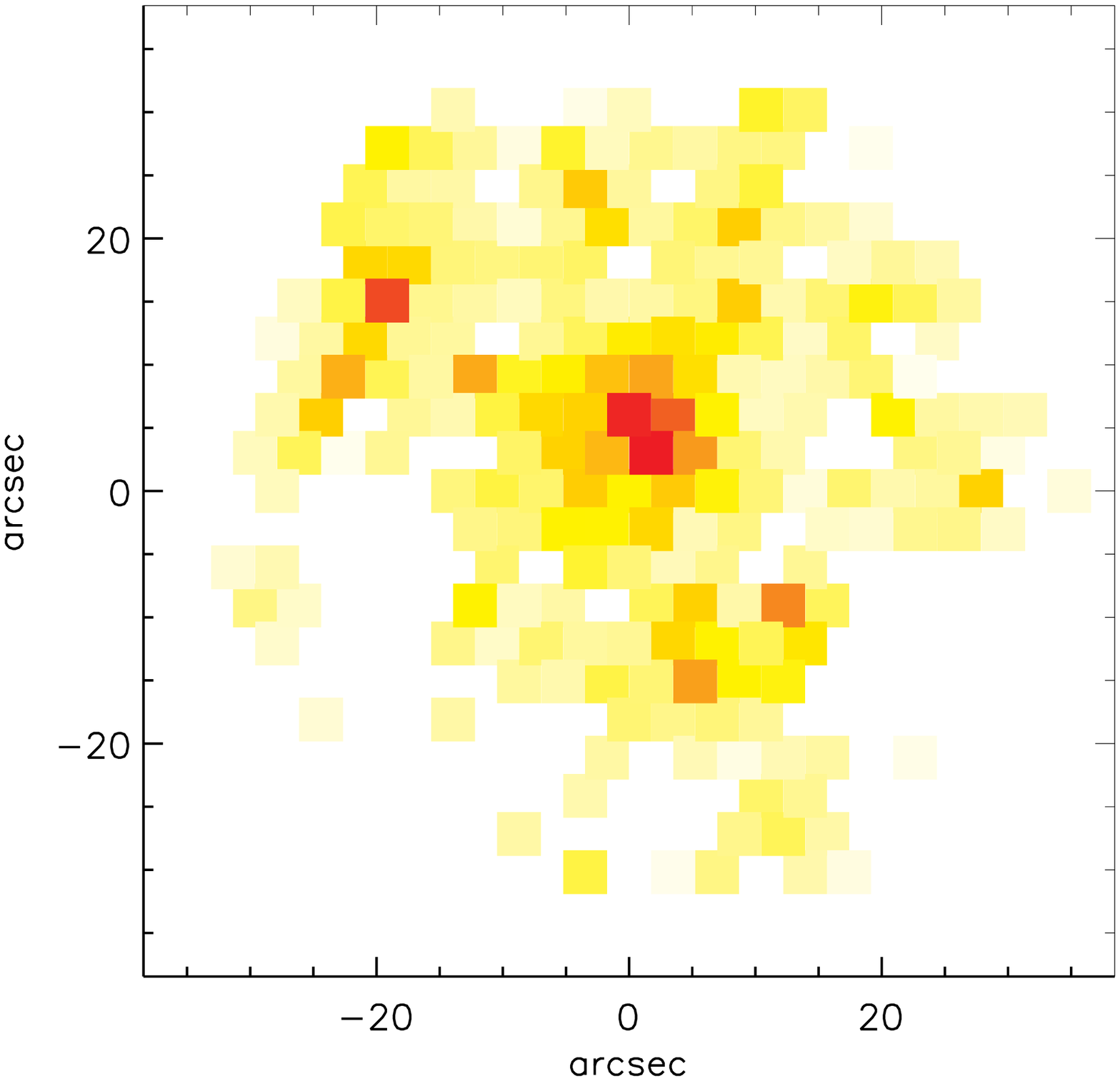}
   \hspace*{-0.25cm}   \includegraphics[height=4.1cm]{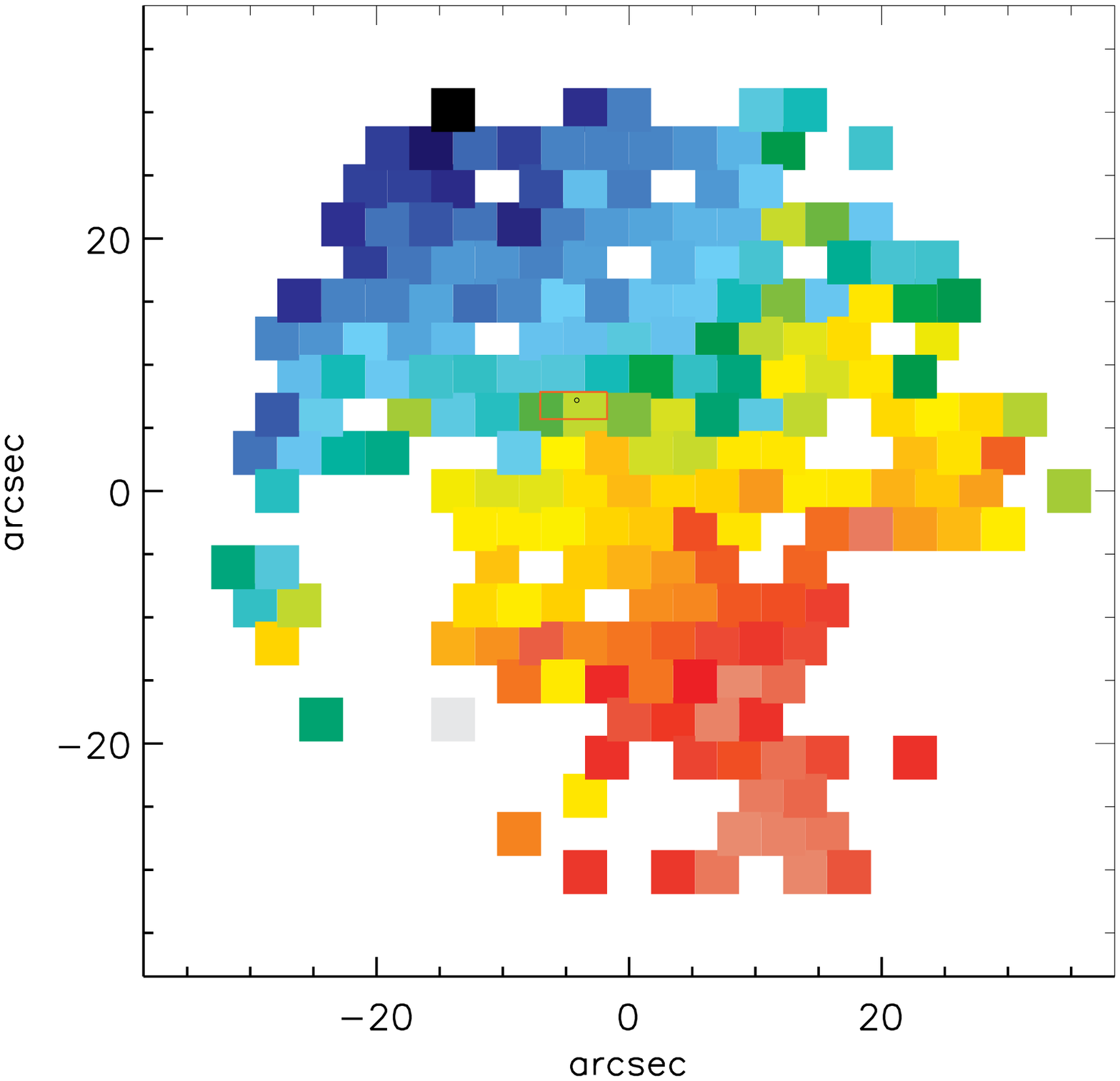}
   \hspace*{-0.25cm}   \includegraphics[height=4.1cm]{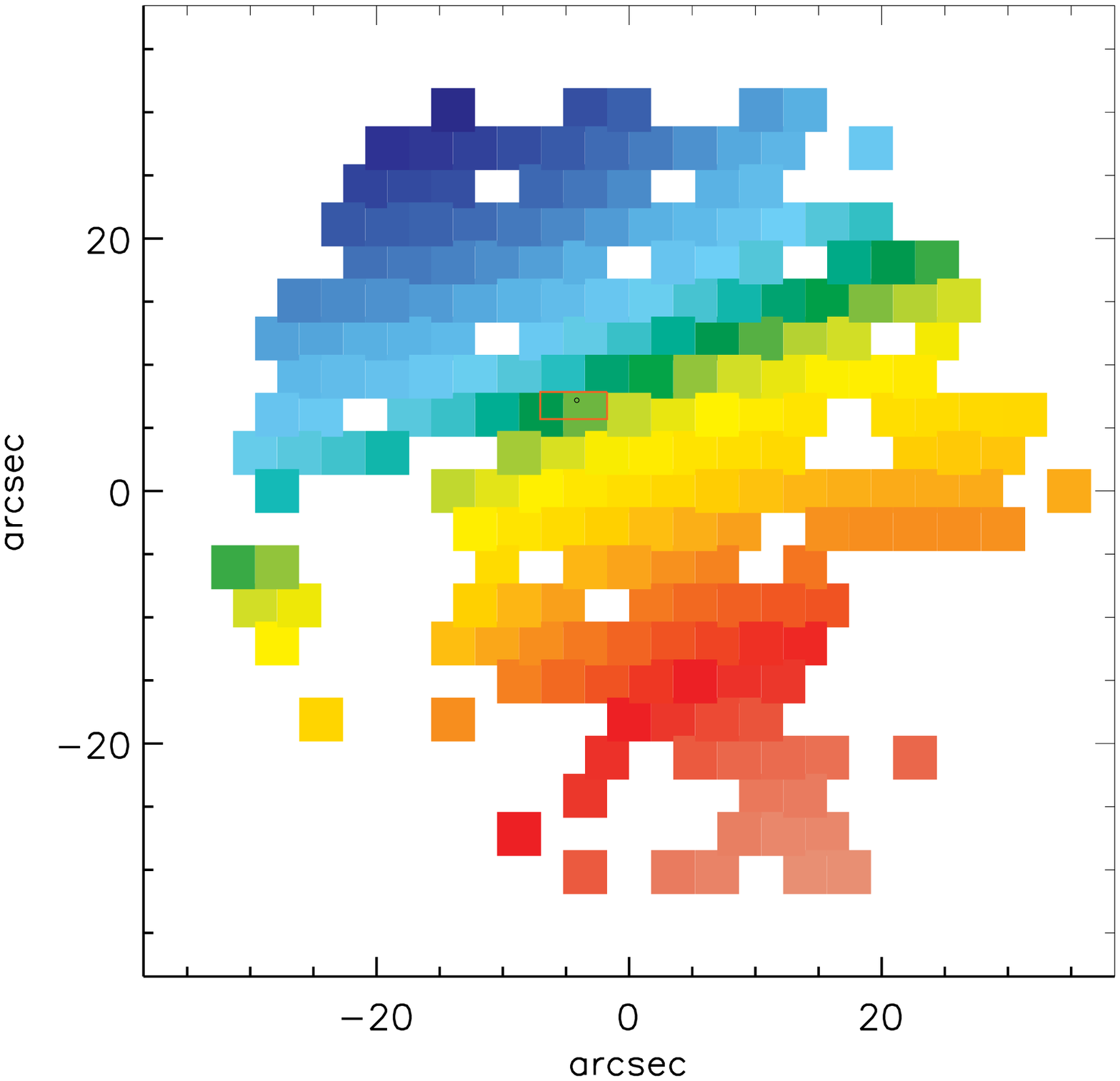} 
 \hspace*{-0.25cm}  \includegraphics[height=3.82cm]{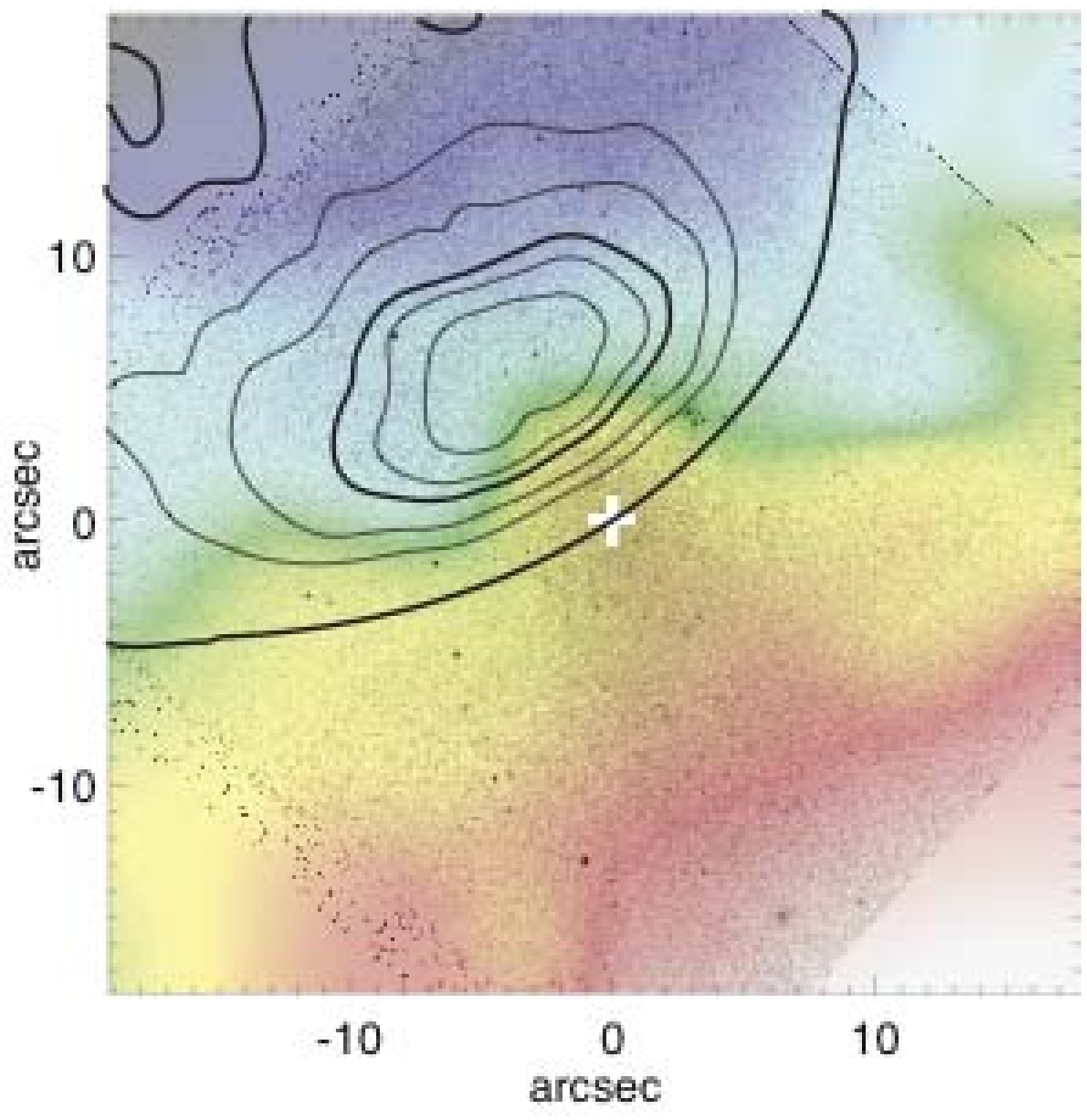}
   \vspace{-0.6cm}
  \end{tabular}
  \end{center}

  \begin{center}
  \begin{tabular}{cc}
  \hspace*{-1.3cm} \includegraphics[height=4.5cm]{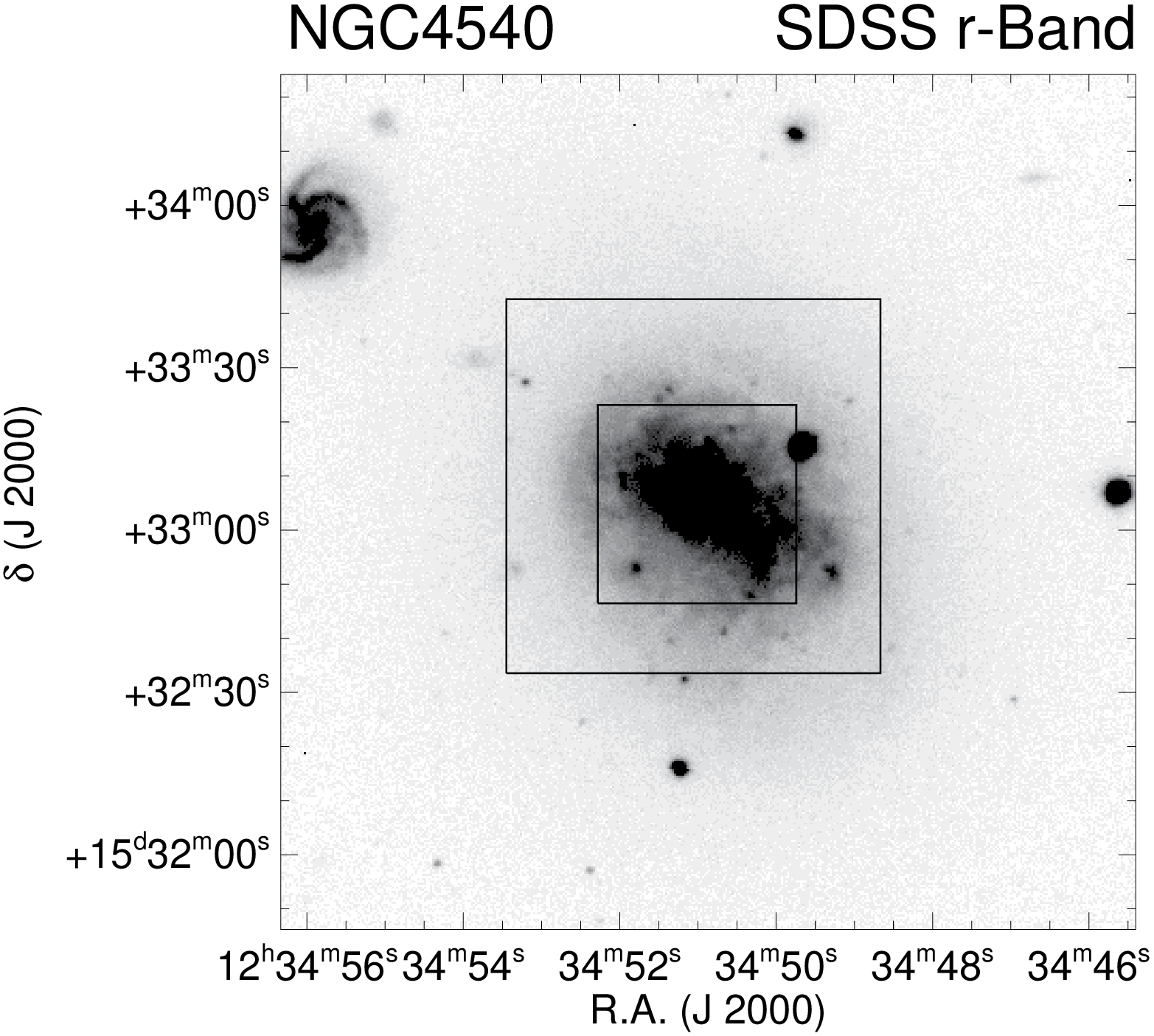}
   \hspace*{-0.25cm} \includegraphics[height=4.1cm]{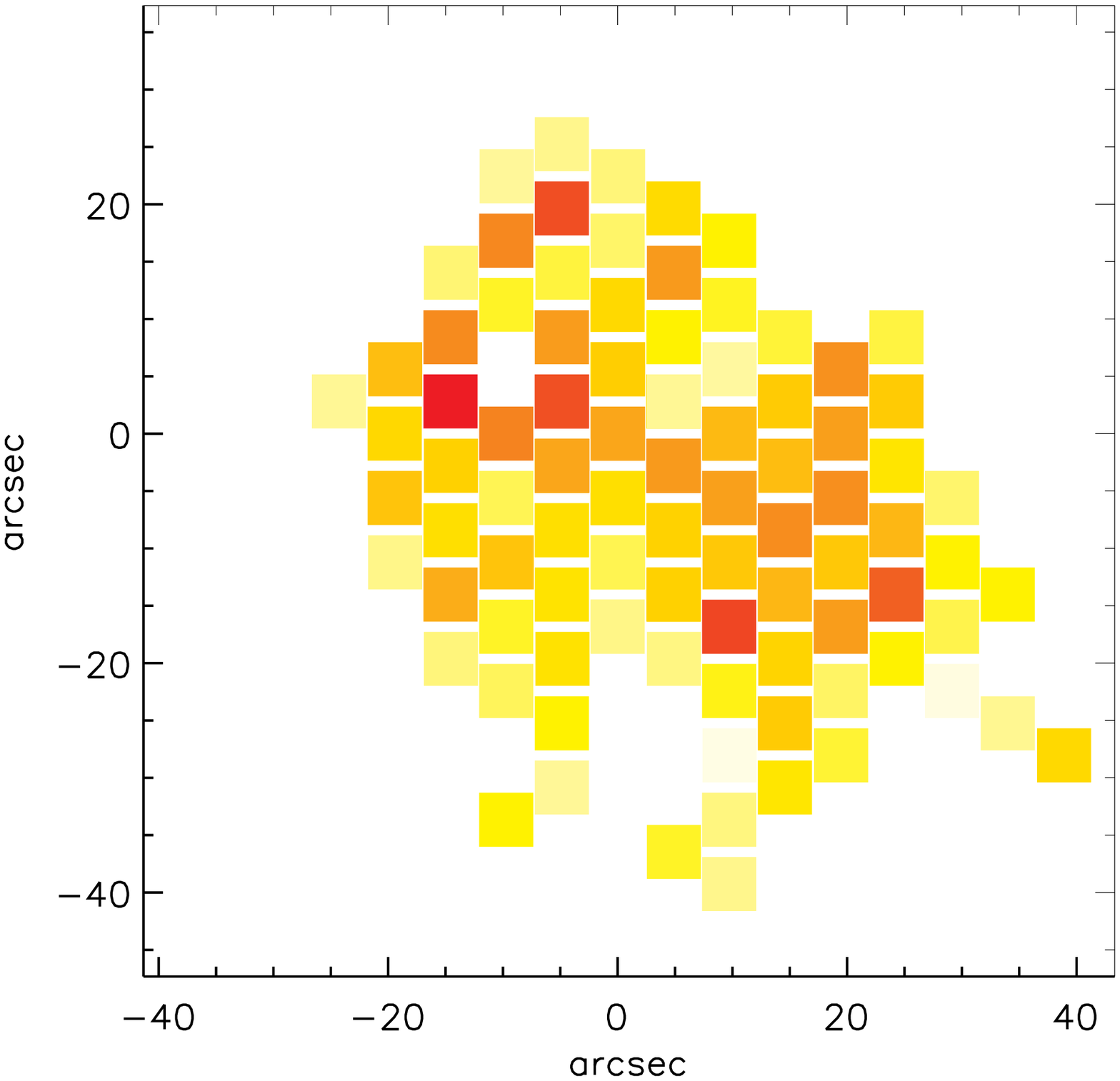}
   \hspace*{-0.25cm}   \includegraphics[height=4.1cm]{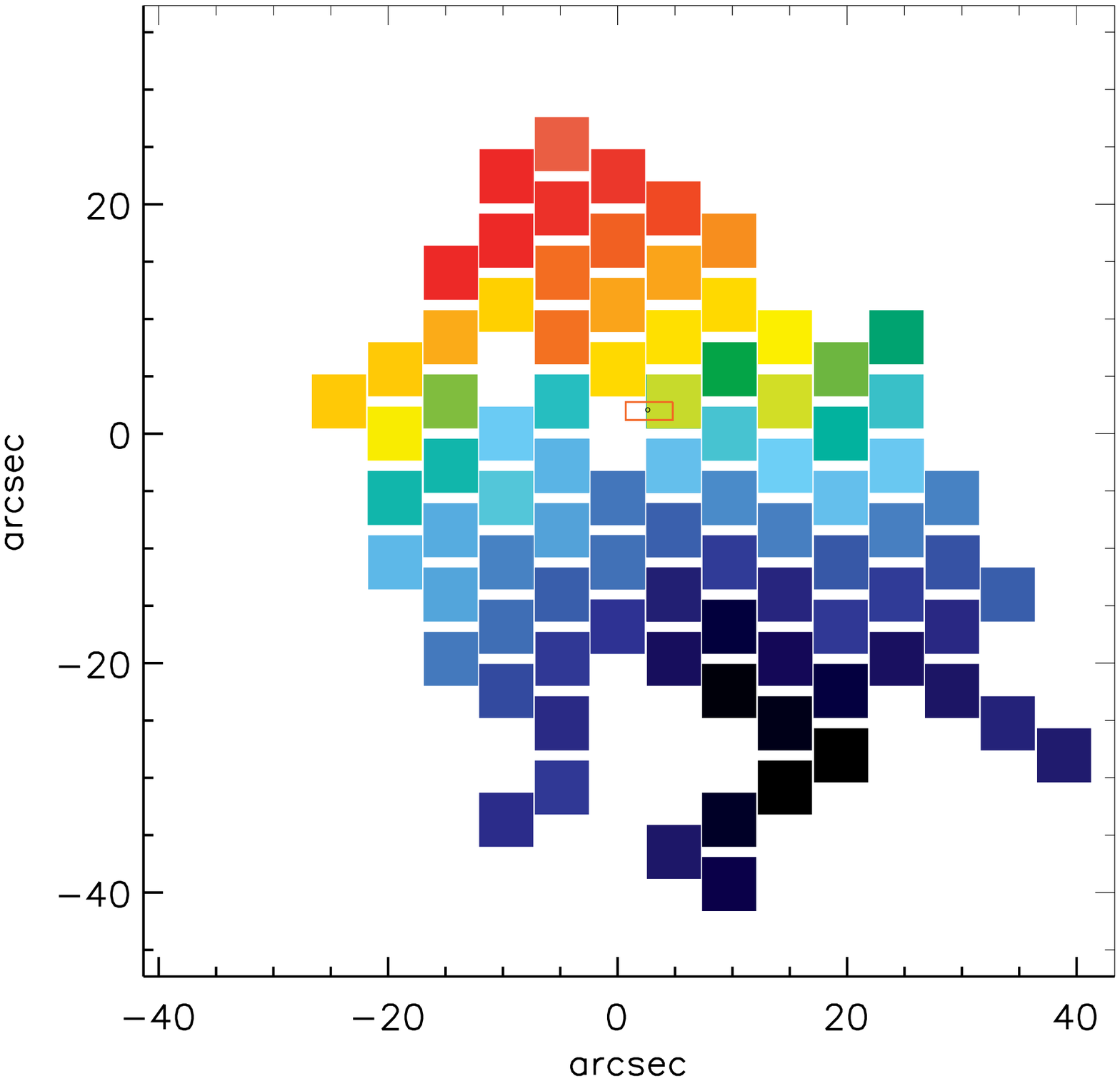}
   \hspace*{-0.25cm}   \includegraphics[height=4.1cm]{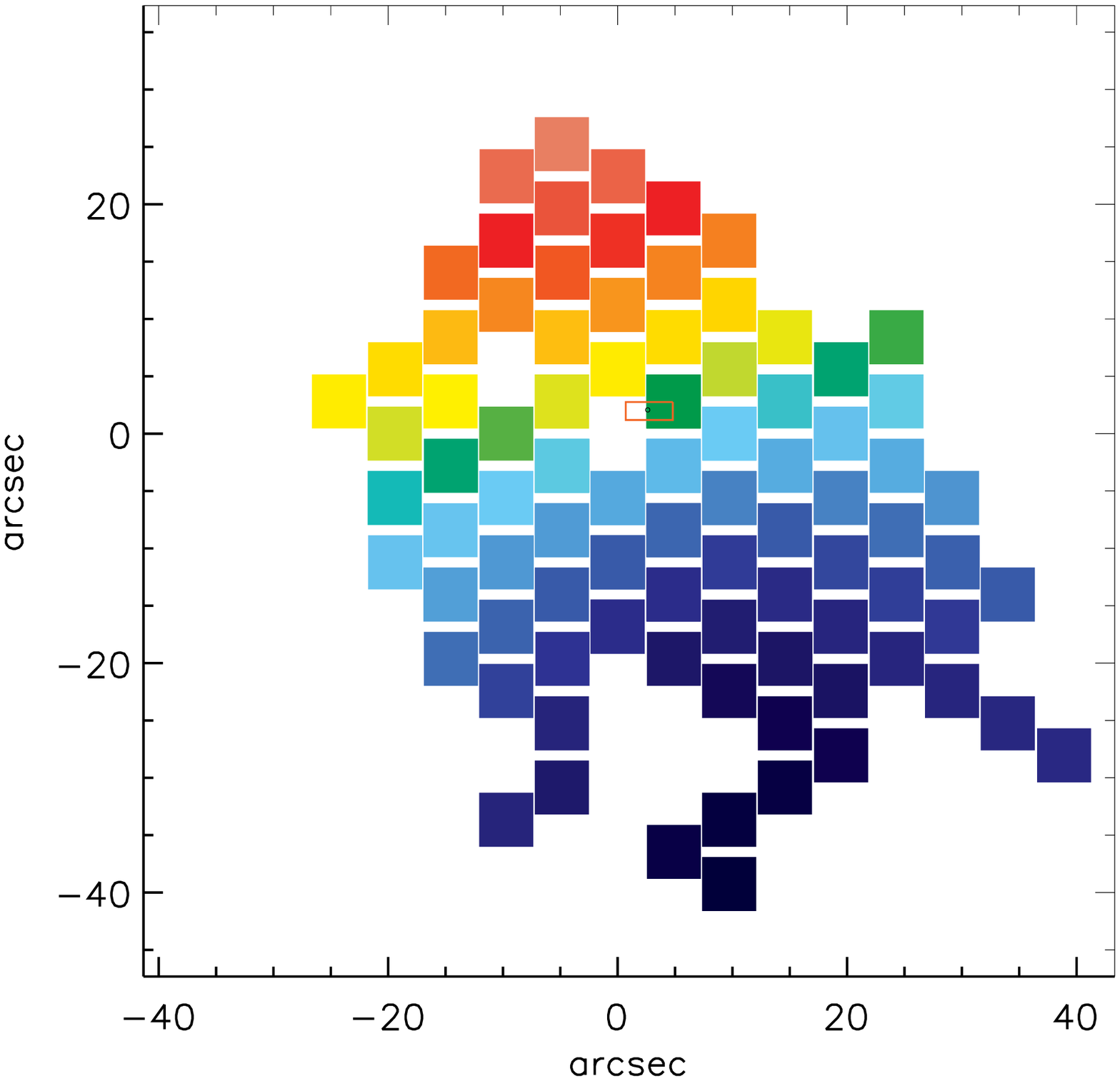} 
  \hspace*{-0.25cm} \includegraphics[height=3.82cm]{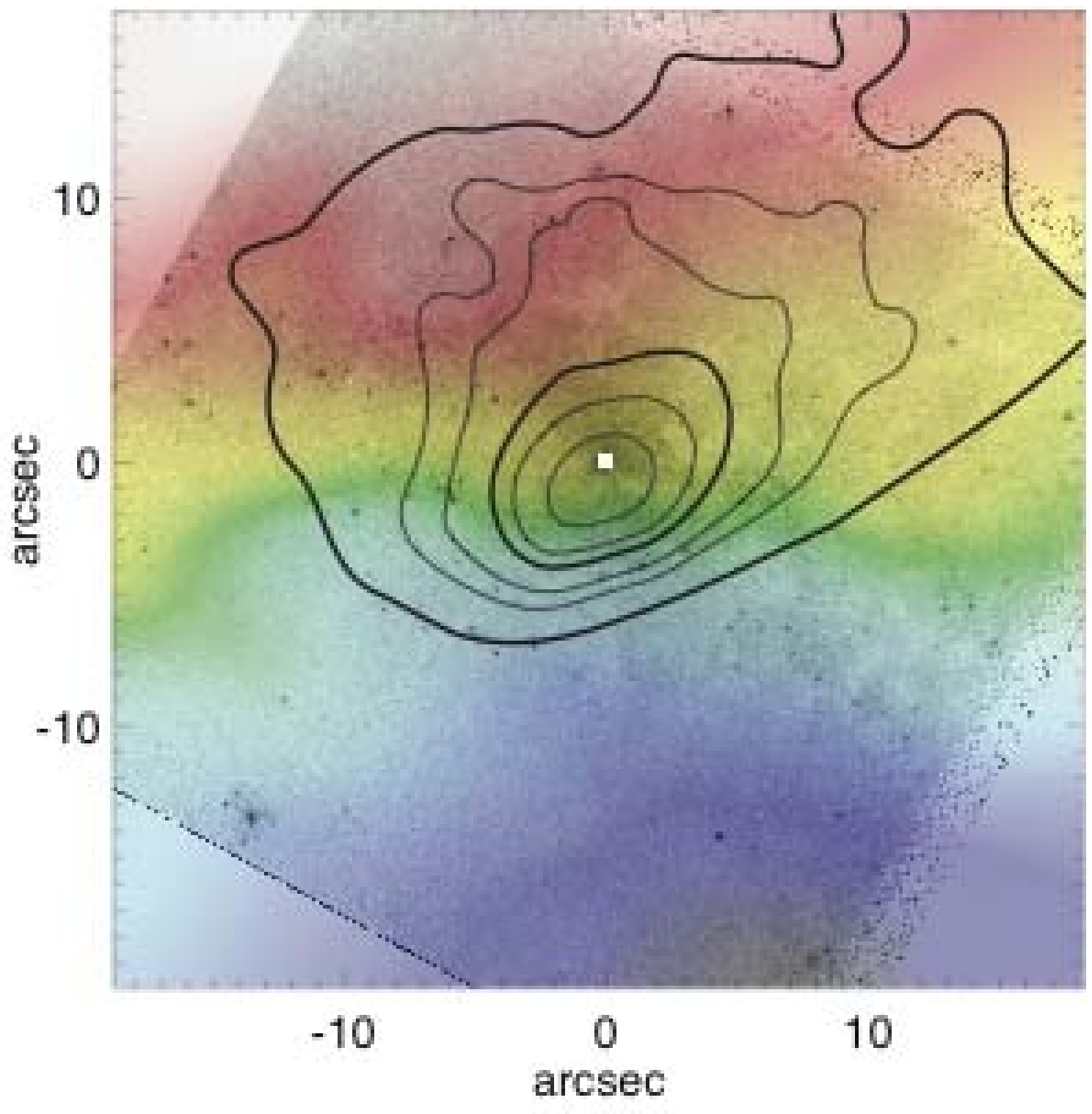}
   \vspace{-0.6cm}
  \end{tabular}
  \end{center}

 \begin{center}
  \begin{tabular}{cc}
  \hspace*{-1.3cm} \includegraphics[height=4.5cm]{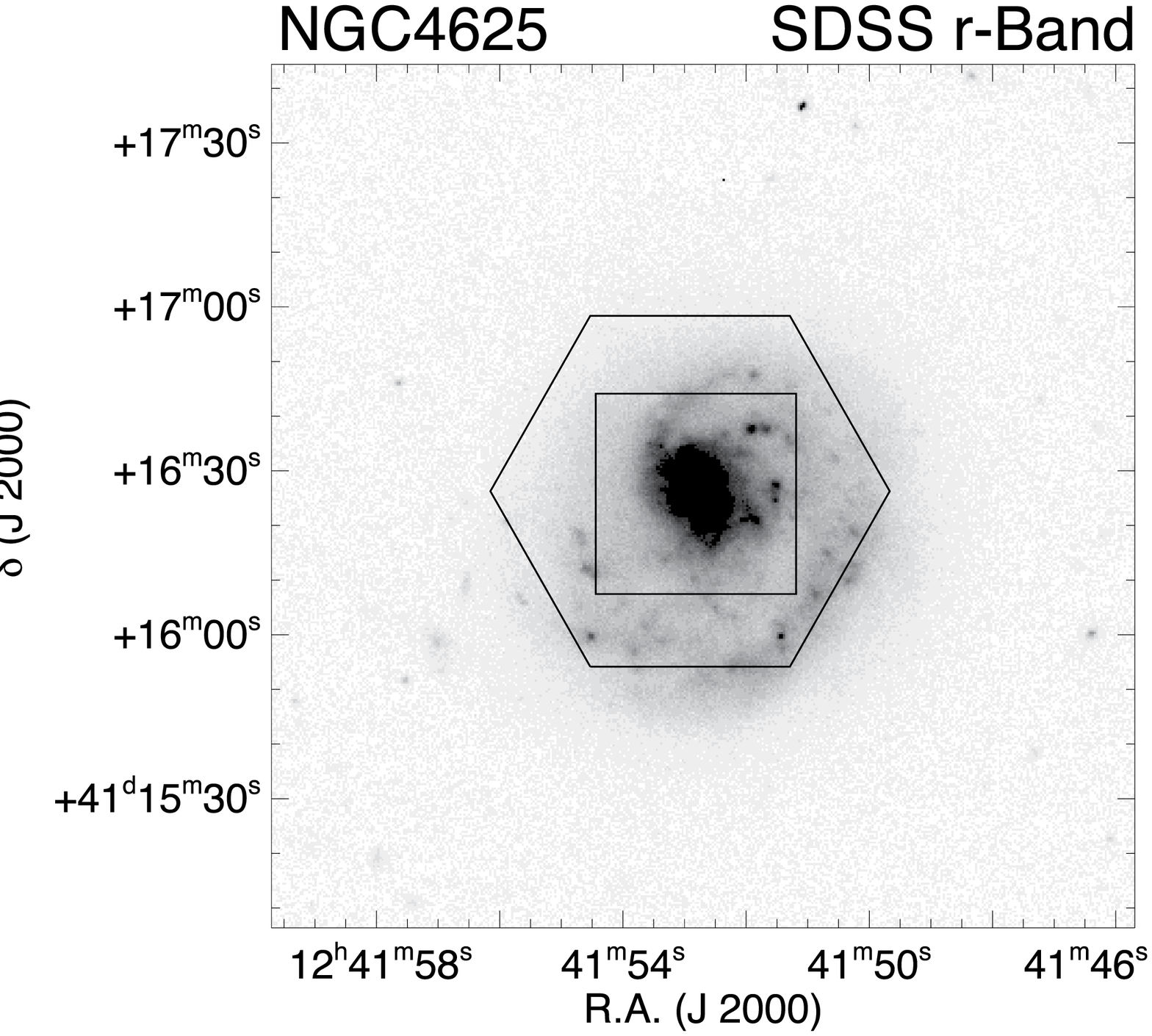}
   \hspace*{-0.25cm} \includegraphics[height=4.1cm]{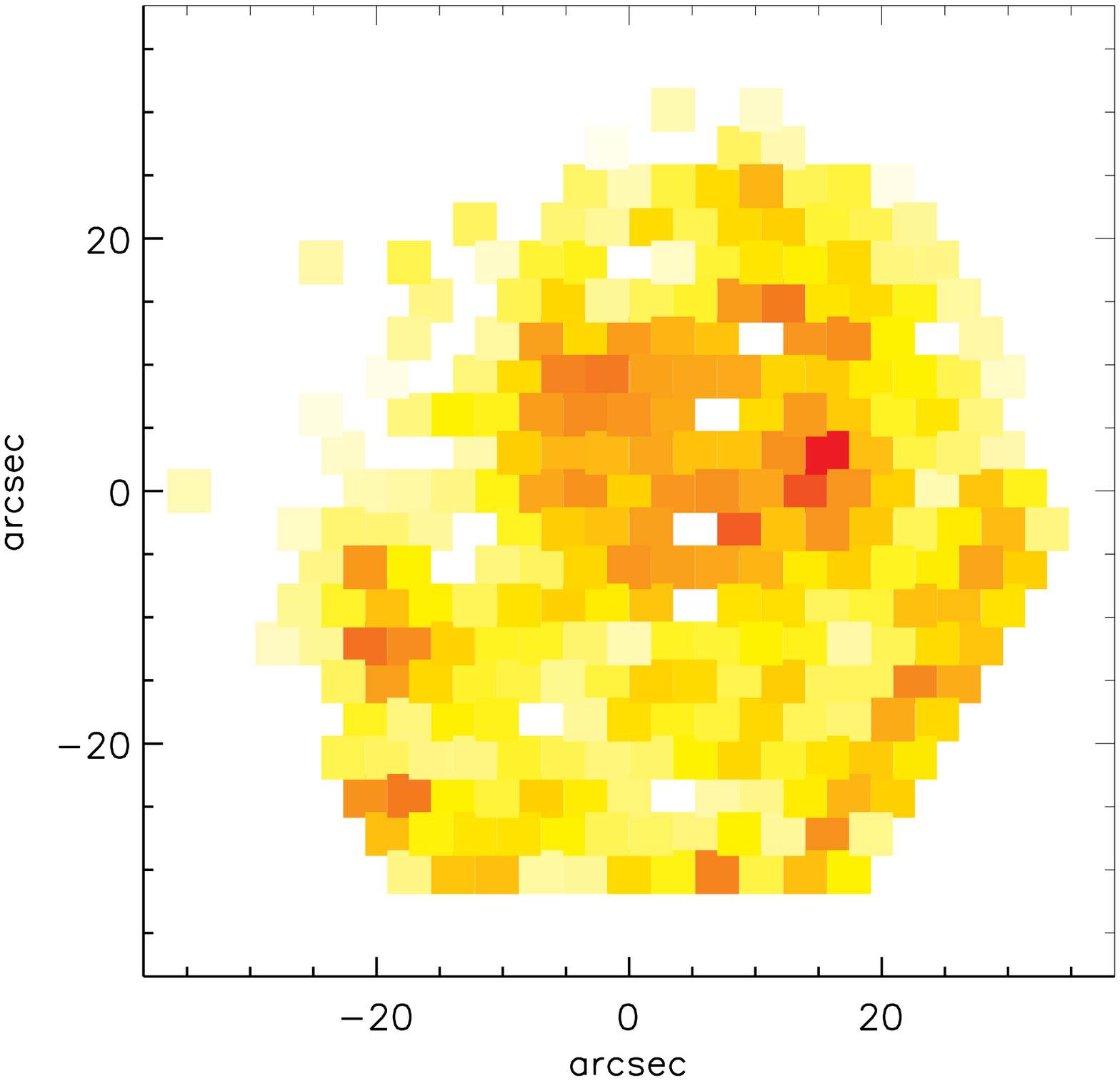}
   \hspace*{-0.25cm}   \includegraphics[height=4.1cm]{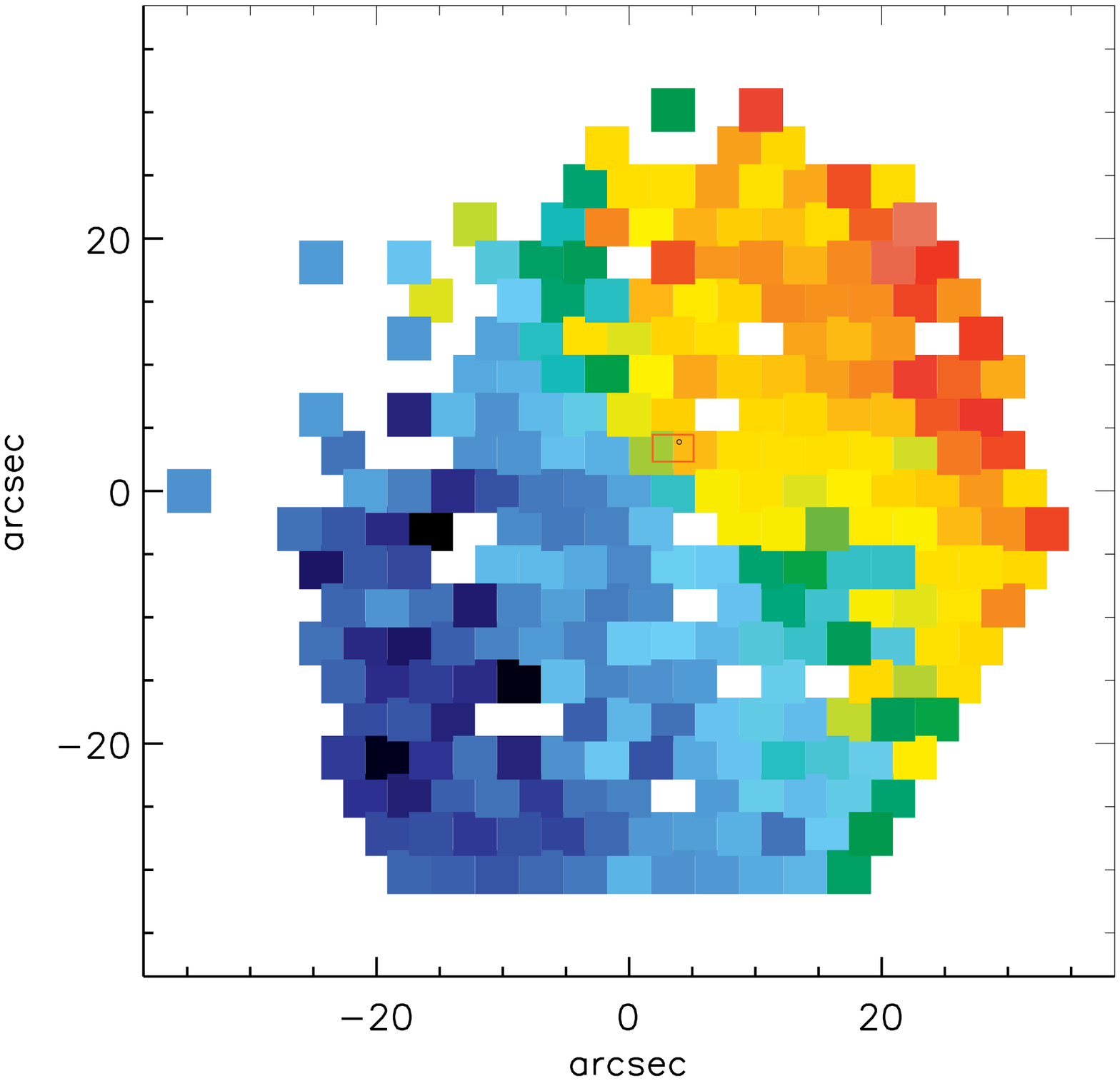}
   \hspace*{-0.25cm}   \includegraphics[height=4.1cm]{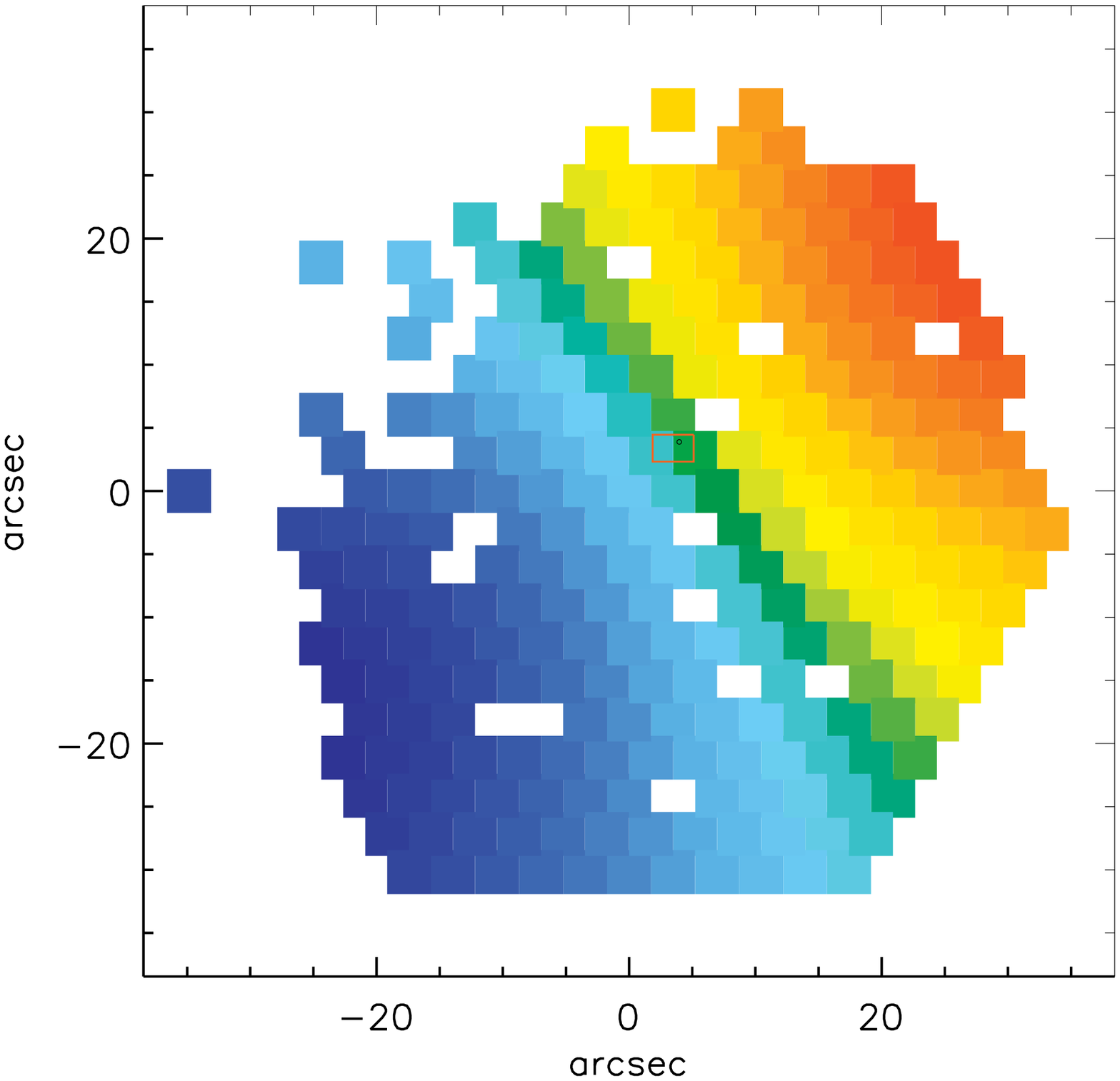} 
  \hspace*{-0.25cm}   \includegraphics[height=3.82cm]{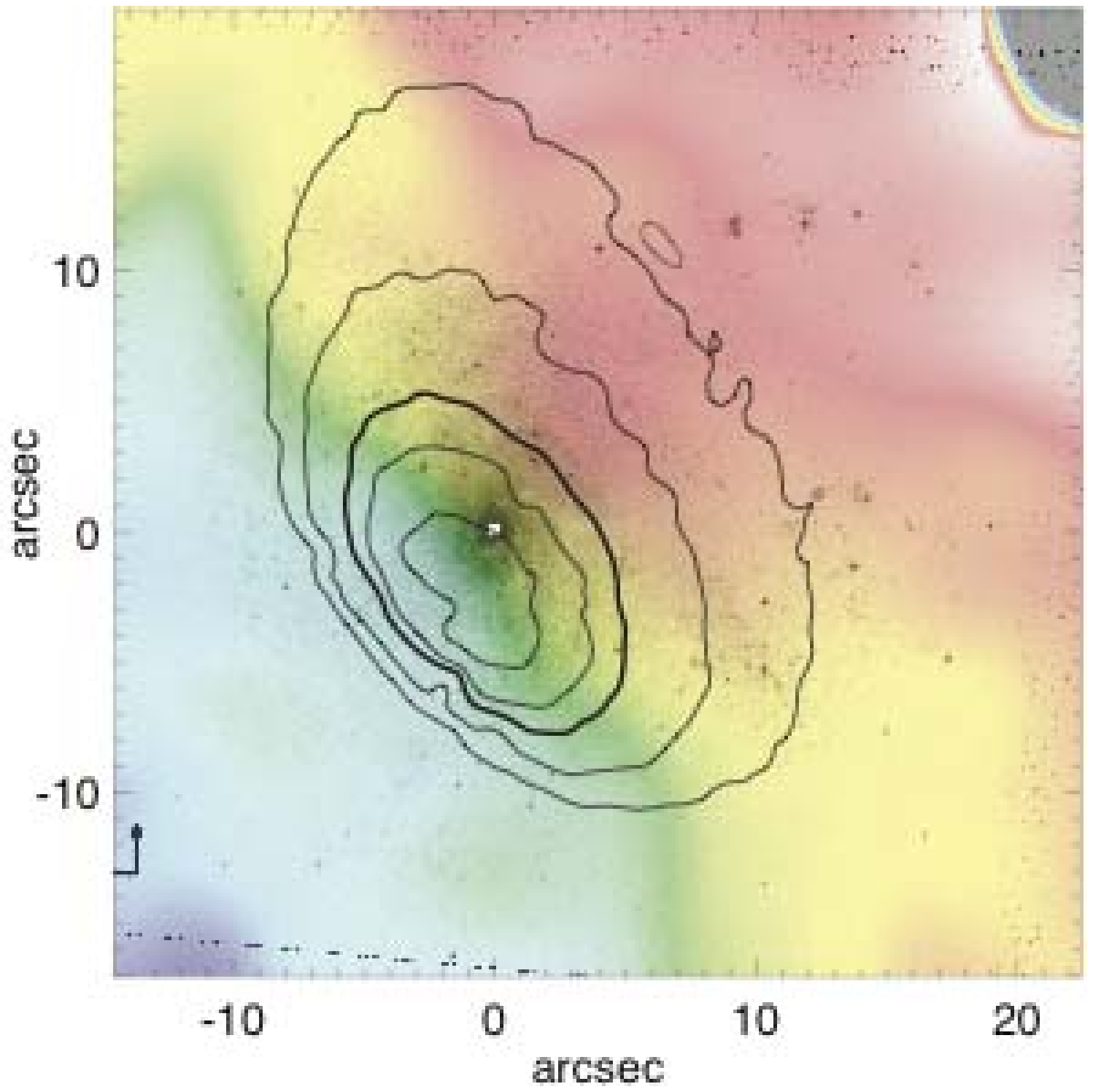}
   \vspace{-0.6cm}
  \end{tabular}
  \end{center}

 \begin{center}
  \begin{tabular}{c}
    \hspace*{-1.3cm} \includegraphics[height=4.5cm]{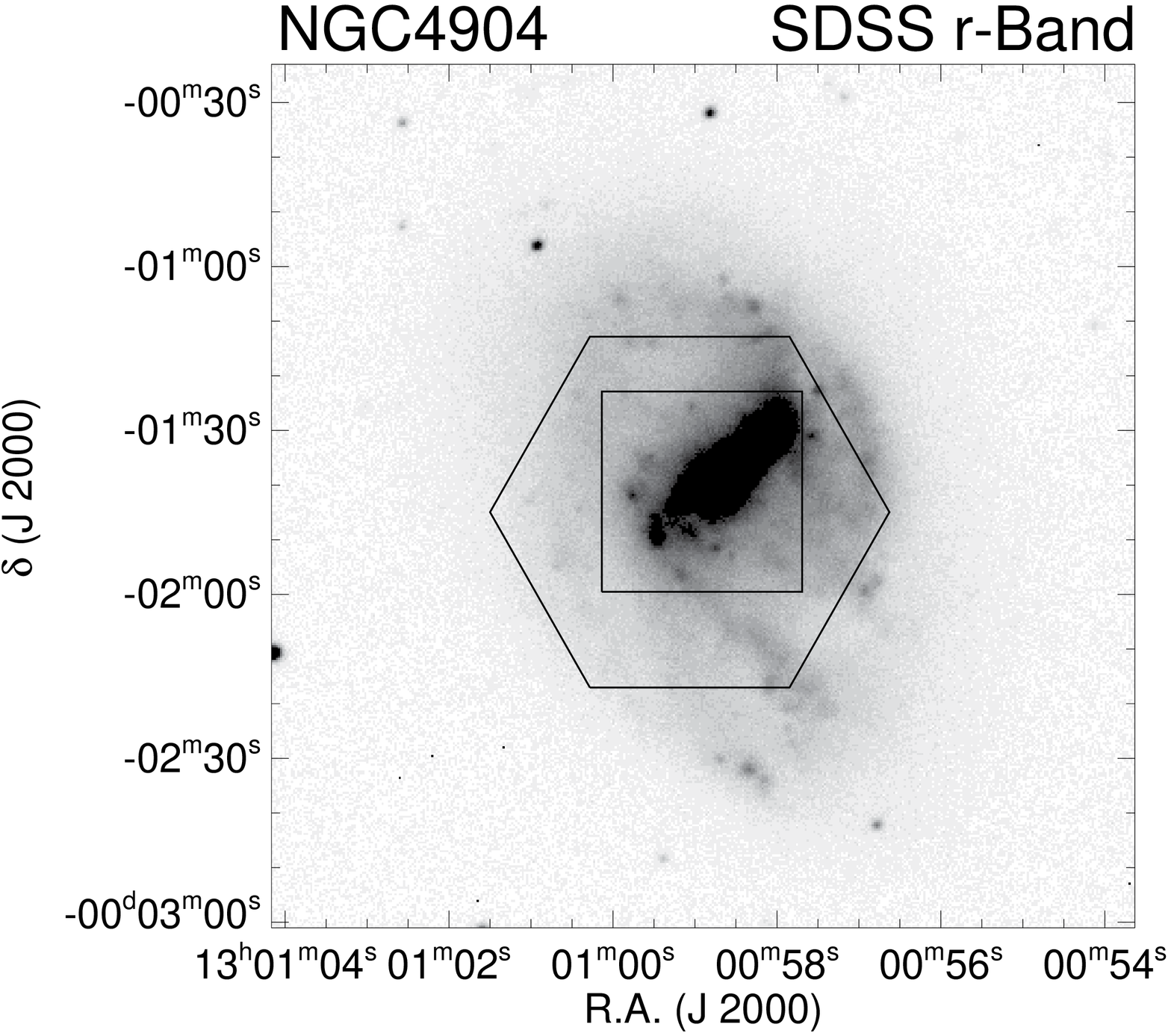}
   \hspace*{-0.25cm} \includegraphics[height=4.1cm]{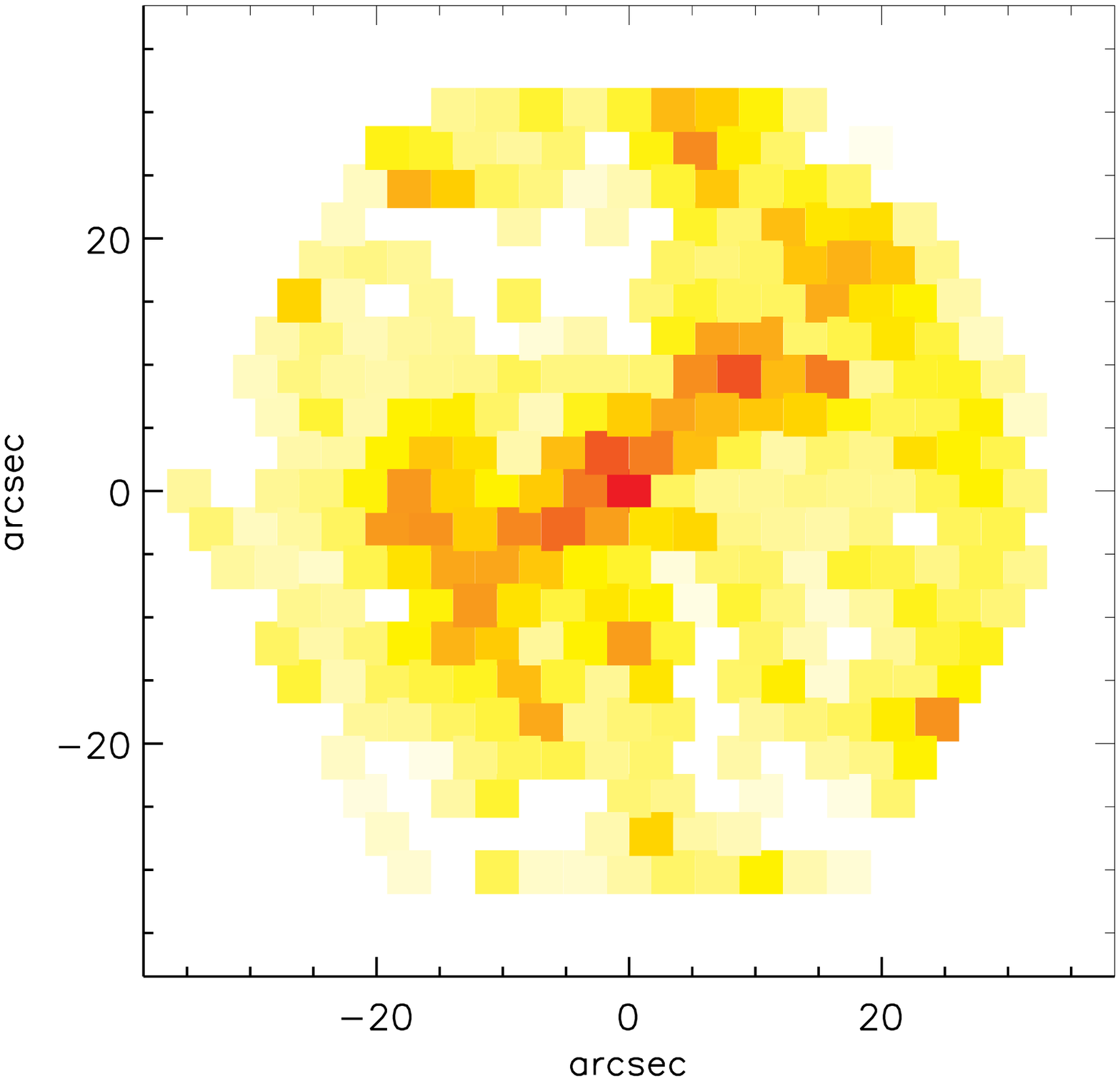}
   \hspace*{-0.25cm}   \includegraphics[height=4.1cm]{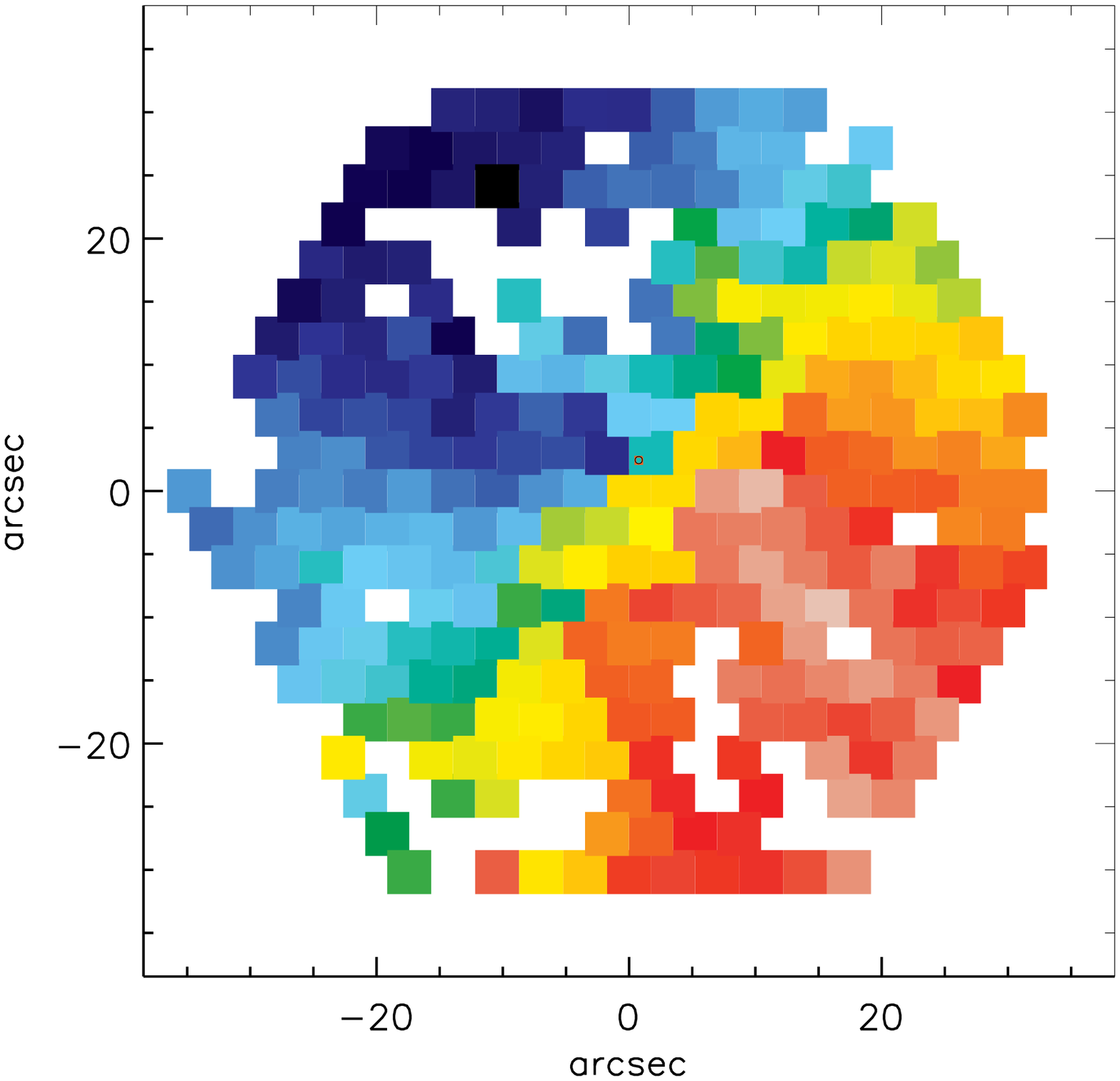}
   \hspace*{-0.25cm}   \includegraphics[height=4.1cm]{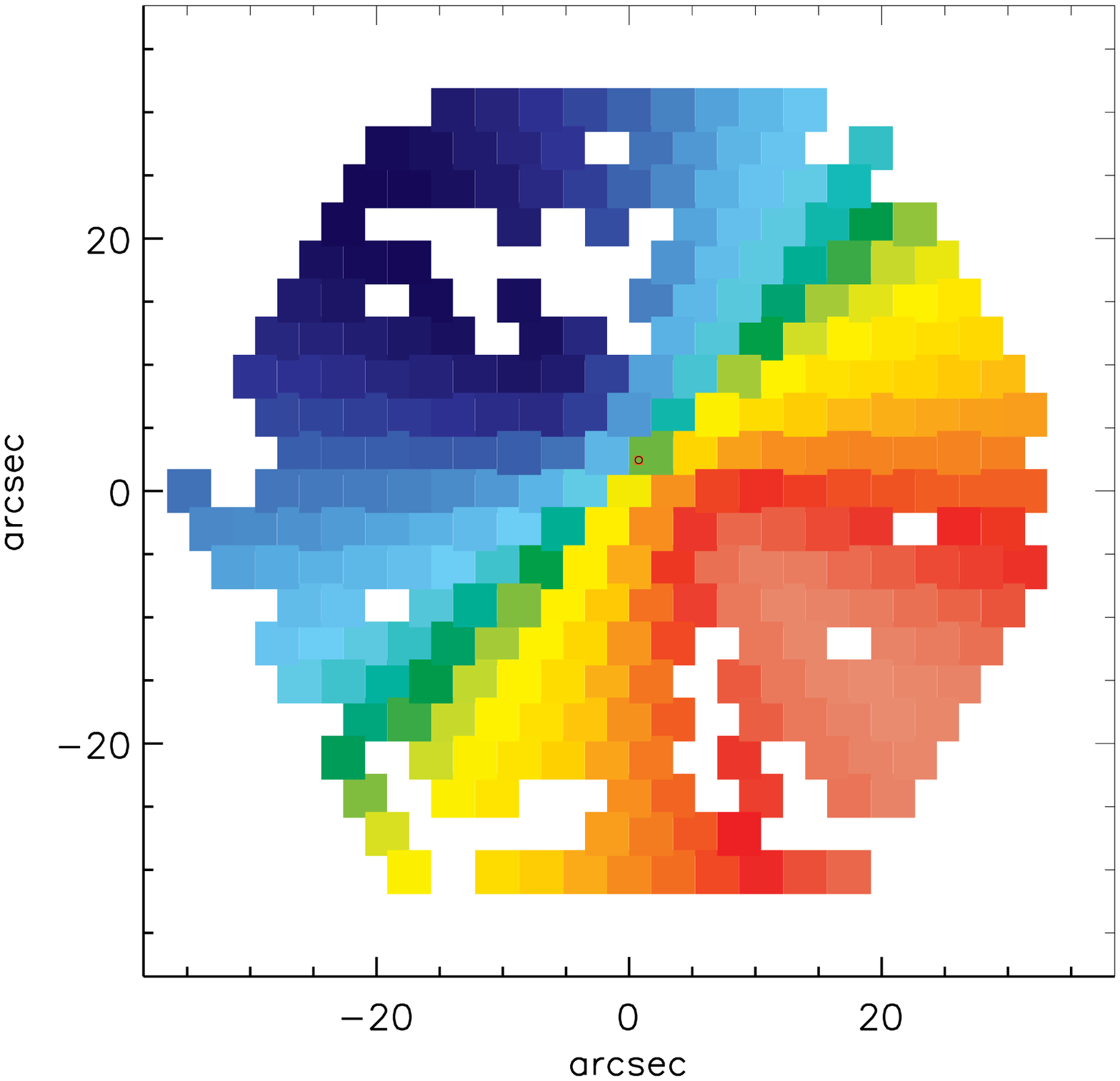} 
  \hspace*{-0.25cm}   \includegraphics[height=3.82cm]{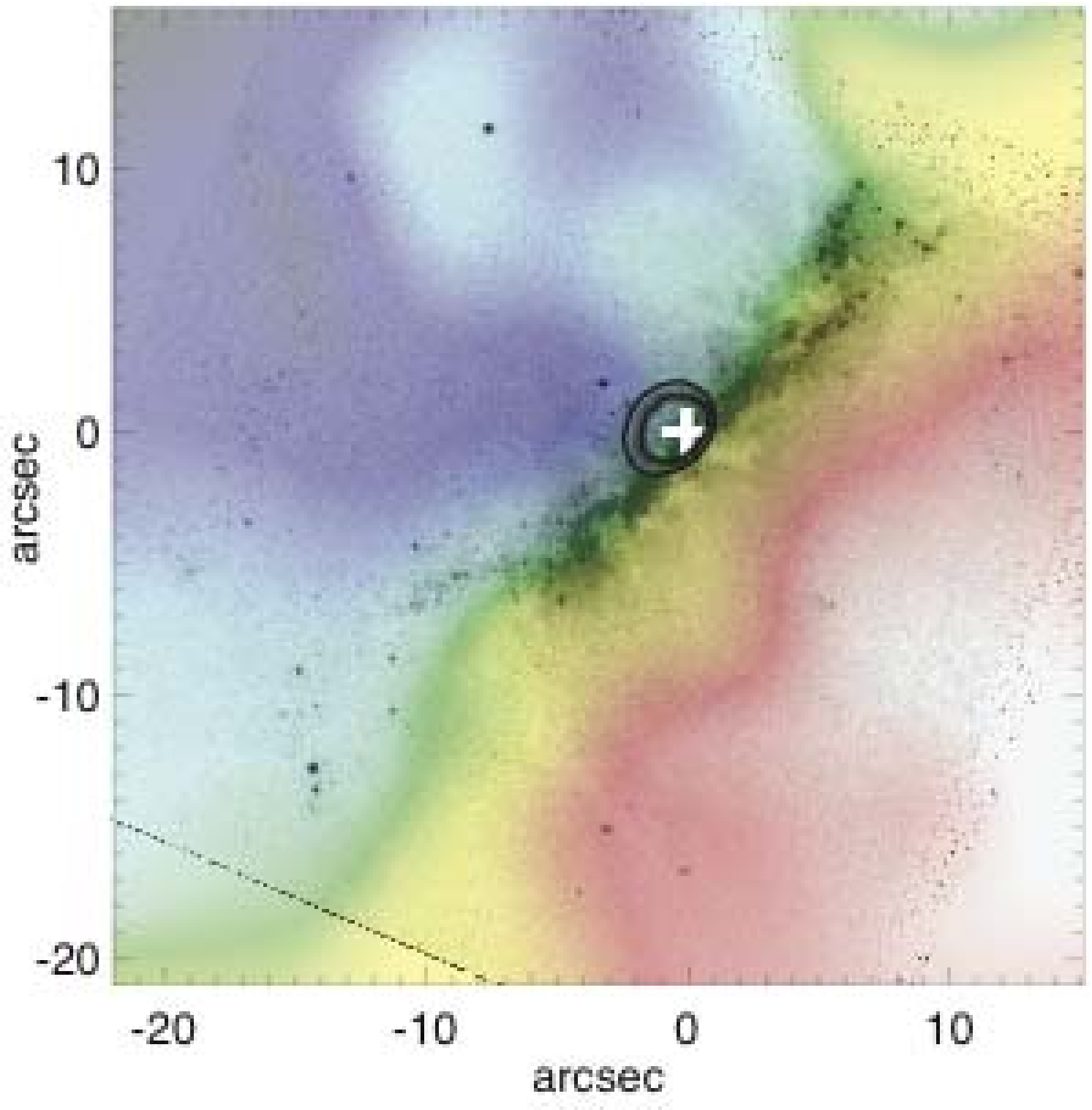}
  \end{tabular}
  \end{center}

  \contcaption{
}
  \end{figure*}

  \begin{figure*}
  \begin{center}
  \begin{tabular}{cc}
    \hspace*{-1.3cm} \includegraphics[height=4.5cm]{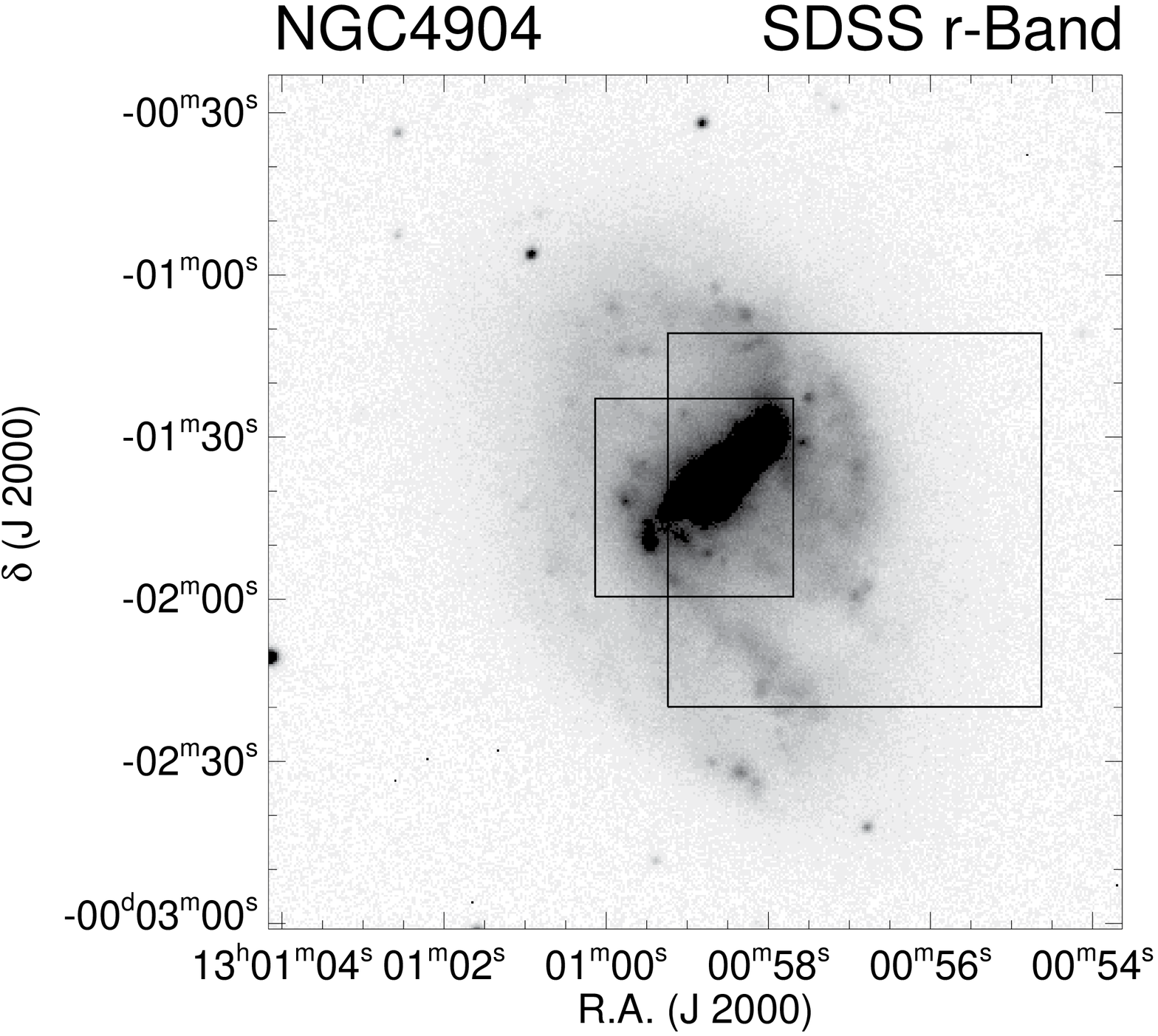}
   \hspace*{-0.25cm}  \includegraphics[width=4.1cm]{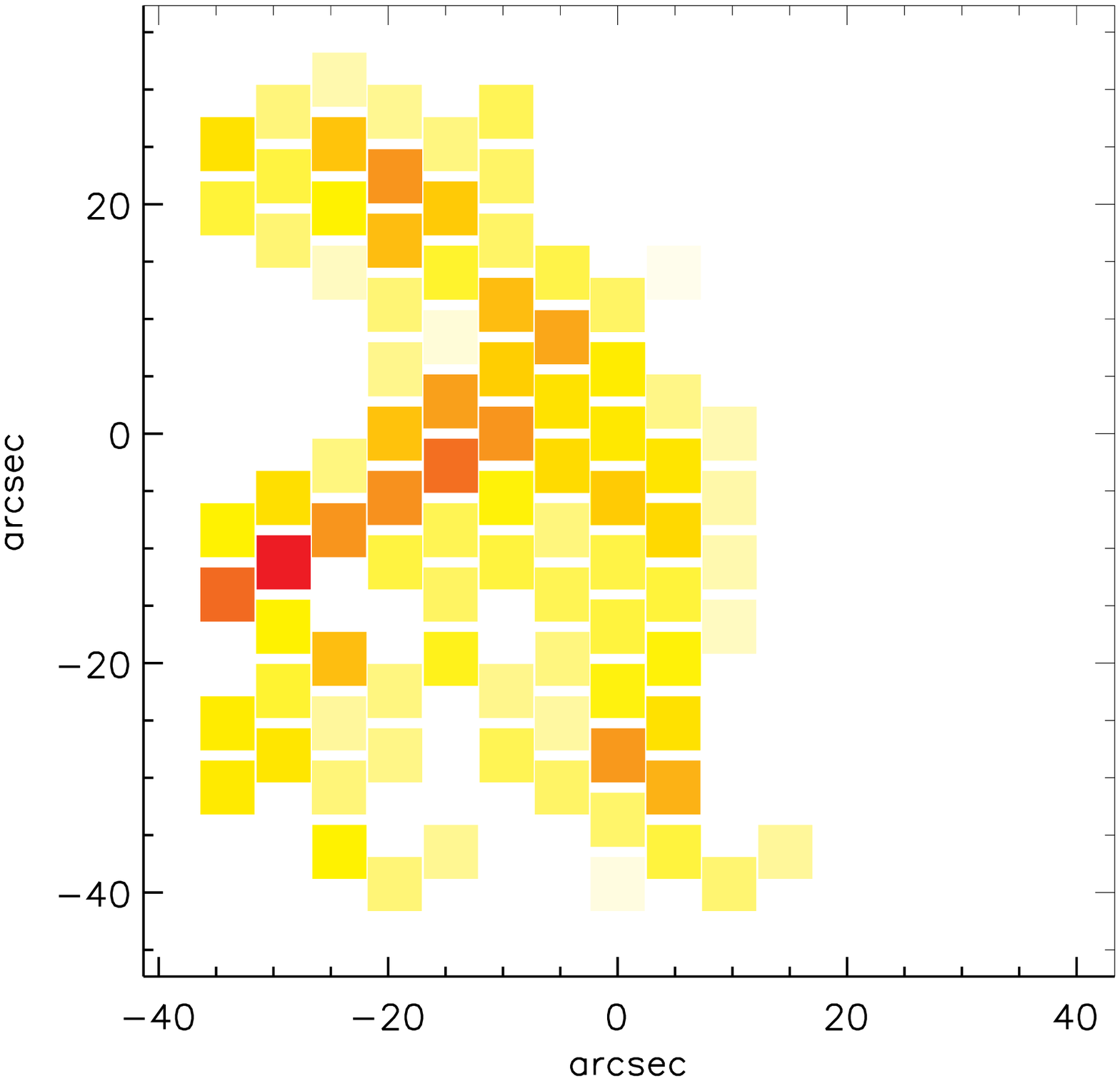}
    \hspace*{-0.25cm}  \includegraphics[height=4.1cm]{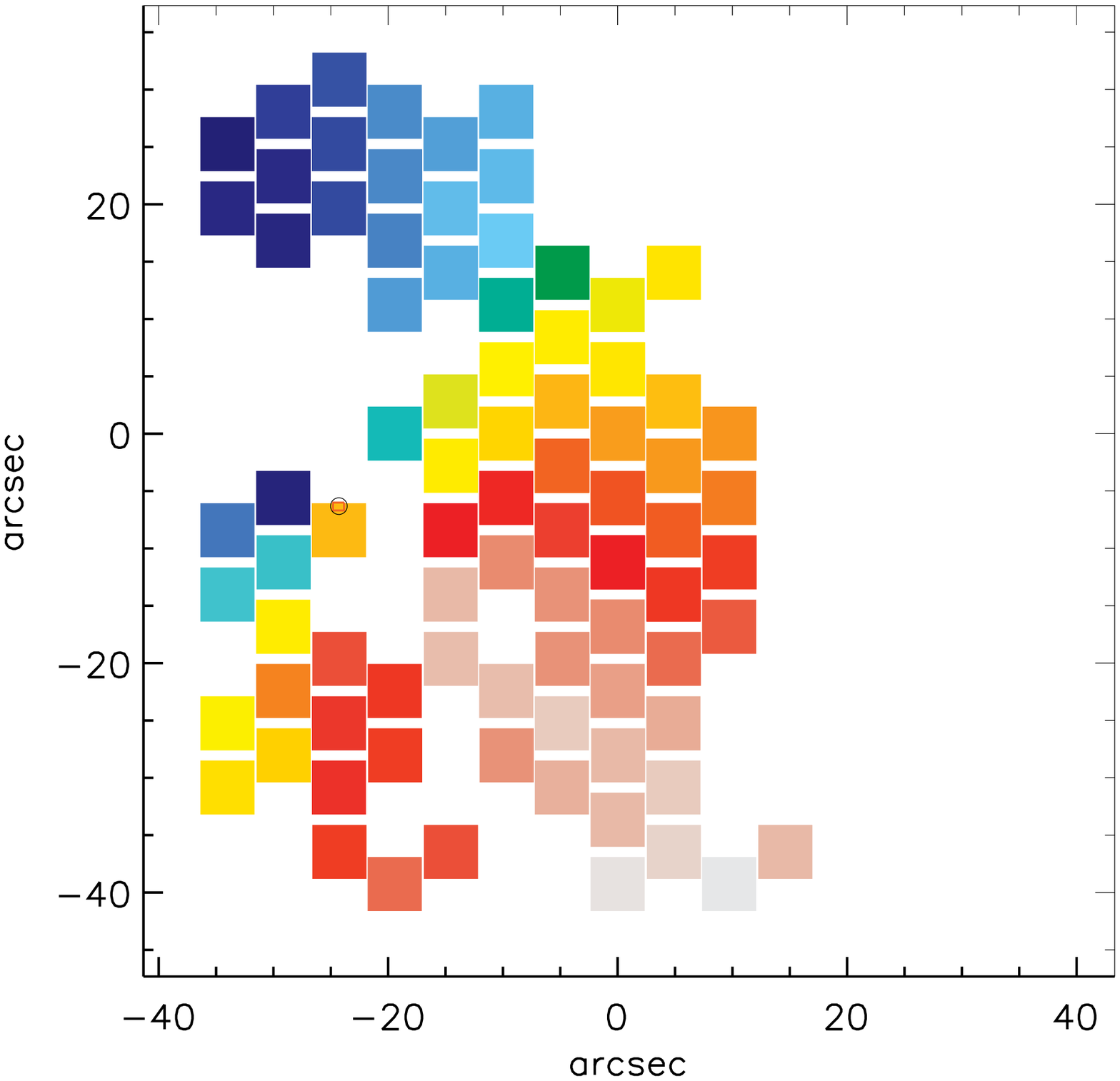}
    \hspace*{-0.25cm}  \includegraphics[height=4.1cm]{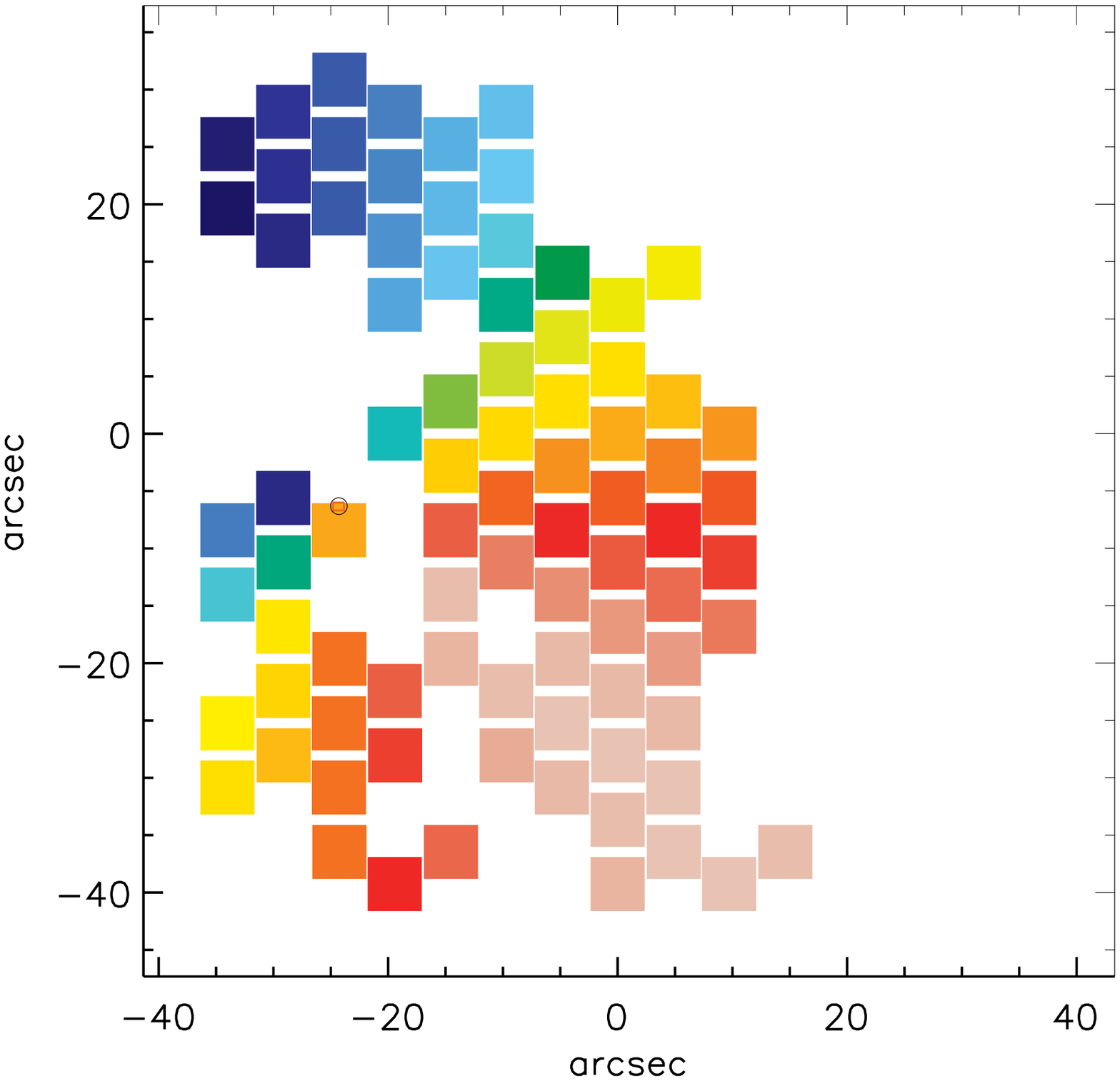}
  \hspace*{-0.25cm}  \includegraphics[height=3.82cm]{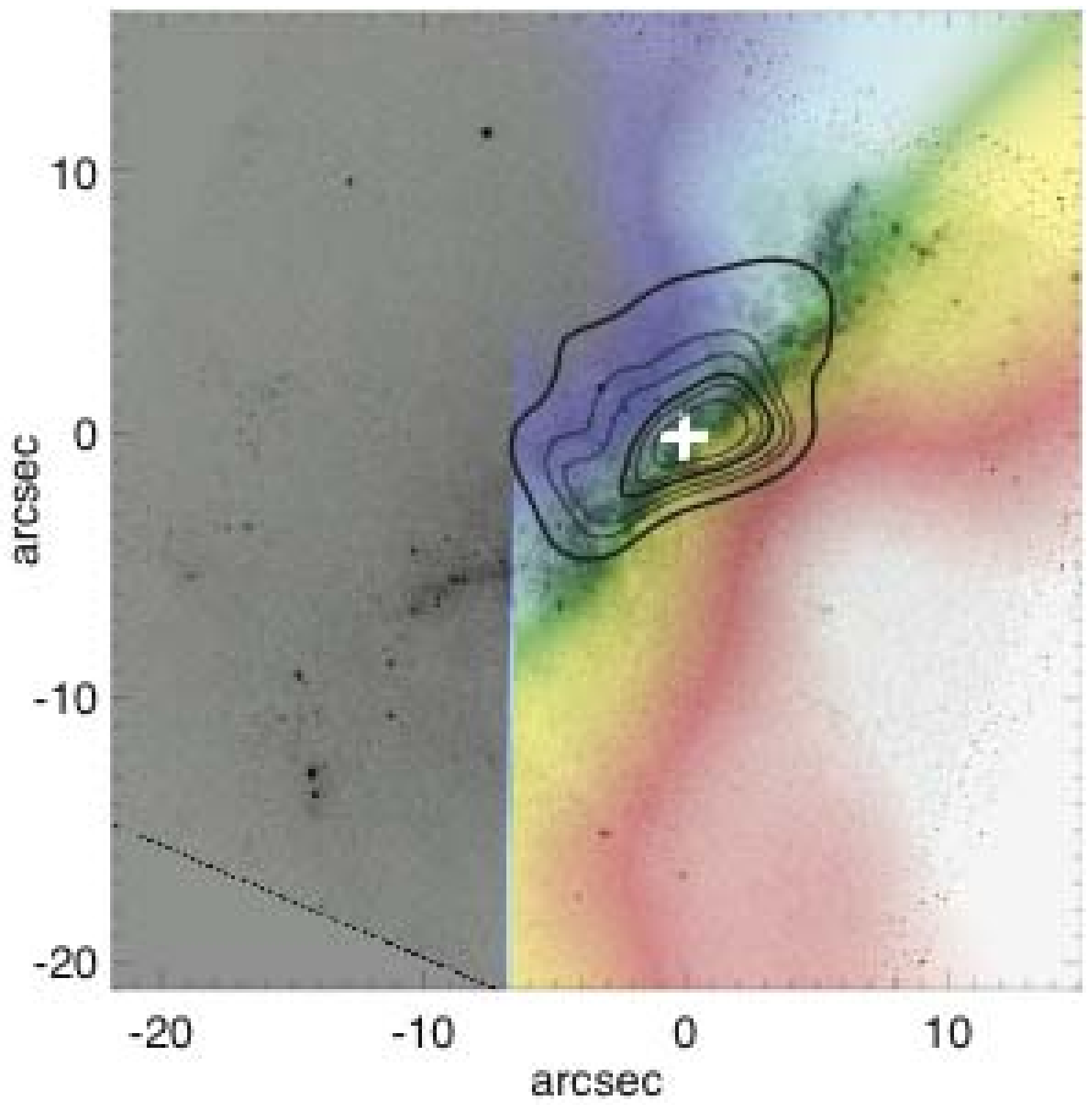}

    \vspace{-0.6cm}
  \end{tabular}
  \end{center}

 \begin{center}
  \begin{tabular}{cc}
   \hspace*{-1.3cm} \includegraphics[height=4.5cm]{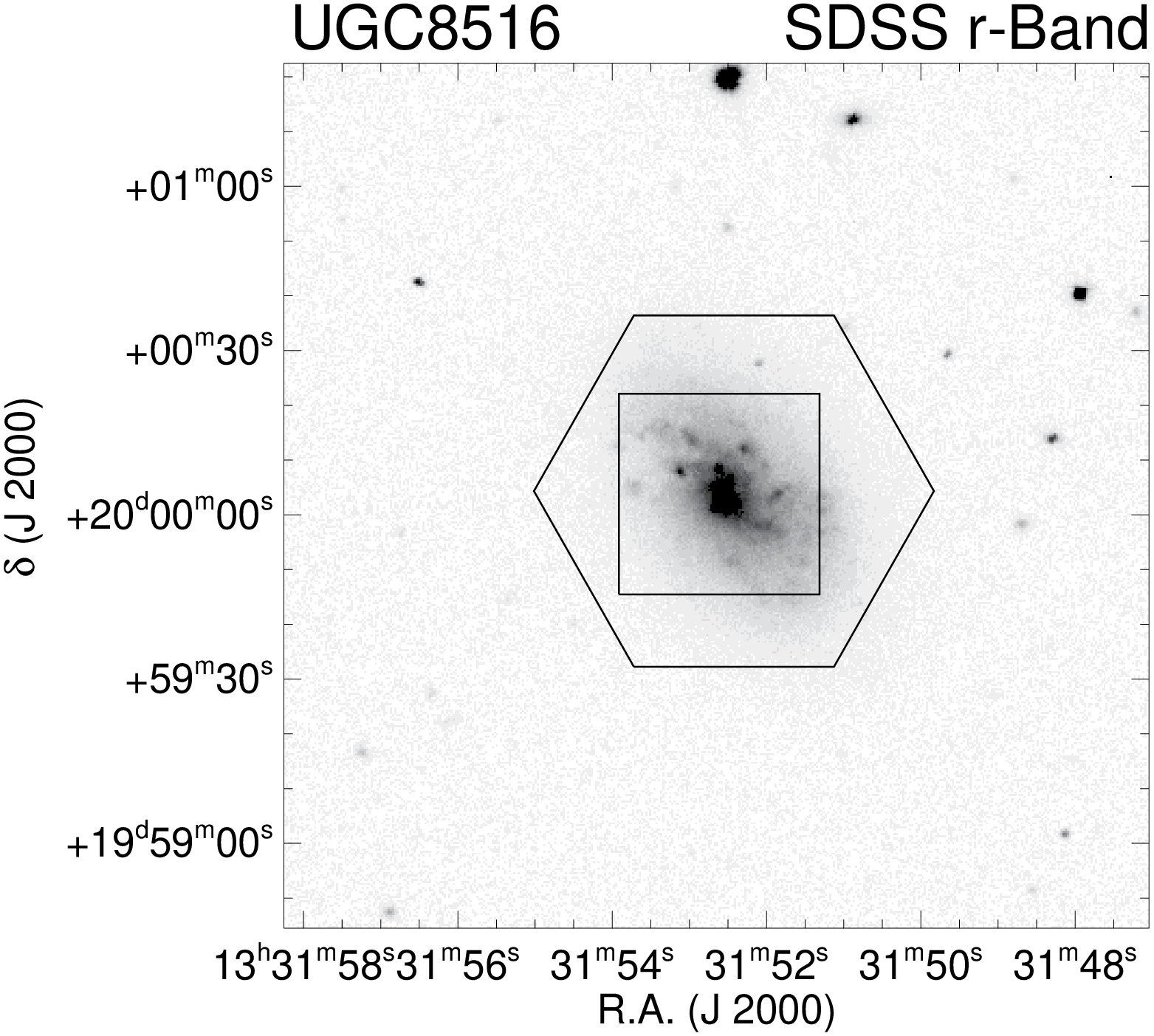}
   \hspace*{-0.25cm} \includegraphics[height=4.1cm]{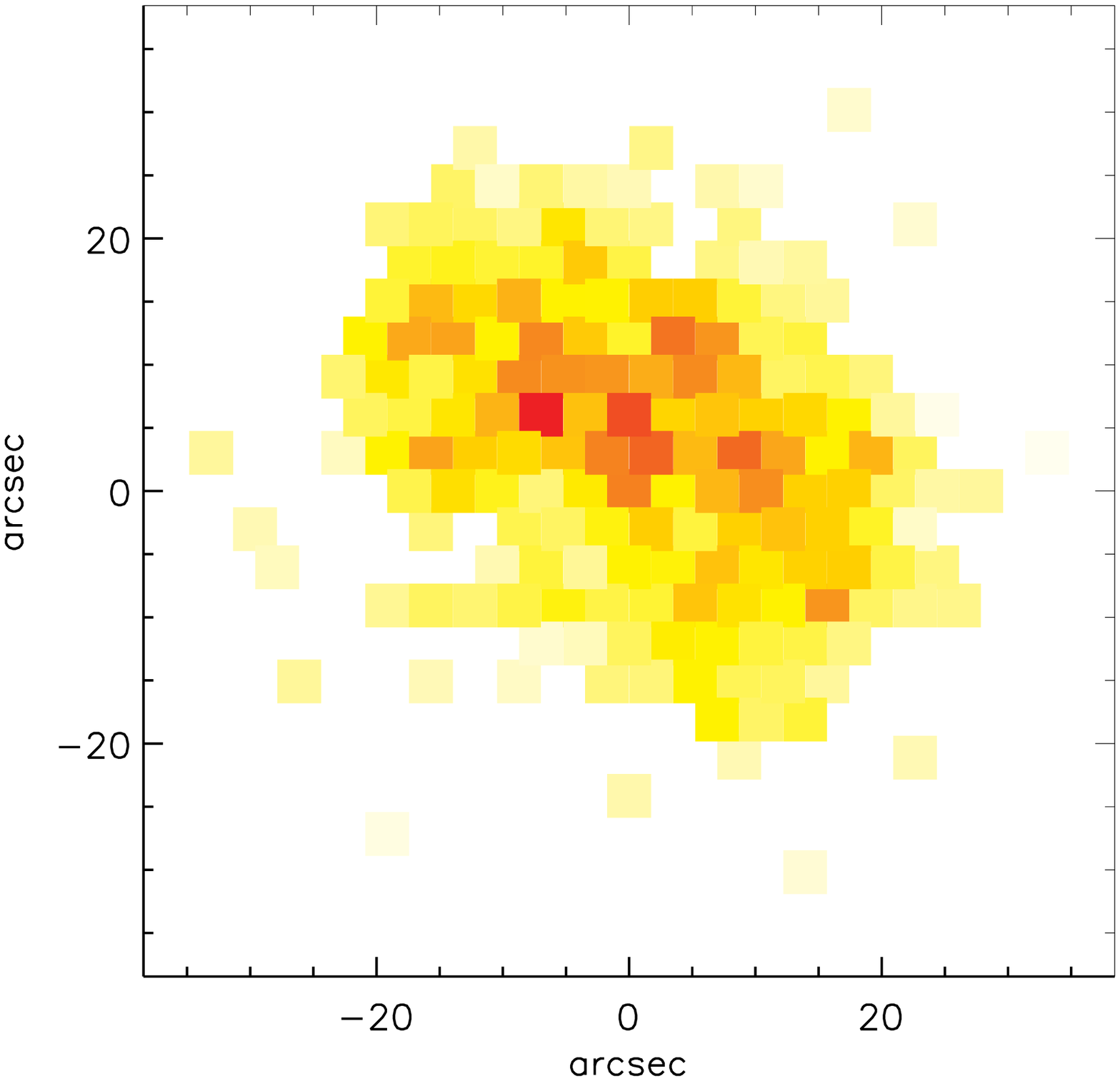}
   \hspace*{-0.25cm}   \includegraphics[height=4.1cm]{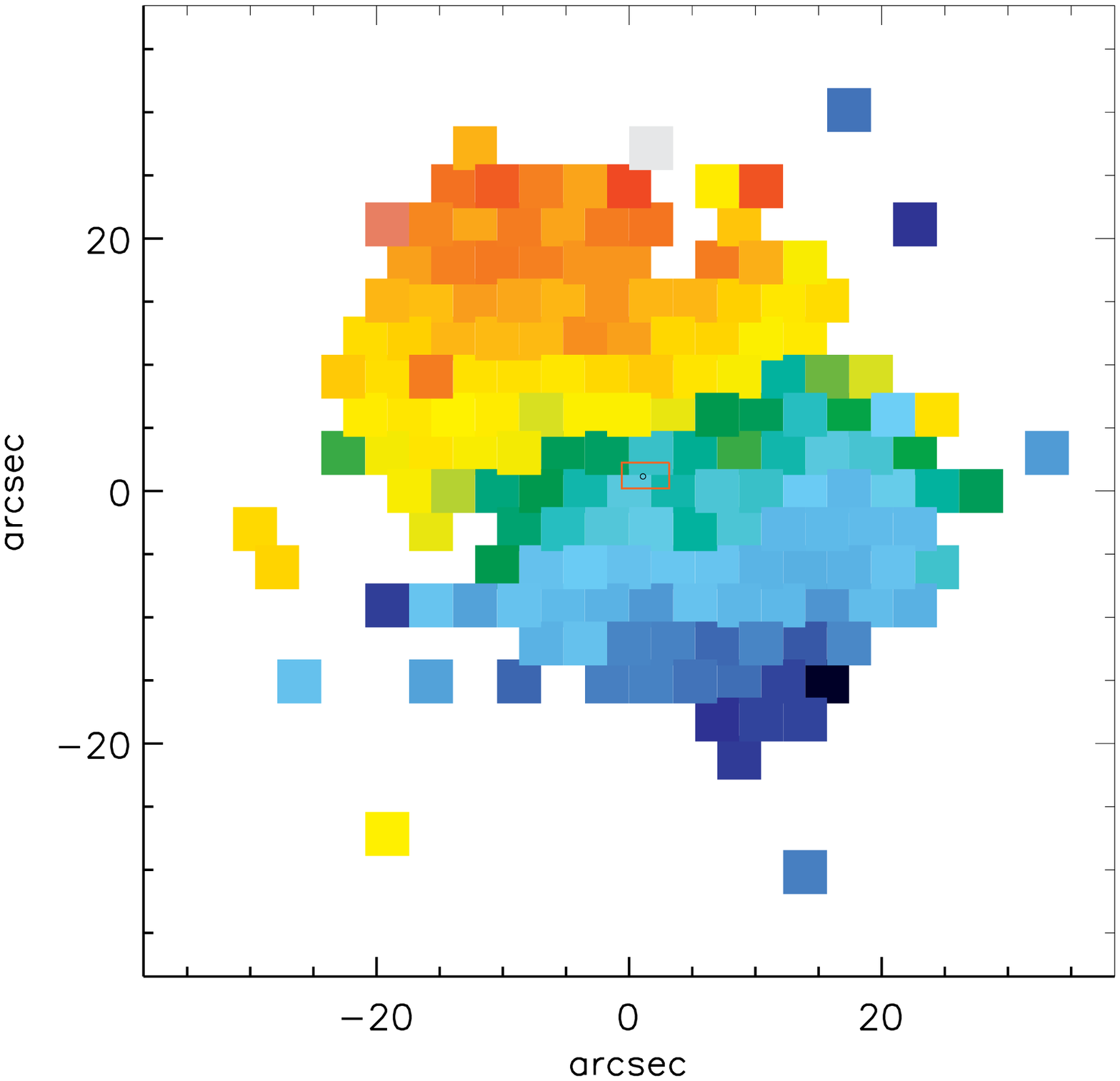}
   \hspace*{-0.25cm}   \includegraphics[height=4.1cm]{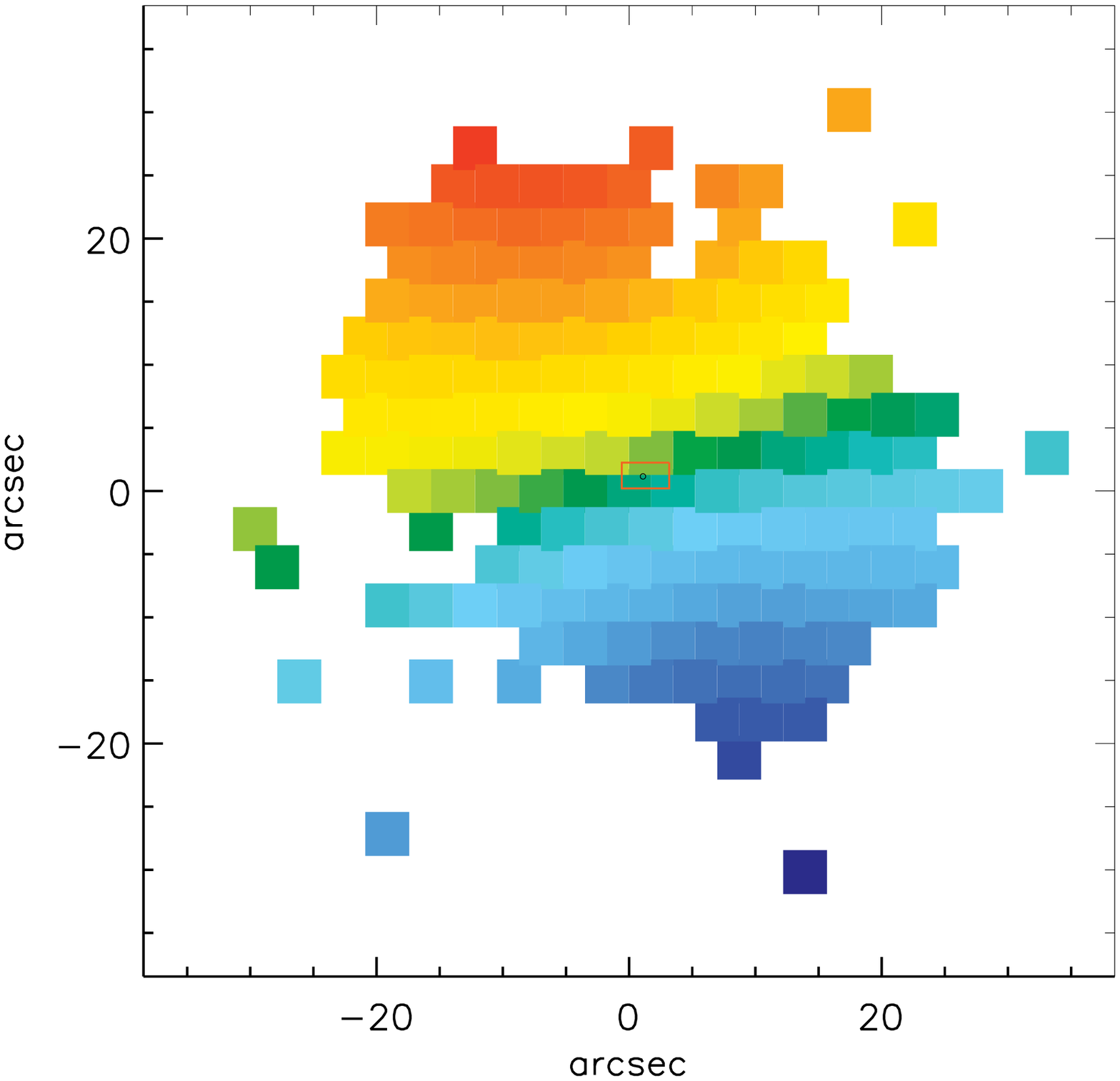} 
 \hspace*{-0.25cm}   \includegraphics[height=3.82cm]{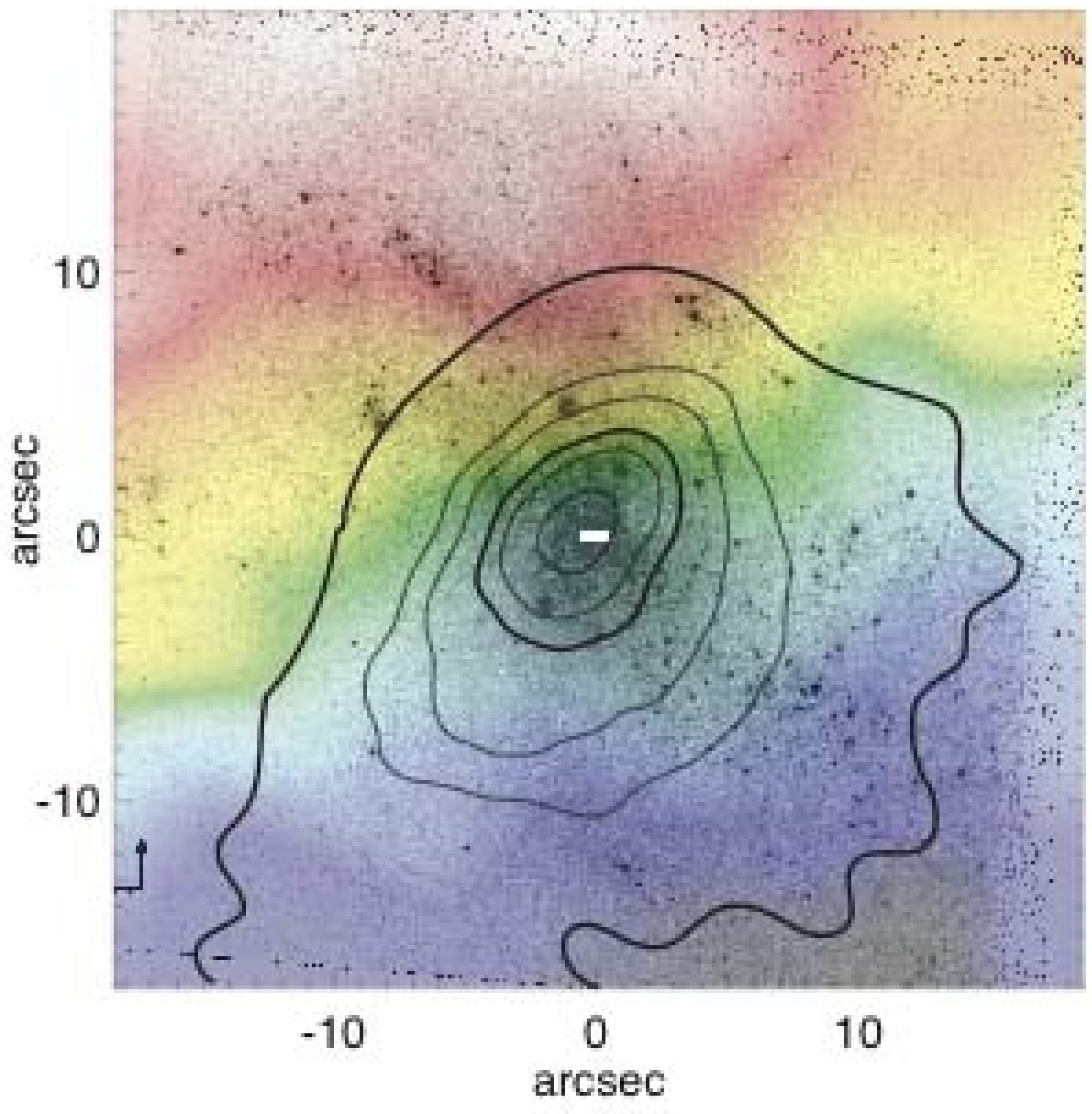}
   \vspace{-0.6cm}
  \end{tabular}
  \end{center}
  
\begin{center}
  \begin{tabular}{c}
   \hspace*{-1.3cm} \includegraphics[height=4.5cm]{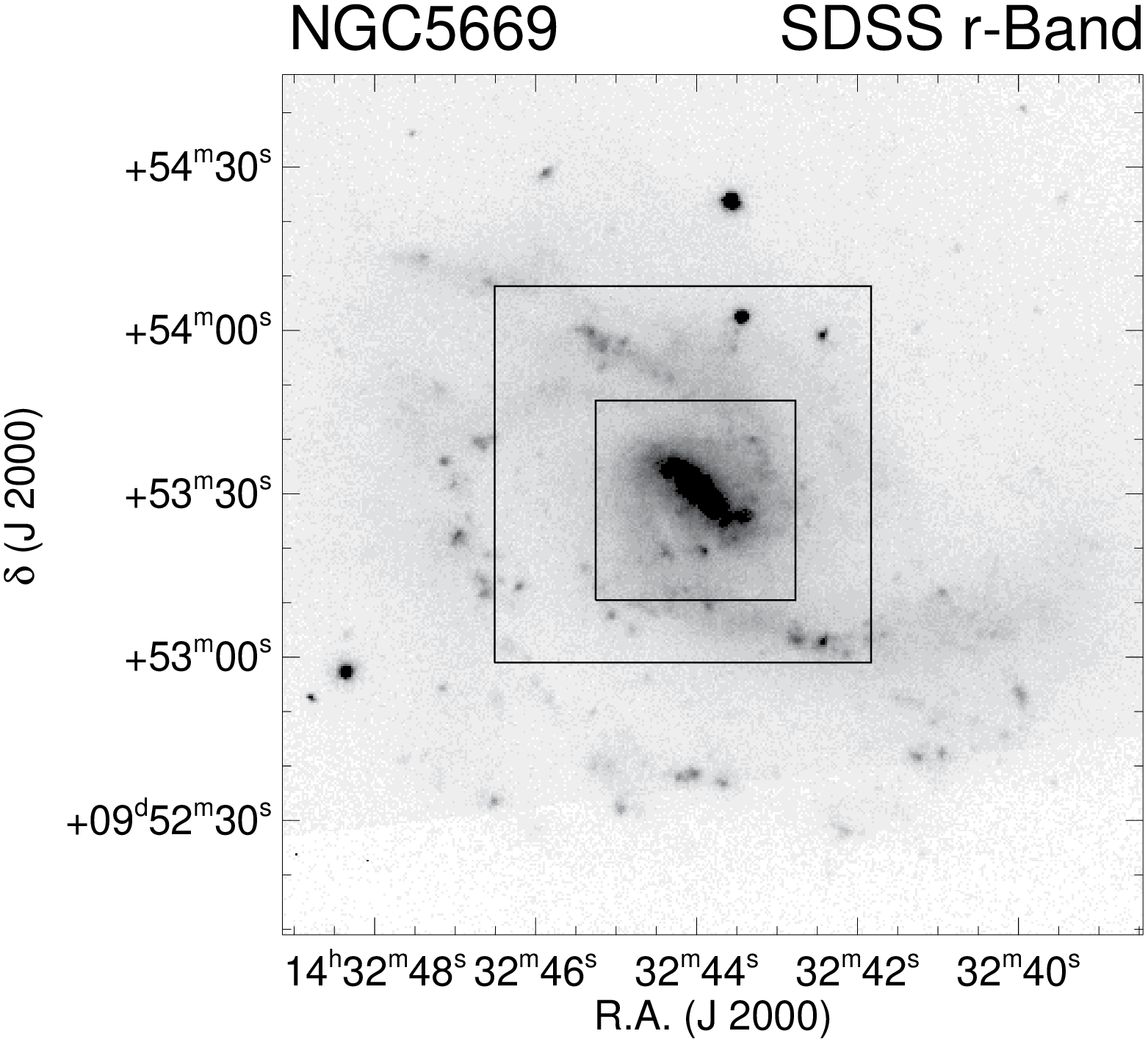}
   \hspace*{-0.25cm} \includegraphics[height=4.1cm]{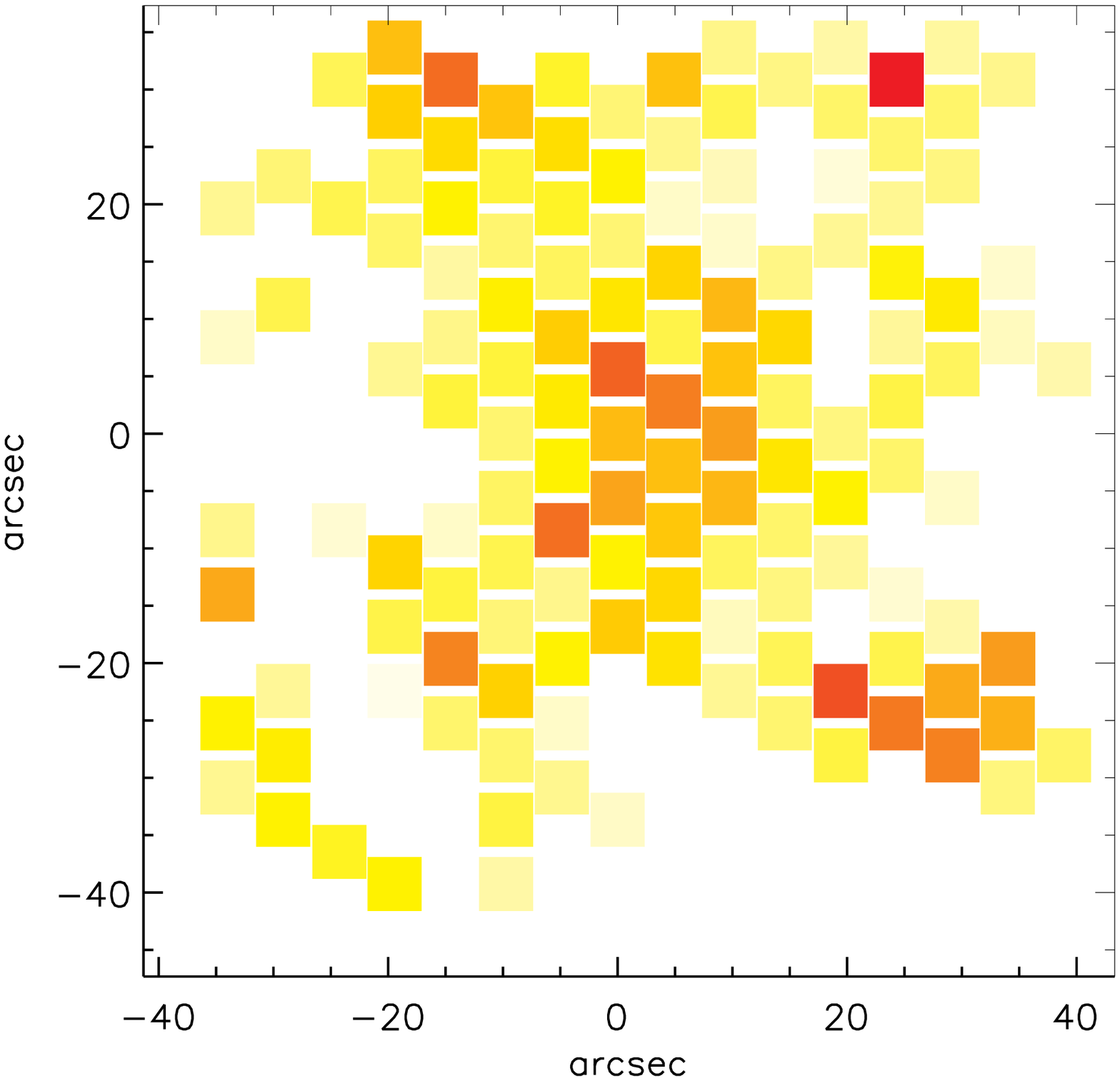}
   \hspace*{-0.25cm}   \includegraphics[height=4.1cm]{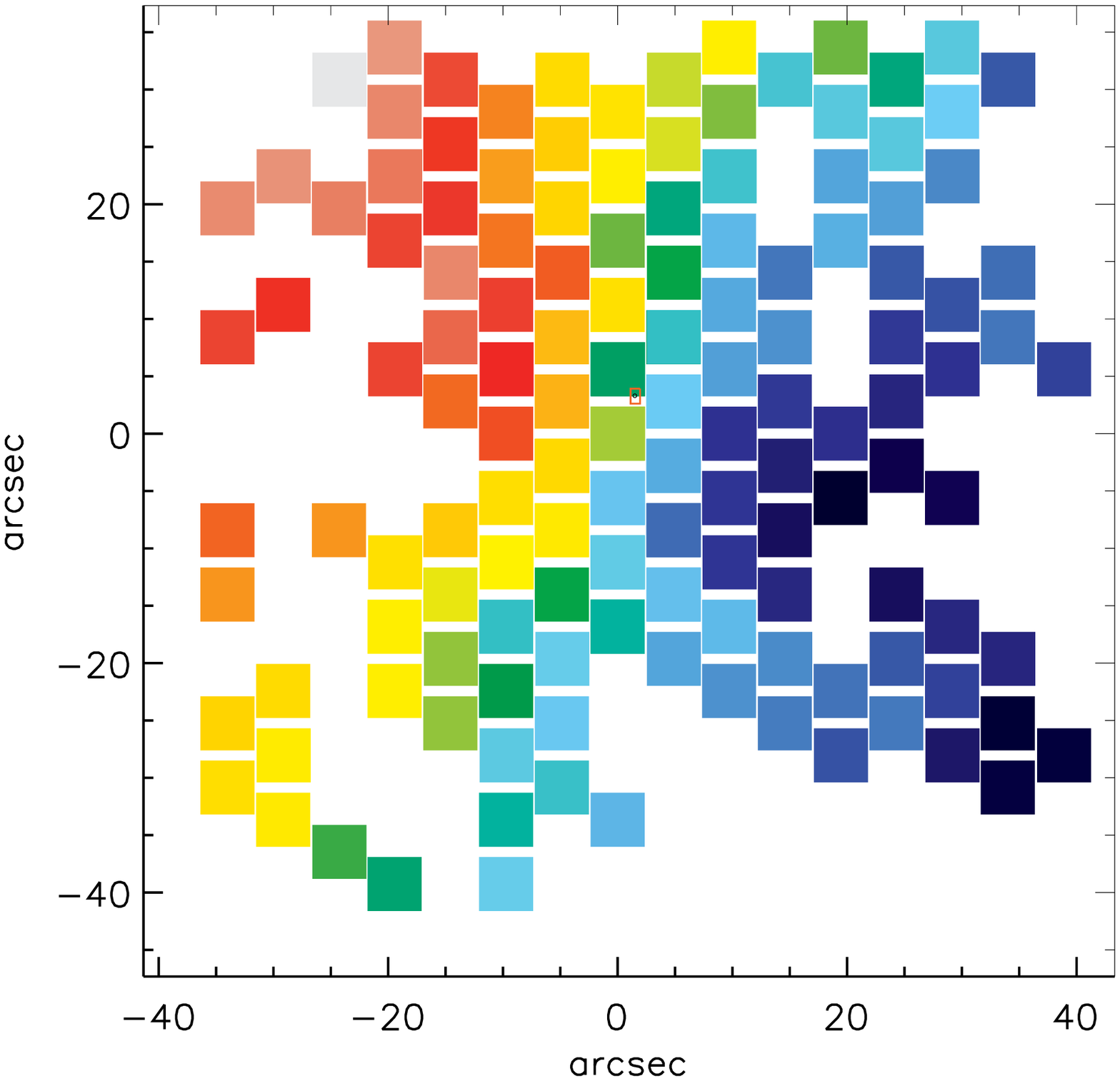}
   \hspace*{-0.25cm}   \includegraphics[height=4.1cm]{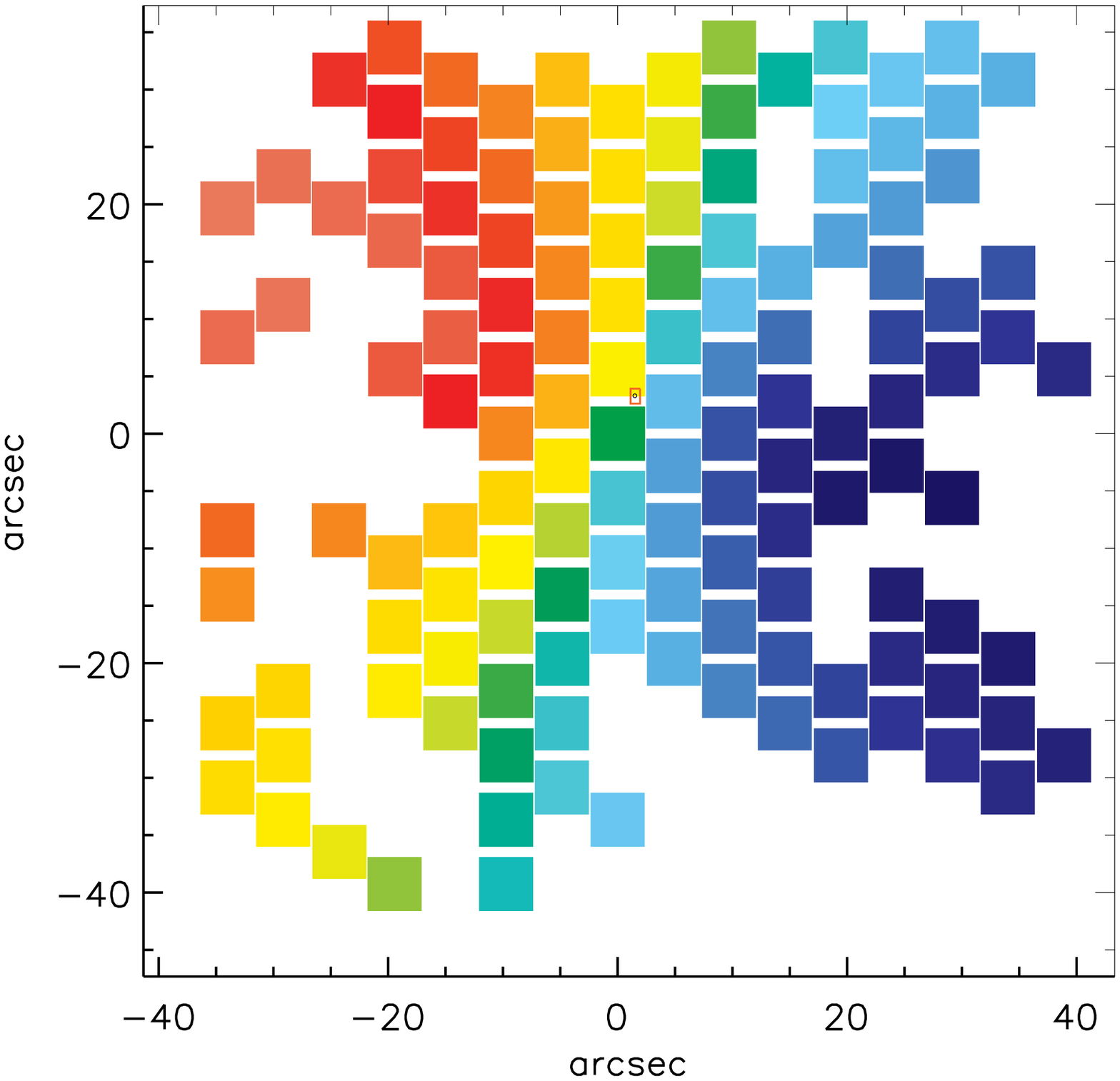} 
  \hspace*{-0.25cm}   \includegraphics[height=3.82cm]{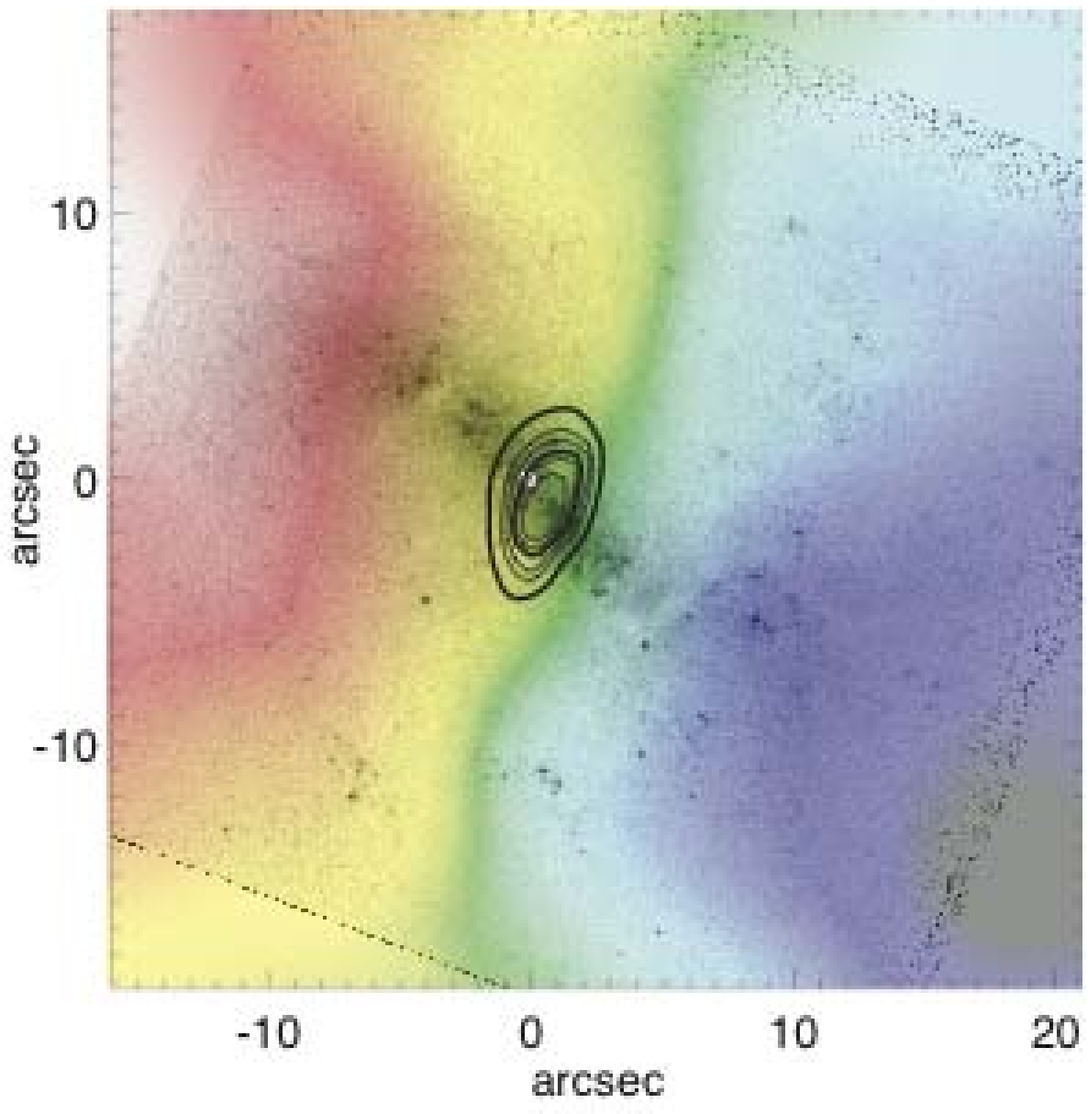}
  \vspace{-0.6cm}
   \end{tabular}
  \end{center}

 \begin{center}
  \begin{tabular}{cc}
  \hspace*{-1.3cm} \includegraphics[height=4.5cm]{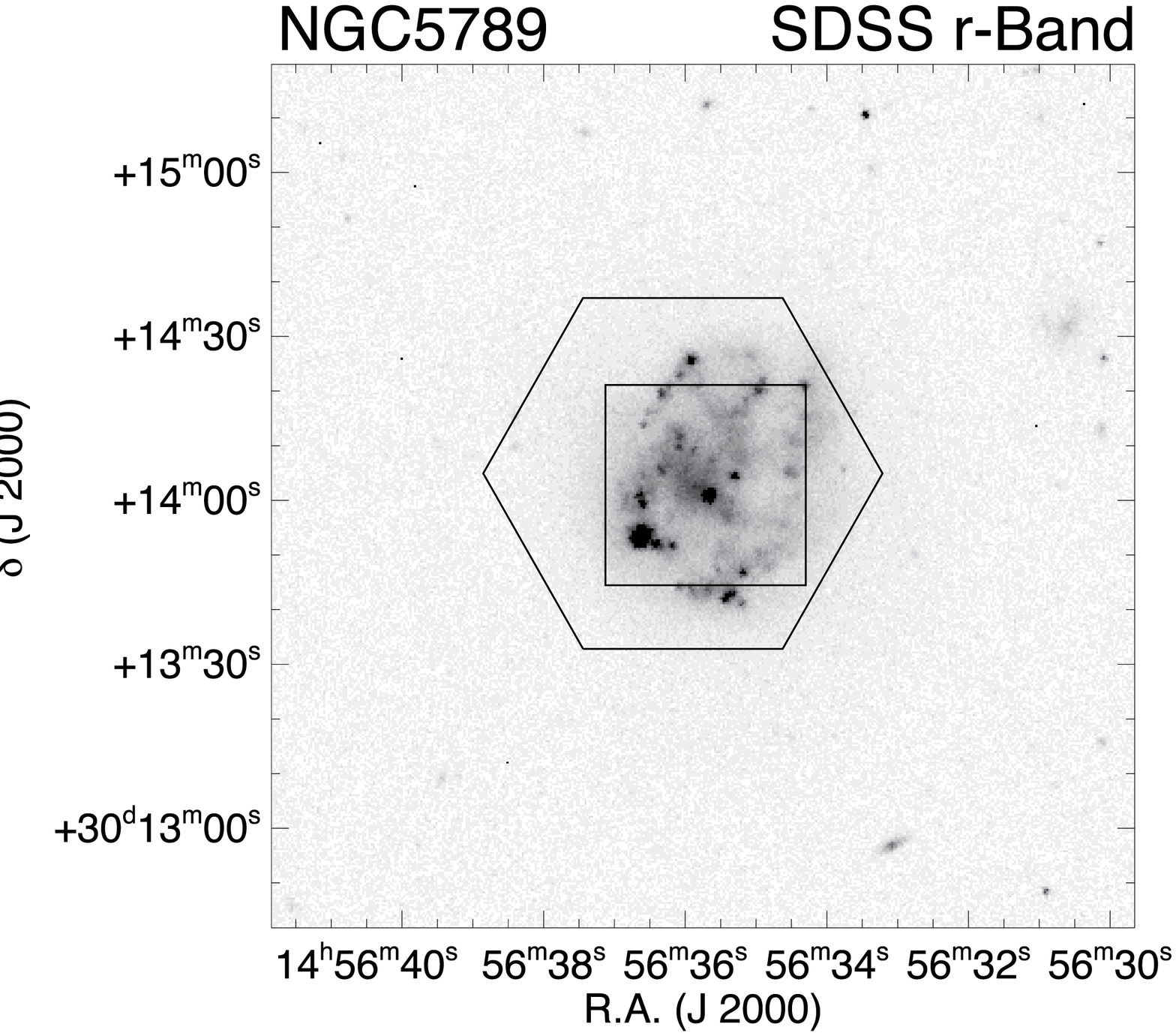}
   \hspace*{-0.25cm} \includegraphics[height=4.1cm]{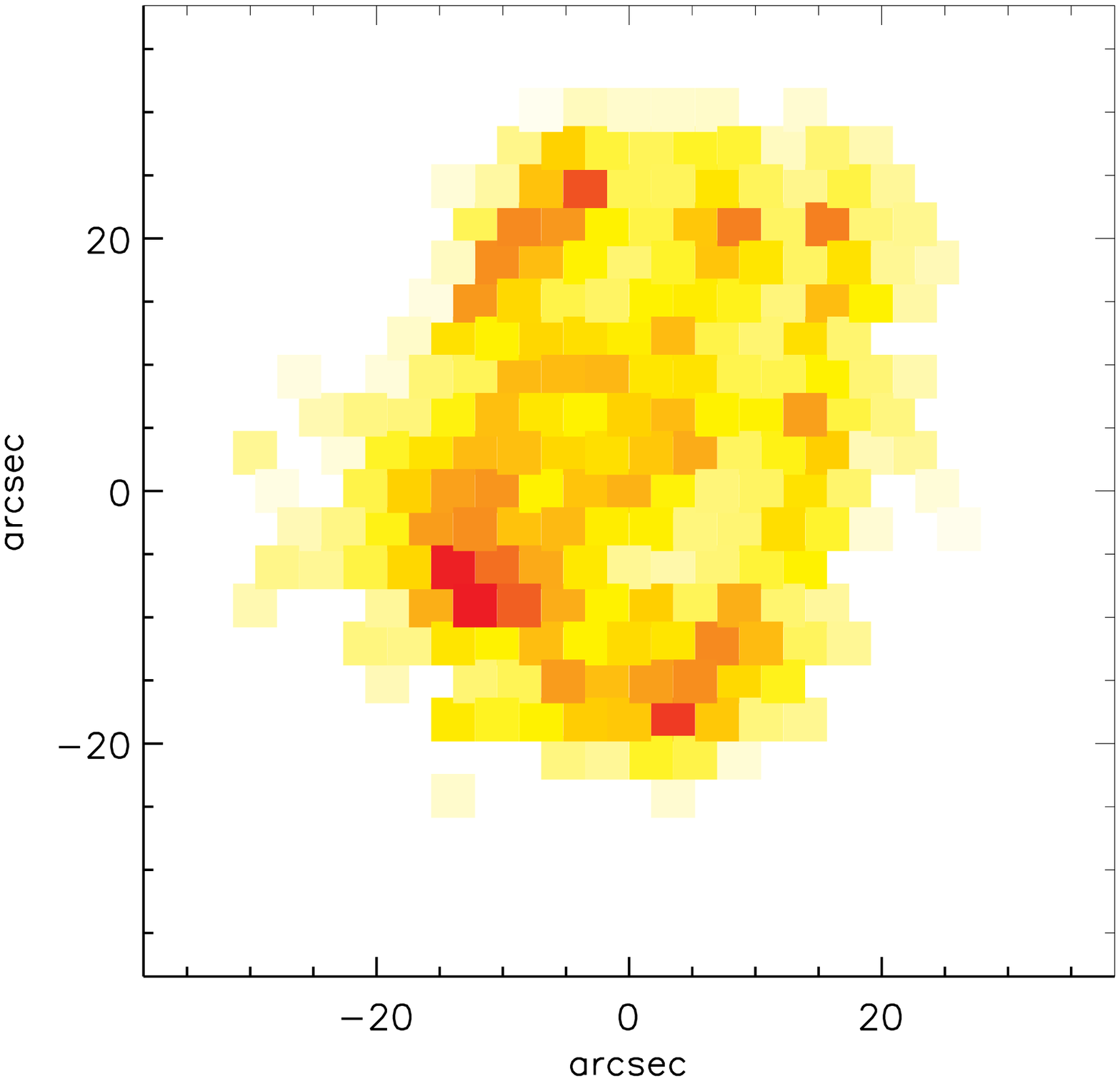}
   \hspace*{-0.25cm}   \includegraphics[height=4.1cm]{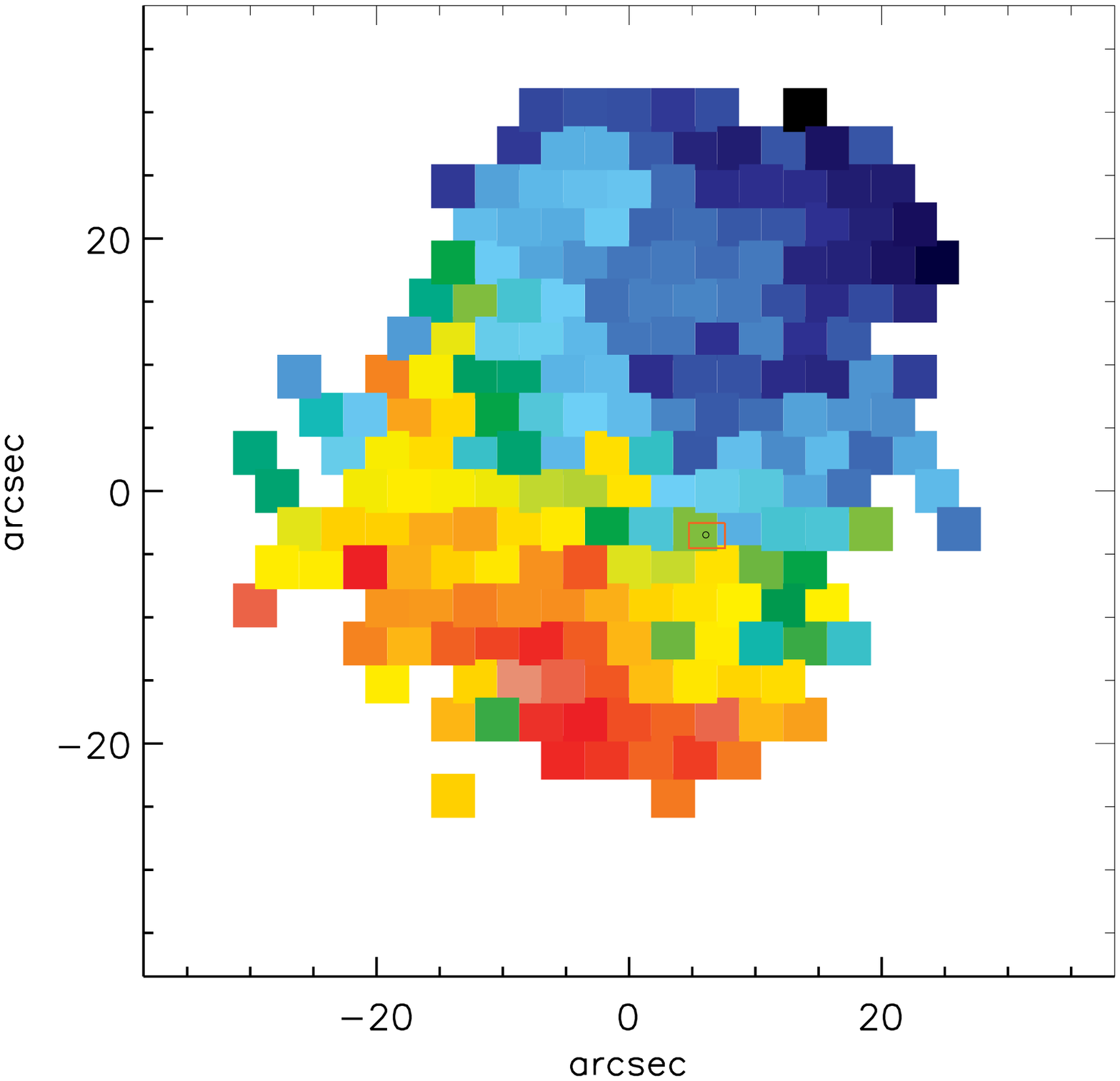}
   \hspace*{-0.25cm}   \includegraphics[height=4.1cm]{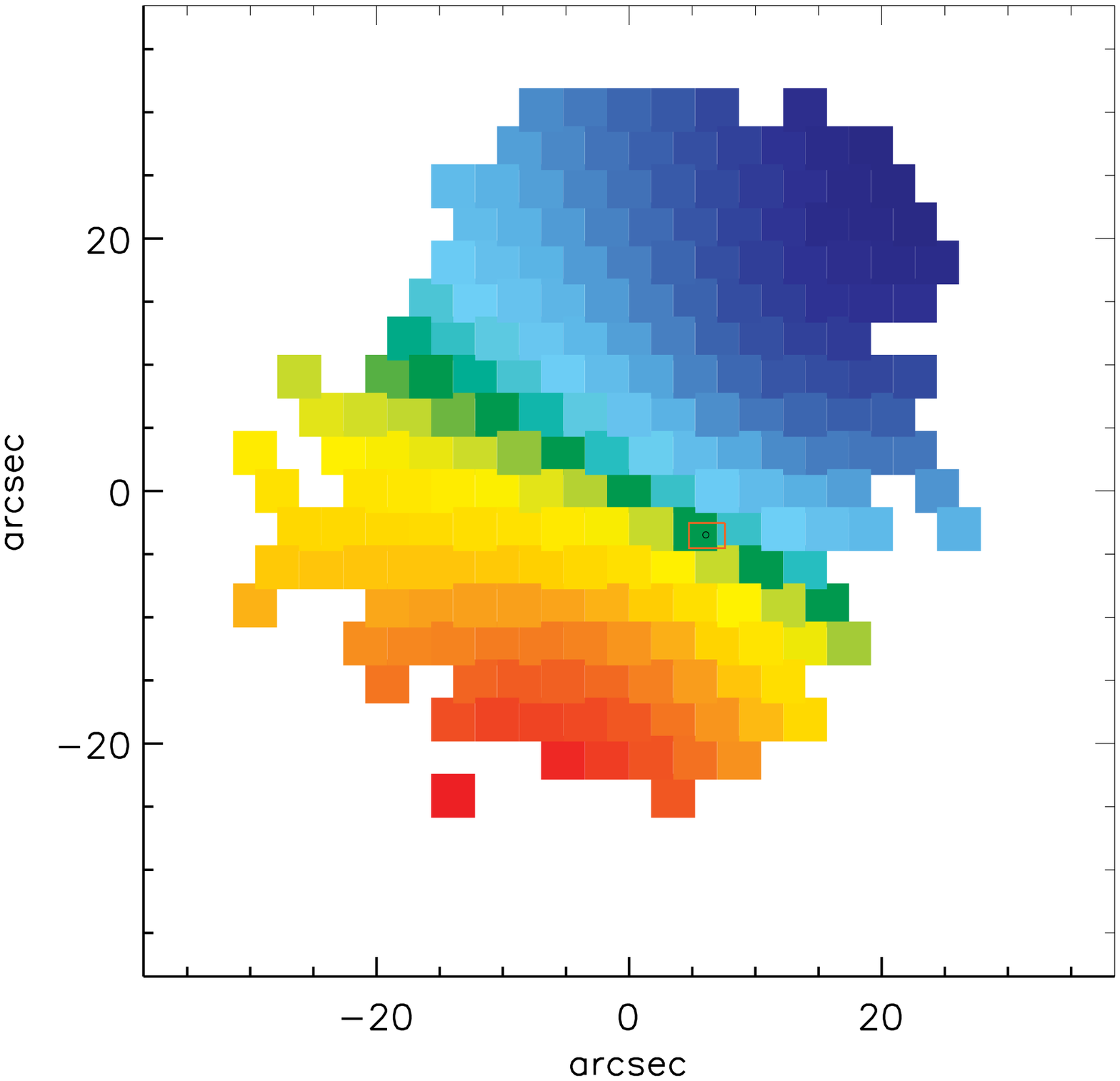} 
  \hspace*{-0.25cm}   \includegraphics[height=3.82cm]{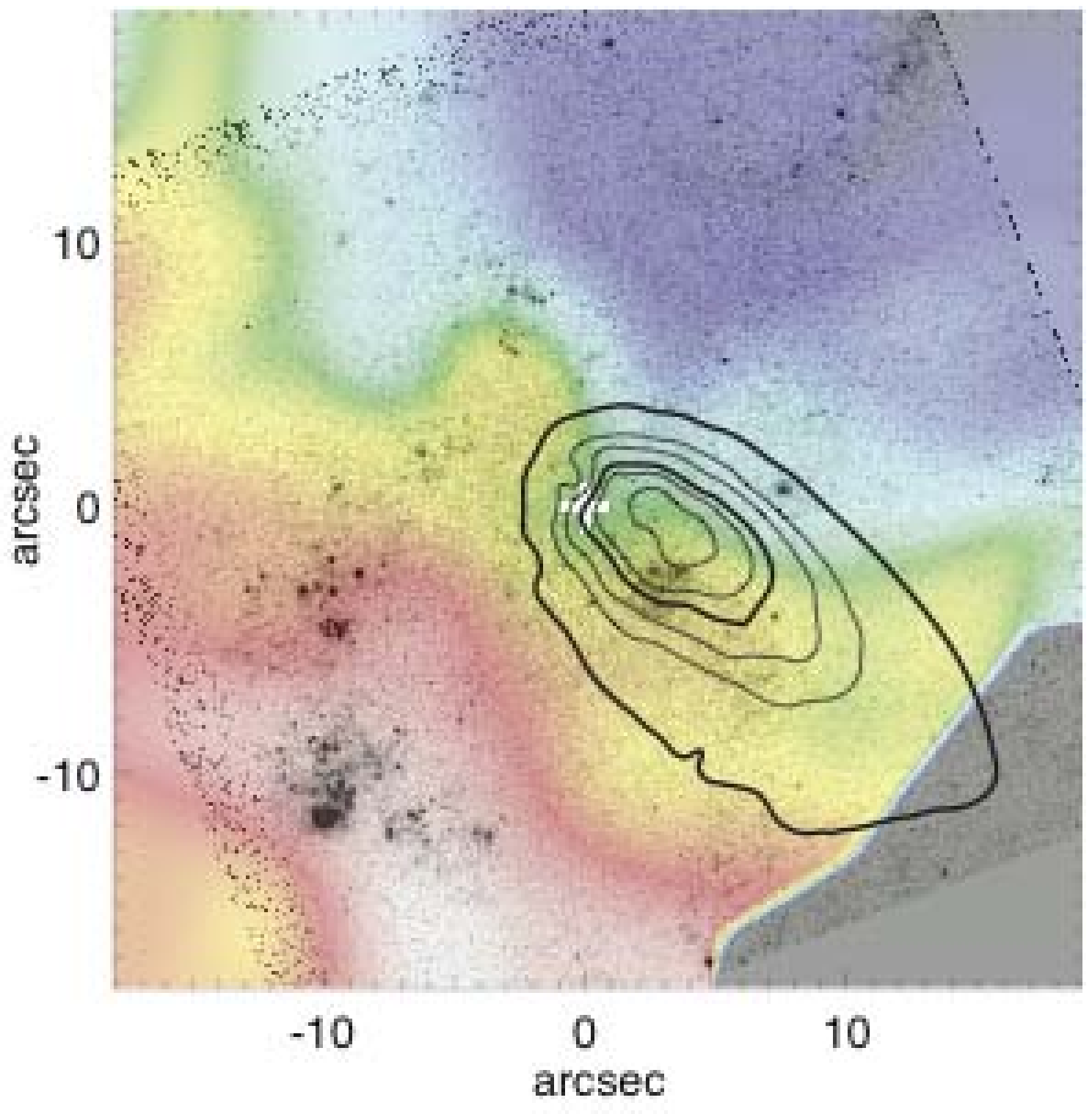}
   \vspace{-0.6cm}
  \end{tabular}
  \end{center}
  
 \begin{center}
  \begin{tabular}{cc}
\hspace*{-1.3cm} \includegraphics[height=4.5cm]{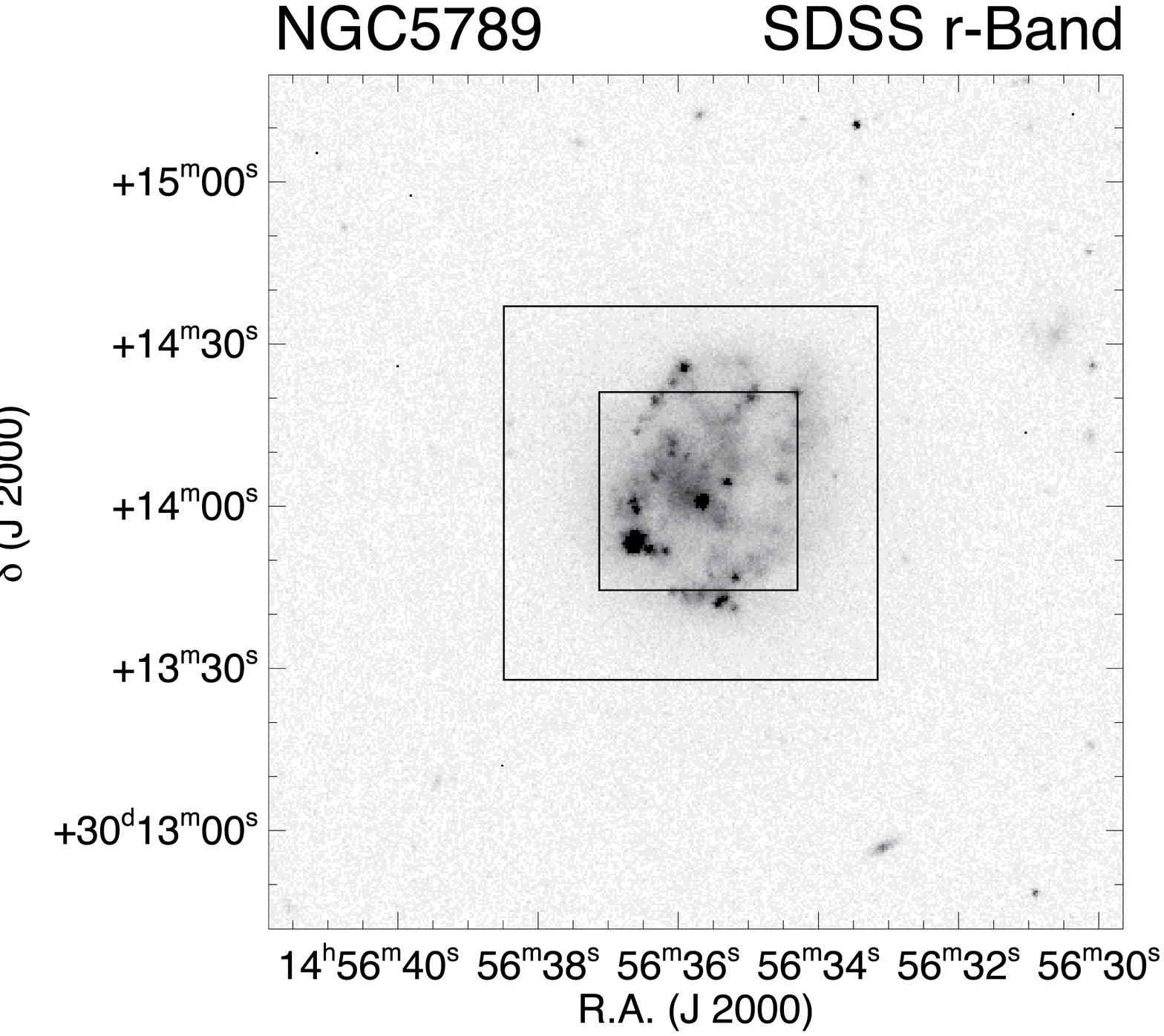}
   \hspace*{-0.25cm} \includegraphics[height=4.1cm]{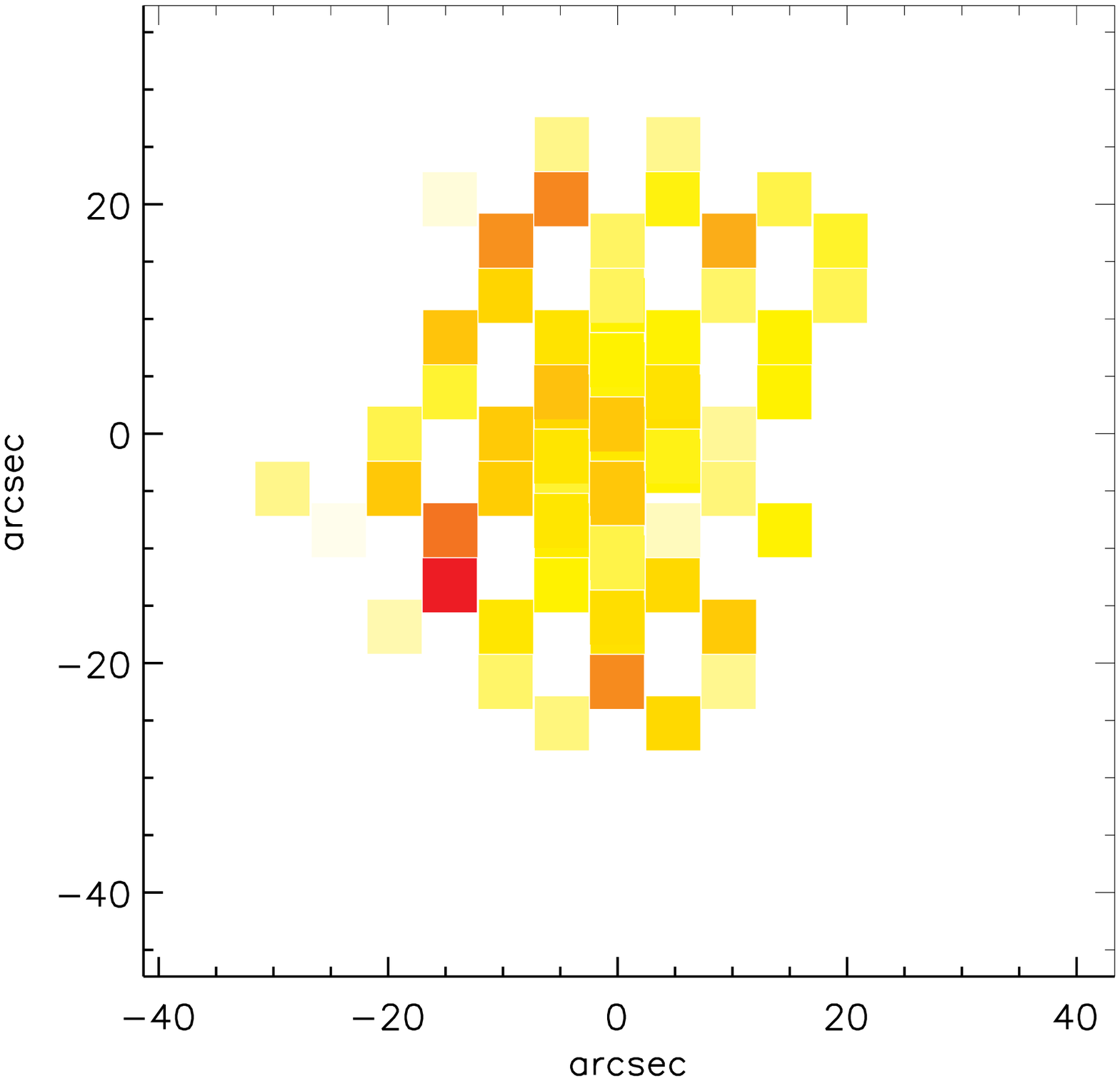}
   \hspace*{-0.25cm}   \includegraphics[height=4.1cm]{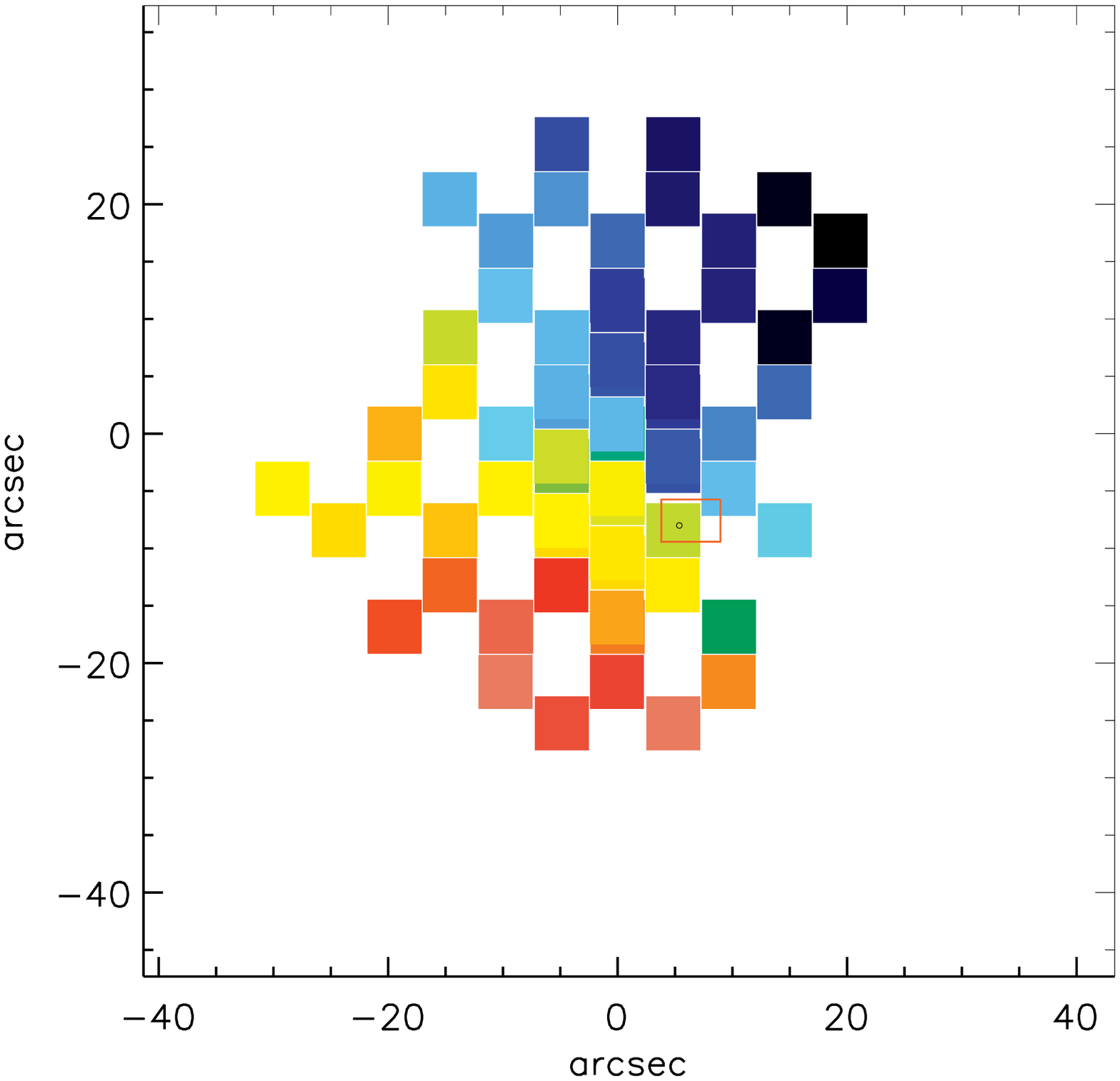}
   \hspace*{-0.25cm}   \includegraphics[height=4.1cm]{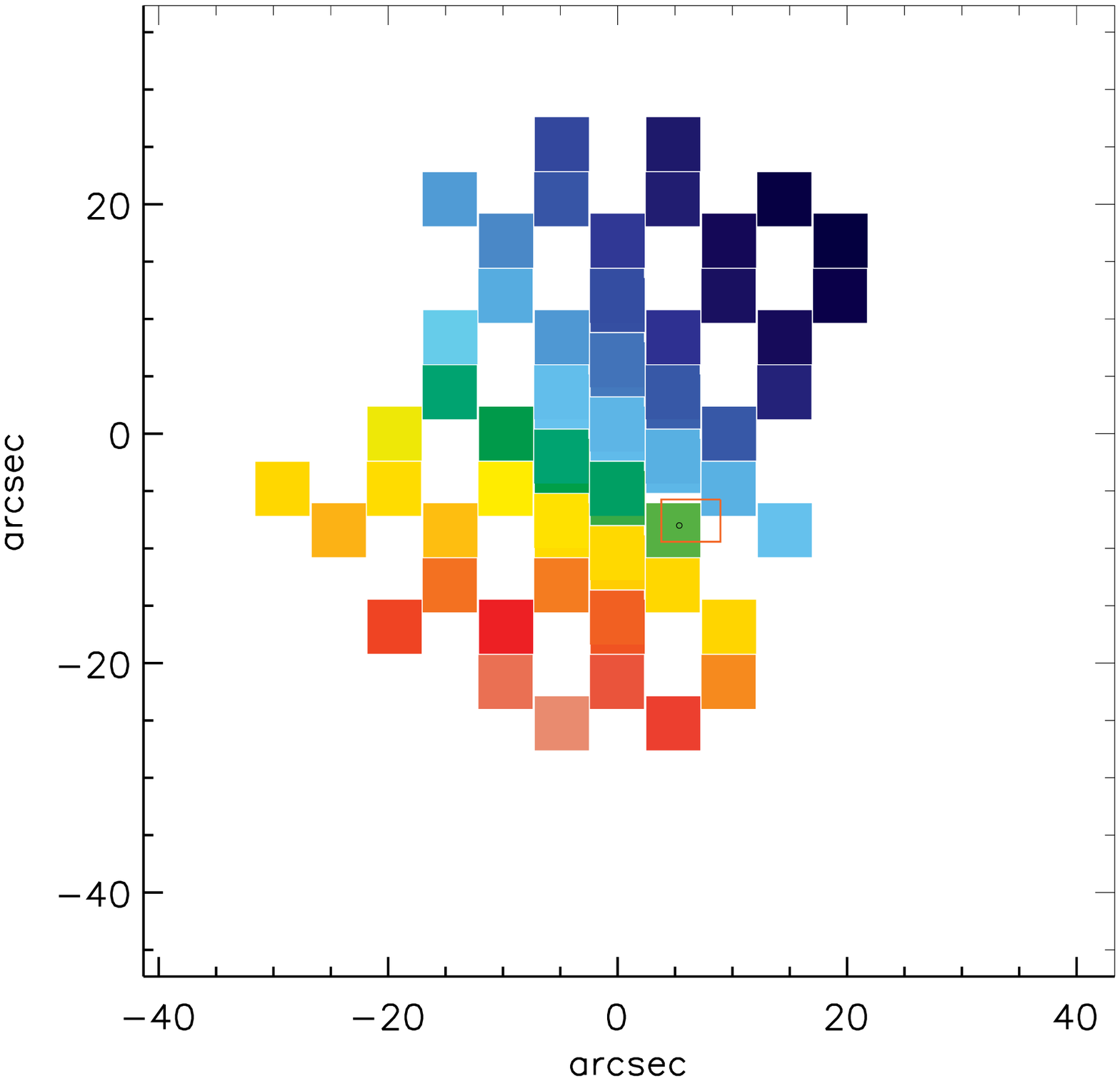} 
  \hspace*{-0.25cm}   \includegraphics[height=3.82cm]{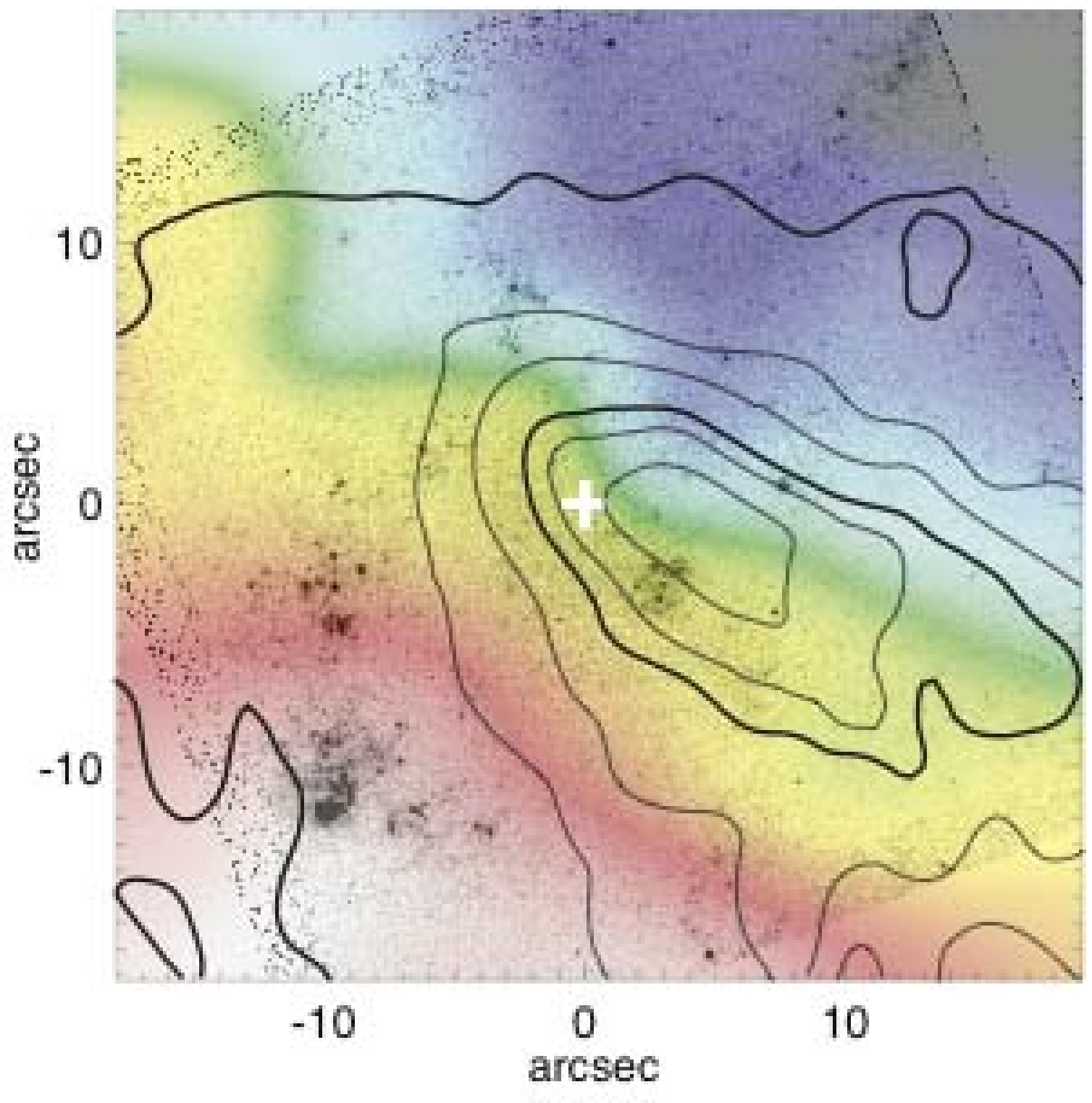}
  \end{tabular}
  \end{center}
  
  \contcaption{
}
  \end{figure*}
    
  \begin{figure*}
 
  \begin{center}
  \begin{tabular}{cc}
  \hspace*{-1.3cm} \includegraphics[height=4.5cm]{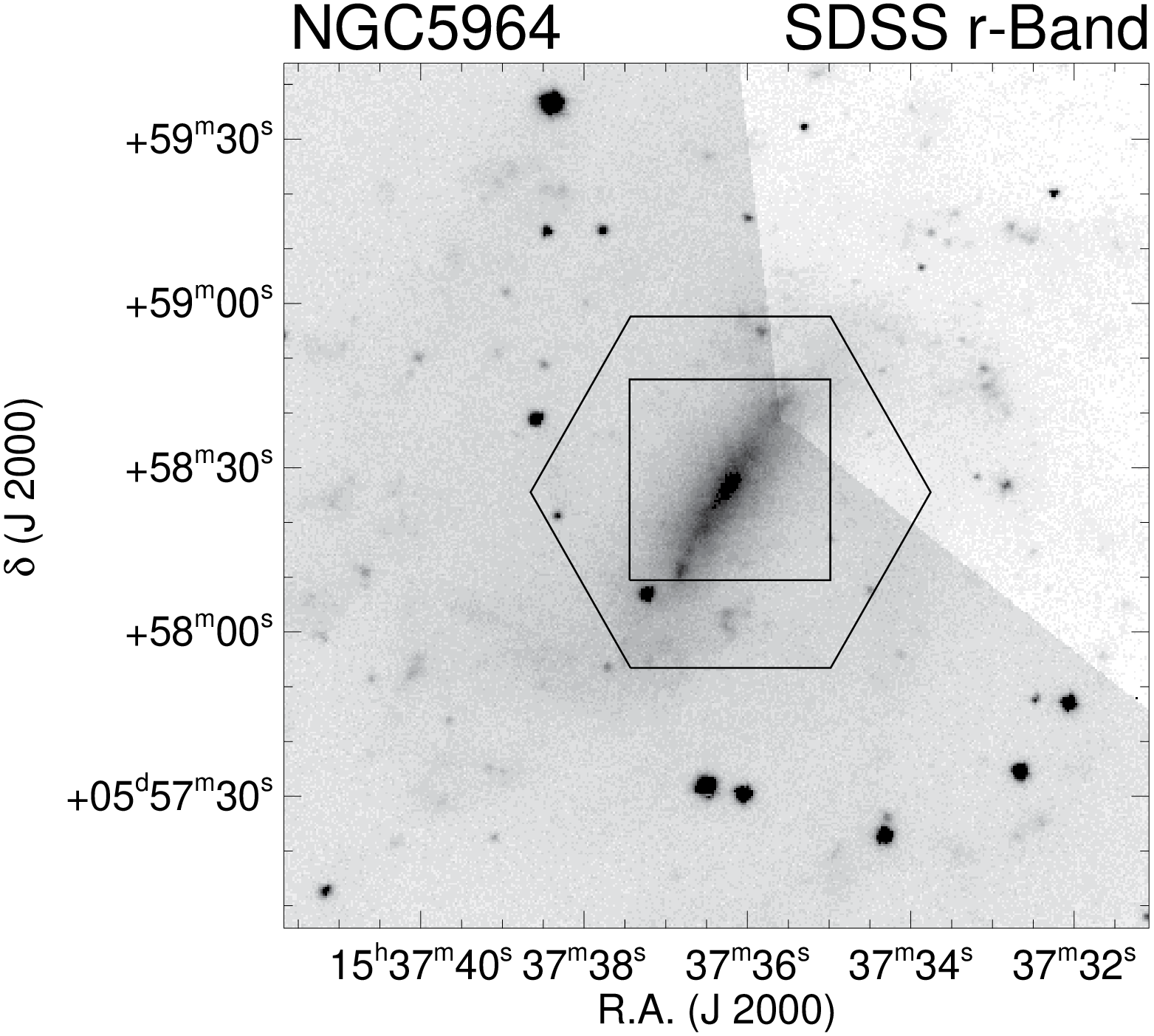}
   \hspace*{-0.25cm} \includegraphics[height=4.1cm]{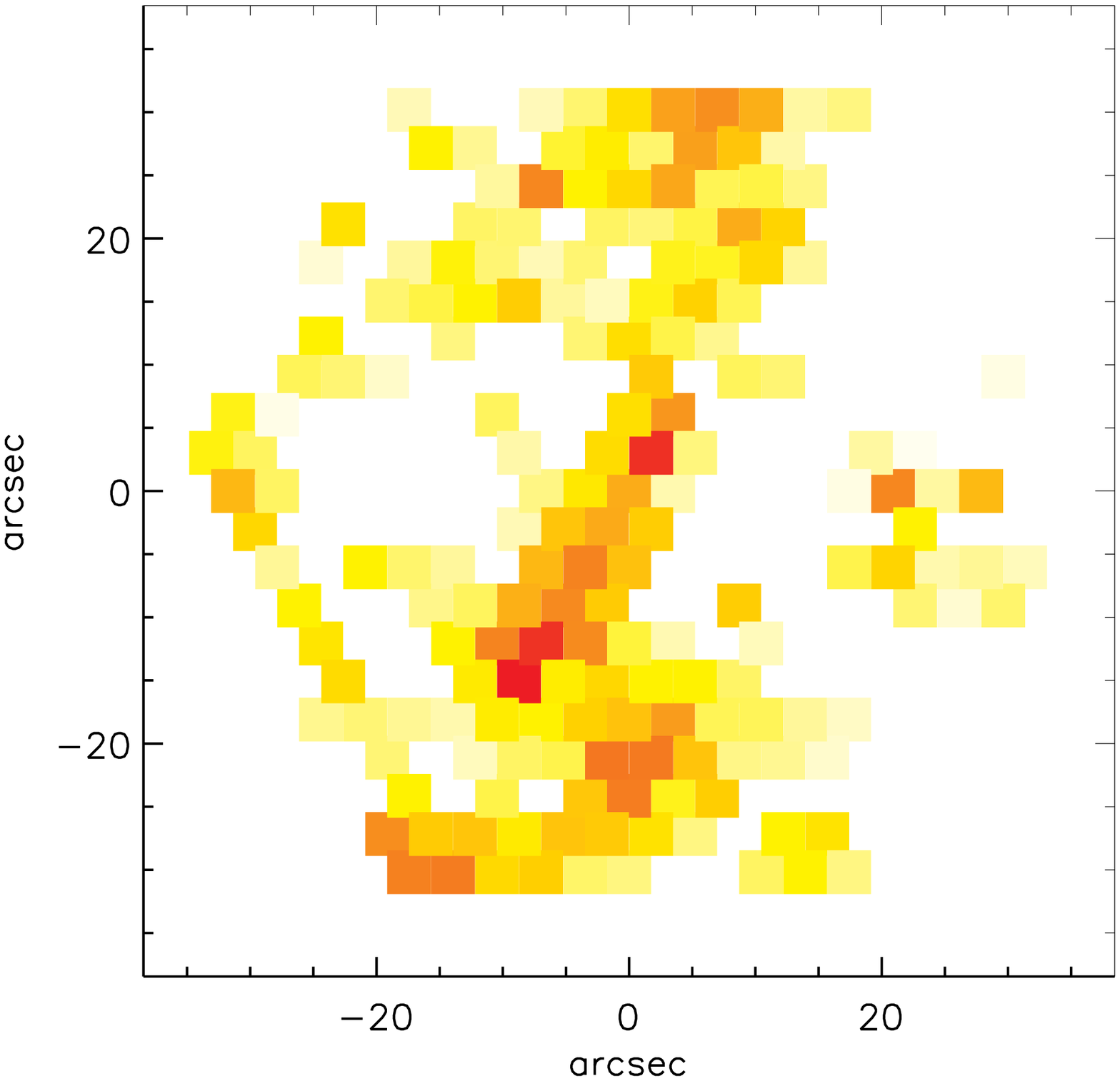}
   \hspace*{-0.25cm}   \includegraphics[height=4.1cm]{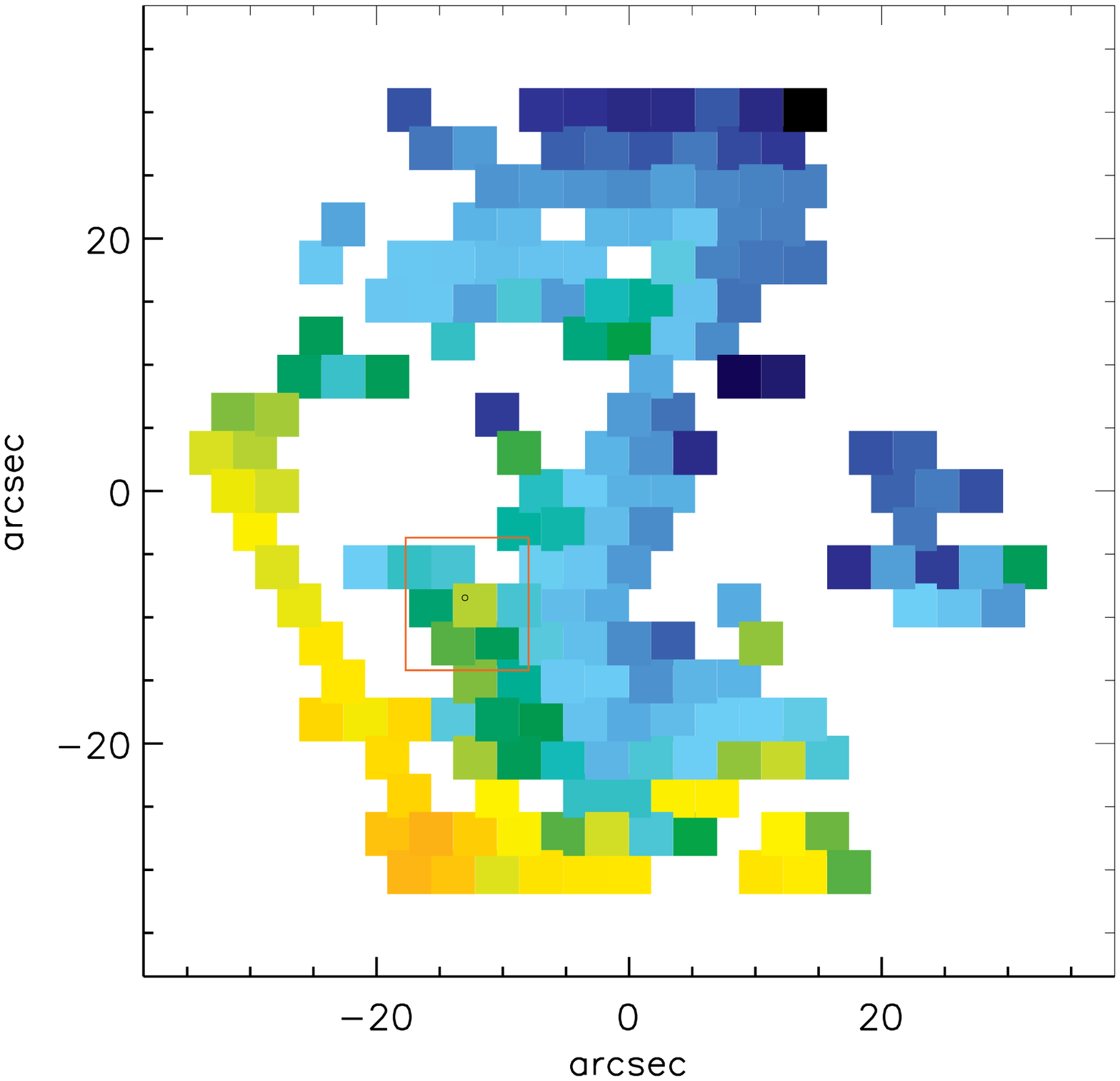}
   \hspace*{-0.25cm}   \includegraphics[height=4.1cm]{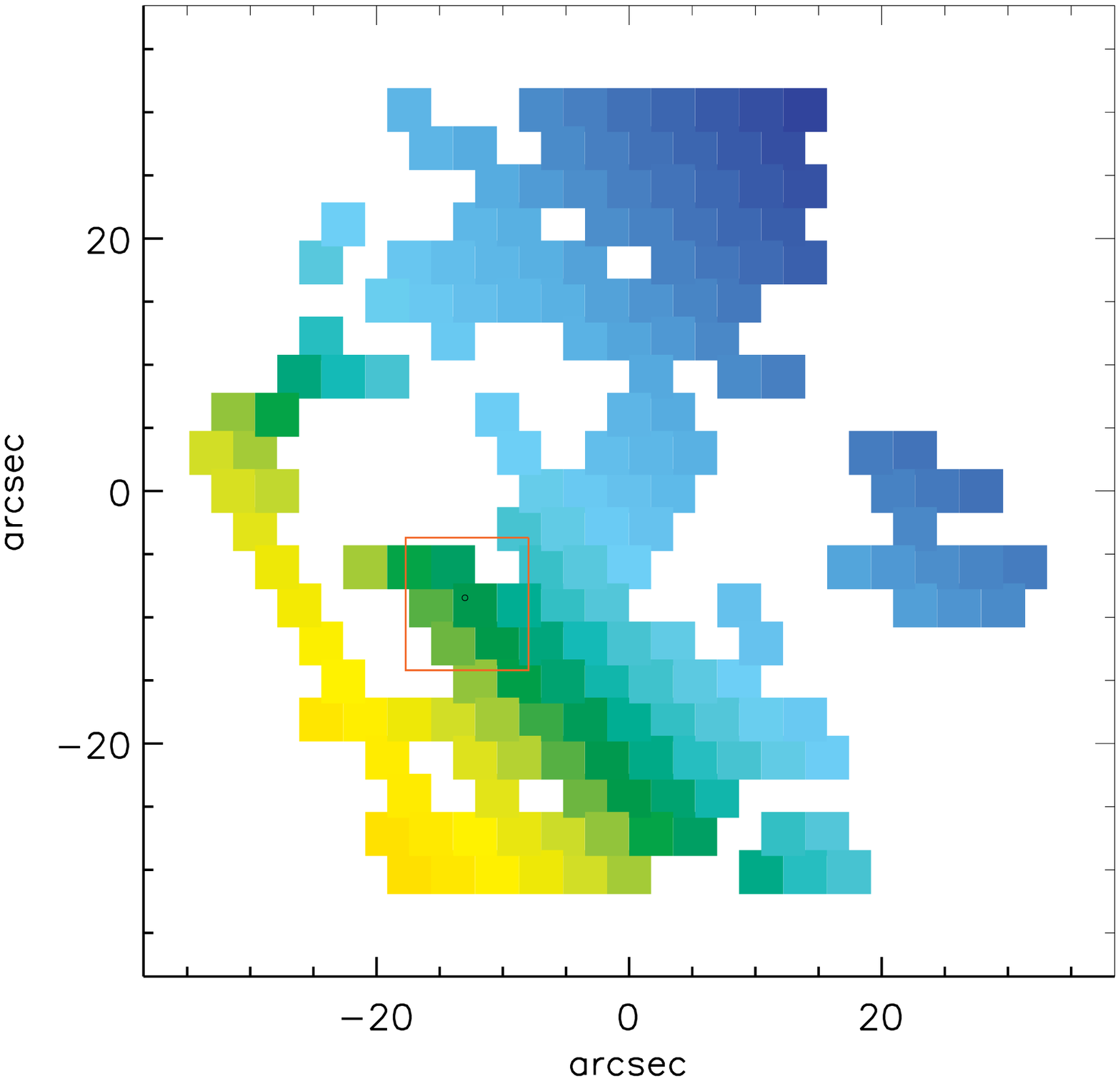} 
 \hspace*{-0.25cm}  \includegraphics[height=3.82cm]{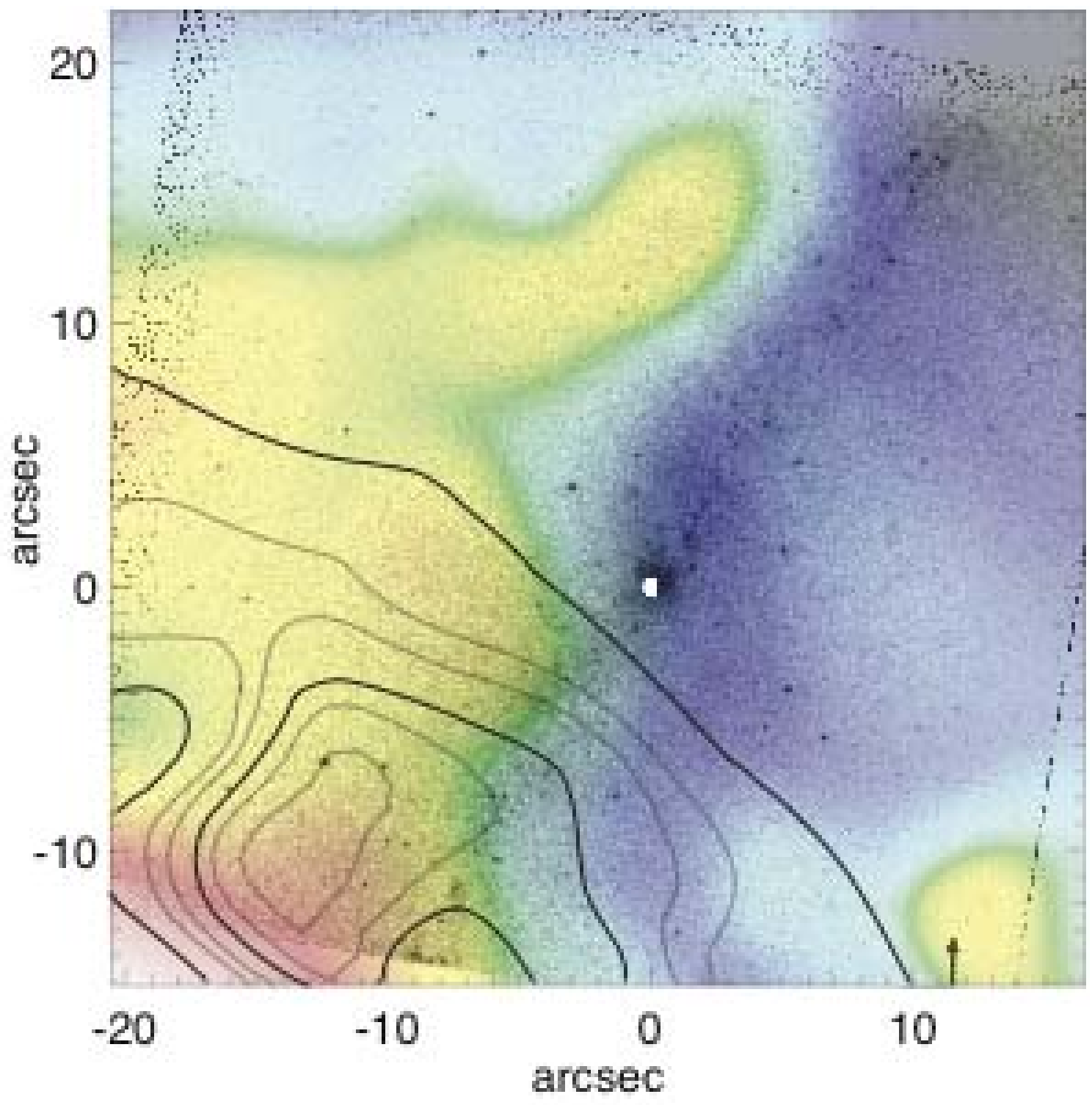}
   \vspace{-0.6cm}
  \end{tabular}
  \end{center}

  \begin{center}
  \begin{tabular}{cc}
  \hspace*{-1.3cm} \includegraphics[height=4.5cm]{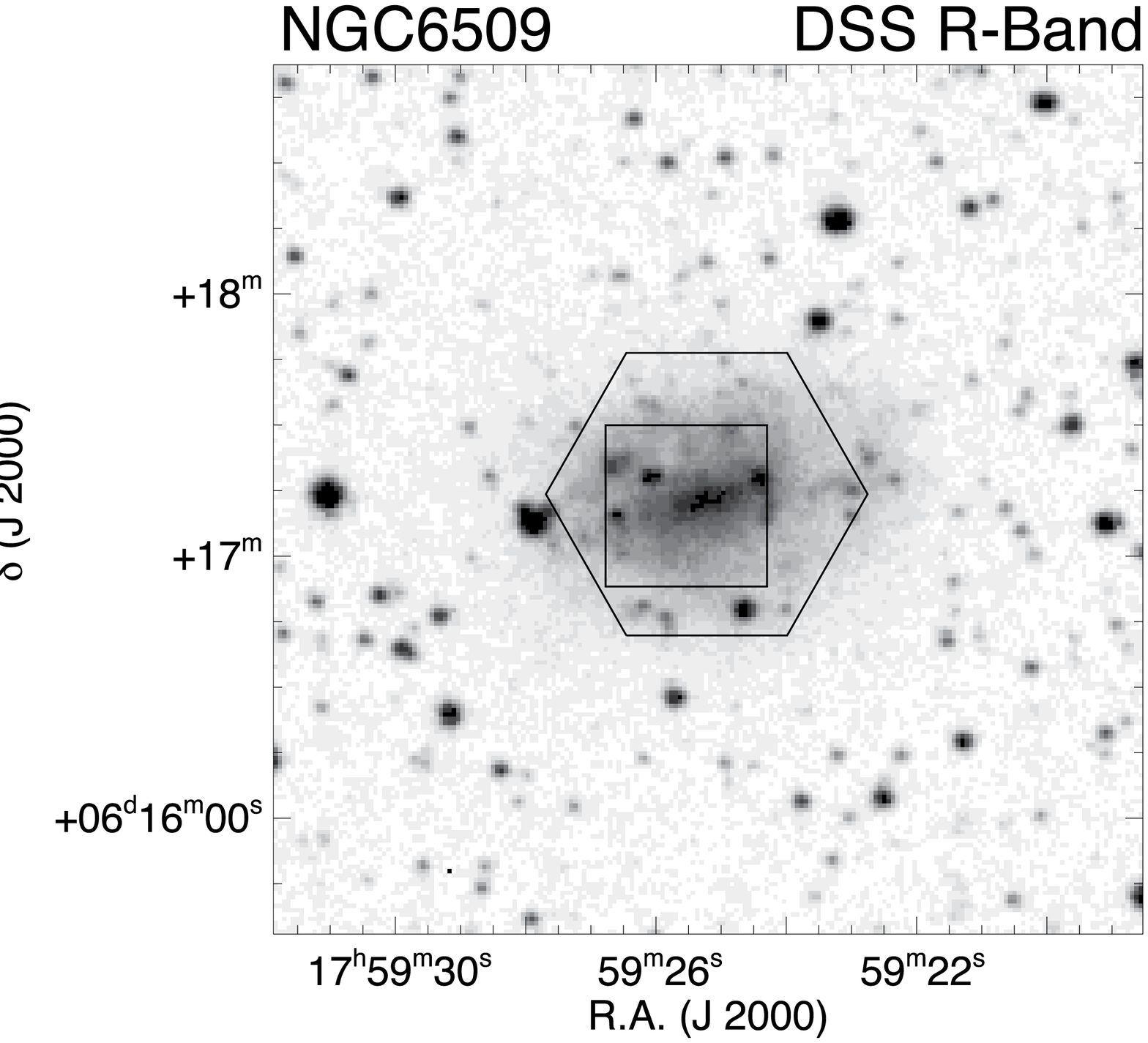}
   \hspace*{-0.25cm} \includegraphics[height=4.1cm]{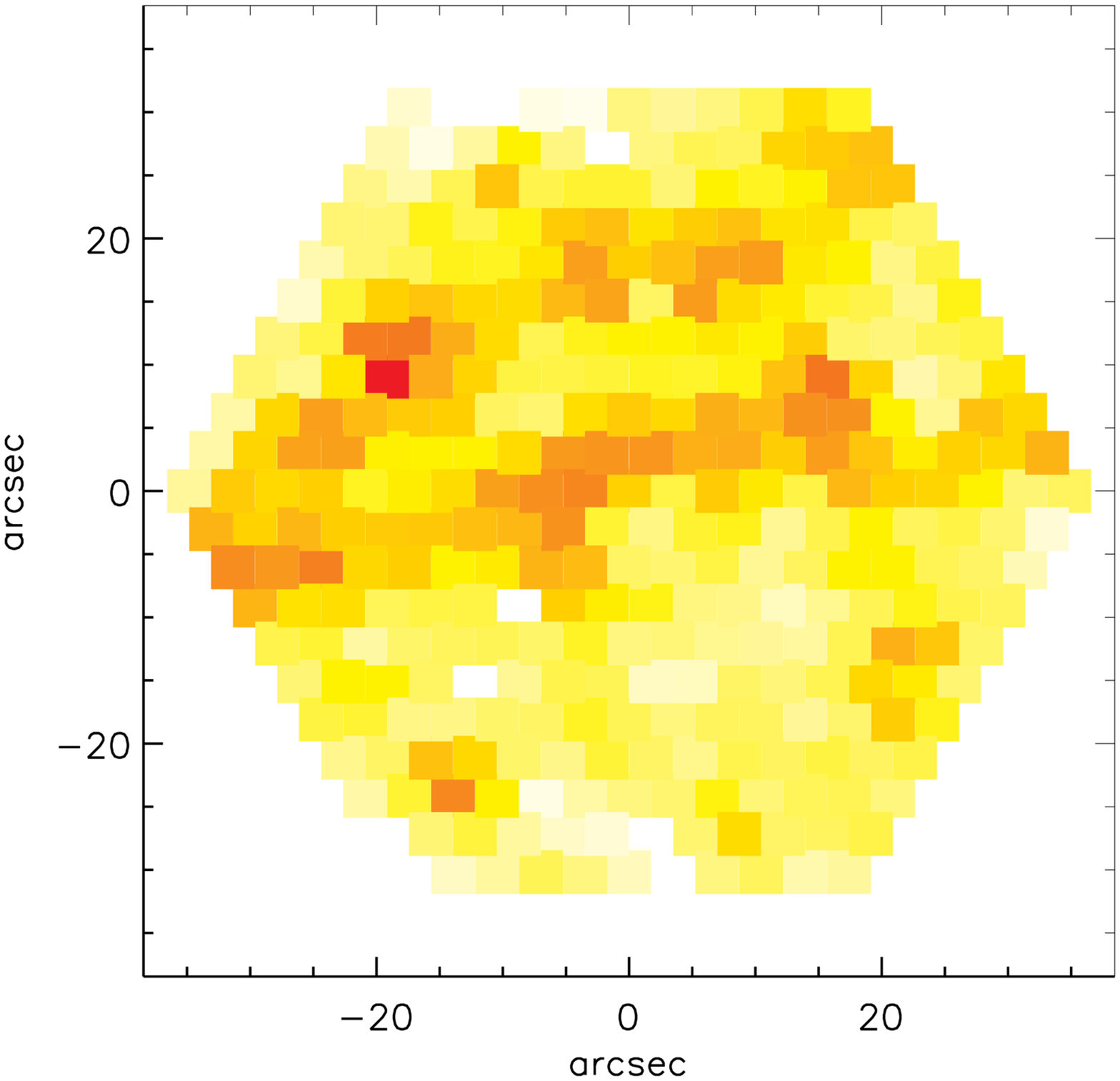}
   \hspace*{-0.25cm}   \includegraphics[height=4.1cm]{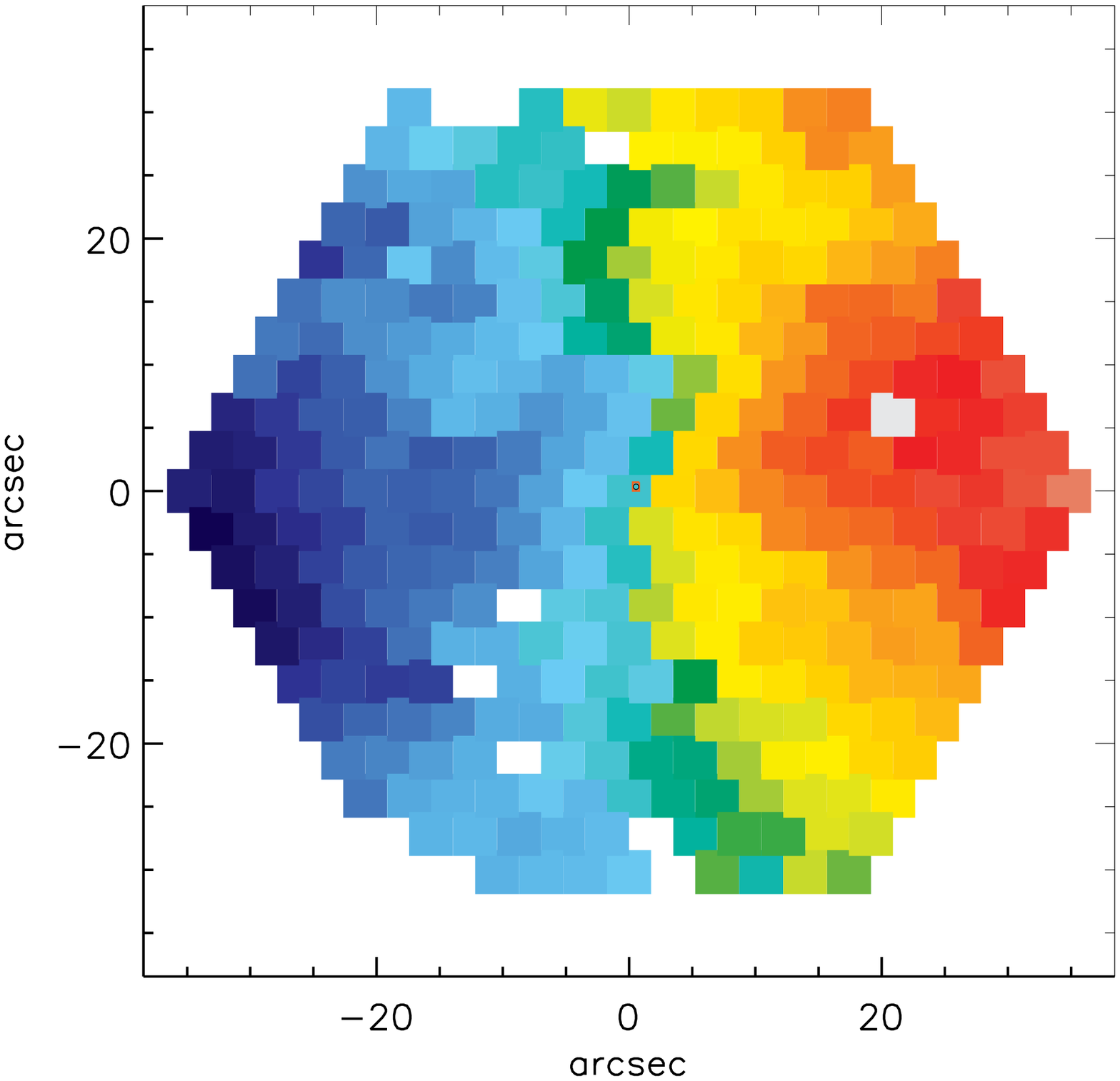}
   \hspace*{-0.25cm}   \includegraphics[height=4.1cm]{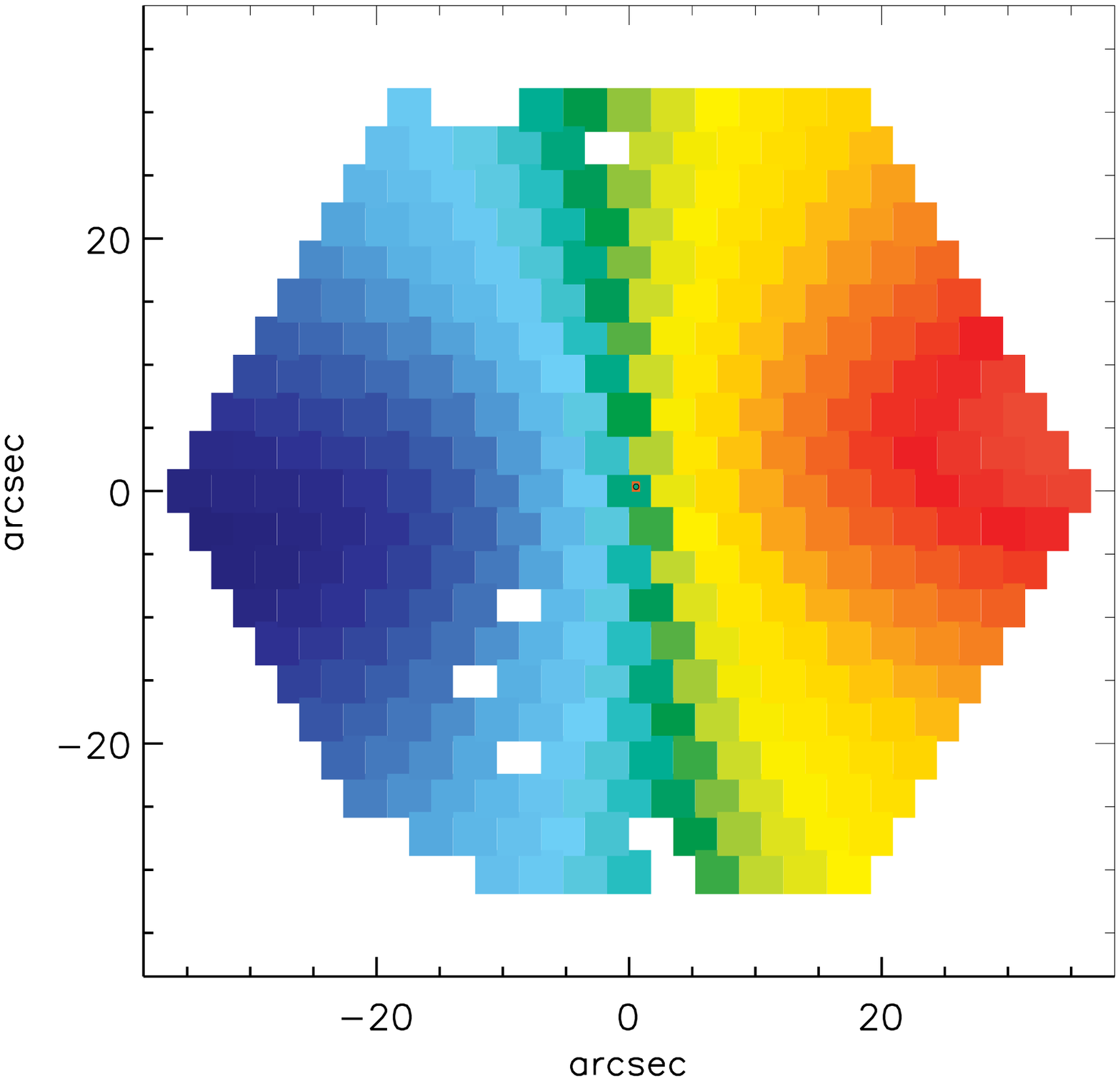} 
  \hspace*{-0.25cm} \includegraphics[height=3.82cm]{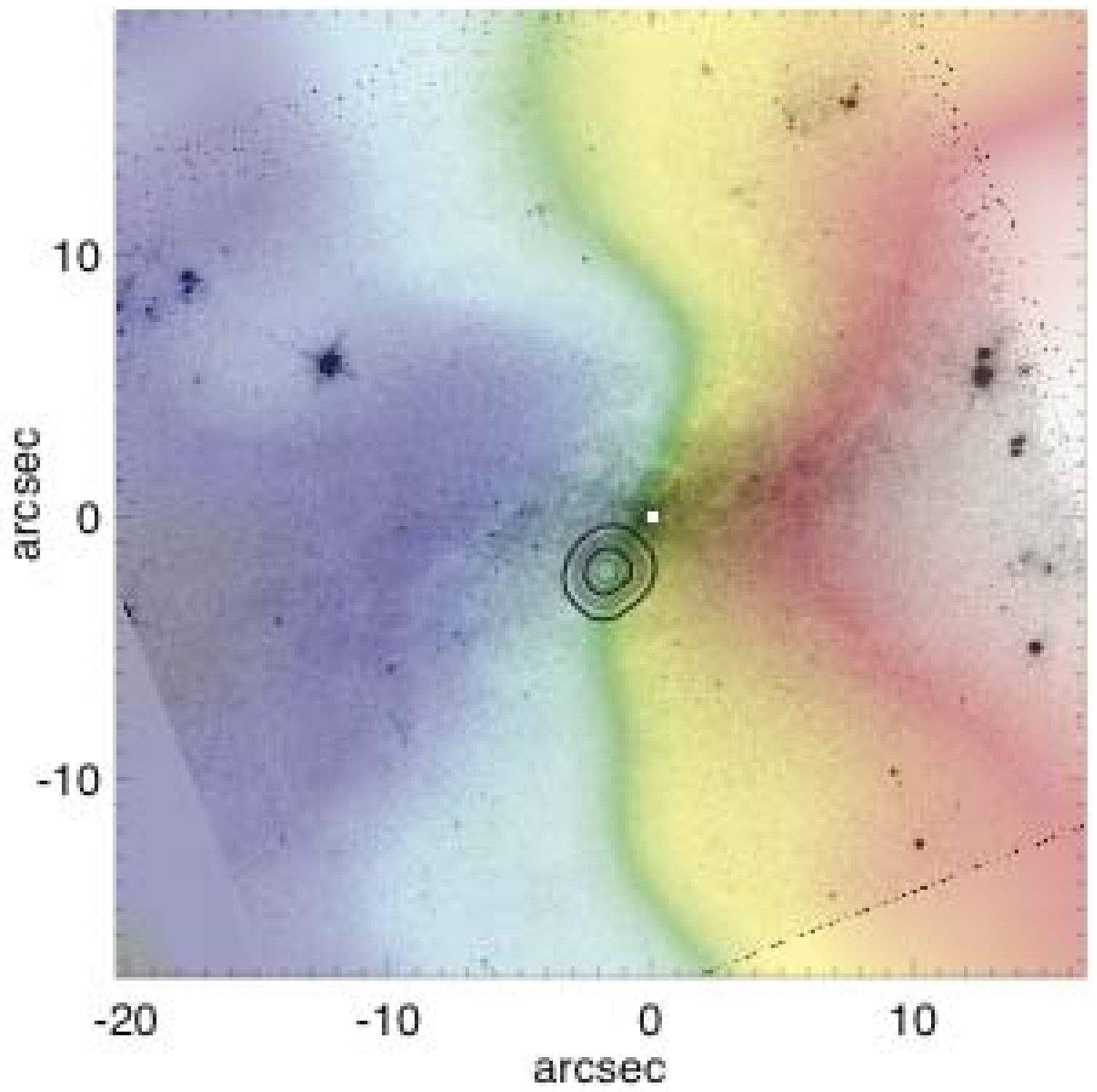}
  \end{tabular}
  \end{center}

  \contcaption{
}
  \end{figure*}
  

A further consistency check can be performed by comparing the dynamical models derived from 
\ppak\ and \spak\ data for NGC\,4904 and NGC\,5789, the two galaxies which have been observed 
with both instruments. As can be seen from Fig.~\ref{fig:mc1}, the observations agree well in both 
cases. The fit results also agree well, both in the location of the KC, and in the overall 
shape of the $\chi^2$ contours, despite marked differences in the spatial sampling. We take this as 
confirmation of both the quality of our data and the robustness of our analysis method.

However, our main uncertainty in the velocity field modelling is the actual complexity of the velocity fields. Our simple 
model does e.g.~not include the effects of streaming motions due to bars or the effects on the velocity 
field due to interactions with a companion galaxy. We have performed an eye-ball check of our trust in the velocity 
field modelling, by looking at the spacing of 
the $\chi^2$ contours of the fit to the dynamical centre (see Fig.~\ref{fig:mc1} column 5), weighing in the presence/absence 
of a bar/companion, the filling factor of Halpha emission, 
the overall regularity of the velocity field and the quality of the fit to the rotation curve.
On that basis we have assigned quality flags 
from 0 (unusable) over 1 (low trust) to 2 (well modelled) to each velocity field. This quality 
assessments number is added as a column in Table~\ref{table:pos}. This exercise has been somewhat sobering as to 
our ability to systematically determine dynamical galaxy centers, as only 7 velocity fields out of 22 obtain 
the label ``well modelled''. Only two of these are neither barred nor show any other sign of problems 
(NGC\,3423, NGC\,3206). Late type disk galaxies tend to be irregular and a 
well-defined center simply does not exist in quite a few of them. The implications of this will be further 
discussed in Section \ref{sec:discussion}.

\subsection{Comparing kinematic centers and photocenters} 

\begin{figure*}
\begin{center}
\hspace*{-0.5cm} \includegraphics[width=9cm]{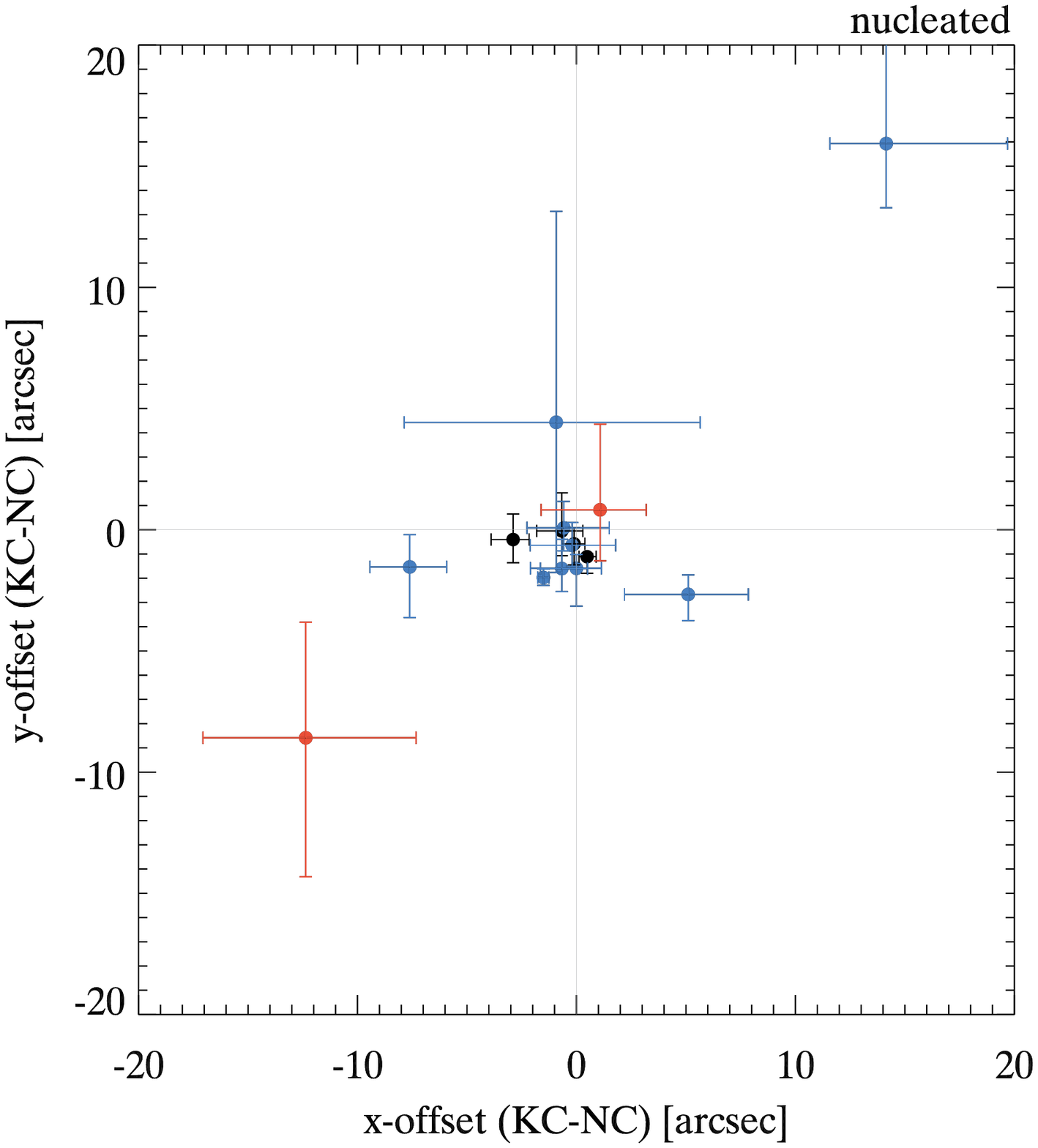}
\hspace*{-1cm} \includegraphics[width=9cm]{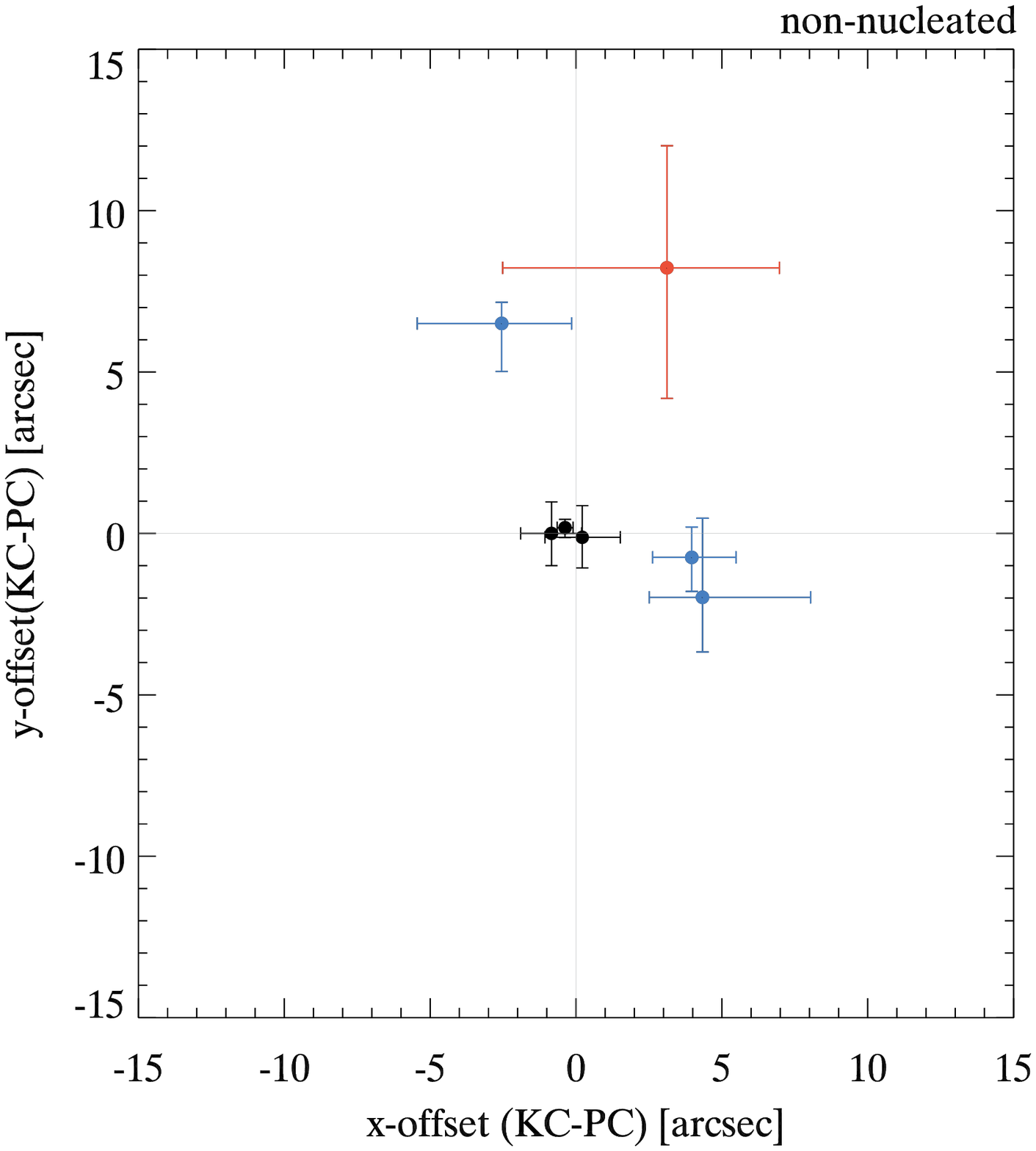}
\end{center}
\caption{   \label{fig:offsets}
{\it Left}: Projected position of all nuclear clusters, relative to the kinematic centers of their respective host galaxy. 
Size of the crosses denotes the $1 \sigma$ errorbar in the kinematic centers.  {\it Right}: Offset of the 
photometric center relative to the kinematic center of the non-nucleated galaxies. Here, the 
size of the crosses denotes the combined $1 \sigma$ error in the kinematic plus photometric centers. 
The colors of the symbols indicate how well the velocity fields can be modelled. 
{\it Black} symbols indicate a model with high fidelity, {\it blue} symbols denote models with low trust, 
and {\it red} symbols denote galaxies that can not be well modelled due to interactions, or non-circular 
motions (see text for details).  }
\end{figure*}

Following image registration as described in \S~\ref{subsec:registration}, we can measure the
absolute coordinates of the best fit KC. Table~\ref{table:pos} 
summarizes the
KC positions, along with those of the PC and NC (if present). 
The latter two were derived from isophotal fits to the WFPC2 image, as described in \citet{boker02}.
All coordinates in Table~\ref{table:pos} 
refer to the J2000 coordinate system of the HST image, which 
has an absolute accuracy with respect to the ICRS system of $\sim 1$\arcsec. In Fig.~\ref{fig:offsets}, 
we compare the relative positions of the NC and the KC as well as the PC 
and the KC for the non-nucleated galaxies.

Generally, we find that there are only a few galaxies (6 out of 20) with a velocity field regular enough 
(trust level 2 in Table \ref{table:pos}) to meaningfully compare the locations of NC, PC and 
KC. For all of these galaxies these locations agree within the uncertainties, although 
some of the galaxies show bars.  On the other hand, for all but one galaxy with offset KC we can always 
identify a reason. NGC\,2552, NGC\,4299, NGC\,5789, have a very fuzzy or patchy 
appearance in \ha\ and in stellar emission. These galaxies may simply not be in regular rotation. 
UGC\,5288, NGC\,4204, NGC\,4496, NGC\,5669, NGC\,5964, NGC\,6509 show a strong bar and the 
velocity field is clearly affected by large scale streaming motions along the bar.ÊNGC\,4625 is lopsided 
from a strong m=1 mode, which seems to displace the KC. There are only two potentially odd cases: in 
NGC4517  there is a large offset between KC and NC. However, this galaxy probably has a nuclear starburst 
(see \ha\ map) that affects the central part of the velocity field, although a possible bar is weak if existent.Ê
UGC\,3574 looks very regular, yet NC and KC show an offset of the size of the 1 $\sigma$ error bar. 
However, in this galaxy the central $5\arcsec$ unfortunately do not show \ha\ emission, which might hamper 
our ability to determine an accurate KC.Ê We are left with no confirmed offset between the NC and the KC.

\subsection{Rotation curves and velocity residuals}

\begin{figure*}
\begin{center}
\includegraphics[width=12cm]{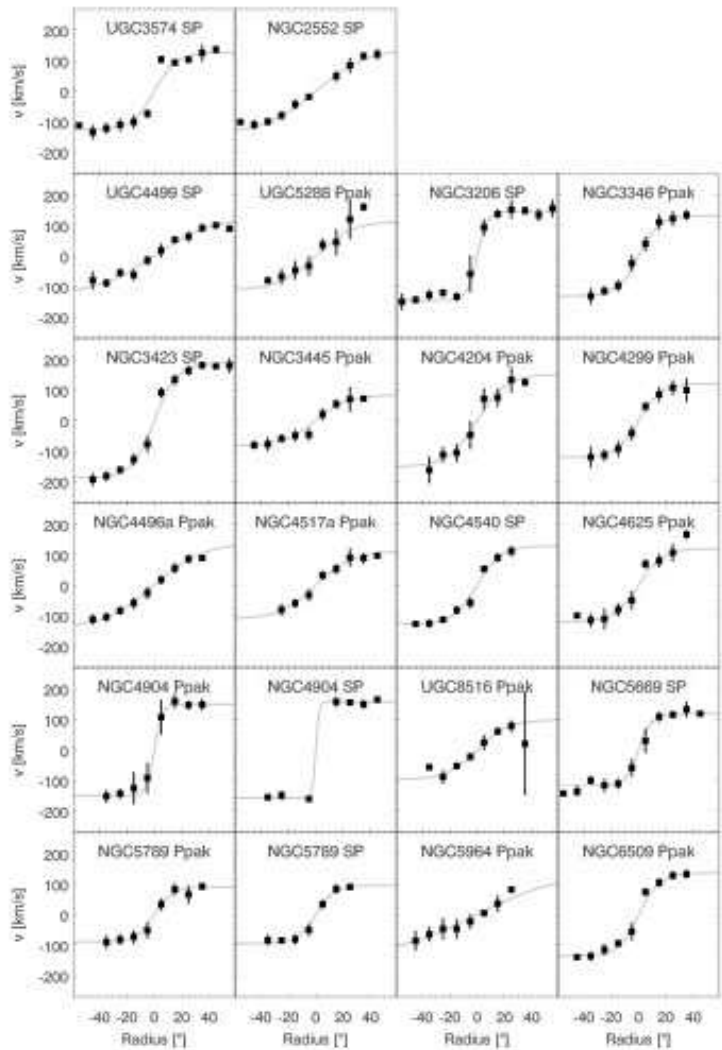}
\end{center}
\caption{ \label{fig:rotcurv}
The derived rotation curves for the galaxies in our sample. The lines overplotted to the data 
points give the best fit to Eq.~\ref{eq:rotcurve}. The abbreviation next to the galaxy name 
indicates whether the galaxy was observed with \spak (SP) or with \ppak.
}

 \end{figure*}

Rotation curves of our sample galaxies are shown in Figure \ref{fig:rotcurv}. As mentioned in Section 
\ref{subsec:kinmod} we find that the functional form given in Eq.~\ref{eq:rotcurve} represents the 
rotation curves of latest type spirals well in general. 
To quantify this statement in terms of the scientific goal of our paper we have used the rotation 
curves of our best models (trust level 2 in Table~\ref{table:pos}) and our bad models (trust level 0). 
There is no obvious trend with nucleatedness (4 out of 6 well-fit galaxies are nucleated and 1 
out of 3 not-well-fit galaxies). 

Those rotation curves which are least consistent with the arctan functional form are dominated by irregularities 
of their host galaxies. The irregular rotation curve of NGC\,4625 can be explained by an 
ongoing interaction, while NGC\,5964 is strongly barred, and UGC\,4499 has a very sparsely 
sampled velocity field. 

The rotation curve shapes of late type galaxies have also been discussed in \cite{swaters09} from 
HI data. The morphological variety of their rotation curves is similar to ours.

\subsection{Dynamical Friction Timescales}
\label{sec:friction}
Based on Chandrasekhar's formula we calculate the dynamical friction timescales for star clusters in our 
galaxies using equation 8.12 of \cite{binney08}:
\begin{equation}
t_{\mathrm{fric}} = \frac{1.17}{ln\Lambda} \frac{r_i^2 v_c}{GM},
\label{eq:dyn_fric}
\end{equation}
where $ln\Lambda$ is the Coulomb logarithm, $v_c$ the typical velocity of the stars in the galaxy at radius 
r$_i$, and $M$ is the mass of the test particle that is going through the much larger mass of the underlying 
galaxy. Although \cite{binney08} derived Eq.~\ref{eq:dyn_fric} for the decay of black hole orbits in a singular isothermal
sphere, they explicitly state that it is approximately correct even for mass distributions other than the singular
isothermal sphere (given that the mass ratio of the subject body to the interior mass of the host $\ll 1$). 
For point like particles, $\Lambda \approx \frac{r_i v^2_{c}}{GM} \gg 1$, while for extended bodies the 
Coulomb logarithm is obtained from N-body simulations, that give values of $ln \Lambda \sim 2 - 7$ 
\citep{spinnato03,penarrubia04}. The calculation of  t$_{\mathrm{fric}}$ requires the typical stellar velocity, 
while we are actually measuring the velocity fields of the ionised gas. The circular velocity of the stars in the 
central region of disk galaxies will be 1.5-2 times slower than the ionized gas (Ganda et al. 2006) due to asymmetric drift.
We scale our measured velocity values to take that into account.

\begin{figure*}
\begin{center}
\hspace*{-0.5cm} \includegraphics[width=9cm]{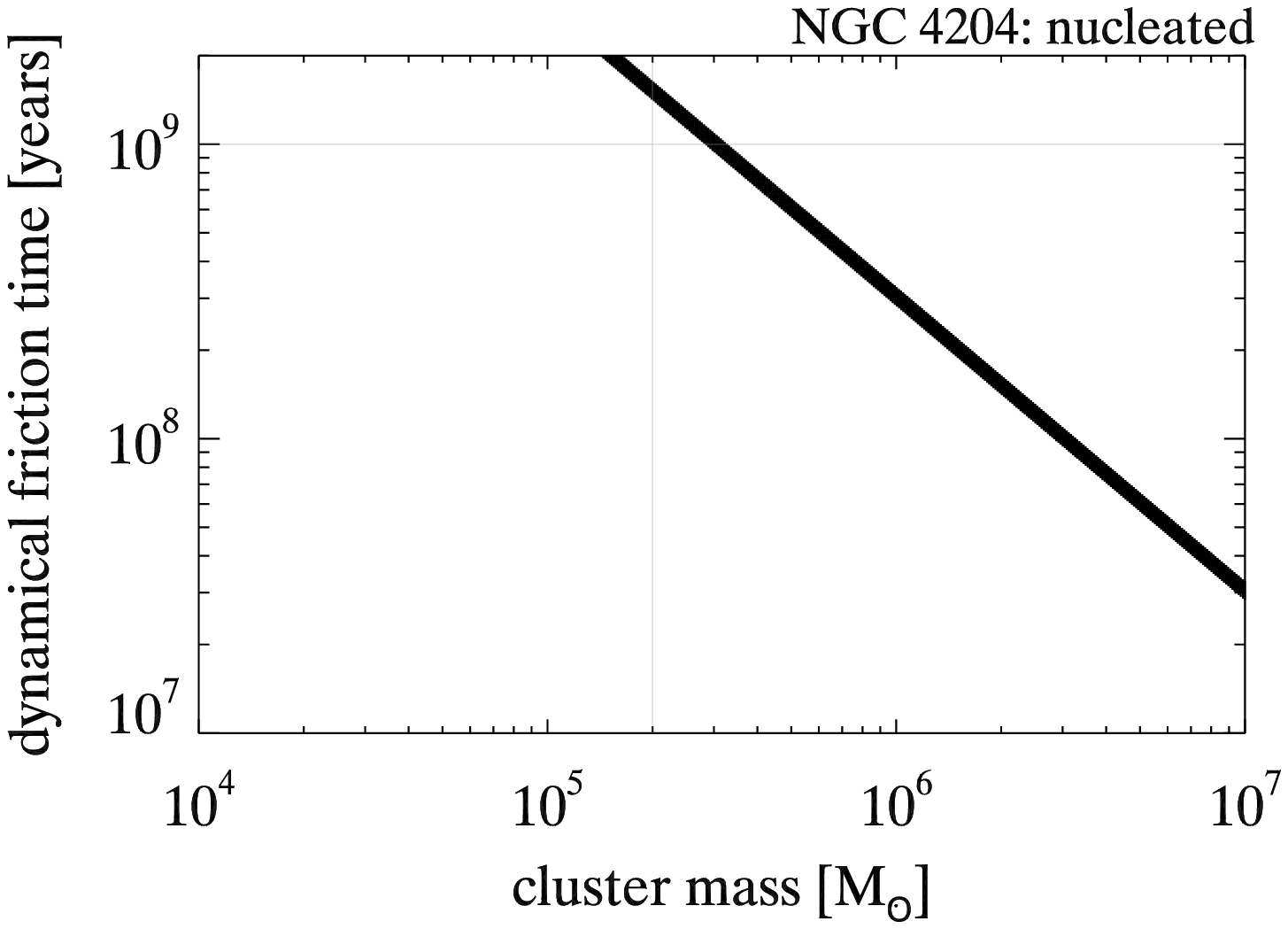}
\hspace*{-1cm} \includegraphics[width=9cm]{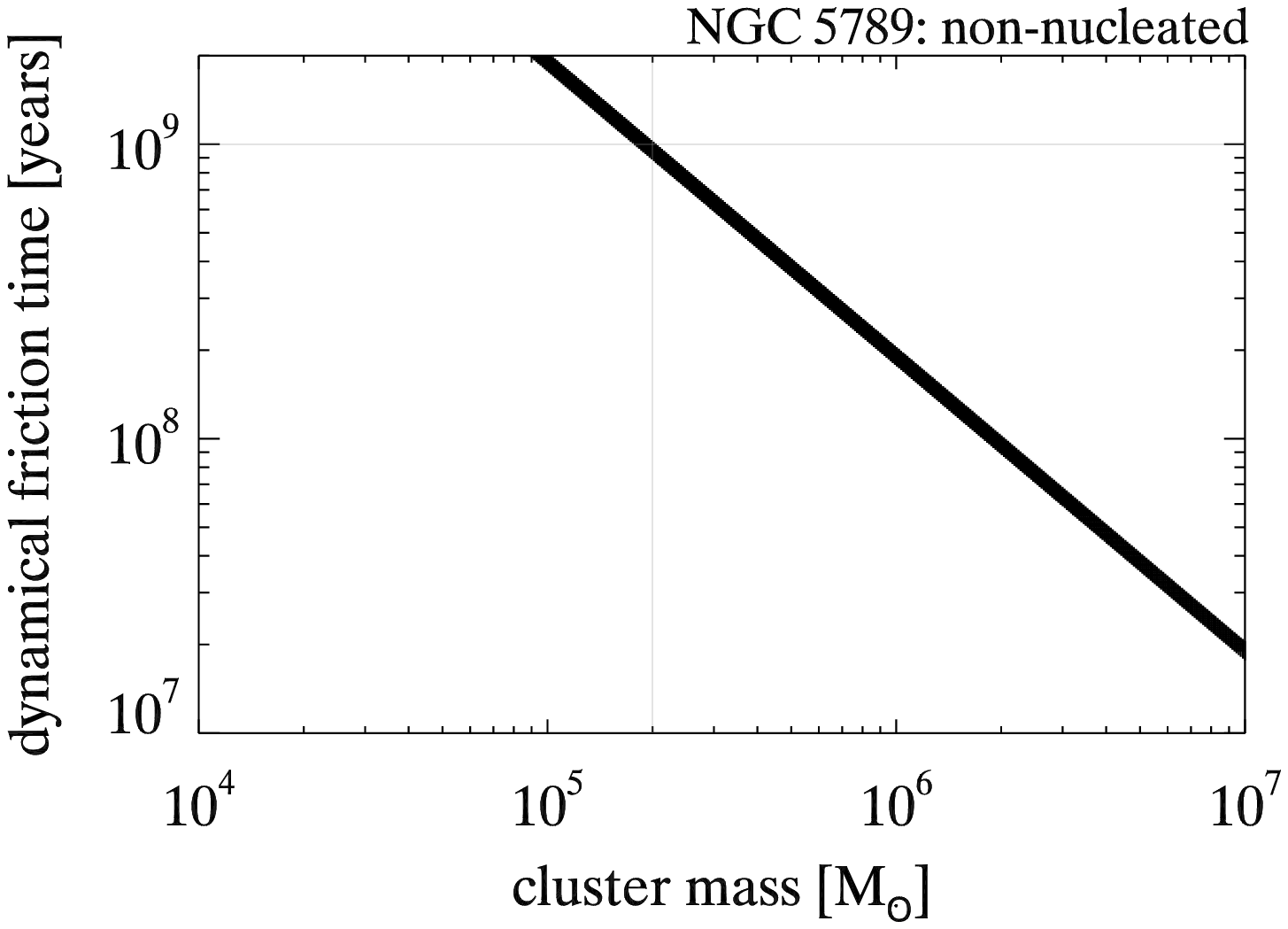}
\end{center}
\caption{Dynamical friction time as a function of cluster mass, for a cluster starting 500pc away from the dynamical center of the galaxy, and
assuming a fixed value of $ln\Lambda = 7$. Two example calculations are shown: a galaxy with a nuclear
cluster (NGC 4204, {\it left}), and a non-nucleated galaxy (NGC 5789, {\it right}). In both cases, a cluster of $\sim 2\times 10^5$\Msun\ would sink into the
dynamical center within two Gyrs.}
\label{fig:dynfric}
\end{figure*}

We calculate the dynamical friction timescale for clusters (taken to be point-like test particles) 
of masses between $10^4$ and $10^7$ \Msun, that start at an initial distance of $\sim 500$pc to the center. 
We find that typically clusters with masses above $2 \times10^5$ \Msun\ will make it to the dynamical 
center within $\le 2\times 10^9$ years (see Figure~\ref{fig:dynfric}). These findings are in very good agreement 
with the recent study by \cite{bekki10}, who performed numerical simulations on dynamical evolution
of disk galaxies and investigated the orbital evolution of star clusters influenced by
dynamical friction against disk field stars. \cite{bekki10} finds that dynamical friction of star clusters against disk
field stars is much more effective in orbital decay of star clusters in comparison with that against galactic halos in disk galaxies. Moreover,
dynamical friction seems to be most effective in disks with disk masses lower than $10^9$ \Msun\ owing to smaller stellar velocity dispersions.
\cite{milosavljevic04} argued that dynamical friction 
timescales are too long to bring massive clusters to the centers of late-type spiral galaxies. However, recently 
\cite{agarwal10} pointed out that \cite{milosavljevic04} only considered migration from a distant location in the 
disk, and indeed clusters that form close to the galactic center can reach the center and merge with the nuclear cluster.
 
\section{Discussion}
\label{sec:discussion}

The main goal of this paper is to verify whether the NC and the KC in late-type, bulgeless galaxies 
always coincide and whether the presence or mass of a NC is related to the kinematic state of its 
host galaxy. As discussed in Section \ref{sec:analysis}, we indeed find that for galaxies with high 
fidelity velocity fields, the NC, the KC and the PC coincide within the errors. A similar result was 
obtained by \citet{trachternach08} who conclude, based on HI rotation curves, that there are no 
systematic offsets between optical and kinematic centers. 

However, \citet{trachternach08} also conclude that, based on HI observations, non-circular motions 
are small, especially in late Hubble types. As has become evident, this is not the case for velocity fields with higher spatial 
resolution and derived from the ionized gas. 

Our derived velocity fields compare well with literature results.
For example, \citet{ganda06} have published \hb\ velocity fields for 
NGC\,3346 and NGC\,3423. While they did not perform 
kinematical modeling of their data, i.e. did not publish rotation curves or the KC location, the 
general shape and orientation of their velocity fields agrees well with those published here. 

Two of our 
galaxies, NGC\,5964 and NGC\,6509, are also part of the recent HI study of bulgeless spirals by \cite{watson10}. 
Comparing our \ha\ velocity fields to their HI data, we find that the position angles and velocity scales are in 
good agreement between atomic and ionised hydrogen. The advantage of the HI maps is their large field of view,
which covers the entire galaxy and thus the complete rotation curve. Our \ha\ velocity fields cover 
a good fraction of the rotation curves, but for some galaxies we are restricted to the central part of the velocity 
gradient and do miss the turnover of the velocity curve, like in the case of NGC\,5964. This results in large 
error bars on the location of the KC. We compare the KC 
position derived from HI \citep{watson10} and \ha\ and find good agreement in the case of NGC\,6509 
but for NGC\,5964 the offset is larger than the measurement error. This is reflected in the fact that the best fit to the
velocity field of NGC\,5964 appears unreliable.

We have used our measurements to derive realistic dynamical friction timescales for all sample galaxies (see Section~\ref{sec:friction}). 
We find that the threshold mass for a star cluster to be able to migrate to the center efficiently from 
a realistic distance ($\sim500$pc) is $1-2 \times 10^5$ \Msun (in very good agreement with the numerical
simulations by \cite{bekki10}). This is indeed the observed lower mass threshold for NCs. 
\cite{boker02} find that there is a lower luminosity cut-off to the NC luminosity function at M$_I = -9$. 
For a known M/L this would imply a lower limit mass cut-off for NCs. \cite{walcher05} published dynamically 
determined M/L ratios for a sample of nine NCs with a mean of M/L$_I$ = 0.6. This would translate into an 
observed lower mass limit for NCs of $\approx 10^5$ \Msun. 

We thus find that our data are entirely consistent with a NC formation scenario in which a massive seed cluster 
forms within $\sim 500$pc from the center of the galaxy, and spirals in to the center due to dynamical friction. This 
scenario has the advantage of automatically accounting for galaxies without a nuclear star cluster, implying 
merely that no suitable seed clusters formed close enough to the center in these disks. 
We also note that this lower mass threshold is one order of magnitude lower 
than the threshold mass of $\sim 10^6$ \Msun\ derived by \cite{pflamm-altenburg09} for efficient accretion of gas into the 
NC, which would lead to repetitive bursts of star formation there. Therefore, it may be that further NC growth 
through in-situ star formation starts out quite slowly, but picks up in pace as the NC grows. Mass growth may 
be further supported by accretion of other massive inspiralling clusters. Soon after its formation the seed cluster 
would then satisfy our definition and observations of a nuclear star cluster, namely being massive, sitting close 
to the center, having undergone recurrent star formation, and being compact. This formation and growth mechanism
is also supported by the fact that NCs are observed to rotate \citep{seth08,seth10}.

The long dynamical friction timescales we infer for massive star clusters that are initially located at significant 
($> 1$ kpc) distances from the galaxy center do not favour models where the NC grows only through accretion 
of globular clusters. We emphasize on the other hand that we have not found a confirmed case where the NC 
sits outside of the KC. We thus cannot rule out in-situ formation scenarios for NCs from our data.

\section{Conclusions}
\label{sec:conclusions}

We have presented two-dimensional \ha\ velocity fields for 20 late-type, disk-dominated 
spiral galaxies, the largest sample of bulgeless disks with high-resolution \ha\
velocity fields to date. 

We fitted the data with kinematic models in order to derive rotation curves and the location of 
the dynamical centers. Most rotation curves are well-fit by the arctan form of \cite{courteau97} 
used for earlier-type spirals. 
We find that the velocity fields span a broad range of morphologies. Some galaxies show regular 
rotation, which allows accurate determination of the kinematic center. However, only 2 out of 20 galaxies 
have a completely regular velocity field.  Many galaxies without bulges, 
but with strong bars show steep rises of their "velocity gradients" ("rotation curves") in the inner part. 
These steep rises are not necessarily due to mass concentrations, but to streaming motions 
along the bar. However, quite a few have some degree of irregular gas motion, which in 
nearly all cases can be either attributed to the presence of a bar or is connected to a rather 
patchy appearance of the \ha\ emission and the stellar light. Thus most galaxies in the sample 
show strong gas motions that cannot be attributed to the overall gravitational potential of the 
galaxy, implying low surface mass densities \citep[compare][]{dalcanton10} and 
implying that many bulgeless galaxies are not in dynamical equilibrium. 

There appears to be no systematic difference in the kinematics of nucleated and non-nucleated 
disks. For galaxies with regular, well-sampled velocity fields, the photometric center, the nuclear 
cluster and the kinematic center coincide within the errors. These centers also coincide for quite a 
few of the not-so-regular galaxies (in total for 13 out of our sample of 20 galaxies). However, we also 
find that nuclear clusters also occur in galaxies with disordered rotation fields. Hence, the large-scale 
velocity field is not a good predictor for the presence or mass of a nuclear cluster. Many formation scenarios for 
nuclear clusters invoke off-center cluster formation and subsequent ``sinking'' of clusters due to dynamical 
friction. We confirm that this scenario is viable for clusters that form within $\sim 500$pc of the center of the 
galaxy, as the dynamical friction timescales inferred from our data are consistent with this scenario. 
More distant globular clusters  do not seem to be promising candidates for nuclear cluster seeds, due to 
their long dynamical friction timescales. On the other hand, we point out that we cannot rule out an in-situ 
formation scenario.

\section*{Acknowledgments}
The authors thank the referee for a constructive report. We thank the MPIA for hospitality and support during a 
very productive visit. NN acknowledges the financial support of ESA through the ESTEC visitor program, 
where part of this work has been performed, and G. and M. H\"aring, who made this visit possible.
NN acknowledges support by the DFG cluster of excellence `Origin and Structure of the Universe'.
This research has made use of the NASA/IPAC Extragalactic Database (NED) which is operated by the 
Jet Propulsion Laboratory, California Institute of Technology, under contract with the National Aeronautics and Space Administration. 
We acknowledge the usage of the HyperLeda database (http://leda.univ-lyon1.fr). This research has made use of 
NASA's Astrophysics Data System Bibliographic Services, as well as of IPAC's Skyview Image 
Display and Analysis Program, developed with support from the National Aeronautics and Space Administration.

\label{lastpage}

\end{document}